\documentclass[a4paper,11pt]{article}
\usepackage{amsfonts, amsmath, amssymb, amsthm}
\usepackage[english]{babel}
\usepackage{booktabs}
\usepackage{bbm}
\usepackage{bm}
\usepackage{eucal}
\usepackage{enumerate}
\usepackage{float}
\usepackage{dsfont}
\usepackage{epsfig}
\usepackage{fontenc}
\usepackage[hmargin = 1.6cm, vmargin = 3cm]{geometry}
\usepackage{lscape}
\usepackage{mathrsfs}
\usepackage{natbib}
\usepackage{rotating}
\usepackage{times}
\usepackage{verbatim}
\usepackage[table]{xcolor}

\DeclareMathOperator{\iid}{\overset{iid}{\sim}}

 \let\oldthebibliography=\thebibliography
 \let\oldendthebibliography=\endthebibliography
 \renewenvironment{thebibliography}[1]{%
   \footnotesize
   \oldthebibliography{#1}%
   \setlength{\parskip}{0mm}%
   \setlength{\itemsep}{0mm}%
 }{\oldendthebibliography}

\newtheorem{theorem}{Theorem}
\newtheorem{definition}{Definition}

\newtheorem{lemma}{Lemma}
\newtheorem{remark}{Remark}

\renewcommand{\thefootnote}{$\dagger$}
\newcommand{\dif}{{\mathrm{d}}}
\newcommand{\KL}{{\mathrm{KL}}}

\begin{document}
\title{Bayesian semiparametric modelling of \\ phase-varying point  processes}
  \author{Bastian Galasso, Yoav Zemel, and Miguel de Carvalho}
  \date{\today}
\maketitle 

\begin{abstract}\footnotesize
We propose a Bayesian semiparametric approach for registration of multiple point processes. Our approach entails modelling the mean measures of the phase-varying point processes with a Bernstein--Dirichlet prior, which induces a prior on the space of all warp functions. Theoretical results on the support of the induced priors are derived, and posterior consistency is obtained under mild conditions. Numerical experiments suggest a good performance of the proposed methods, and a climatology real-data example is used to showcase how the method can be employed in practice.
\end{abstract}
\let\thefootnote\relax\footnotetext{Bastian Galasso is PhD Candidate, Department of Mathematics, Pontificia Universidad Cat\'olica de Chile, Chile (\textit{bigalass@mat.uc.cl}). Yoav Zemel is Research Associate, Statistical Laboratory, University of Cambridge, United Kingdom (\textit{zemel@statslab.cam.ac.uk}). Miguel~de Carvalho is Reader, School of Mathematics, University of Edinburgh, United Kingdom (\textit{miguel.decarvalho@ed.ac.uk}).}
\textsc{key words:} Bernstein--Dirichlet prior; Fr\'echet mean; Phase
variation; Point processes; Random Bernstein polynomials; Wasserstein
distance.

\section{Introduction}
A prototypical characteristic in the analysis of a random function $X(t)$---that distinguishes it from classical multivariate analysis---is that it potentially exhibits two distinct layers of stochastic variability.  Amplitude variation is encapsulated in the fluctuations of $X \equiv X(t)$ around its mean function $\mu(t)$, and can be probed by linear tools, perhaps most prominently the covariance operator of $X$ and the subsequent Karhunen--Lo\`eve expansion.  Phase variation amounts to variability in the argument $t$, usually modelled by a random warp function $T$ defined on the domain of definition of $X$, so that one observes realisations (discretised over some grid) from the random function $\widetilde X(t)=X(T^{-1}(t))$ instead of $X(t)$.  In short, phase variation is randomness in the $t$-axis, whereas amplitude variation pertains to stochasticity in the $X$-axis.

Typically, one is interested in inferring properties of the original function $X$, rather than those of $\widetilde X$.  In such situations phase variation can be thought of as a nuisance parameter, and failing to account for it may result in a severely distorted statistical analysis:  the mean function and Karhunen--Lo\`eve expansion of $\tilde X$ are smeared and less informative than those of $X$.  Consequently, one needs to undo the warping effect of the phase variation by constructing estimators $\widehat{T}$ for the warp functions, and composing them with the observed realisations from $\widetilde X$, a procedure known as registration, or alignment, of the functions.  The registered functions $\widetilde X_i\circ\widehat{T_i}=X_i\circ T_i^{-1}\circ \widehat{T_i}$ are then treated as distributed approximately as $X$, allowing for their use in probing the law of $X$.  For a textbook treatment of phase variation, we refer to the books by \cite{ramsay2002applied,ramsay2005functional};  one may also consult the review articles \cite{marron2015functional} and \cite{wang2016}.

\begin{figure}
\centering
\includegraphics[scale = .8]{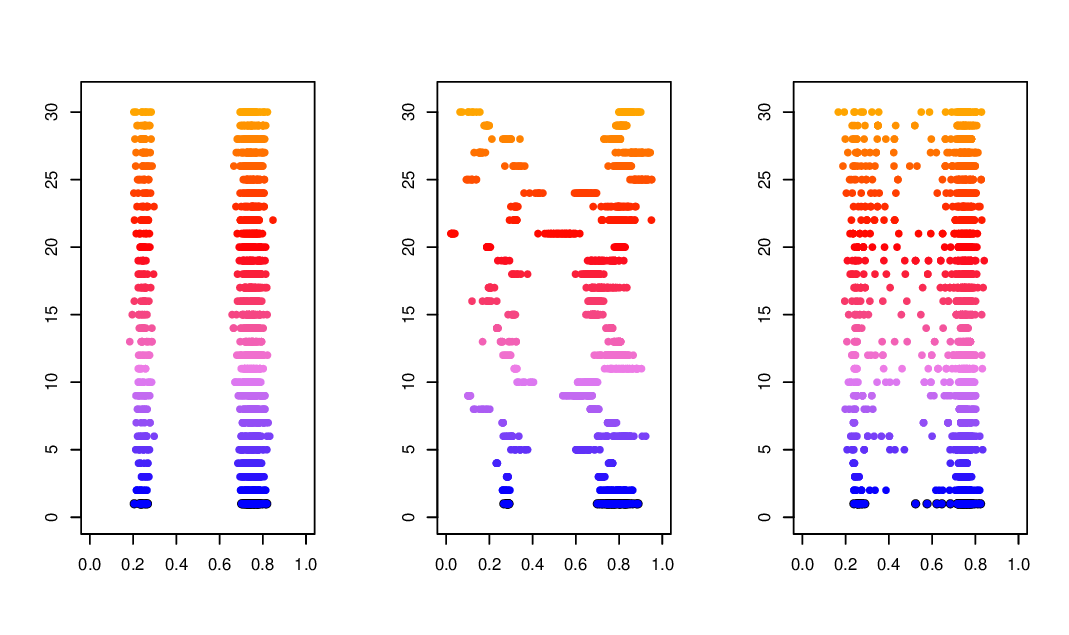}
\vspace{-.5cm}
\caption{\footnotesize Left: Realisations of the original point process. Middle: Phase-varying versions. Right: Registered versions obtained using the method proposed in the manuscript. Details on the underlying processes can be found in Section~ \ref{numerics}.}
\label{fig1}
\end{figure}

In this paper, we propose a Bayesian method for registering phase-varying point processes. Our paper is aligned with recent developments focused on modelling phase and amplitude variation of complex objects that are not functional data per se, yet still carry infinite-dimensional traits.  An intriguing example is that of point processes, appearing as spike trains in neural activity \citep[e.g.,][]{wu2014analysis}, where phase variation can be viewed as smearing locations of peaks of activity.  See Figure~\ref{fig1} for an example of such phase-varying point processes (and Section~\ref{numerics} for more details on the underlying processes). Such data can be transformed into functional data by smoothing and considering density functions \citep{wu2013functional}, but can be also be dealt with directly, replacing the ambient space $L^2$ used for functional data by a space of measures. Indeed, \cite{panaretos2016} formalise the problem and show how the Wasserstein metric of optimal transport arises canonically in the point process version of the problem. Here we propose a Bayesian model that is flexible in being nonparametric, whilst at the same adapted to the warping problem in a point process context, in the sense that our priors for the warp functions obey the same classical phase variation assumption in functional data analysis (see Subsection~\ref{bnp_pvp}). From a conceptual viewpoint, our model can be regarded as a semiparametric Bayesian version of \cite{panaretos2016}, but by putting directly a prior on the space of all random measures on the unit interval it allows for straightforward inference from posterior outputs---both in terms of credible bands for warp functions, and credible intervals for registered points. By modelling the mean measure of each phase-varying point process with a random Bernstein polynomial \citep{petrone1999a,petrone1999b}, we are able to show that the support of the induced priors for the warping functions and collections of registered points is `large' in the sense made precise in Subsections~\ref{support}--\ref{consistency}. Posterior consistency is established under a proviso that is asymptotically equivalent to that of \cite{panaretos2016}, but our large sample results only require the number of points in each process to increase.

Section~\ref{methods} develops details of our approach, in Section~\ref{numerics} we report numerical experiments, and Section~\ref{application} includes a climatology real-data example. Concluding remarks are given in Section~\ref{conclusion}. Proofs of results characterising the prior and limiting posterior can be found in the appendix (Section~\ref{sec:appendix}).  An R implementation of our method is available from  GitHub (\verb+https://github.com/bgalasso/Rmpp+).  To streamline the presentation, supplementary materials providing further figures, simulations, and implementation details are given in Sections~\ref{sec:supp-proofs}--\ref{sec:supp-app}.

\section{Random Bernstein polynomial-based registration of multiple point processes}\label{methods}
\subsection{Random Bernstein polynomials}
Random Bernstein polynomials were introduced by~\cite{petrone1999a,petrone1999b} and are defined as
\begin{equation}\label{Bern:dist}
B(x \mid k,G) = \sum_{j=0}^k G\left(\frac{j}{k}\right){k \choose j} x^j(1-x)^{k-j},
\end{equation}
\noindent
where $G$ is a random function on $[0,1]$ and $k$ is a (positive) integer-valued random variable. When $G$ is a distribution function, so is $B(t \mid k,G)$, and if in addition $G(0)=0$ then $B(t\mid k,G)$ has a density given by
\begin{equation}\label{b}
  b(x \mid k,G) = \sum_{j=1}^k w_{j,k} \beta(x \mid j,k-j+1), 
\end{equation}
where $w_{j,k} = G(j/k) - G((j-1)/k)$ and $\beta(x \mid a,b)$ is a Beta density function with parameters $a,b >0$. Since $G(1)=1$ it follows that $(w_{1,k},\ldots,w_{k,k})$ is in the unit simplex $S_{k} = \{(w_1,\ldots,w_k) \in [0,1]^k \; : \; \sum_{j=1}^k w_j = 1\}$; if $G$ has a continuous density $g$, then $b(x \mid k, G)$ approximates $g$ uniformly as $k \to \infty$ (see Lemma~\ref{lem:BernDer}). Following \cite{petrone1999a,petrone1999b} we have the next definition.
\begin{definition}\label{BPprior}
The probability measure $\pi$ induced by $B$ in \eqref{Bern:dist}, on the set $\Delta$ of all continuous distribution functions defined on $[0,1]$, is called Bernstein prior with parameters $(k,G)$. In symbols, $\pi \equiv \pi(k,G)$.
\end{definition}
Further details on random Bernstein polynomials can be found in \citet[][Section~5.5]{ghosal2015}.  To avoid unnecessarily burdening notation, measure-theoretical considerations will be kept to a minimum (including the measures with respect to which expected values are defined).

\subsection{Bayesian semiparametric inference for phase-varying point processes}\label{bnp_pvp}
Let $\Pi$ be a point process in $[0, 1]$, with finite second moment ($E\{(\Pi[0,1])^2\}<\infty$), and denote its mean measure by $\lambda(\cdot) = E\{\Pi(\cdot)\}$. Estimation of $\lambda$ is straightfoward when one has access to multiple realisations $\{\Pi_1, \ldots, \Pi_n\}$ from $\Pi$, with $\widehat{\lambda}$ asymptotically normal \citep[Proposition~4.8]{karr1991point}.  Suppose, however, that one instead observes a sample $\{\widetilde\Pi_1, \ldots, \widetilde\Pi_n\}$ with
\[
\widetilde \Pi_i=T_{i \#}\Pi_i,
\]
where $T_{i \#}\Pi_i(\cdot)=\Pi_i\{T_i^{-1}(\cdot)\}$ denotes the push-forward of $\Pi_i$ through $T_i$, for all $i$.  In other words, if a given realisation of $\Pi_i$ is the collection of points $\{x_{i, j}\}_{j = 1}^{m_i}$, then one observes the deformed collection $\{\widetilde x_{i, j}\}_{j = 1}^{m_i} \equiv \{T_i(x_{i, j})\}_{j = 1}^{m_i}$ , for all $i$.  Here, $\{T_1, \ldots, T_n\}$ is a sequence of random warp functions, that is, increasing homeomorphisms on $[0, 1]$.  A target of interest will be on learning about the warp functions, so to register the point processes. To achieve this goal we model the (conditional) mean measures of the phase-varying point processes with a Bernstein--Dirichlet prior, which induces a prior on the space of all warp functions. The conditional mean measure of the warped version  $\widetilde{\Pi}_i$ given $T_i$ is denoted by $\Lambda_i(\cdot) = E\{\widetilde{\Pi}_i(\cdot) \mid T_i\}$. We impose the rather standard assumptions that $E\{T_i(x)\} = x$ (unbiasedness) for all $x\in[0,1]$, and that the collection $\{T_1, \ldots, T_n\}$ is independent of $\{\Pi_1, \ldots, \Pi_n\}$; the assumptions of unbiasedness and monotonicity of warp functions are \textit{sine qua non} in the phase variation literature, often accompained with additional conditions \citep[e.g.,][]{tang2008, wang2016}. In words, unbiasedness is tantmount to requiring the average time change $E\{T(x)\}$ to be the identity: on average, the ``objective'' time-scale should be maintained, so that time is not sped up or slowed down. In fact, unbiasedness and monotonicity are key for identifiability.

To learn about $F_i(x) = \int_0^x \Lambda_i(\dif t)$, for $x \in [0, 1]$, we set the prior 
\begin{equation}
  F_i(x) = B(x \mid k_i, G_i),\quad x \in [0, 1],
  \label{rbp}
\end{equation}
where $\{k_1, \ldots, k_n\}$ is a sequence of independent integer-valued random variables and $\{G_1, \ldots, G_n\}$ is a sequence of independent random measures. In a more concrete specification of \eqref{rbp}, we proceed as follows. Let $\{\widetilde x_{i, j}\}_{j = 1}^{m_i}$ be the points corresponding to $\widetilde \Pi_i$, and for $i = 1, \dots, n$ set 
\begin{equation}\label{warpver}
  \begin{split}
    \widetilde x_{i, j} \mid F_i \sim F_i, \quad j = 1, \ldots, m_i, \quad F_i(x) = B(x \mid k_i,G_i), \\  
    G_i \mid \alpha \sim \text{DP}(\alpha, G^*), 
    \quad k_i \sim \text{Unif}\{1, \ldots, k_{\text{max}}\}.
  \end{split}
\end{equation}
Here `DP' stands for Dirichlet process \citep{ferguson1973}, with precision parameter $\alpha > 0$ and centering distribution $G^* = E(G_i)$. To complete the model specification we set $G^* = \text{Beta}(a_{0}, b_{0})$ and 
$\alpha \sim \text{Gamma}(\textsl{a}_{0}, \textsl{b}_{0})$, for $i = 1, \ldots, n$. More sophisticated versions of  \eqref{warpver} are, of course, possible by e.g., specifying different precision and centering for the DP per each point process; for simplicity, we will focus on \eqref{warpver}. Below, we assume that the $\{G_i\}$ and $\{k_i\}$ are independent.  Moreover, by a slight abuse of notation we identify $F_i$ with $\Lambda_i$ and more generally, a measure $\mu$ with its distribution function $F_\mu(x)=\mu\{[-\infty,x]\}$.

Now, $\{F_1, \ldots, F_n\}$, specified as in \eqref{rbp}, can be used to induce a prior $F$ on the mean measure $\lambda$ of the random point process $\Pi$ and on the warp maps $T_i$. The prior $F$ will be centred around the structural mean $\lambda$ in the Fr\'echet mean sense that $\lambda$ is the closest to $F$ in expectation, that is, $E_{\lambda}\{d^2(\lambda, F)\} \leq E_{\lambda}\{d^2(\gamma, F)\}$, for all diffuse measures $\gamma$ on $[0, 1]$. An obvious question that arises is what metric $d$ should one use, but the Wasserstein distance \citep{santambrogio2015, panaretos2019} has been shown to be the canonical metric for phase-varying point processes by \citet[][Section~3]{panaretos2016}. It is defined by
\begin{equation}\label{metrics}
d(\mu,\nu)
= \inf_{Q \in \Gamma(\mu,\nu)}\sqrt{\int_0^1 \{Q(x)-x\}^2\mu(\dif x)},
\end{equation}
where $\Gamma(\mu,\nu)$ is the collection of functions $Q: [0, 1] \to [0, 1]$ such that $Q_{\#} \mu = \nu$. (If $\mu$ is not diffuse, then $\Gamma(\mu,\nu)$ may be empty and the definition of $d$ needs to be modified, but we will only have to deal with diffuse measures in the sequel.) Since Fr\'echet averaging with respect to Wasserstein distance amounts to averaging of quantile functions \citep{agueh2011barycenters}, the prior on $F$ is induced from the prior on $\{{F}_1,\ldots,{F}_n\}$ as the probability law of 
\begin{equation}\label{frechet:mean}
F(x) = \left(\frac{1}{n}\sum_{i=1}^n F_i^{-1}\right)^{-1}(x), \quad x \in [0, 1].
\end{equation}
The random Bernstein polynomial-induced prior on each $T_i$ defines the optimal transport map of $F$ onto $F_i$  \citep{santambrogio2015}: 
\begin{equation}
  \label{warp}
  T_i = F_i^{-1} \circ F.
\end{equation}
Since $F_1,\ldots,F_n$ are independent, identically distributed and increasing distribution functions, it follows that the $T_i$ are homeomorphisms with $E\{T_i(x)\} = x$. Indeed, by construction it can be shown that $T_1(x) + \cdots + T_n(x) = nt$ for every $x$, $T_1,\ldots,T_n$ are identically distributed given $F$ and so, $E(T_i \mid F) = E(T_{i'} \mid F)$ for every $i \neq i'$, and taking expectation in both sides, we have that $E(T_i) = E(T_{i'})$;  therefore, 
\begin{equation}\label{unbias}
nE\{T_i(x)\} =
  E\{T_1(x)\} + \cdots + E\{T_n(x)\} =
  E\{T_1(x) + \cdots + T_n(x)\} =
  nx,
\end{equation}
and thus it follows that $E\{T_i(t)\} = t$, for $i=1,\ldots,n$.

The random Bernstein polynomial-induced priors on the registered point processes is constructed by pushing them forward through the registration maps
\begin{equation}\label{regist}
\Pi_i = T_{i\#}^{-1} \widetilde{\Pi}_i, \quad i=1,\ldots,n.
\end{equation}

The posterior sampling for the warping maps and registered points is then conducted as follows. Let $F_{i,[1]},\ldots,F_{i,[M]}$ be posterior samples from $F_i$, for $i = 1,\ldots,n$, which can be obtained via a Gibbs sampler as described in the supplementary materials (Section~ 2). Then, for each $j = 1,\ldots,M$ we get $F_{[j]} = (\sum_{i=1}^n F_{i,[j]}^{-1}/n)^{-1}$ and so, $T_{i,[j]} = F_{i,[j]}^{-1} \circ F_{[j]}$ and $\Pi_{[j]} = T_{i,[j] \; \#}^{-1} \widetilde{\Pi}_i$. Finally, pointwise estimation for mean measure, warp functions, and registered points are given by the posterior means,
\begin{equation}\label{postmeans}
\widehat{F} = \frac{1}{M} \sum_{j=1}^M F_{[j]}, \quad \widehat{T}_i = \frac{1}{M}\sum_{j=1}^M T_{i,[j]}, \quad \widehat{\Pi}_i = \frac{1}{M}\sum_{i=1}^M \Pi_{i,[j]}.  
\end{equation}
Credible intervals or pointwise credible bands can be also directly obtained from the relevant quantiles of the corresponding posterior outputs.

\subsection{Kolmogorov--Smirnov, Wasserstein, and Kullback--Leibler supports of induced priors}\label{support}
As it will be shown below, full support of the relevant parameters in our setup holds, under conditions on the support of the law of the $k_i$ and on that of $w_{1, k_i}, \dots, w_{k_i, k_i} \mid k_i$. Extending the assumptions in \cite{petrone1999a}, we assume that the prior probability mass function $\rho_i$ of $k_i$ is positive, that is $\rho_i(k) > 0$ for $i = 1, \dots, n$ and all $k$, and that $w_{1, k_i}, \dots, w_{k_i, k_i} \mid k_i$ has a family of conditional densities $l_i(w_{1, k_i}, \dots, w_{k_i, k_i} \mid k_i) > 0$, for every $(w_{1, k_i}, \dots, w_{k_i, k_i}) \in S_{k_i}$ and for every sequence of independent integer valued random variables $\{k_1, \ldots, k_n\}$. Define the supremum norm 
\[
\|F - H \|_{\infty} = \sup_{x \in [0,1]} |F(x) - H(x)|.
\]
Below, $\mathscr{F} \equiv (F_1, \dots, F_n)$ denotes the joint Bernstein prior and $N_i \equiv \Pi_i([0, 1]) > 0$ is the total number of points in the $i$th point process, for $i = 1, \dots, n$. 

\begin{theorem}\label{th1} 
Let $F_1,\dots,F_n \iid \pi$ with Fr\'echet--Wasserstein mean $F$, and with induced priors $T_i$ and $\Pi_i$ as defined in \eqref{warp} and \eqref{regist}. For any continuous strictly increasing $\mathbb{F}_1, \dots, \mathbb{F}_n \in \Delta$, with Fr\'echet--Wasserstein mean $\mathbb{F}$, transport maps $\mathbb{T}_i = \mathbb{F}_i^{-1} \circ \mathbb{F}$, and registered discrete measures $P_i = \mathbb{T}_{i\#}^{-1} \widetilde \Pi_i$, and for any $\varepsilon > 0$ the following events occur with positive probability:
\begin{equation*}
\begin{array}{lll}
(a)~\{\mathscr{F}: \|F_{j} - \mathbb{F}_{j}\|_{\infty} < \varepsilon, j = 1,\dots,n\}, & \hspace{1cm} &                                                                                   (b)~\{\mathscr{F}: \|F - \mathbb{F}\|_{\infty} < \varepsilon\}, \\
(c)~\{\mathscr{F}: \|T_i - \mathbb{T}_i\|_{\infty} < \varepsilon\}, & \hspace{1cm} & 
(d)~\{\mathscr{F}: d(\Pi_i / N_i, P_i / N_i) < \varepsilon\},
\end{array}  
\end{equation*}
for $i = 1, \dots, n$. 
\end{theorem}
Claims (a), (b), and (c) in the Theorem~\ref{th1} respectively state that the joint Bernstein prior, the Fr\'echet--Wasserstein mean, and the warp functions have large Kolmogorov--Smirnov support. Claim (d) states that the registered point processes have large Wasserstein support. The proof actually shows that the intersection of these four events (a)--(d) has positive probability. Whilst the latter properties may not look surprising ex-post, as their proofs show they are not straightforward facts.

The characterisation of the Kullback--Leibler (KL) support is more challenging. By definition, a density $f$ is said to possess the KL property relatively to a prior $\pi$ if for any $\varepsilon > 0$ one has that $\pi\{H: \KL(F, H) < \varepsilon\} > 0$, where 
\[
\mbox{KL}(F, H)
=\int_0^1 h(x)\log\frac{h(x)}{f(x)} \, \dif x,
\]
with $F$ and $H$ denoting the distribution functions respectively corresponding to $f$ and $h$.

When $\pi$ is a random Bernstein polynomial prior as per Definition \ref{BPprior}, any density $f$ possesses the KL propery \citep[Theorem 2]{petrone2002}. The following theorem inspects the permanence of the Kullback--Leibler property on the functionals of interest, and it shows that the property is preserved for Fr\'echet--Wasserstein mean and the warping functions.

\begin{theorem}\label{thm2}
Let $F_1,\dots,F_n \iid \pi$ with Fr\'echet--Wasserstein mean $F$ and with transport maps $T_i=F_i^{-1}\circ F$ as defined in \eqref{warp}.  For any $\varepsilon>0$ and strictly increasing $\mathbb{F}_1, \dots, \mathbb{F}_n \in \Delta$ with densities $\mathbbm f_i$ that are continuous on $(0,1)$, Fr\'echet--Wasserstein mean $\mathbb{F}$ and transport maps $\mathbb{T}_i = \mathbb{F}_i^{-1} \circ \mathbb{F}$, $\mathbb F$ also has a density $\mathbbm f$ and:
\begin{enumerate}
\item[(a)] If $\int_0^1\mathbbm f(x)\log \mathbbm f(x)\dif x<\infty$ then $\KL(F,\mathbb F)<\varepsilon$ with positive probability.
\item[(b)] If each $\mathbbm f_i$ is strictly positive on $(0,1)$, then with positive probability $\KL(T_i,\mathbb T_i)<\varepsilon$ for all $i=1,\dots,n$.
\end{enumerate}
\end{theorem}
\begin{remark}\normalfont
The densities $\mathbbm f_i$ can be unbounded or approach zero near 0 or 1. The condition $\int_0^1\mathbbm f(x)\log \mathbbm f(x) \dif x <\infty$ in (a) is very weak and is satisfied when $\mathbbm f$ is a beta density with arbitrary (positive) parameters.  This condition is, in fact, necessary; if it fails to hold, then $\KL(F,\mathbb F)=\infty$ almost surely.  The assumptions on the densities can be further relaxed to $\mathbbm f_i$ having finitely many discontinuity points on $[0,1]$, and for part (b) $\mathbbm f_i$ may vanish on finitely many points on $[0,1]$.  We refrained from this level of generality for the purpose of clarity and because the current version includes the most important case of beta distributions.
\end{remark}

Theorem~\ref{thm2} shows that under mild conditions, the Fr\'echet--Wasserstein mean and the warping functions possess the Kullback--Leibler property with respect to the prior on $F$ induced from $F_1,\dots,F_n$ via \eqref{frechet:mean}. We now study the large-sample behaviour of the posterior.

\subsection{Posterior consistency}\label{consistency}
Contrarily to \citet[Theorem~1]{panaretos2016}, our asymptotic theory does not require $n \to \infty$; indeed we only require that $m_i \to \infty$, with $i=1,\ldots,n$, for any finite $n$. Yet note that the consequence is that under this assumption one is only able to approximate warping functions of the type $\mathbb{T}_i = \mathbb{F}_i^{-1} \circ \mathbb{F}$,  for all $i$, where $\mathbb{F}$ is the Fr\'echet--Wasserstein mean of $\mathbb{F}_1, \dots, \mathbb{F}_n$. This proviso is less and less restrictive as $n$ increases, and it is asymptotically compatible with that of \cite{panaretos2016}, as indeed if the $\mathbb{T}_i$ are independent and identically distributed---rather than fixed as assumed in Theorem~\ref{thm2}---then it follows that as $n \to \infty$,
\begin{equation*}
  \frac{1}{n} \sum_{i = 1}^n \mathbb{T}_i(x) \underset{p}{\to} E\{\mathbb{T}_1(x)\} = x.
\end{equation*}
Below, the posteriors induced by~\eqref{frechet:mean} and~\eqref{warp} should be understood respectively as the laws of ${F}$ and ${T}_i$ conditional on $\{\widetilde x_{i, j}\}_{j = 1}^{m_i}$. The following result holds.
\begin{theorem}\label{thm:consistency}
Under the same conditions as in Theorem~\ref{thm2}, if $m_i \to \infty$ for $i=1,\ldots,n$, then the posteriors induced by~\eqref{frechet:mean} and~\eqref{warp} are respectively Kolmogorov consistent at $\mathbb{F}$ and $\mathbb{T}_i$, for all $i$.
\end{theorem}
This result closes the large sample properties of our methods; we next focus on assessing their finite-sample properties.

\begin{figure}[H]
  \centering
  \begin{minipage}{0.31\linewidth}
    \includegraphics[scale = 0.37]{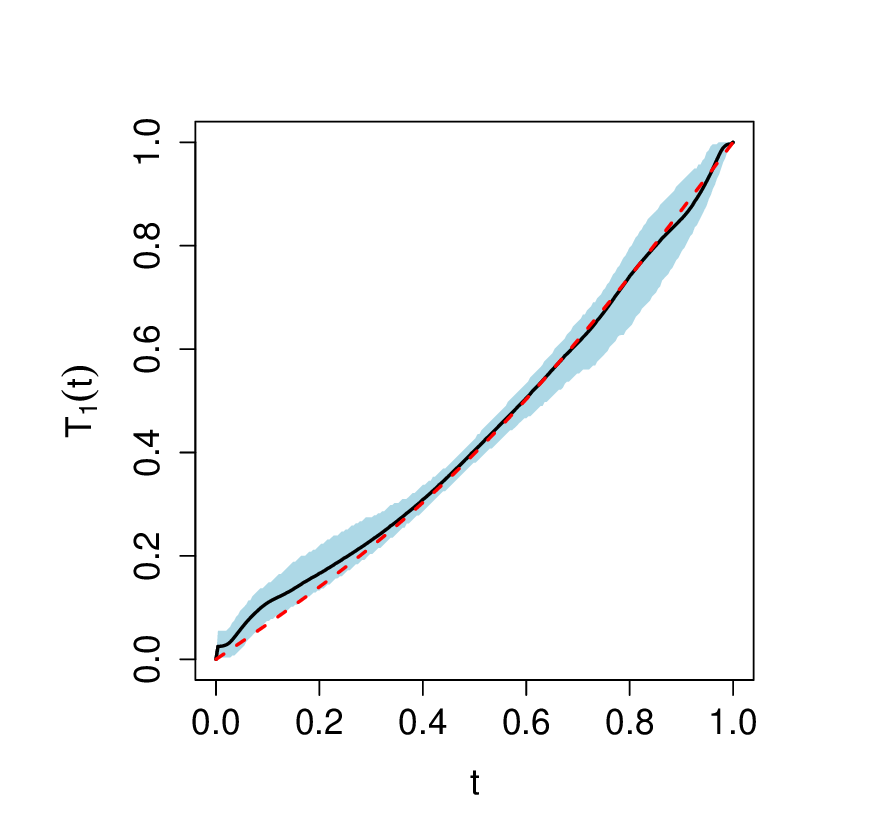}
    \includegraphics[scale = 0.37]{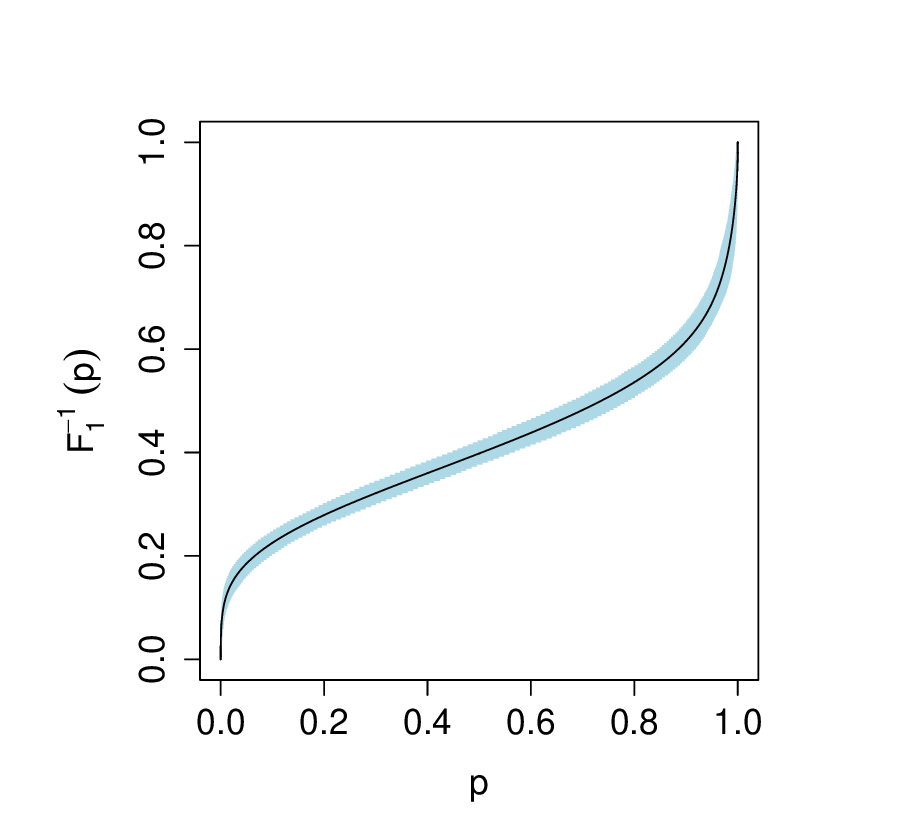} 
  \end{minipage} \hspace{0.2cm} 
  \begin{minipage}{0.31\linewidth}
     \includegraphics[scale = 0.37]{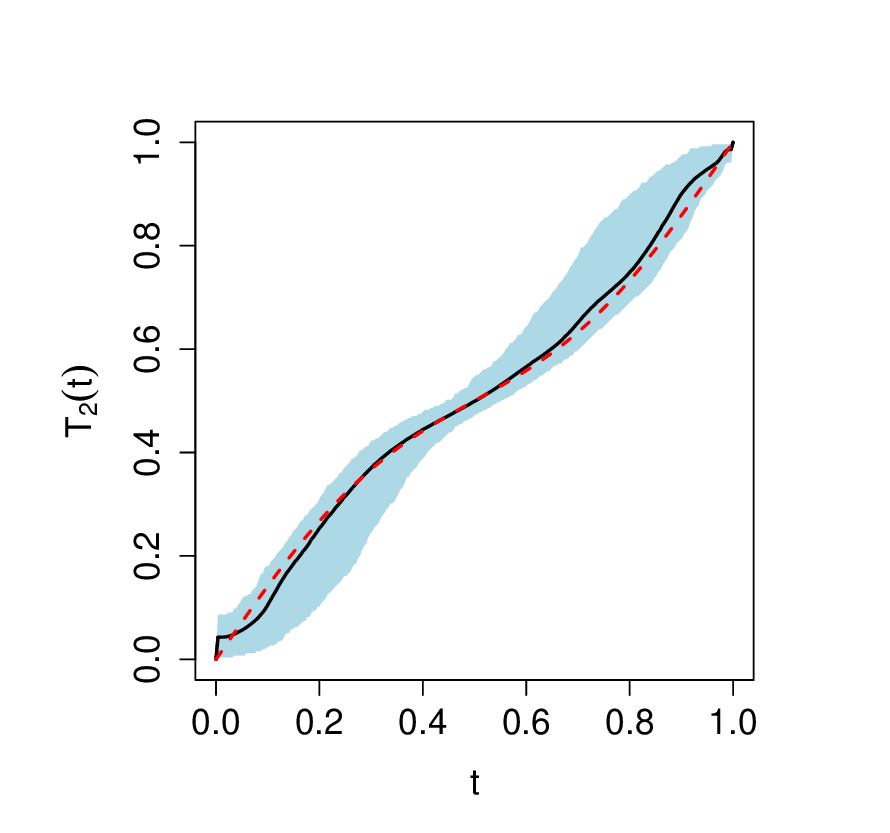}
     \includegraphics[scale = 0.37]{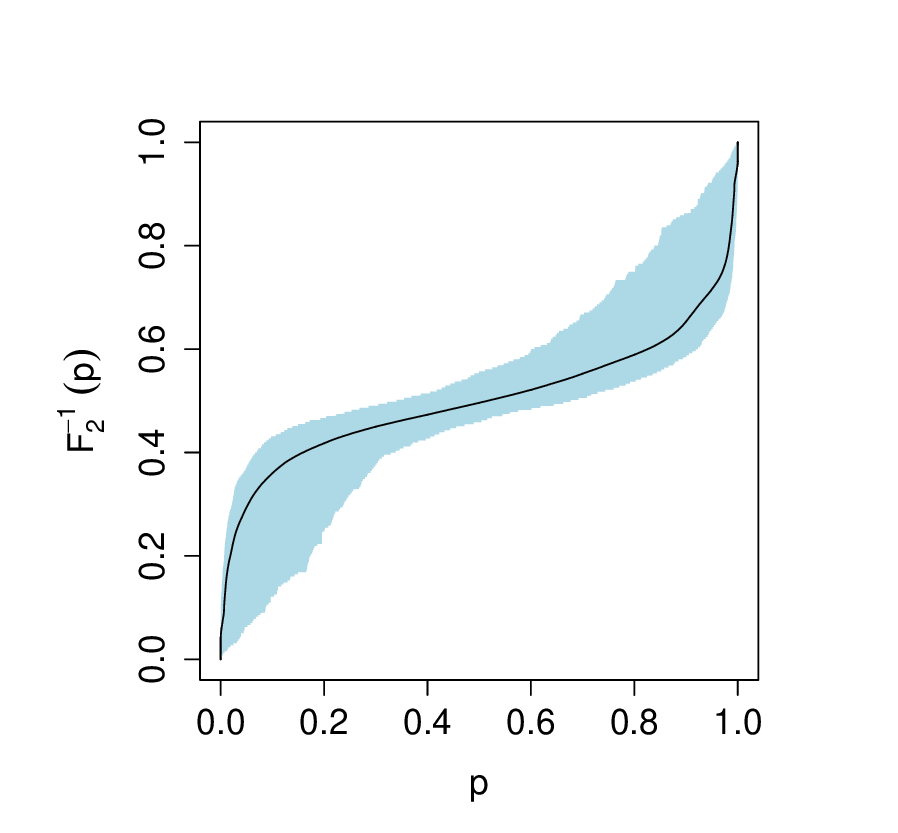} 
  \end{minipage} \hspace{0.2cm}
  \begin{minipage}{0.31\linewidth}
    \includegraphics[scale = 0.37]{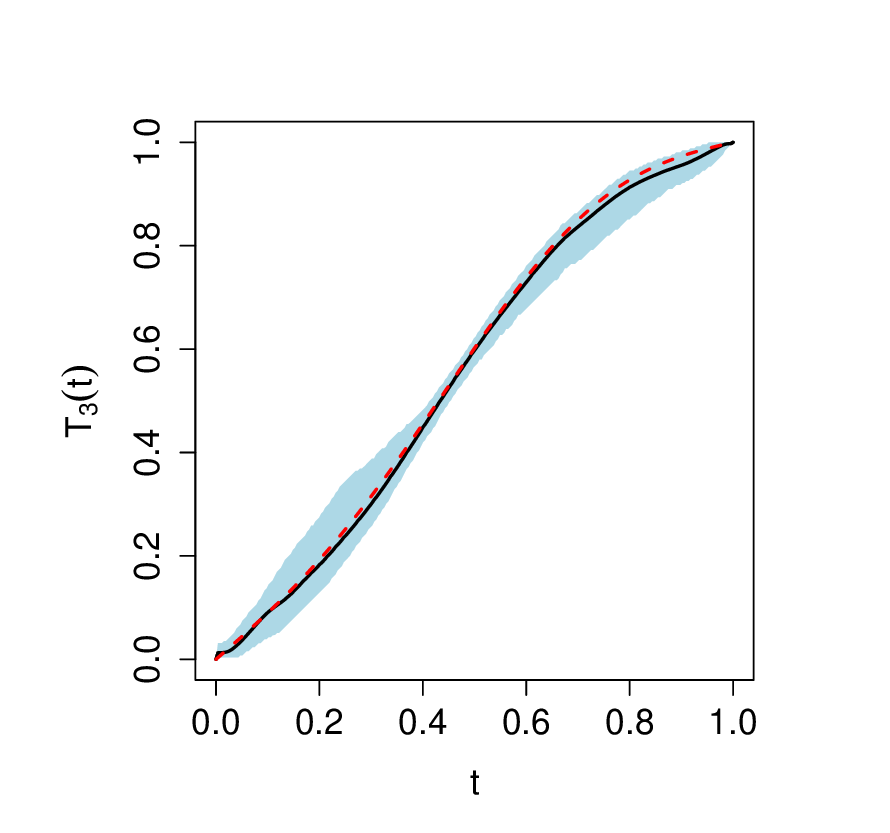}
    \includegraphics[scale = 0.37]{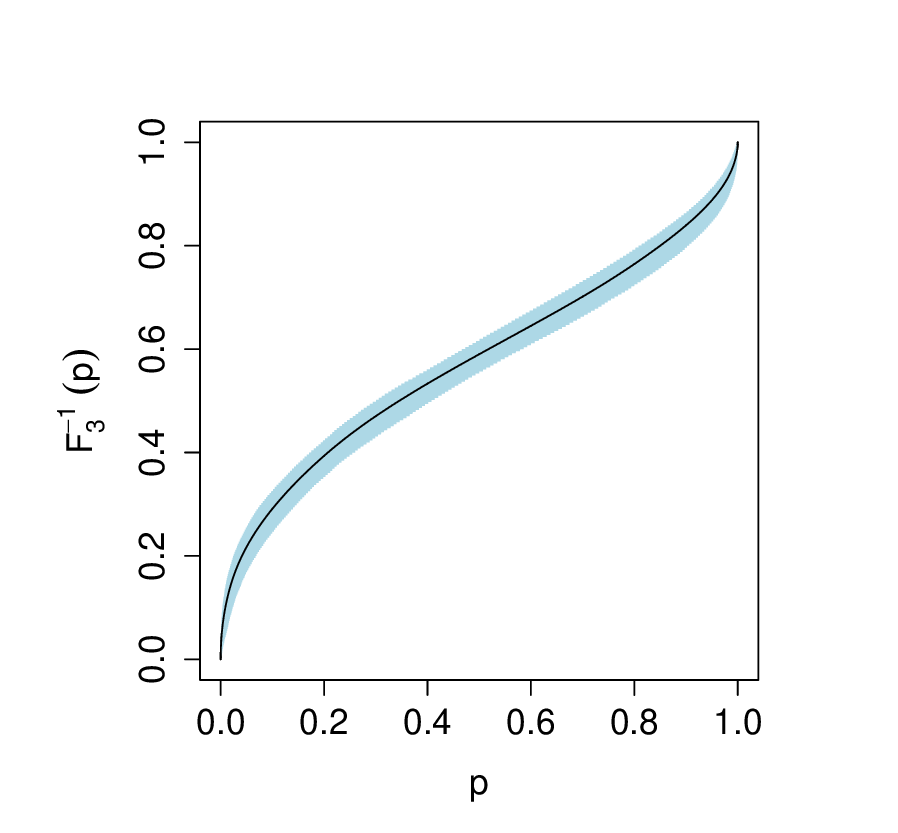} 
  \end{minipage}
\caption{\footnotesize Above: True (dashed red) and estimated (solid black) warp functions along with credible bands.  The estimators are constructed as the posterior mean of the induced prior as~\eqref{warp}. Below: Corresponding quantile function estimates along with credible bands.}
\label{fig2a}
\end{figure}

\section{Numerical experiments and computing} \label{numerics}
\subsection{Small $n$, large $m$} \label{sc1}
\begin{figure}
  \centering
  \includegraphics[scale = 0.35]{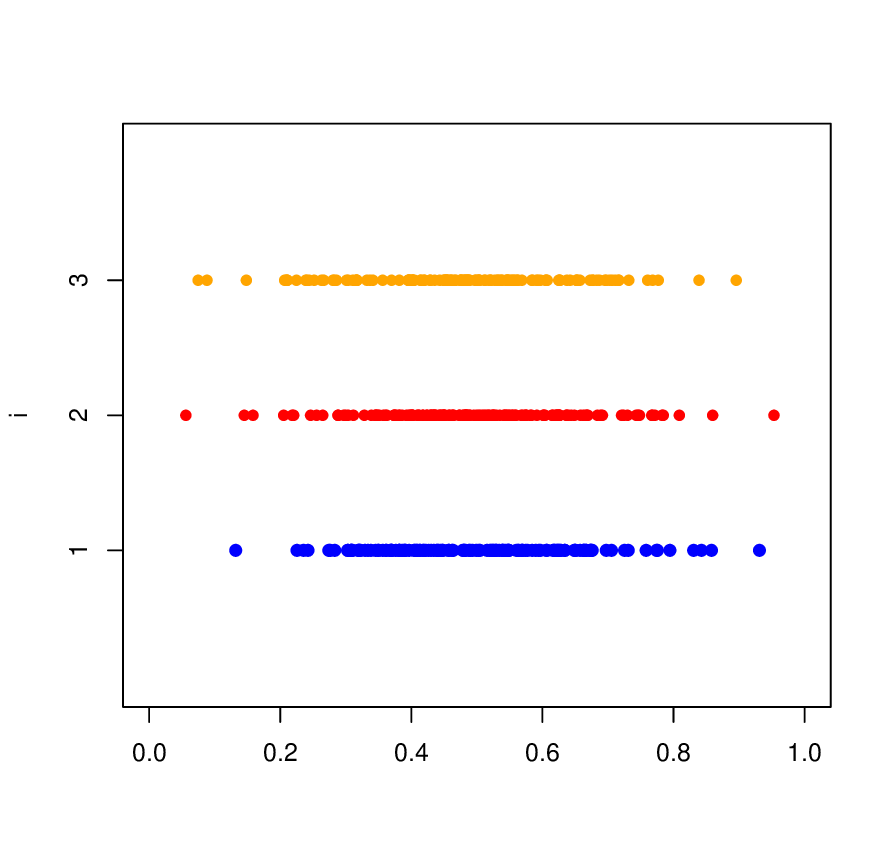}
  \includegraphics[scale = 0.35]{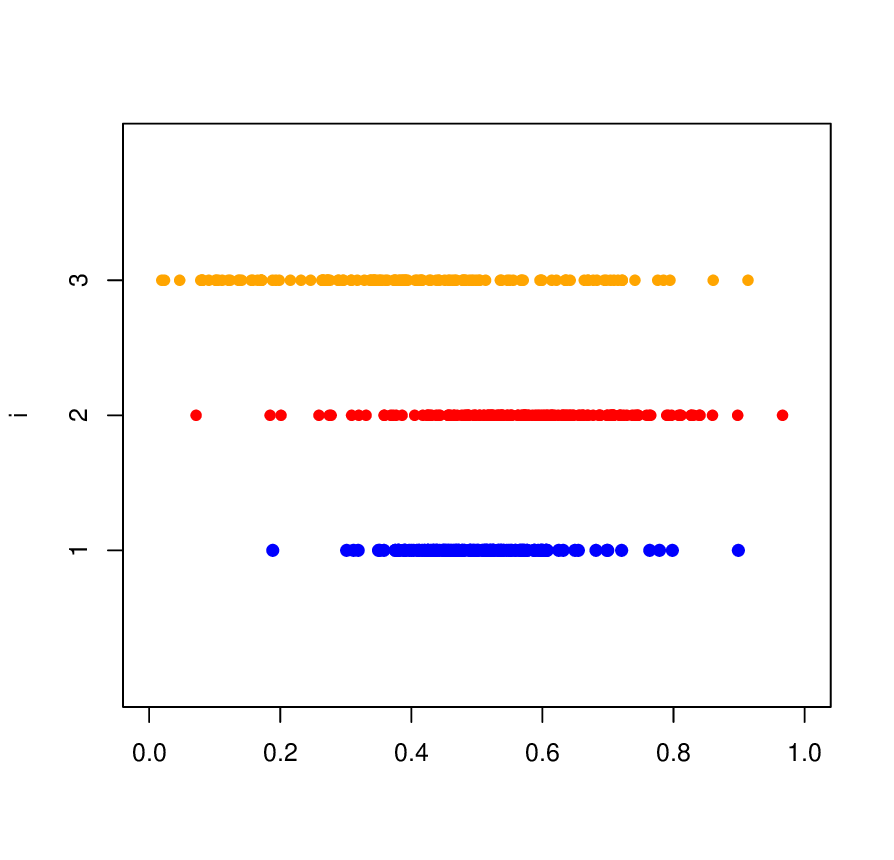}
  \includegraphics[scale = 0.35]{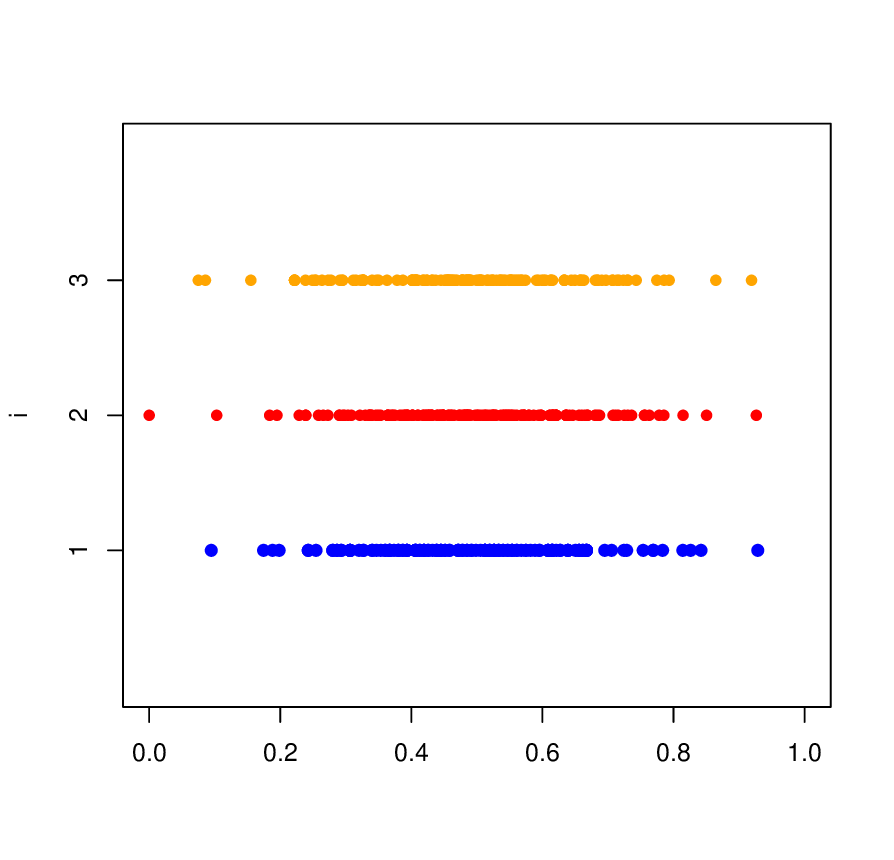}
\caption{\footnotesize Left: Realisations of the original point process from the setup of \S~\ref{sc1} in the small $n$, large $m$ regime. Middle: Corresponding phase-varying versions.  Right: Registered versions.}\label{fig1a}
\end{figure}

As our asymptotic theory does not require $n \to \infty$, we start by assessing performance of the proposed methods in a small $n$, large $m$ setting.  We generate random samples $x_{i,1},\ldots, x_{i,m_i} \mid m_i$, from
\begin{equation*}
\lambda(x) = \Phi(x \mid 0.5, (0.15)^2), \quad m_i \sim \text{Poisson}(L),
\end{equation*}
for $i = 1,2,3$, with $L = 150$ and $\Phi(x \mid \mu, \sigma^2)$ denoting the Normal distribution function. Then the warped data $\tilde{x}_{i,j} = T_i(x_{i,j})$ are obtained using
\begin{equation}\label{warp1}
\left\{
\begin{array}{rcl}
T_i(x) & = & x + \left(a_i - \frac{1}{2}\right)\text{sin}(b_ix\pi)(b_i\pi)^{-1}, \quad i=1,2,\\
 & & \\
T_3(x) & = & 3t - T_1(x) - T_2(x),
\end{array}\right.
\end{equation}
where $a_1,a_2 \stackrel{\text{iid}}{\sim}  \text{Unif}([0,1/4]\cup[3/4,1])$ and $b_1, b_2 \stackrel{\text{iid}}{\sim}  \text{Unif}\{1,2\}$. By construction these warp maps are in line with the model assumptions: each $T_i$ is an increasing homeomorphism of $[0,1]$, and $E\{T_i(x)\}=x$ follows from the fact that $E(a_i)=1/2$. See Figure~\ref{fig1a} (left and middle) for an instance of realisations of the original point process along with phase-varying versions obtained by warping the data as in \eqref{warp1}.

The proposed semiparametric approach in Section~\ref{methods} can be implemented with the aid of the R package Rmpp, which implements a version of the algorithm in \citet[][p.~383]{petrone1999a}; see Subsection~\ref{subsec:supp-gibbs} in the supplementary materials for details. Figure~\ref{fig2a} shows the estimators of each of the three warp maps through the posterior mean of the induced prior defined in~\eqref{warp}, along with their credible bands and the true warp maps.

From Figure~\ref{fig2a} it can be observed that our estimators are reasonably in line with the true warp functions. Thus, the method recovers quite well the original point processes, as can be seen when comparing the left and right panels of Figure~\ref{fig1a}. Given \eqref{warp} the width of the credible bands of the warps maps depicted in Figure~\ref{fig2a} is determined by the width of the credible bands of the quantile function and of the Fr\'echet mean. To put differently, the fact that the bands of the warp maps are narrow on some regions is simply a consequence of the fact that the credible bands for the corresponding quantile functions are themselves narrow, on some other region.

A Monte Carlo study was conducted in this setting based on $B = 50$ simulated datasets. We apply our method to each, and then calculate the Monte Carlo $L^2$-Wasserstein distance mean (WDM) by
\begin{equation}\label{wrm}
  {\text{WDM}}  = 
  \frac{1}{B} \sum_{b = 1}^B \sum_{i=1}^{n} d(\widehat{\Pi}_i^{[b]},\Pi_i^{[b]}),
\end{equation}
where the superscript $[b]$ denotes the corresponding object computed from the $b$th simulated dataset, in order to give a performance of our methods when $n$ is small ($n=3$) and the $m_i$'s are large. We obtained a WDM of 0.01274.  When taking $L=75$ instead of 150 the ${\text{WDM}}$ is 0.01697, in accordance with the intuition that this value decreases with $L$.  For the sake of comparison, if in \eqref{wrm} we use $\widetilde{\Pi}_i$ instead of $\widehat\Pi_i$, the WDM becomes 0.0793 and 0.08 with $L=75$ and $L=150$, respectively. Boxplots of $d(\widehat{\Pi}_i^{[b]},\Pi_i^{[b]})$ are given in the supplementary materials (Subsection~\ref{supp:supportingOutput}), for all $i$. In the supplementary materials (Subsection~\ref{supp:misspec}) we also include an additional simulation study suggesting satisfactory performance of the methods under misspecification, with data being warped via biased warp maps (i.e., $E(T) \neq t$).

Whilst the performance of both the Bernstein polynomial estimator and the kernel-based estimator is remarkable, there are situations where both estimators suffer from  extrapolation issues on some subintervals; see the following subsection.  Mitigating these nontrivial effects is an important avenue for future research.

\begin{figure}\centering
  \hspace{-.8cm}
  \begin{minipage}{0.33\linewidth}
    \includegraphics[scale = 0.46]{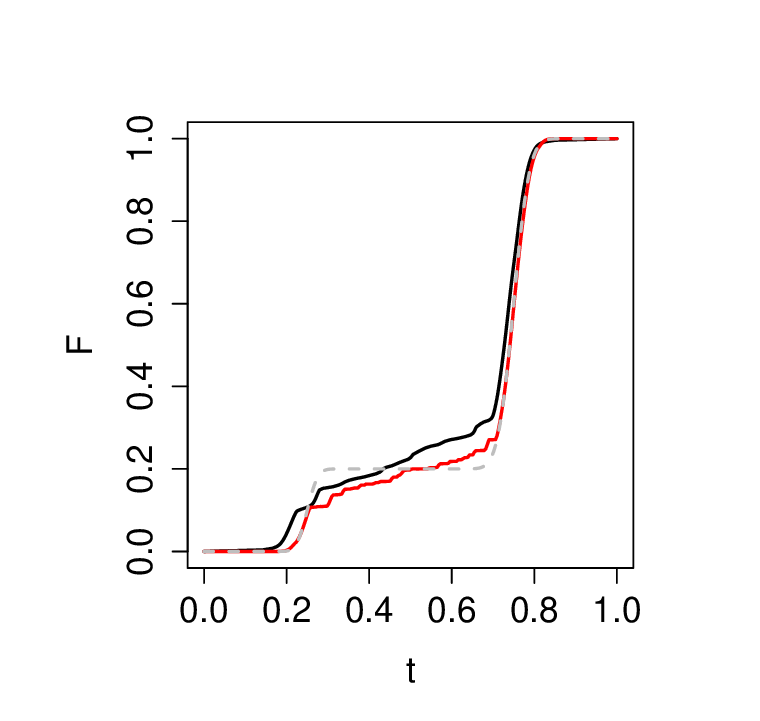}
  \end{minipage}
  \begin{minipage}{0.33\linewidth}
    \includegraphics[scale = 0.46]{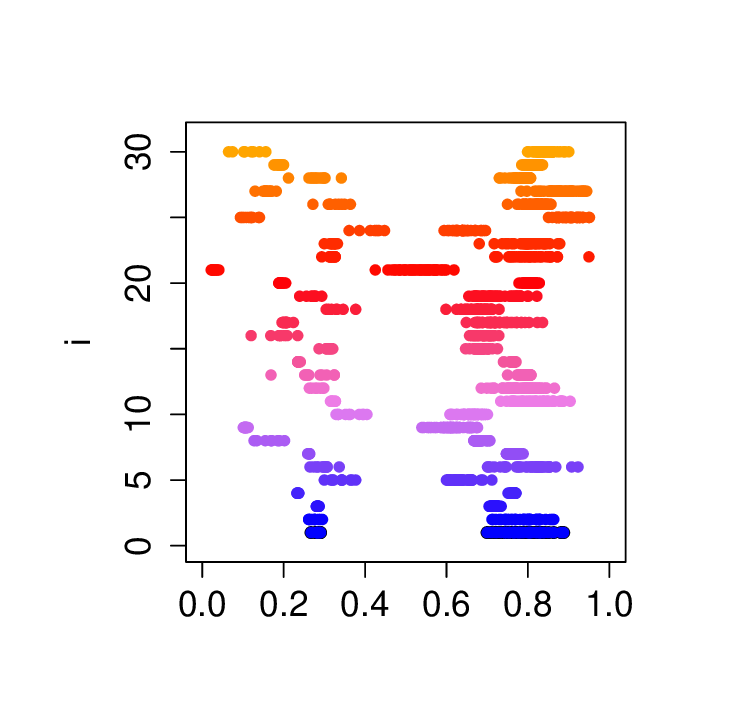}
  \end{minipage}
  \begin{minipage}{0.33\linewidth}
    \includegraphics[scale = 0.46]{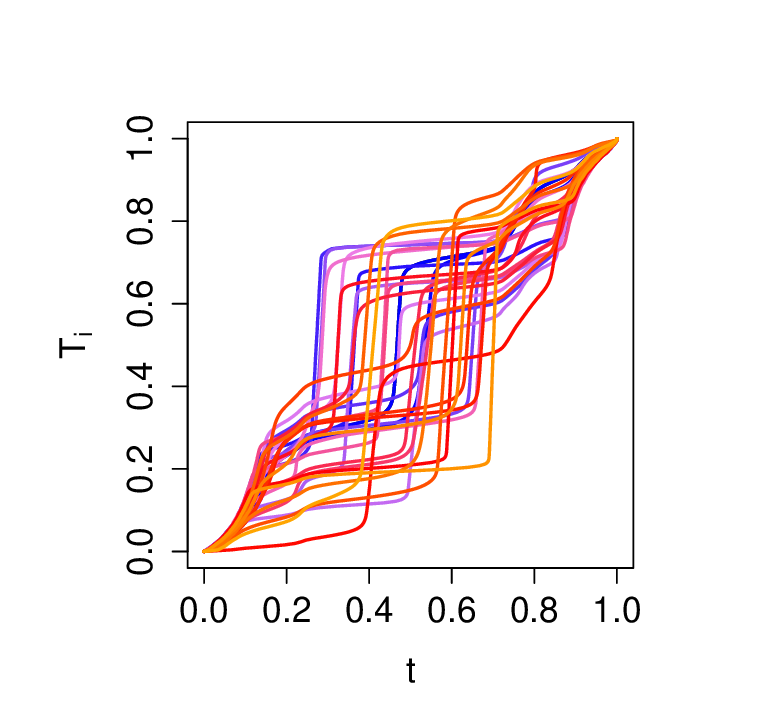}
  \end{minipage}
\caption{\footnotesize Left: Posterior Bernstein polynomial Fr\'echet mean (solid black), kernel smoothing Fr\'echet mean (solid red) and original Fr\'echet mean (grey dashed line). Middle: Phase/varying point process. Right: Posterior mean Bernstein polynomial warp functions colored according to the same palette as in Figure~\ref{fig1}.} 
\label{fig2}
\end{figure}

\subsection{Large $n$, small $m$} \label{sc2}
For comparison with \cite{panaretos2016} we now assess performance over a large $n$ setup. We generate random samples $x_{i,1}, \ldots,x_{i,m_i}\mid m_i$, from  
\begin{equation*}
\lambda(t) = 0.2\, \phi(t \mid 0.25,0.02^2) + 0.8 \, \phi(t \mid 0.75,0.03^2),
\quad m_i \sim \text{Poisson}(L), \quad i = 1,\ldots,n = 30,
\end{equation*}
with $\phi(t \mid \mu, \sigma^2)$ denoting the normal density function and $L =50$. The warped data $\widetilde{x}_{i, j} = T_i(x_{i, j})$ are obtained using 
\begin{equation*}
T_i(t) \stackrel{D}{=} U \, \zeta_{K_1}(t) + (1-U) \, \zeta_{K_2}(t),  \qquad 
  \zeta_k(t) = 
  \begin{cases} \displaystyle 
    t, & k = 0, \\ \displaystyle
    t - \frac{\sin(\pi t k)}{|k|\pi}, & \text{otherwise},
  \end{cases}
\end{equation*}
where $U \sim  \text{Unif}(0,1),$ $K_j \stackrel{D}{=} V_1V_2$ with $V_1 \sim \text{Poisson(3)}$ and $P(V_2 = -1 )= P(V_2 = 1) = 1/2$.

\begin{figure}
  \centering
    \includegraphics[scale = 0.5]{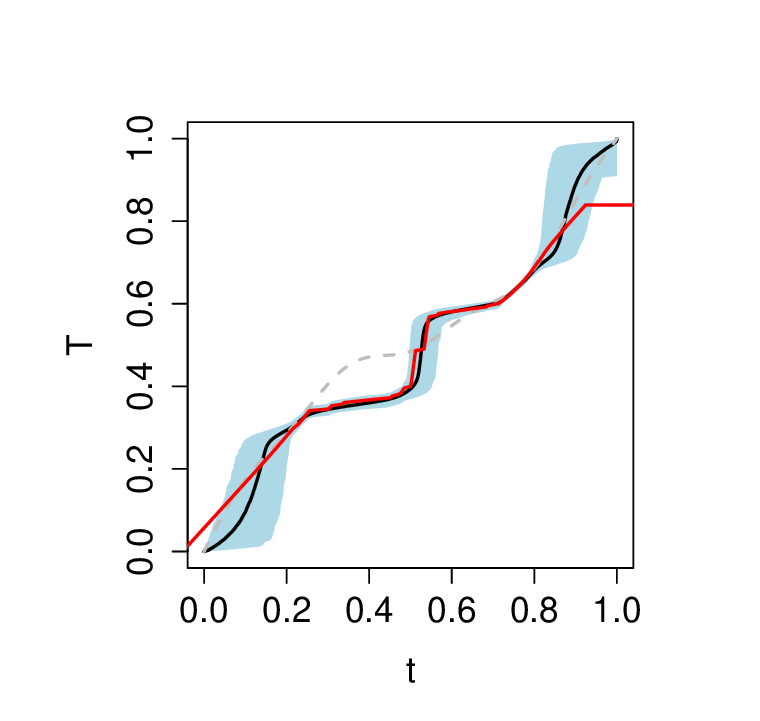}
    \includegraphics[scale = 0.5]{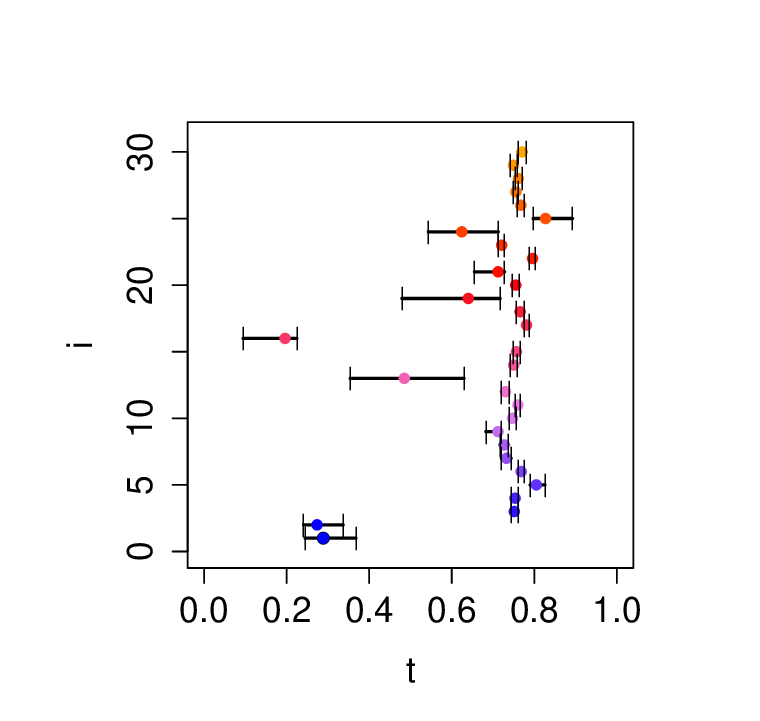}
\caption{\footnotesize Left: Posterior mean Bernstein polynomial warp function (solid black) and corresponding credible band, kernel smoothing warp function estimate  (solid red), and original warp function (dashed grey) for $i = 5$. Right: Credible intervals for randomly selected registered points for each registered point process.}
\label{fig3}
\end{figure}

We start by illustrating our method on this setup on a single run-experiment; a Monte Carlo study was also conducted in this setting along the same lines as in Subsection~\ref{sc1} and it will also be reported below. A realisation of the original point process can be found in Figure~\ref{fig1}. After estimating $F_1, \dots, F_n$ using random Bernstein polynomials we obtain using \eqref{postmeans} the posterior mean Fr\'echet mean depicted in Figure~\ref{fig2}.  The posterior mean is quite similar to the kernel-based estimator of \cite{panaretos2016}, and both are similar to the true Fr\'echet mean. Figure~\ref{fig2} also includes posterior inference for the warp functions. To examine the inference for warp functions in a greater level of detail Figure~\ref{fig3} presents the posterior mean Bernstein polynomial warp function along with credible bands for $i = 5$. As it can be observed from the latter figure, our estimator follows closely that of \cite{panaretos2016}, and is reasonably in line with the original warp function; similar evidence holds for the remainder values of $i$ (see Figure~\ref{FNE:3}). As expected, both estimators have however more difficulty in recovering the true value in the center of the unit interval but this is due to an extrapolation issue as there tends to be much less data on that region. Figure~\ref{fig3} also showcases that our method is more appropriate for bounded domains than the kernel as it takes full advantage of knowledge on the interval where the point processes lie.

Whilst the theoretical claims in Subsection~\ref{sc2} extend those of \cite{panaretos2016}---in the sense that under extra conditions they support the use of the methods even under a small $n$ large $m$ setting---numerical experiments in the supplementary materials suggest that the pointwise performance of our methods is tantamount to that of \cite{panaretos2016}.
Figure~\ref{fig3} presents additionally credible intervals for randomly selected registered points for each registered point process.  Observe that wider intervals are associated to points falling on the interval separating the two `clusters' of points.  

\section{Application: tracking phase variation of annual peak temperatures}\label{application}
\begin{figure}[H]
  \centering
  \hspace{-1cm}
  \includegraphics[scale = 0.5]{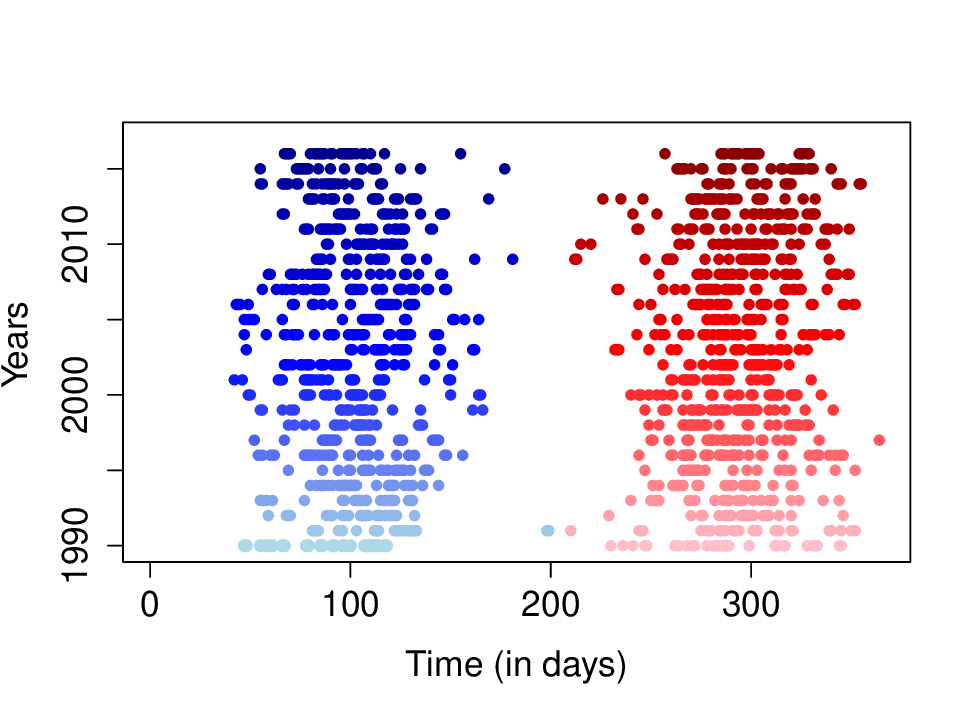} \\
  \hspace{-1cm}
  \includegraphics[scale = 0.4]{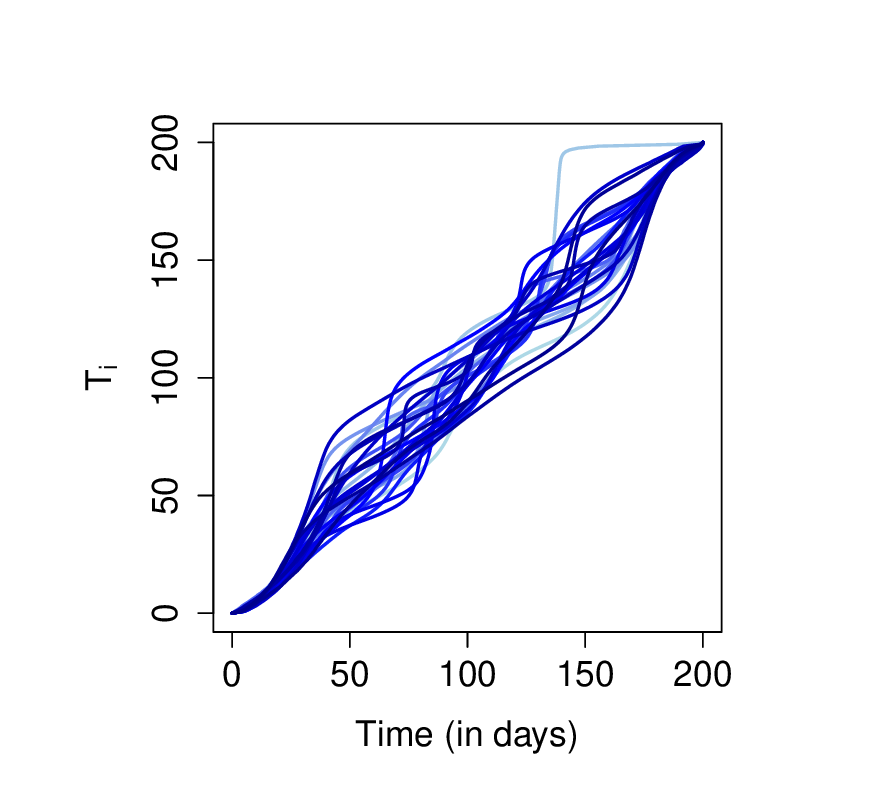}
  \hspace{-1cm}
  \includegraphics[scale = 0.4]{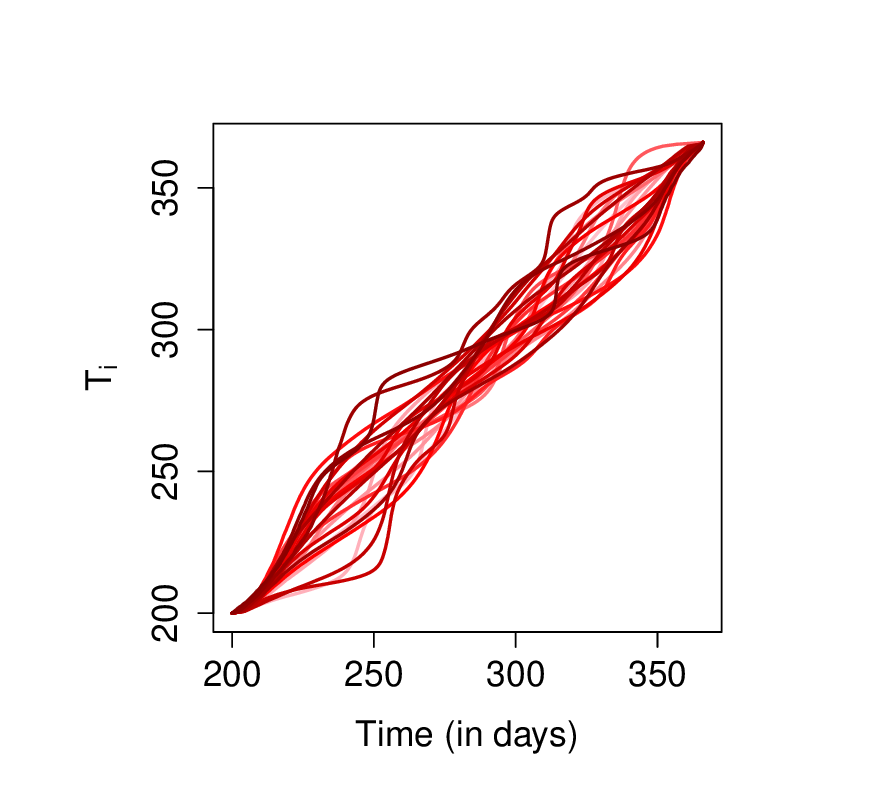}
  \caption{\footnotesize Above: Point processes of annual peaks for peaks above (red) and below (blue) the thresholds. Below---Left and Right: Corresponding posterior mean warp functions in the same palette of colors.}
  \label{fig:pp}
\end{figure}

We now showcase how our method can be used for tracking the phase variation of annual peak temperatures, that is, temperatures above or below a threshold. Peaks of temperature are related with a variety of hazardous events---including heat-related mortality, destruction of crops, wildfires---and have a direct impact on a wealth of economic decisions---such as demand for fuel and electricity. A better understanding of the variation of the regularity of these peaks is thus of the utmost importance from an applied perspective. A main target of our analysis will be on assessing the variation of the onset of temperature peaks, as well as quantifying how atypical is a certain year's pattern of such peaks. Our analysis has points of contact with the subject of shifts in seasonal cycles (e.g., late start of spring, or growing seasons), which is of wide interest in biology and climatology \citep[e.g.,][]{menzel1999, schwartz2006}. To illustrate how the method can be used for such purpose we gathered data from ``\emph{National Centers for Environmental Information of National Oceanic and Atmospheric Administration (NOAA)}'' (https://www.ncdc.noaa.gov/), that consist of average daily air temperatures (in $^\circ$F, rounded to the nearest integer) of Santiago (Chile) from April, 1990 to March, 2017. Let $\widetilde{x}_{i, j}$ be the temperature on day $i$, year $j$. Below, we focus on the point processes of annual peaks over threshold, $\{\widetilde{x}_{i, j}^+ \geq u_j^+\}$, and annual peaks below threshold, $\{\widetilde{x}_{i, j}^- \leq u_j^-\}$; in practice we set the thresholds $u_j^+$ and $u_j^-$ using the 95\% and 5\% quantiles of temperature over year $j$, and this results in $m_1^+, \dots, m_n^+$ and $m_1^-, \dots, m_n^-$ ranging from 19 to 32. The supplementary material includes a sensitivity analysis based on the 97.5\% and 2.5\% quantiles; the main empirical findings  are tantamount to the ones presented here. In Figure~\ref{fig:pp} we present the point processes of interest along with the corresponding warping functions for peaks above the threshold ($T_j^+$) and peaks below the threshold ($T_j^-$). For the analysis of annual peaks over threshold, we fully support the warping functions between the minimum and maximum times corresponding to the pooled exceedances above the threshold; we proceed analogously for the analysis of annual peaks below the threshold. Here the interest is on the highest and lowest temperatures per year, which is tantamount to considering a fixed threshold per year (i.e.~$u_j^+$ and $u_j^-$). Yet the analysis can be easily extended to the situation where the goal is on modeling conditionally high/low temperates (i.e.~taking into account seasonal variation) via a time-varying threshold that can be obtained, for example, via quantile regression \citep{koenker2005}.

To interpret Figure~\ref{fig:pp} we first focus on annual peaks below the threshold, for which there are at least two patterns of points that readily look unusual to the naked eye: 1991, for which there was an atypical cold weather event almost taking place in the summer; and 2010, given that lower temperatures peaked later on a concetrated period. The fact that these patterns of points look unusual agrees with what can be observed from the corresponding warping functions, that are among the ones that further deviate from the identity;  cf.\ Figures~\ref{fig:supp-pp} and \ref{warps:cold} in the supplementary material. In terms of peaks above the threshold, note how the antepenultimate pattern of points started much later than all the remainder, thus meaning that higher temperatures peaked much later than expected.

\begin{figure}[H]\centering
  \hspace{-.8cm}
  \begin{minipage}{0.33\linewidth}
    \includegraphics[scale = 0.43]{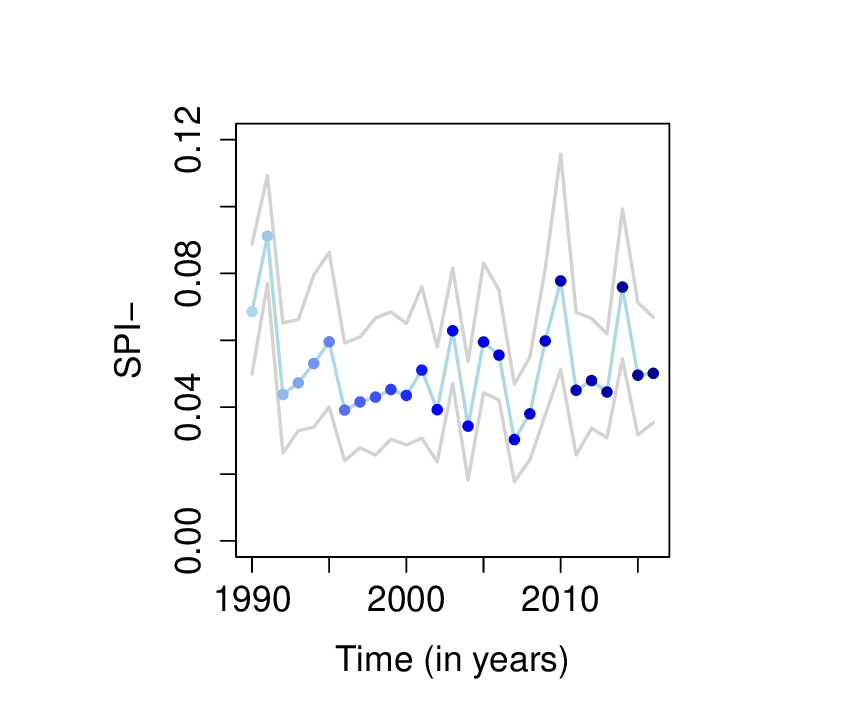}
  \end{minipage}\hspace{0.2cm}
  \begin{minipage}{0.33\linewidth}
    \includegraphics[scale = 0.43]{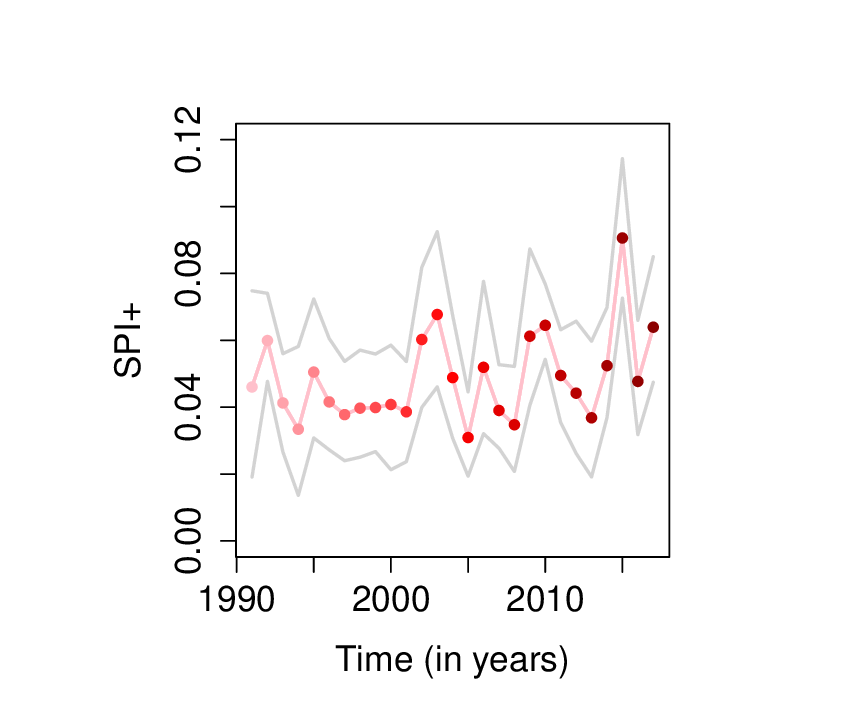}    
  \end{minipage}\hspace{0.2cm}
  \begin{minipage}{0.33\linewidth}
    \includegraphics[scale = 0.43]{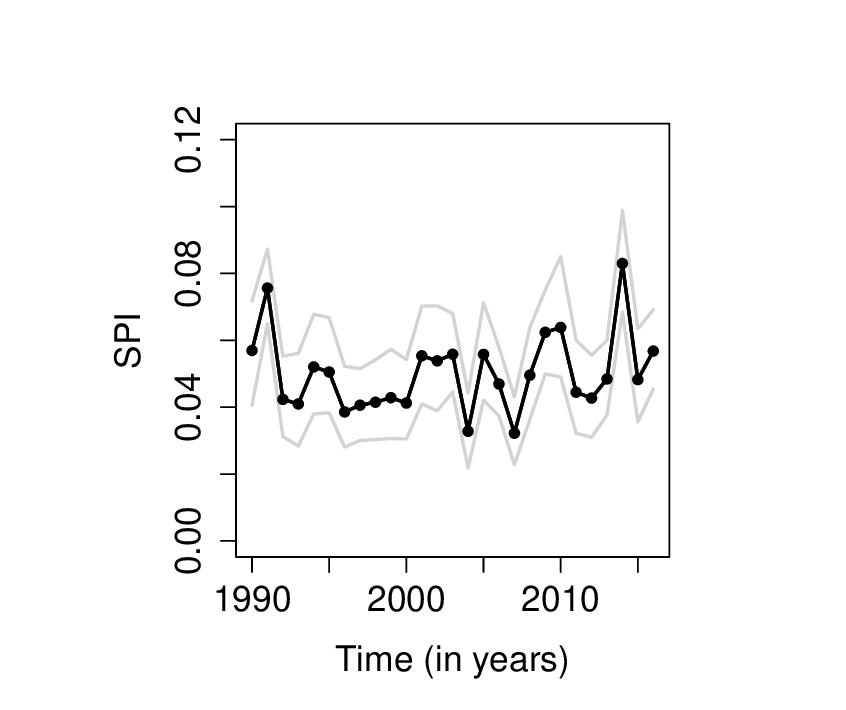}
   \end{minipage}\vspace{-.2cm}
  \caption{\footnotesize Posterior mean \textsc{spi} (scores of peak irregularity), as defined in \eqref{eq:spi}, along with credible intervals, for below threshold (Left), above  threshold (Middle), and global (Right).}
  \label{fig:spi}
\end{figure}

To assess how atypical is the climatological pattern of onset of peaks, we define the following measures to which we refer as scores of peak irregularity (\textsc{spi}), and for temperatures above and below a threshold are respectively defined as 
\begin{equation}
  \label{eq:spi}
  \textsc{spi}^+  = \int_{0}^1 |T_j^+(t) - t| \, \dif t,  \quad 
  \textsc{spi}^- = \int_{0}^1 |T_j^-(t) - t| \, \dif t;
\end{equation}
to combine peaks over and below a threshold, we also define a global $\textsc{spi} = (\textsc{spi}^+ + \textsc{spi}^-) / 2$. Figure~\ref{fig:spi} depicts the scores of peak irregularity over time for peaks above and below a threshold. To shed light on interpretation of Figure~\ref{fig:spi} we note that if the climatological pattern of the onset of peaks above or below the threshold was always the same, then all \textsc{spi}s would be equal to zero. The ranking of the \textsc{spi}s, on the other hand, quantifies which patterns of onset of peaks are the most anomalous---with the largest \textsc{spi}, for the most atypical year. Figure~\ref{fig:spi} is coherent with what was expected given the comments above surrounding Figure~\ref{fig:pp} on the patterns of points that looked immediately atypical, and on the shape of the corresponding warping functions.  

\section{Closing remarks}\label{conclusion}
We propose a semiparametric Bayesian approach for the purpose of separating amplitude and phase variation in point process data.  This paradigm has the advantage of providing a straighforward construction of credible sets via the posterior distribution, and  in particular, we are able to quantify the uncertainty in learning not only the structural mean measure $\lambda$, but also the warping functions $T_i$ and the latent point processes $\Pi_i$. The Bernstein--Dirichlet prior interweaves elegantly with the Wasserstein geometry of optimal transport.  Indeed, its favourable support properties \citep[as established by][]{petrone2002} carry over to the induced priors on the structural mean measure $\lambda$ and all sufficiently regular warping functions, allowing to obtain Bayesian consistency in a genuinely infinite-dimensional setup.

A interesting question would be how to extend this work to the case of spatial point process supported on e.g., $[0,1]^D$ with $D>1$, as explored by \cite{boissard2015distribution} and \cite{zemel2017frechet}; a natural extension of our paper to this setup would entail modelling the mean measures of the corresponding spatial point processes via multivariate Bernstein polynomials \citep{zheng2009}. The computation of the empirical Fr\'echet--Wassertein mean can no longer however be done in closed form, requiring numerical schemes \citep{peyre2019computational}. From a statistical viewpoint, another avenue for future research would be on modelling the phase variation of point processes conditionally on a covariate, by resorting to predictor-dependent versions of the Bernstein--Dirichlet prior \citep{barrientos2017}.

\subsection*{Acknowledgement}
\begin{footnotesize}
BG was partially supported by the graduate scholarship 21140901 from the Chilean NSF (CONICYT), YZ was supported by Swiss National Science Foundation Early Postdoc.Mobility Fellowship \# 178220, and MdC was partially supported by FCT (Funda\c c\~ao para a Ci\^encia e a Tecnologia, Portugal) through the projects PTDC/MAT-STA/28649/2017 and UID/MAT/00006/2019. 
\end{footnotesize}

\section{Appendix}\label{sec:appendix}
\subsection{Auxiliary lemmata}
\renewcommand{\theequation}{\thesection.\arabic{equation}}
We begin by stating a number of auxiliary lemmata that will be useful to deriving our main results.  Lemma~\ref{lem:weakKol} is often known as P\'olya's theorem \citep[][Theorem~11.2.9]{lehmann2006}.  Lemma~\ref{lemma1} states that inversion is continuous in supremum norm \citep[][Lemma~11.2.1]{lehmann2006}.  Lemma~\ref{lem:BernDer} discusses sufficient conditions for (local) uniform convergence of the Bernstein polynomial density; see \citet[][Lemma E.3]{ghosal2015} for a related result under further smoothness assumptions on $\mathbbm f$. As the proof of Lemma~\ref{lem:BernDer} shows, the uniform converges holds on any set bounded away from the discontinuity points of $\mathbbm f$. Proofs of Lemmata~\ref{lem:weakKol}--\ref{lem:BernDer} are available from the supplementary materials.
\begin{lemma}\label{lem:weakKol}
Let $\mathbb F$ be a continuous distribution function and let $F_n$ be a sequence of distribution functions that converge weakly to   $\mathbb F$.  Then $\|F_n-\mathbb F\|_\infty\to0$.
\end{lemma}
\begin{lemma}\label{lemma1} Let $\mathbb F:[0,1]\to[0,1]$ be continuous, strictly increasing and with $F(0)=0$, $F(1)=1$.  Then $\mathbb F^{-1}$ is also continuous and strictly increasing, and for any $\varepsilon > 0$ there exists $\delta > 0$ such that for any continuous strictly increasing $H:[0,1]\to[0,1]$:
\begin{enumerate}
\item If $\|\mathbb F - H\|_{\infty} < \delta$, then $\|\mathbb F^{-1} - H^{-1}\|_{\infty} < \varepsilon$.
\item If $\|\mathbb F^{-1} - H^{-1}\|_{\infty} < \delta$, then $\|\mathbb F - H\|_{\infty} < \varepsilon.$
\end{enumerate}
\end{lemma}
\begin{lemma}\label{lem:BernDer}
Let $\mathbb F:[0,1]\to \mathbb R$ be differentiable with derivative $\mathbbm f$ that is continuous on $(0,1)$.  Then for any $a>0$, $b(x \mid k, \mathbb F)$ as defined in \eqref{b} converges to $\mathbbm f$ uniformly on $[a,1-a]$.  If $\mathbbm f$ is continuous on $[0,1]$, then $b(x\mid k,\mathbb F)\to \mathbbm f$ uniformly on $[0,1]$.
\end{lemma}

\subsection{Proofs of main results}
\textit{Proof of Theorem~1.}
  \begin{enumerate}
  \item[(a)] The proof follows from Theorem~3 in~\cite{petrone1999b}, combined with the fact that by assumption $F_1,\ldots,F_n \iid \pi$. Indeed,
    \begin{equation*}
      \pi^{(n)}\{\mathscr{F}: \; \|F_{j} - \mathbb{F}_{j}\|_{\infty} < \varepsilon, \; j=1,\ldots,n\} = \prod_{j=1}^n \pi\{F_j: \|F_{j} - \mathbb{F}_j\|_{\infty} < \varepsilon\} > 0.
    \end{equation*}
   \item[(b)] From (a) and Lemma~\ref{lemma1} it follows that 
     \begin{equation}\label{cons}
       \pi^{(n)}\{\mathscr{F}: \|F_i^{-1} - \mathbb{F}^{-1}\|_{\infty} < \eta, i = 1, \dots, n\}> 0, \quad \eta > 0. 
     \end{equation}
  Also, note that 
  \begin{equation}\label{eq:obs}
    \|F^{-1} - \mathbb{F}^{-1}\|_{\infty} =
    \bigg\|\frac{1}{n} \sum_{i = 1}^n 
    F_i^{-1} - \mathbb{F}^{-1} \bigg \|_{\infty} \leq 
    \frac{1}{n} \sum_{i = 1}^n \|F_i^{-1} - \mathbb{F}^{-1}\|_{\infty}. 
  \end{equation}
From \eqref{eq:obs} and Lemma~\ref{lemma1}, it follows that to have $\|F - \mathbb{F}\|_{\infty} < \varepsilon$ it would suffice having $\|F_i^{-1} - \mathbb{F}^{-1}\|_{\infty} < \delta$ for all $i$, thus implying that 
\begin{equation*}
  \pi^{(n)}\{\mathscr{F}: \|F - \mathbb{F}\|_{\infty} < \varepsilon\}
  \geq  
  \pi^{(n)}\{\mathscr{F}: \|F_i^{-1} - \mathbb{F}^{-1}\|_{\infty} < \delta\} > 0.
\end{equation*}
\item[(c)] Lemma~\ref{lemma1} and the assumption $\mathbb{F}_i$ is (uniformly) continuous on $[0,1]$ imply that $\mathbb{F}_i^{-1}$ is also uniformly continuous, for $i = 1, \dots, n$. Given $\eta > 0$, let $\delta > 0$ such that $|t - s| \leq \delta \Rightarrow |\mathbb{F}_i^{-1}(t) - \mathbb{F}_i^{-1}(s)| \leq \eta$, for $i = 1, \dots, n$. From (a) and (b) it respectively follows that
\[
  \pi^{(n)}\{\mathscr{F}: \|F_i^{-1} - \mathbb{F}^{-1}_i\|_{\infty} \leq \eta, \, 
  i = 1, \dots, n\} > 0, \quad 
  \pi^{(n)}\{\mathscr{F}: \|F - \mathbb{F}\|_{\infty} \leq \delta\} > 0. 
\]
Thus, $\pi^{(n)}\{\mathscr{F}: |F(x) - \mathbb{F}(x)| \le \delta, \, x \in [0, 1]\} > 0$, and this implies that the event  
\begin{equation}\label{eq:preps}
  \begin{cases}
    F_i^{-1}(F(x)) \leq F^{-1}_i(\mathbb{F}(x) + \delta) \leq 
    \mathbb{F}_i^{-1}(\mathbb{F}(x) + \delta) + \eta \leq 
    \mathbb{T}_i(x) + 2 \eta, \\
    F_i^{-1}(F(x)) \geq F^{-1}_i(\mathbb{F}(x) - \delta) \geq 
    \mathbb{F}_i^{-1}(\mathbb{F}(x) - \delta) - \eta \geq 
    \mathbb{T}_i(x) - 2 \eta, \\
  \end{cases}
\end{equation}
occurs with positive probability, for $i = 1, \dots, n$. This thus yields that
\begin{equation*}
  \pi^{(n)}\{\mathscr{F}: \|T_i - \mathbb{T}_i\|_{\infty} \leq 2 \eta\} > 0, \quad i = 1, \dots, n.
\end{equation*}

\item[(d)] The strategy of the proof is similar to that of \citet[][p.~798]{panaretos2016}. We start by noting that $T_i^{-1} \circ \mathbb{T}_i \in \Gamma(\Pi_i / N_i, P_i / N_i)$ as a consequence of
\begin{equation*}
  \Pi_i = T_{i \; \#}^{-1} \widetilde \Pi_i = (T_i^{-1} \circ \mathbb{T}_{i})_{\#} P_i, \quad i = 1, \dots, n.
\end{equation*}
It thus follows that 
\begin{equation*}
  \begin{split}
  d^2(\Pi_i / N_i, P_i / N_i) 
  \leq \int_0^1 \{(T_i^{-1} \circ \mathbb{T}_{i})(x) - x\}^2 \, \frac{\Pi_i(\dif x)}{N_i}
  \leq \|\{T_i^{-1} \circ \mathbb{T}_{i} - x\}^2\|_{\infty}.
  \end{split}
\end{equation*}
To complete the proof just note that (c) implies that for all $i$
\begin{equation*}
  \pi^{(n)}\{\mathscr{F}: 
  \|T_i^{-1} \circ \mathbb{T}_{i} - x\|_\infty < \varepsilon\}
  =   \pi^{(n)}\{\mathscr{F}: 
  \|T_i^{-1} - \mathbb{T}_{i}\|_\infty < \varepsilon\} > 0,   
\end{equation*}
from where the final result follows.
\end{enumerate}

\textit{Proof of Theorem~\ref{thm2}.} The derivatives of the induced priors \eqref{frechet:mean} and \eqref{warp} will be required for the proofs, and are respectively
  \begin{equation*}
    f(x) = n\left( \sum_{i=1}^n \frac1{f_i(T_i(x))}  \right)^{-1}
    , \quad T_i'(x) = \frac{f(t)}{f_i(T_i(x))}, \quad i = 1, \dots, n
    ,\qquad f_i=F_i'.
  \end{equation*}
  \begin{enumerate}
  \item[(a)] Let $\mathbbm{f}_i$ be the density corresponding to $\mathbb{F}_i$ and $\mathbbm f$ that corresponding of $\mathbb{F}$. Then
\begin{equation} \label{boun}
|f(x) - \mathbbm{f}(x)|
= n\left| \left(\sum_{i=1}^n \frac 1{f_i(T_i(x))}\right)^{-1}  
 - \left(\sum_{i=1}^n \frac 1{\mathbbm{f}_i(\mathbb{T}_i(x))} \right)^{-1} \right|.  
\end{equation}
We first assume that $\inf \mathbbm f_i\ge2l>0$ for all $i$, and consequently $\inf \mathbbm f>2l$ as well.  For $g:[0,1]\to\mathbb R$ and $1/2>a>0$ denote $\|g\|_{\infty,a}=\sup_{x\in[a,1-a]}|g(x)|$.  We shall show that the event
\[
\Omega_{a,\rho}
=\{f_i\ge l \ \&\  \|f_i - \mathbbm{f}_i\|_{\infty,a}<\rho,i=1,\dots,n\},
\]
has positive probability for all $a,\rho>0$.  Let $k_i$ be large so that $\|b(x \mid k_i,\mathbbm{f}_i) - \mathbbm{f}_i\|_{\infty,a}<\rho/2$ (using Lemma~\ref{lem:BernDer}), set $k=\max_i k_i$ and denote $b(x \mid k,\mathbbm{f}_i)=\sum_{j=1}^k w_{i,j}\beta(x \mid j,k-j+1)$.  The set of polynomials with slightly perturbed coefficients
\[
\mathcal P_{i,\delta}
=
\left\{p=\sum_{j=1}^k w'_{i,j}\beta(x \mid j,k-j+1):
(w_{i,1}', \dots, w_{i,k}') \in S_k \text{ with } |w'_{i,j} - w_{i,j}|<\delta, \text{for all }j \right\}
\]
has positive probability under the Bernstein polynomial prior, for all $\delta>0$, as a consequence of \citet[][p.~84]{petrone2002} because the set where $(w'_{i,1},\dots,w'_{i,k})$ lies is open in the unit simplex.  Moreover, each $p\in \mathcal P_{i,\delta}$ satisfies
\[
\|p - b(x \mid k, \mathbbm{f}_i)\|_\infty 
\le \delta k\max_{1\le j\le k} \sup_x\beta(x \mid j,k-j+1)
<\infty
\]
because $1\le j\le k$.  Thus for small enough $\delta$, $\|p - b(x \mid k,\mathbbm{F}_i)\|_{\infty,a}<\rho/2$.  Since the $F_i$'s are independent, there is a positive probability that $f_i\in \mathcal P_{i,\delta}$ for all $i$, which implies that $\|f_i - \mathbbm{f}_i\|_{\infty,a}<\rho$ for all $i$.  Moreover, as $\mathbbm f_i\ge 2l$, $w_{i,j}\ge2l/k$ and if $\delta<l/k$ this yields $w'_{i,j}\ge l/k$ and thus $b(x\mid k,\mathbb F_i)\ge l$.  Hence $\Omega_{a,\rho}$ has positive probability.

Fix $\epsilon>0$; we wish to show that $\|F_i - \mathbb F_i\|_\infty\le\epsilon$ holds on $\Omega_{a,\rho}$ for appropriate $a,\rho>0$.  Let $1/2>a>0$ such that $\mathbb F_i(a)<\epsilon/3$ and $\mathbb F_i(1-a)>1-\epsilon/3$, and let $\rho<\epsilon/3$.  When $\Omega_{a,\rho}$ holds, we have
\begin{align*}
1\ge 
F_i(1-a)
=F_i(a)
+\int_a^{1-a}f_i(x)dx
&\ge F_i(a)  +\int_a^{1-a}\mathbbm f_i(x)dx - \rho(1-2a)\\
&=F_i(a) + \mathbb F_i(1-a) - \mathbb F_i(a) - \rho(1-2a).
\end{align*}
Thus $-\epsilon\le F_i(a) -  \mathbb F_i(a)\le 1-\mathbb F_i(1-a)  + \rho(1-2a)<2\epsilon/3$.  For $x\le a$ we have
\[
-\epsilon \le F_i(x) - \mathbb F_i(x)
\le F_i(a) - \mathbb F_i(x)
\le 1-\mathbb F_i(1-a)  + \rho(1-2a) + \mathbb F_i(a) - \mathbb F_i(x)
\le \epsilon.
\]
Thus $|F_i - \mathbb F_i|\le\epsilon$ on $[0,a]$ and by a similar argument the same holds on $[1-a,1]$.  For $x\in[a,1-a]$ observe that
\[
|F_i(x) - \mathbb F_i(x)|
\le |F_i(a) - \mathbb F_i(a)|
+\int_a^{x}|f_i(y) - \mathbbm f_i(y)|dy
\le |F_i(a) - \mathbb F_i(a)| + \rho
<\epsilon.
\]
Conclude that $\|F_i - \mathbb F_i\|_\infty\le\epsilon$.  As in the proof of Theorem~\ref{th1} we have as a conseqeuence that for sufficiently small $a$ and $\rho$, on $\Omega_{a,\rho}$ $\|F_i^{-1}\circ F - \mathbb F_i^{-1}\circ \mathbb F\|_\infty<\epsilon$.  Fix $a,\rho_2\in(0,1/2)$.  Let $c_i=\min(\mathbb F_i^{-1}(\mathbb F(a)),1-\mathbb F_i^{-1}(\mathbb F(1-a)))$ and $a_1=\min_ic_i/2$.  Since $\mathbbm f_i$ is uniformly continuous on $[a_1,1-a_1]$, there exists $\delta_2>0$ such that $|\mathbbm f_i(x) - \mathbbm f_i(y)|\le \rho_2$ for all $x,y\in[a_1,1-a_1]$ such that $|x-y|\le \delta_2$; without loss of generality $\delta_2\le a_1$.  Choose small $a_1>a_2,\rho>0$ such that on $\Omega_{a_2,\rho}$, $\|F_i^{-1}\circ F - \mathbb F_i^{-1}\circ \mathbb F\|_\infty<\delta_2$.  Then on $\Omega_{a_2,\rho}$
\[
\|f_i\circ F_i^{-1}\circ F
-\mathbbm f_i\circ F_i^{-1}\circ F\|_{\infty,a}
\le \|f_i - \mathbbm f_i\|_{\infty,a_1}
\le \rho
\]
and
\[
\|\mathbbm f_i\circ F_i^{-1}\circ F
-\mathbbm f_i\circ \mathbb F_i^{-1}\circ \mathbb F\|_{\infty,a}
\le \sup_{x,y\in[a_1,1-a_1],|x-y|\le \delta_2}
|\mathbbm f_i(x) - \mathbbm f_i(y)|
\le \rho_2.
\]
This means that for any $\rho,\rho_2,a>0$ there is positive probability that for all $i=1,\dots,n$
\[
\|f_i\circ F_i^{-1}\circ F
-\mathbbm f_i\circ \mathbb F_i^{-1}\circ \mathbb F\|_{\infty,a}
\le \rho+\rho_2,
\]
and since $\Omega_{a_2,\rho}$ implies also that $f_i,\mathbbm f_i\ge l$, it follows that for all $a,\epsilon>0$ there is positive probability that $\|f-\mathbbm f\|_{\infty,a}<\epsilon$.  Now write
\[
\KL(F,\mathbbm F)
=\int_{x\in[a,1-a]} \mathbbm f\log\frac{\mathbbm f}f
+\int_{x\notin[a,1-a]} \mathbbm f\log\frac{\mathbbm f}f
=\KL_1 + \KL_2.
\]
The definition of $\Omega_{a,\rho}$ implies that on this event $f\ge l$.  Hence
\[
\KL_2
=\int_{x\notin[a,1-a]} \mathbbm f\log \mathbbm f
-\int_{x\notin[a,1-a]} \mathbbm f\log f
\le \int_{x\notin[a,1-a]} \mathbbm f\log \mathbbm f
-[1-F(1-a)+F(a)]\log l,
\]
and this vanishes as $a\to0$ because $\int\mathbbm f\log \mathbbm f<\infty$.  Hence we can pick $a>0$ such that $\KL_2<\epsilon$.  To bound $\KL_1$ notice that when $\epsilon<l\le\inf \mathbbm f$, and $\|f-\mathbbm f\|_{\infty,a}\le \epsilon$, $\|\log\frac{\mathbbm{f}}{f}\|_{\infty,a}\le \log\frac {\inf \mathbbm f}{\inf \mathbbm f-\epsilon}\le \log\frac l{l-\epsilon}$.  Thus, for all $\epsilon>0$ we have with positive probability
\[
\KL(F,\mathbbm F)
\le \epsilon + \log\frac l{l-\epsilon}.
\]
As this vanishes when $\epsilon\to0$, the proof is complete under the assumption that $\inf \mathbbm f_i>0$ for all $i$.  This assumption can be relaxed as in \citet[][p.~85]{petrone2002}\footnote{beware that they denote $\KL(F,\mathbbm F)$ by $\KL(\mathbbm F,F)$}: take any $\mathbbm f$ as in the statement of the theorem and define $\mathbbm{f}^a(x)=\max(\mathbbm{f}(x),a)/A$, where $A=\int_0^1\max(\mathbbm{f}(x),a)dx\in[1,1+a]$.  Then $\mathbbm{f}\le A\mathbbm{f}^a$ and consequently $\int_0^1\mathbbm f^a(x)\log \mathbbm f^a(x)dx<\infty$.  Applying the theorem to $\mathbbm{f}_1=\dots=\mathbbm{f}_n=\mathbbm{f}^a$ we deduce the KL property for $\mathbbm{f}^a$.  Now, as $\mathbbm{f}\le A\mathbbm{f}^a$ we have \citep[Lemma~5.1]{ghosal1999consistent}
\[
\KL\bigg(\int h,\int \mathbbm{f}\bigg)
\le (A+1)\log A + A\left[\KL\left(\int h,\int \mathbbm{f}^a\right) + \sqrt{\KL\left(\int h,\int \mathbbm{f}^a\right)}\right].
\]
As $a\searrow0$, $A\searrow1$.  If we choose $a>0$ such that $A<2$ and $(A+1)\log A<\varepsilon/3$, and then $\delta>0$ such that $\delta+\sqrt{\delta}<\varepsilon/3$ then
\[
\left\{h:\KL\left(\int h,\int \mathbbm f\right)\le \varepsilon\right \}
\supseteq
\left\{h:\KL\left(\int h,\int \mathbbm f^a\right)\le \delta\right\},
\]
and the latter has positive prior probability.  This completes the proof.

\item[(b)] Again begin with the assumption that $\inf\mathbbm f_i>0$ for all $i$.  Let $T_i'(x) = f(x) / f_i(T_i(x))$ and $\mathbb{T}_i'(x) = \mathbbm{f}(x)/\mathbbm{f}_i(\mathbb{T}_i(x))$, and note that 
\[
|T_i'(x) - \mathbb{T}_i'(x)|
\le \frac{|f(x) - \mathbbm{f}(x)|}{f_i(T_i(x))}  + \mathbbm{f}(x)\left|\frac{1}{f_i(T_i(x))} - \frac1{
\mathbbm{f}_i(\mathbb{T}_i(x))}\right|.
\]
For all $a,\epsilon>0$, since $\mathbbm f$ is bounded on $[a,1-a]$, the same idea as in part (a) shows that with positive probability $\|T_i'-\mathbb T_i'\|_{\infty,a}<\epsilon$.  Write again
\[
\KL(T_i,\mathbbm T_i)
=\int_{x\in[a,1-a]} \mathbbm T_i'\log\frac{\mathbbm T_i'}{T_i'}
+\int_{x\notin[a,1-a]} \mathbbm T_i'\log\frac{\mathbbm T_i'}{T_i'}
=\KL_1 + \KL_2.
\]
These two terms can be made small as in part (a) because $\mathbbm T_i'\le n$. 

To relax the condition $\inf \mathbbm f_i>0$ we use a similar idea as for part (a) but the argument is more subtle.  Fix $a>0$ and define
\[
A_i=
\int_0^1\max(\mathbbm T_i'(x),a) \, \dif x
,\qquad
h_i^a(x)
=\max(\mathbbm T_i'(x),a)/A_i,
\qquad H_i^a(x)
=\int _0^xh_i^a(t) \, \dif t.
\]
For brevity we omit the dependence of $h_i$, $H_i$ and $A_i$ on $a$.  Clearly $H_i$ is strictly increasing, differentiable almost surely with derivative bounded below by $a/A_i$, $H_i(0)=0$ and $H_i(1)=1$.  Moreover $h_i$ is continuous and strictly positive on $(0,1)$ because so is $\mathbbm T_i'$.  We shall view $H_i$ as transport maps from a Fr\'echet mean to well-behaved measures; first we need to fix the issue that they do not necessarily average to the identity by adding another transport map that corrects the discrepancy.

By assumption
\[
A_i
\le\int_0^1(\mathbbm T_i'(x)+a)dx
=\mathbbm T_i(1) - \mathbbm T_i(0) + a
= 1+a
\]
and similarly $A_i\ge1$.  Thus we can choose $a>0$ small such that $(1+a)/A_i\le 1+1/(2n)$ for all $i=1,\dots,n$.  Define the correction function
\[
H_{n+1}(x)
=(n+1)x - \sum_{i=1}^n H_i(x).
\]
Then $H_i$, $i=1,\dots,n+1$ average to the identity.  Since $\mathbbm T_i$, $i=1,\dots,n$ average to the identity, whenever they are differentiable (that it, Lebesgue almost everywhere since they are nondecreasing) we have $\sum_{i=1}^n \mathbbm T_i'(x)=n$.  Hence $H_{n+1}$ is differentiable almost surely with derivative
\[
n+1 - \sum_{i=1}^n h_i(x)
\ge n+1 - \sum_{i=1}^n \frac{\mathbbm T_i'(x)}{A_i} - \frac {na}{A_i}
\ge n+1 - n\frac{1+a}{A_i}
\ge n+1 - n(1+\frac1{2n})
=\frac12.
\]
Now consider the distribution functions $\mathbb G_i=H_i^{-1}$, $i=1,\dots,n+1$ and let $\mathbb G$ denote the identity.  Then $\mathbb G_i$ have Fr\'echet mean $\mathbb G$ with densities bounded above by $\max(2,A_i/a)$ and below by $1/(n+1)$.  Therefore, by the previous part of the proof $H_i^a=\mathbb G_i^{-1}\circ\mathbb G$ is in the KL support of the induced Bernstein polynomial prior.  Since $\mathbbm T_i'\le A_ih_i^a$ almost surely we have \citep[Lemma~5.1]{ghosal1999consistent}
\begin{align*}
\KL(S,\mathbbm{T}_i)
&\le (A_i+1)\log A_i + A_i[\KL(S,H_i^a) + \sqrt{\KL(S,H_i^a)}]
\\&\le (a+2)\log (a+1) + (a+1)[\KL(S,H_i^a) + \sqrt{\KL(S,H_i^a)}].
\end{align*}
As $(H_i^a)'$ is continuous and strictly positive on $(0,1)$, $\KL(S,{H_i}^a)$ can be made as small as we wish with positive probability.  The fact that $a>0$ is arbitrary completes the proof.
\end{enumerate}

\textit{Proof of Theorem~\ref{consistency}}. Under the given assumptions the prior on $F_i$ satisfies the Kullback--Leibler property \citep[Theorem~2]{petrone2002} at $\mathbb F_i$ and consequently the sequence of posteriors are weakly consistent for each $\mathbb F_i$.  The operations
\[
(F_1,\dots,F_n)\mapsto (F_1^{-1},\dots,F_n^{-1})\mapsto \left[F^{-1}=\frac1n\sum_{i=1}^n F_i^{-1}\right]\mapsto F,
\]
are continuous in the supremum norm around $(\mathbb F_1,\dots,\mathbb F_n)$ by Lemma~\ref{lemma1}, \eqref{eq:obs} and again Lemma~\ref{lemma1}.  Taking into account the equivalence of the supremum norm with weak convergence (Lemma~\ref{lem:weakKol}), conclude that the operation $(F_1,\dots,F_n)\mapsto F$ is weakly continuous around $(\mathbb F_1,\dots,\mathbb F_n)$.  Since each $F_i$ is weakly consistent for $\mathbb F_i$, this yields that $F$ is weakly (in fact, Kolmogorov) consistent for $\mathbb F$.

Weak (and Kolmogorov) consistency of $T_i$ to $\mathbb T_i$ follows in the same way, since in Equation~\eqref{eq:preps} it has been established that
\[
(F_i^{-1},F)\mapsto F_i^{-1}\circ F
\]
is continuous in supremum norm around $(\mathbb F_i,\mathbb F)$.

{\centering

{\huge{SUPPLEMENTARY MATERIALS}

}

}

\section{Proofs of auxiliary lemmata}\label{sec:supp-proofs}
\begin{proof}[Proof of Lemma~\ref{lem:weakKol}]~
Since $\mathbb F$ is continuous $F_n\to \mathbb F$ pointwise.  Let $\epsilon>0$ and let $x<y$ such that $\mathbb F(x)\le\epsilon$ and $\mathbb F(y)\ge1-\epsilon$.  Since $\mathbb F$ is uniformly continuous on $[x,y]$ there exists a finite grid $x=x_1<\dots<x_k=y$ with $\mathbb F(x_i)\ge \mathbb F(x_{i+1})-\epsilon$ for all $i\le k-1$.  For $n$ large $|F_n(x_i)-\mathbb F(x_i)|\le\epsilon$ for all $i$ so that
\[
\sup_{z\in [x_i,x_{i+1}]}F_n(z)-\mathbb F(z)\le F_n(x_{i+1}) - \mathbb F(x_i)\le 
|F_n(x_{i+1}) - \mathbb F(x_{i+1})| + |\mathbb F(x_{i+1}) - \mathbb F(x_i)|
\le 2\epsilon.
\]
In the same way
\[
\sup_{z\notin[x,y]}|F_n(z)-\mathbb F(z)|\le 2\epsilon,
\qquad \sup_{z\in [x_i,x_{i+1}]} \mathbb F(z)-F_n(z)\le \mathbb F(x_{i+1}) - F_n(x_i)\le 2\epsilon,
\]
and we conclude that $\|F_n-\mathbb F\|_\infty\le2\epsilon$ for $n$ sufficiently large.
\end{proof}
\begin{proof}[Proof of Lemma~\ref{lemma1}]~
Since $\mathbb F$ is bijective, it has an inverse $\mathbb F^{-1}$.  The latter is nondecreasing and, being a bijection, must also be continuous and with $\mathbb F^{-1}(0)=0$, $\mathbb F^{-1}(1)=1$.  Let $p\in (0,1)$, and let $x\in (0,1)$ such that $\mathbb F(x)=p$.  For $\epsilon\in(0,1-p)$ we have $F_n(x+\epsilon)\to \mathbb F(x+\epsilon)>p$, which means that $x+\epsilon\ge F_n^{-1}(p)$ for $n$ large.  Similarly, $x-\epsilon\le F_n^{-1}(p)$ for any $\epsilon\in(0,p)$ and all $n$ large.  This implies that $F_n^{-1}(p)\to x=\mathbb F^{-1}(p)$ for all $p\in(0,1)$.  Since
\[
0\le 
F_n^{-1}(0)
\le F_n^{-1}(p)
\stackrel{n\to\infty}\to \mathbb F^{-1}(p)
\stackrel{p\to0}\to \mathbb F^{-1}(0)=0,
\]
it also follows that $F_n^{-1}(0)\to \mathbb F^{-1}(0)$.  Similarly $F_n^{-1}(1)\to \mathbb F^{-1}(1)$ and we conclude that $F_n^{-1}\to \mathbb F^{-1}$ pointwise on $[0,1]$.  By Lemma~\ref{lem:weakKol} the convergence is uniform.  Convergence of sequences is equivalent to the statement of the lemma because the supremum norm defines a metric space.

Part b) is shown in the same way, since $(\mathbb F^{-1})^{-1}=\mathbb F$.  There is a slight complication though because $F_n^{-1}$ is only defined on $[F_n(0),F_n(1)]$ which may be a strict subinterval of $[0,1]$.  Let $x=\mathbb F^{-1}(p)$ for $x,p\in(0,1)$.  Then $F_n^{-1}(p-\epsilon)\to \mathbb F^{-1}(p-\epsilon)>x$ for $\epsilon>0$ small, which means in particular that $F_n(0)\le p-\epsilon$ and $F_n^{-1}(p-\epsilon)$ is defined, and also that $F_n(x)\le p-\epsilon$ for $n$ large.  The inequality $F_n(x)\ge p+\epsilon$ is shown in the same way and we conclude that $F_n\to \mathbb F$ pointwise, and hence uniformly on $[0,1]$ by Lemma~1.
\end{proof}
\begin{proof}[Proof of Lemma~\ref{lem:BernDer}]~
Since $\mathbb F$ is differentiable, there exists $x_j^*\in[j,j+1]/k$ such that 
\[
b(x \mid k, \mathbb F)
=B(x \mid k - 1, \mathbbm f)
+\sum_{j=0}^{k-1}\left[\mathbbm f\left(x_j^*\right) - \mathbbm f\left(\frac j{k-1}\right)\right]\binom {k-1}j x^j(1-x)^{k-1-j}.
\]
Notice that $|x^*_j - j/(k-1)|\le j/(k-1) - j/k \le 1/(k-1)\to0$ uniformly in $j$ and as $\mathbbm f$ is uniformly continuous on $[a-1/k,1-a+1/k]$ for all $k>1/a$, the sum at the right-hand side vanishes uniformly in $x\in[a,1-a]$ as $k\to\infty$.  If $\mathbbm f$ is continuous on $[0,1]$ then it is uniformly continuous there and the sum at the right-hand side vanishes uniformly in $x\in[0,1]$. Since $B(x \mid k-1,\mathbbm f)$ converge to $\mathbbm f$ uniformly, this completes the proof.
\end{proof}

\section{Posterior sampling and computing}
\subsection{Gibbs sampling}\label{subsec:supp-gibbs} 
Posterior sampling for each conditional mean measure, $\Lambda(\cdot) = E\{\widetilde{\Pi}(\cdot) \mid T\}$, was conducted according to the following hierarchical structure:
\begin{equation*}
  \begin{split}
k \sim \rho, \quad 
G \sim \text{DP}(\alpha,G^*), \quad  
y \mid k,G \sim  G, \quad 
\widetilde x \mid k,G,y \sim p(\cdot \mid k,y),
\end{split}
\end{equation*}
where $\widetilde x = \{\widetilde x_j\}_{j = 1}^m$ are the raw warped data, $y = \{y_j\}_{j = 1}^m$ are auxiliary latent indicators, and
\begin{equation*}
p(x \mid k,y) = \sum_{j=1}^k \beta(x \mid j,k-j+1)\mathds{1}_{\{(j-1)/k < y_i \leq j/k\}}.
\end{equation*}
The posterior distribution can be computed using a Gibbs sampler with full conditionals being given by \citep[][Section~5.5]{ghosal2015}:
\begin{enumerate}
\item For $k$:
\begin{equation*}
           k \mid \widetilde{x}, y \sim \rho(k \mid \widetilde{x}, y) \propto \rho(k) \prod_{i=1}^m \beta(\widetilde{x}_i \mid z(y_i,k), k -z(y_i,k) + 1),
\end{equation*}
where $z(y,k)=j$ if $(j-1)/k < y \leq j/k$.
      
\item For $y_i$:
\begin{equation*}
y_i \mid k,\widetilde{x},y_{-i} \sim \sum_{j \neq i} q_{i,j}\delta_{y_j} + q_{i,0}G_{b,i},
\end{equation*}
where $y_{-i} = (y_1,\ldots,y_{i-1},y_{i+1},\ldots,y_m)$ and
\begin{equation*}
q_{i,j} \propto \begin{cases} \alpha \, b(\widetilde{x}_i \mid k, G), & j = 0,\\
\beta(\widetilde{x}_i \mid z(y_j,k), k -z(y_j,k) + 1), & 1 \leq j \leq k, j \neq i,
\end{cases}
\end{equation*}
with
\begin{equation*}
\dif G_{b,i}(y \mid k,Y_i) \propto g(y)\, \beta(\widetilde{x}_i \mid z(y,k), k -z(y,k) + 1).
\end{equation*}
\end{enumerate}
\subsection{Parallel computing on the cloud}
The simulation studies reported in the paper and in this supplement were conducted using a virtual machine instance on the Google Cloud Platform (\texttt{cloud.google.com}) running Linux SO with 8 vCPU and 32 GB RAM. Parallel computing was implemented with the R package \textit{parallel} so to speed up the computations.

\section{Further numerical experiments}
\subsection{Supporting outputs}\label{supp:supportingOutput}
\begin{figure}
\centering
  \includegraphics[scale=0.8]{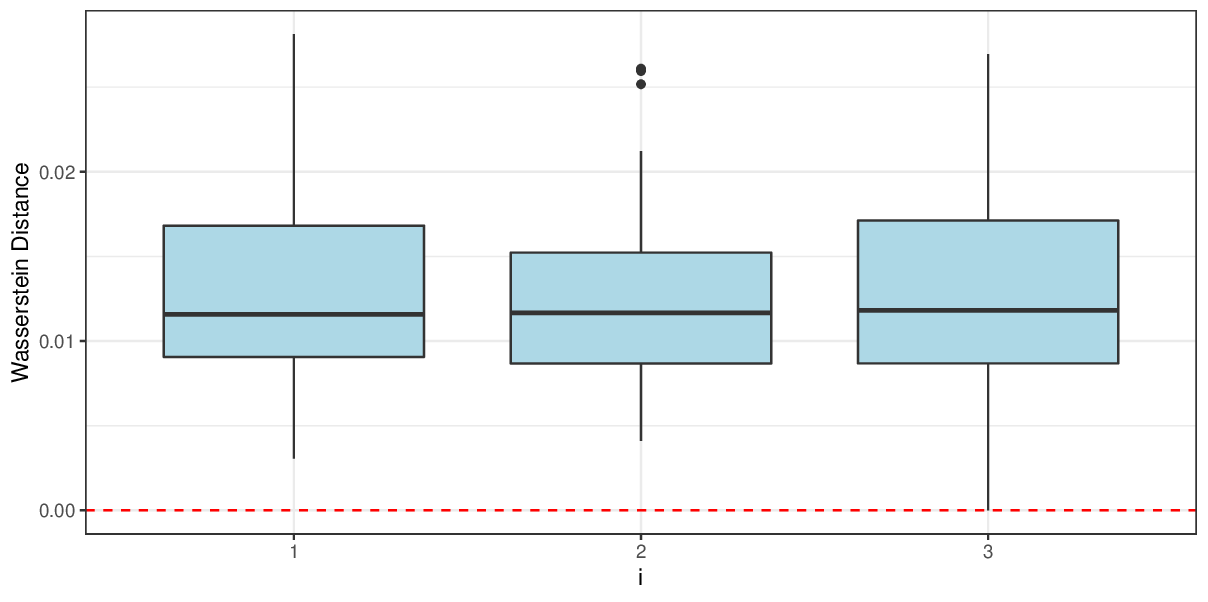}
 \caption{\footnotesize Boxplots of the $L^2$-Wasserstein distance between the original processes $\Pi_i^{[b]}$ and the registered ones $\widehat{\Pi}_i^{[b]}$.  Here $b$ ranges from 1 to $B=50$ and $i=1,2,3$ correspond to the three panels.}
 \label{FNE:1}
\end{figure}

\begin{figure}
\centering
 \includegraphics[scale=0.8]{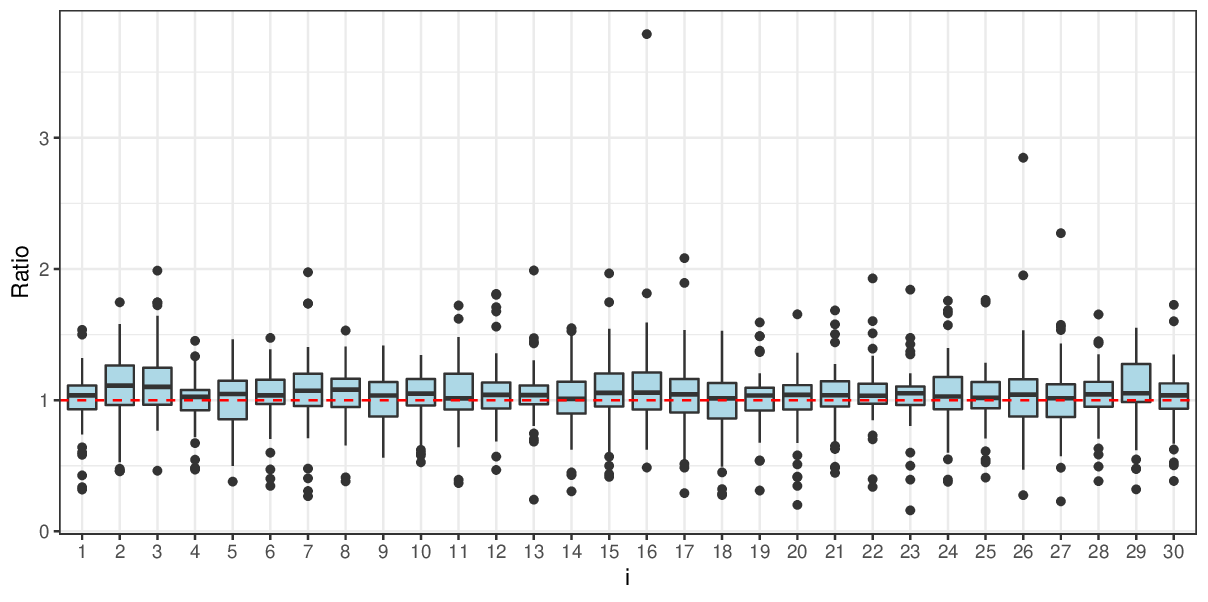}    
 \caption{\footnotesize Comparison of our Bayesian registration with the kernel-based registration of \cite{panaretos2016}.  Each boxplot contains the ratio $d(\widehat{\Pi}_i^{[b,\text{Bayes}]},\Pi_i^{[b]})/d(\widehat{\Pi}_i^{[b,\text{Kernel}]},\Pi_i^{[b]})$ for all $i\in\{1,\dots,30\}$.}
\label{FNE:2} 
\end{figure}

\begin{figure}
\begin{center}
 \includegraphics[scale=0.180]{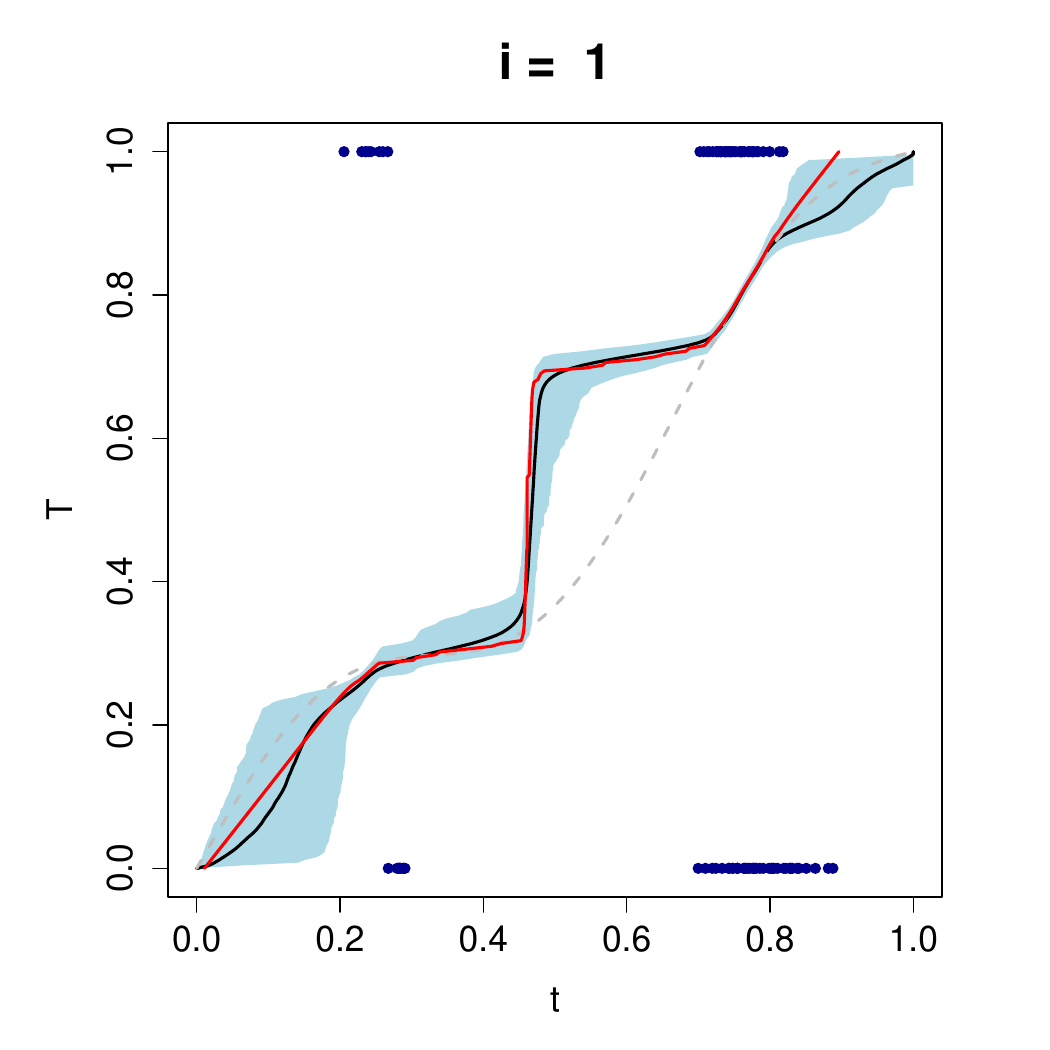}
 \includegraphics[scale=0.180]{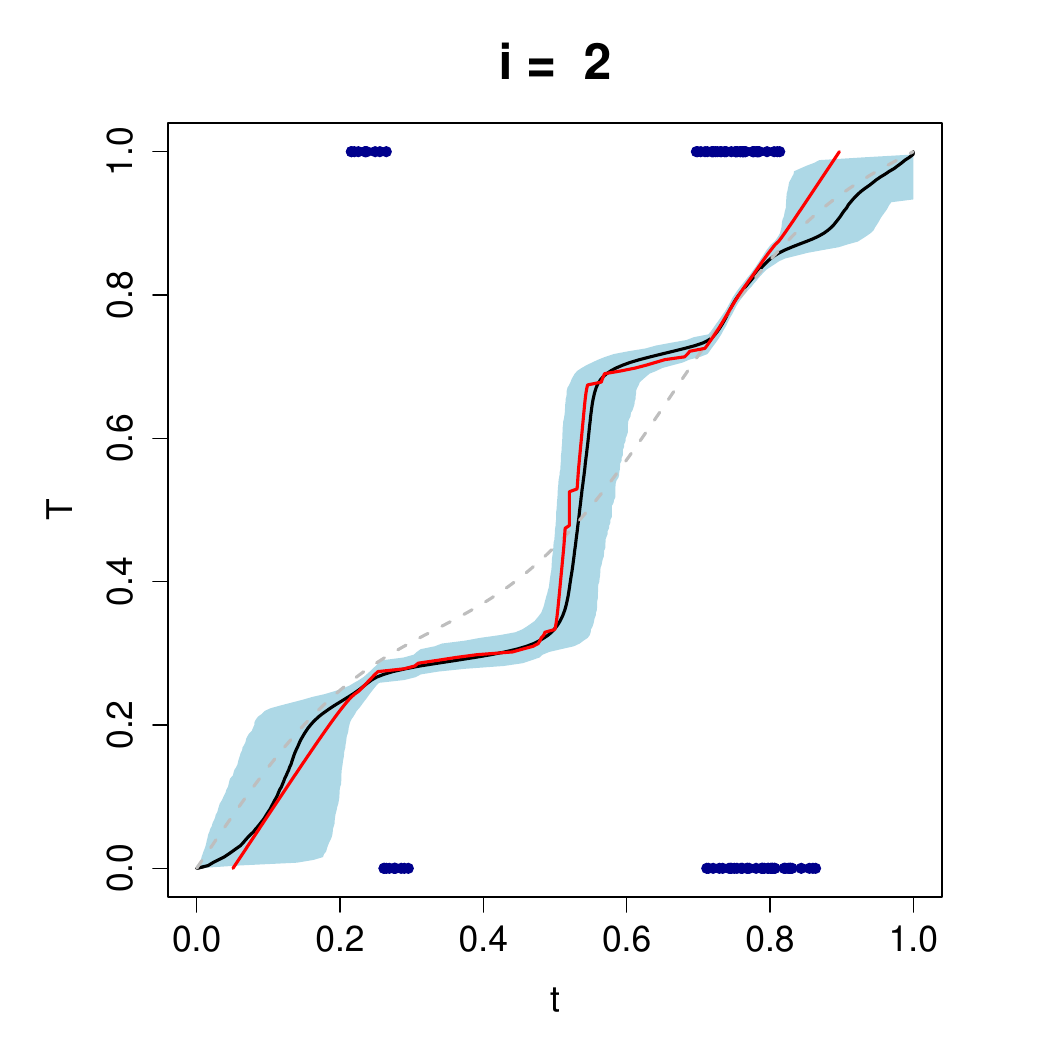}
 \includegraphics[scale=0.180]{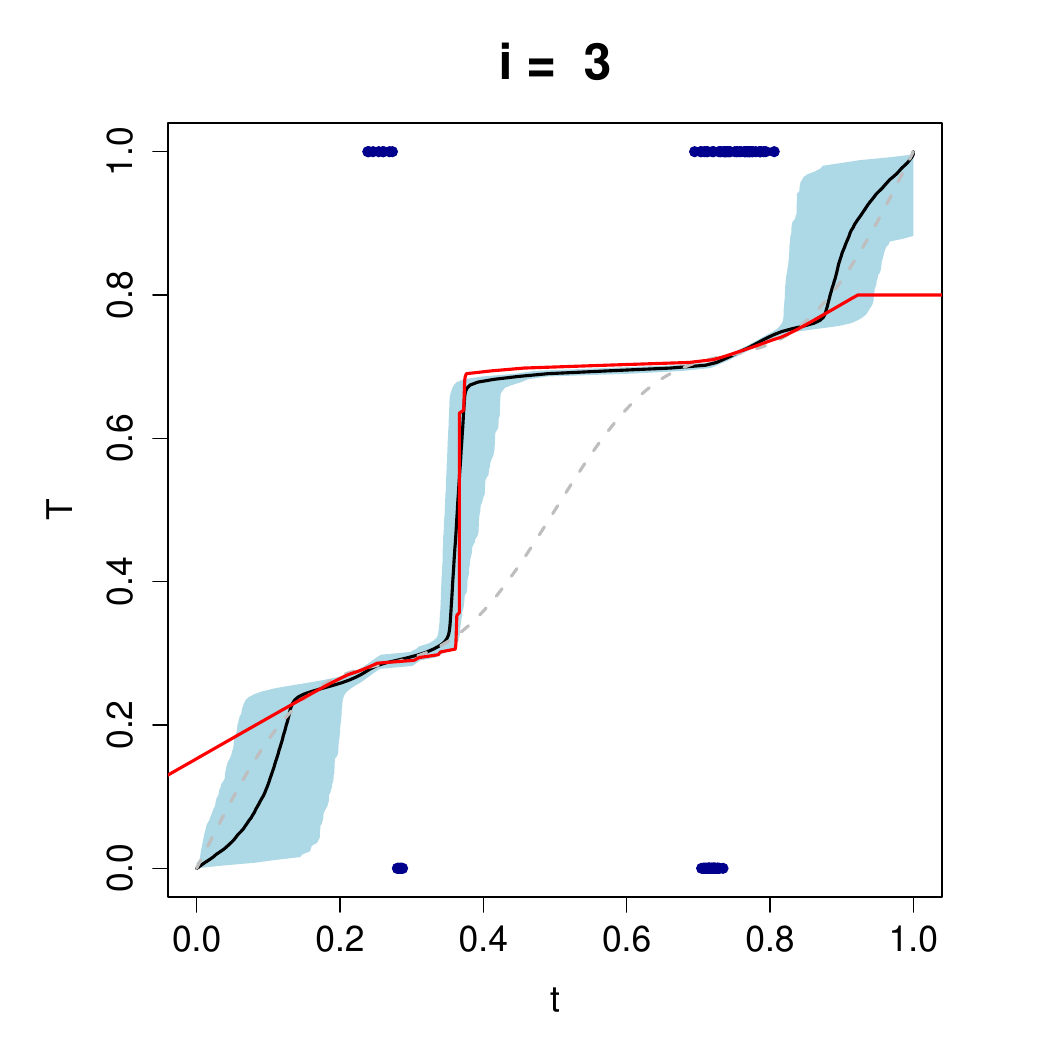}
 \includegraphics[scale=0.180]{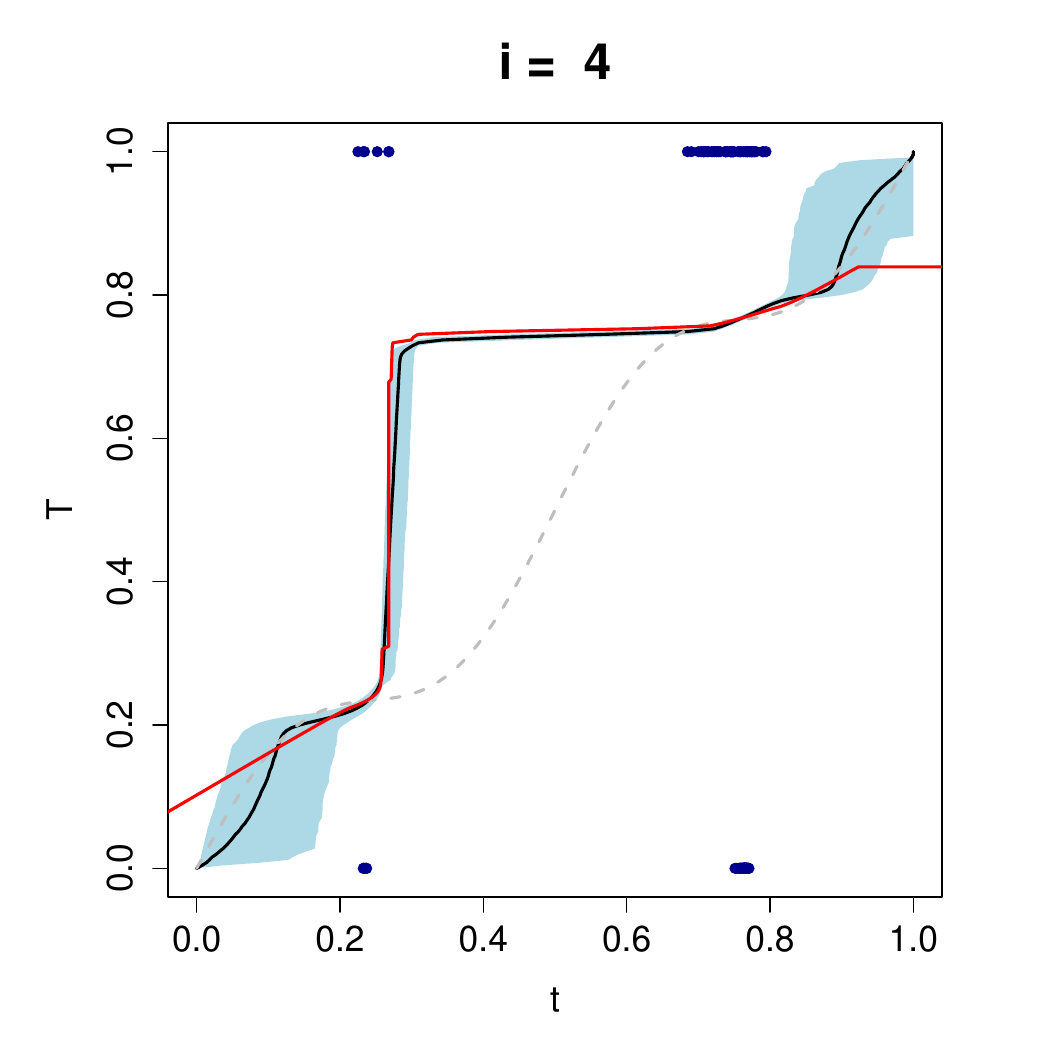}
 \includegraphics[scale=0.180]{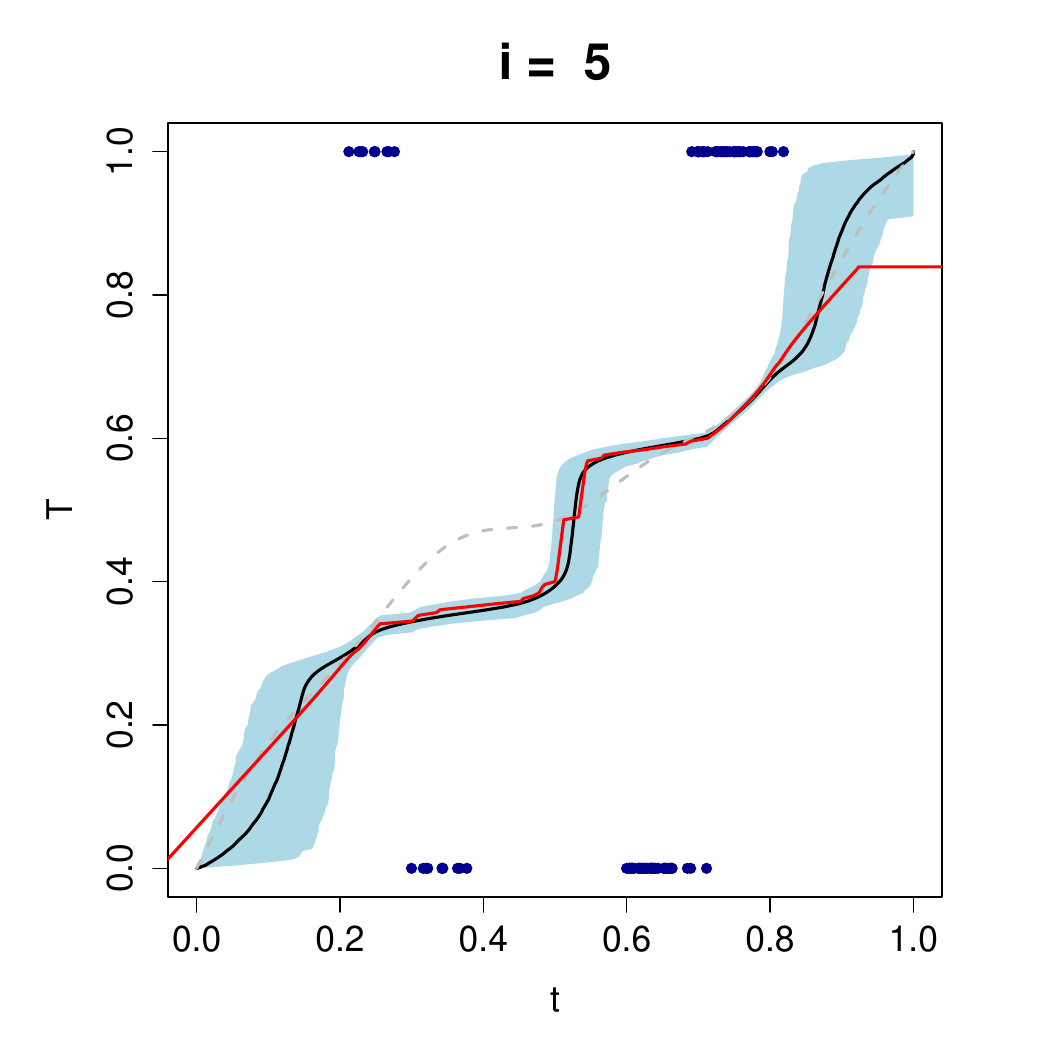}    \\
 \includegraphics[scale=0.180]{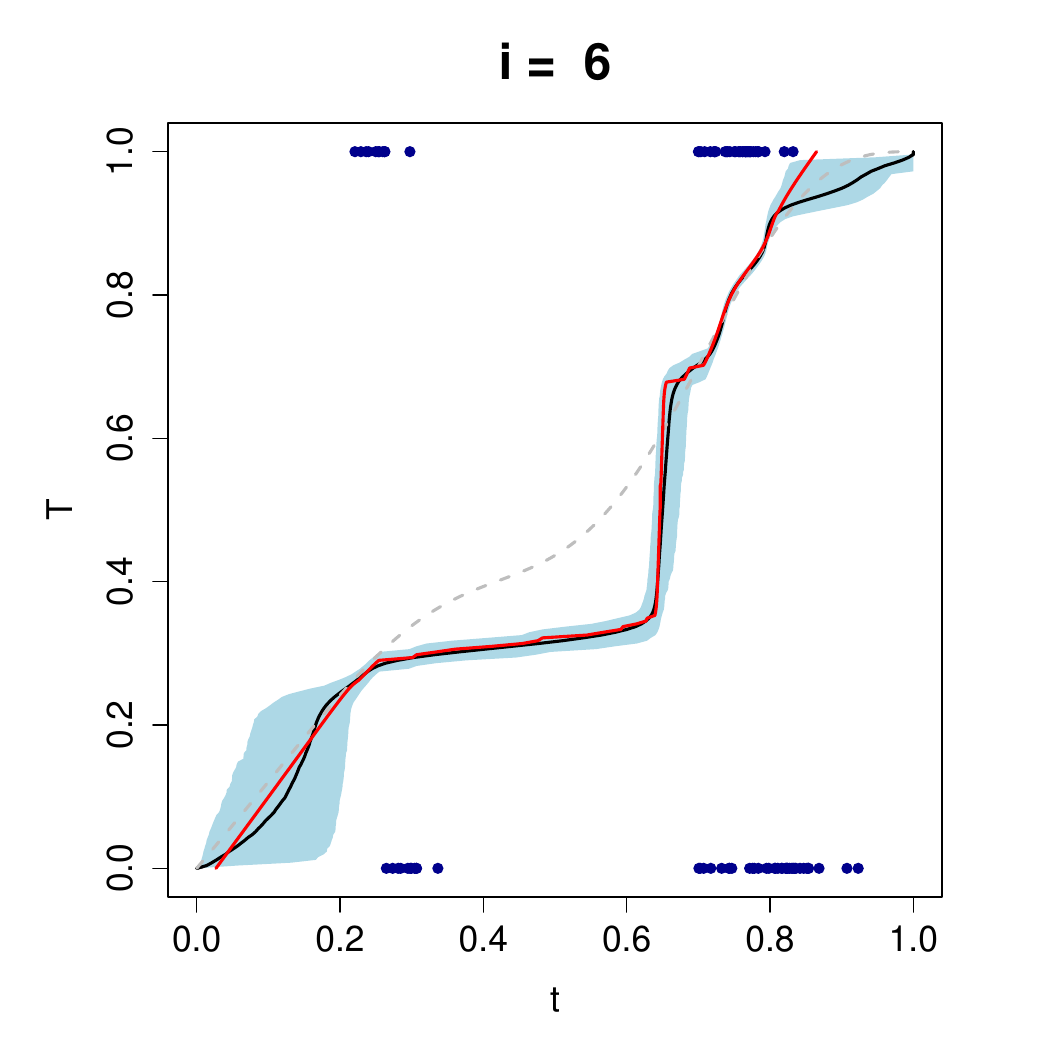}
 \includegraphics[scale=0.180]{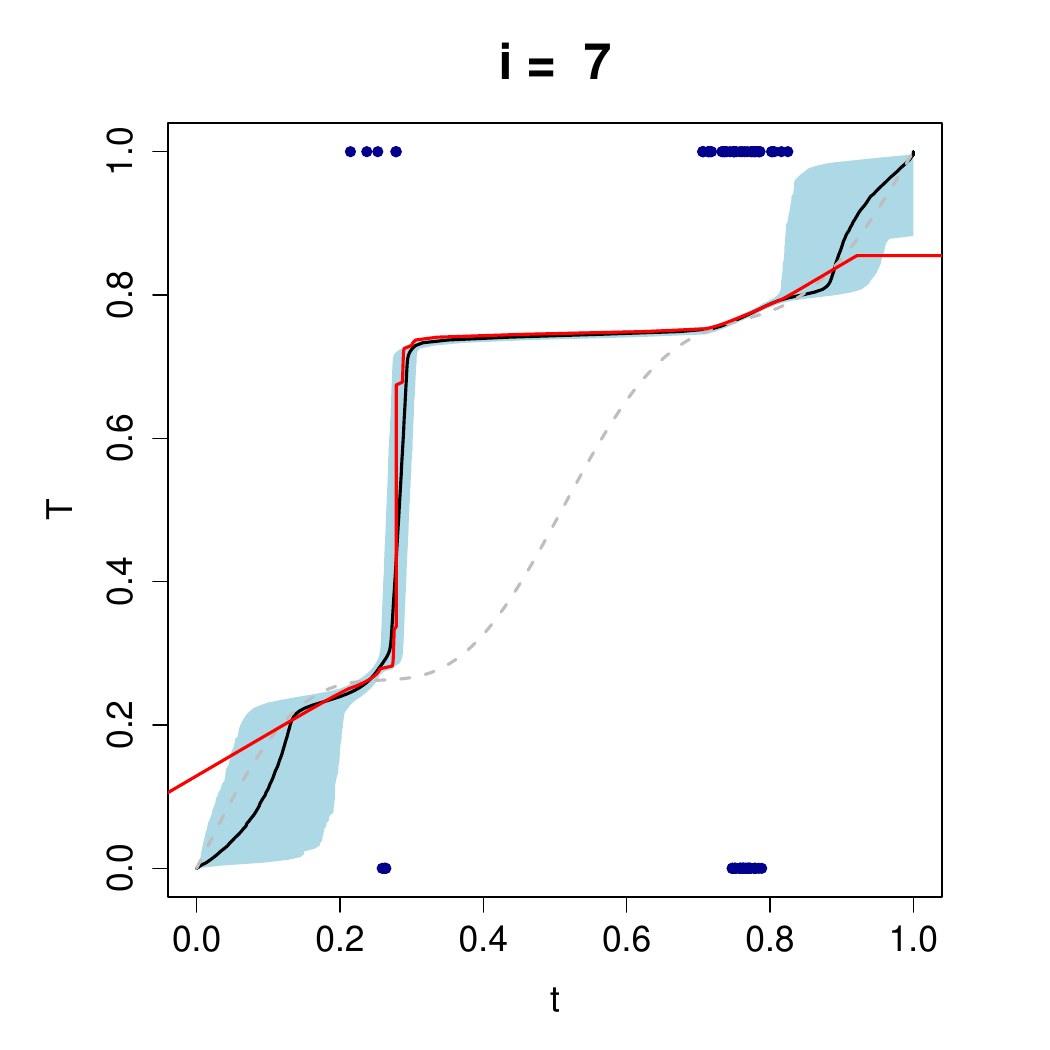}
 \includegraphics[scale=0.180]{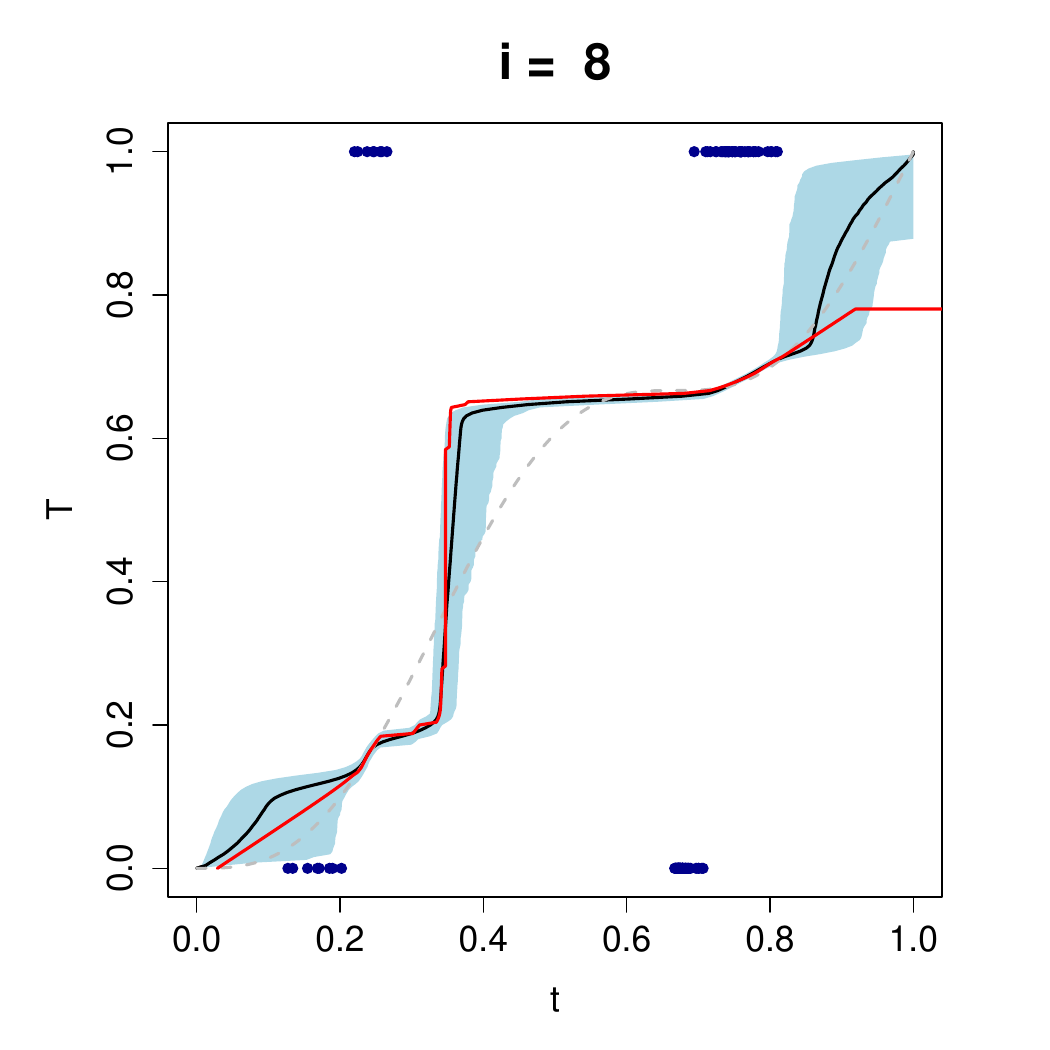}
 \includegraphics[scale=0.180]{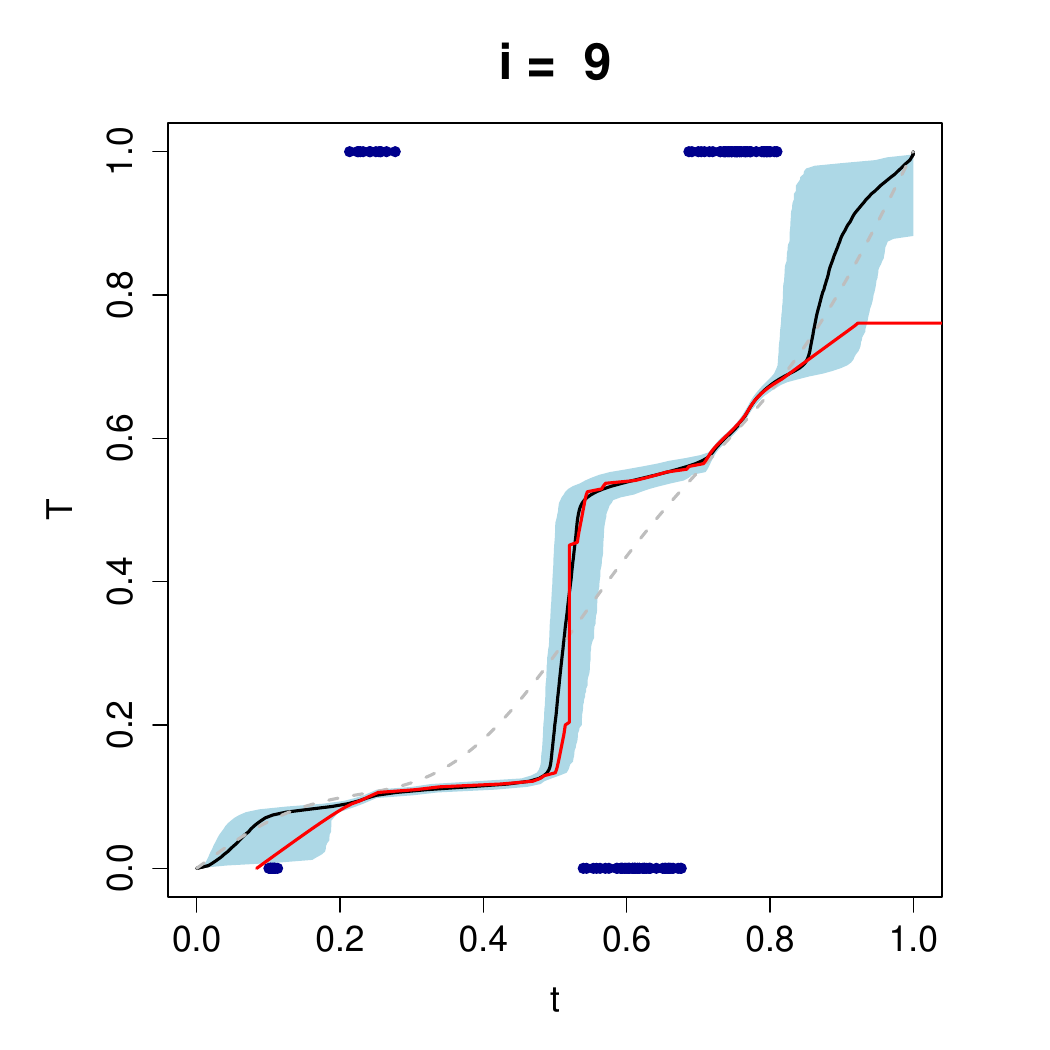}
 \includegraphics[scale=0.180]{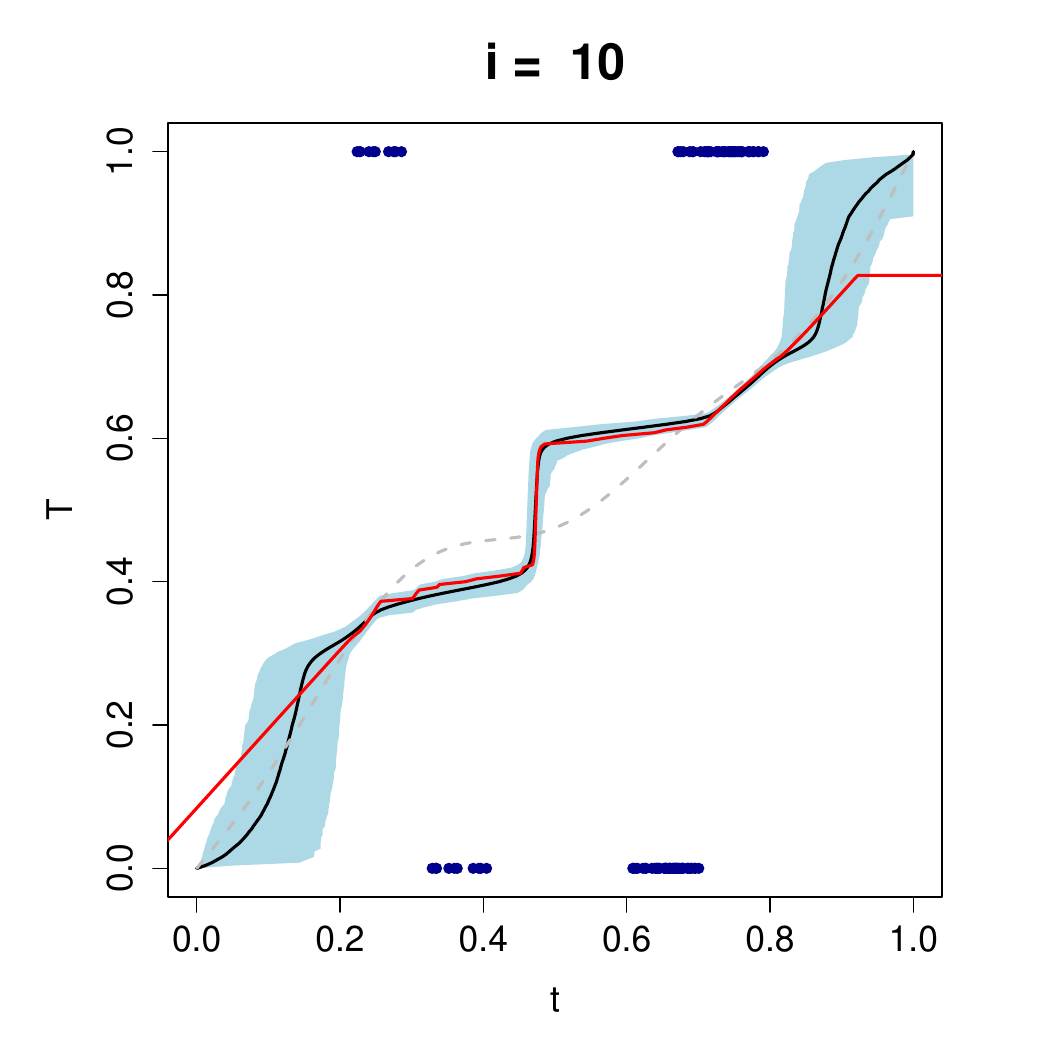}    \\
 \includegraphics[scale=0.180]{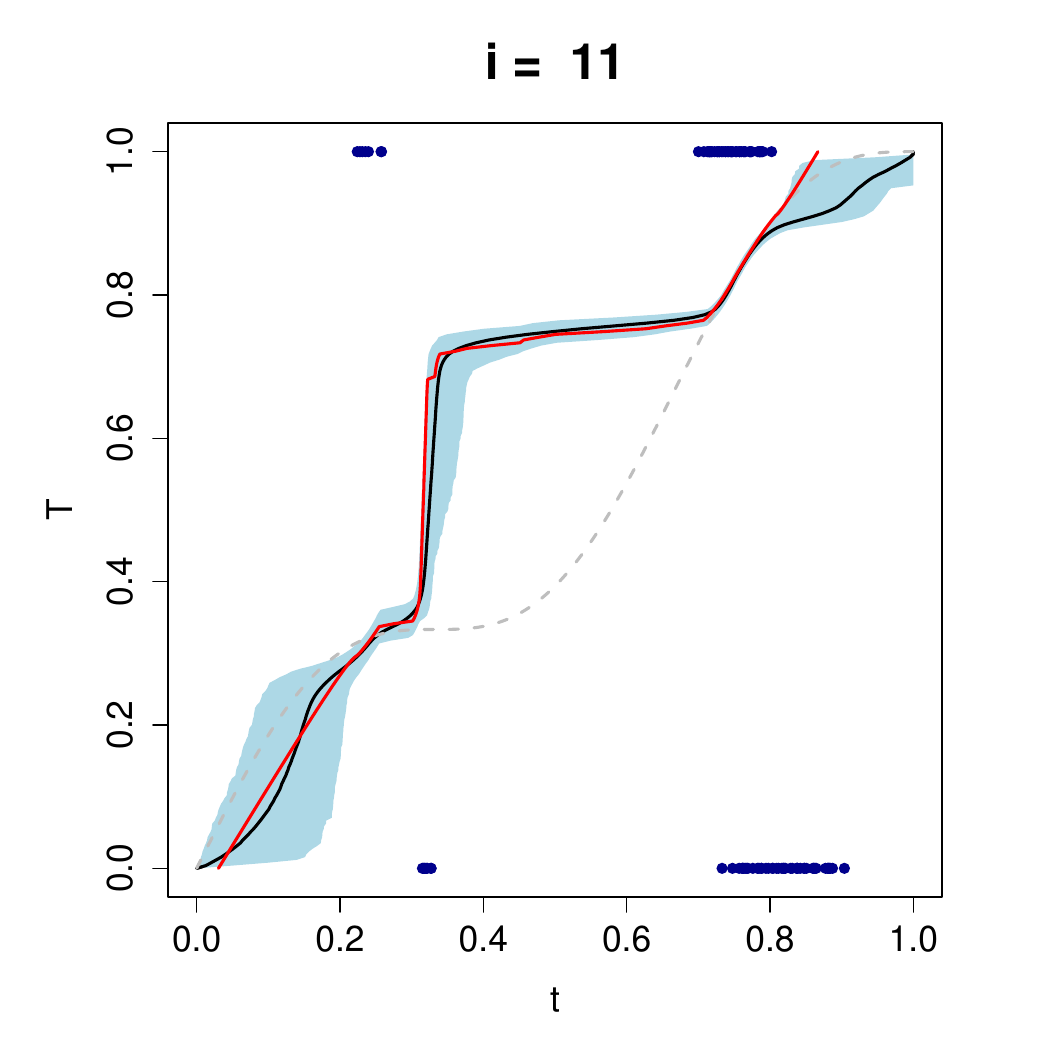}
 \includegraphics[scale=0.180]{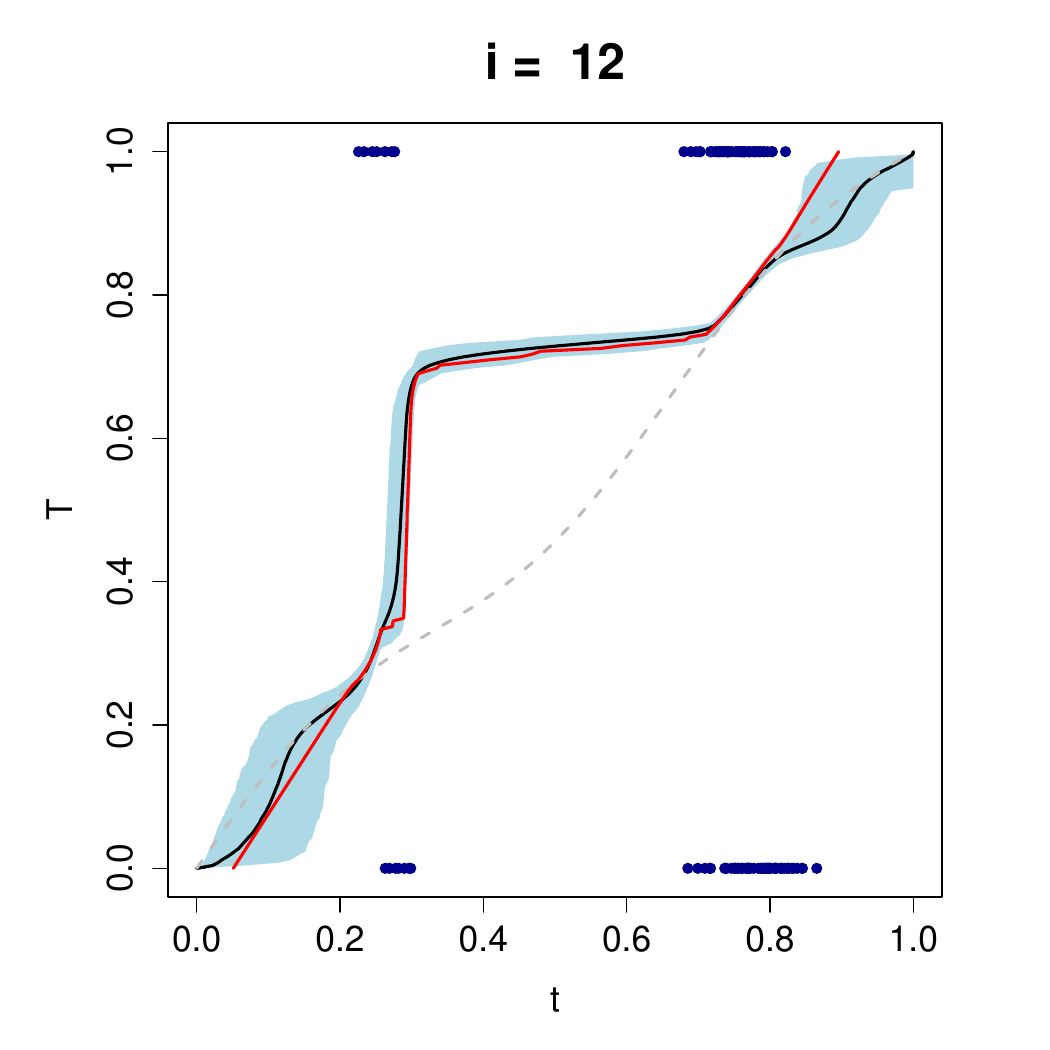}
 \includegraphics[scale=0.180]{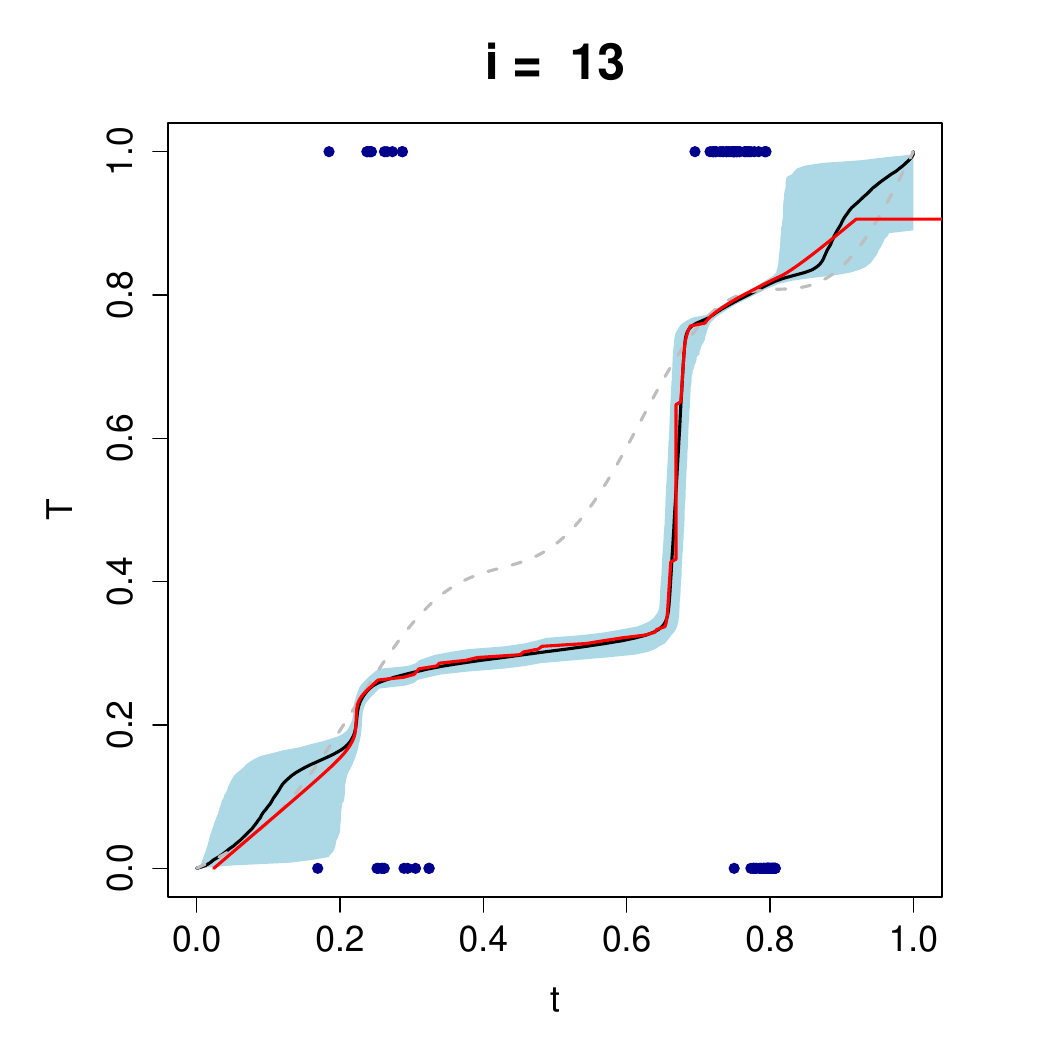}
 \includegraphics[scale=0.180]{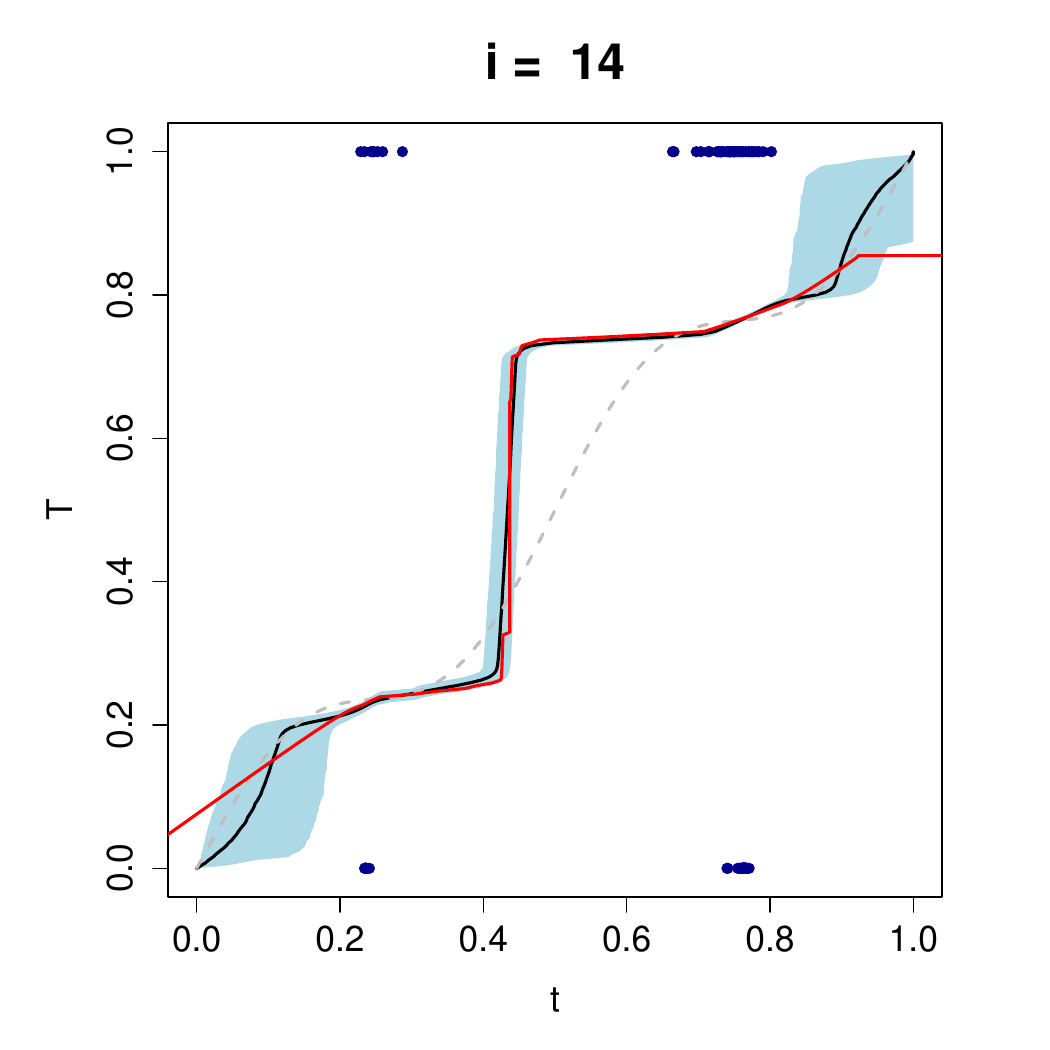}
 \includegraphics[scale=0.180]{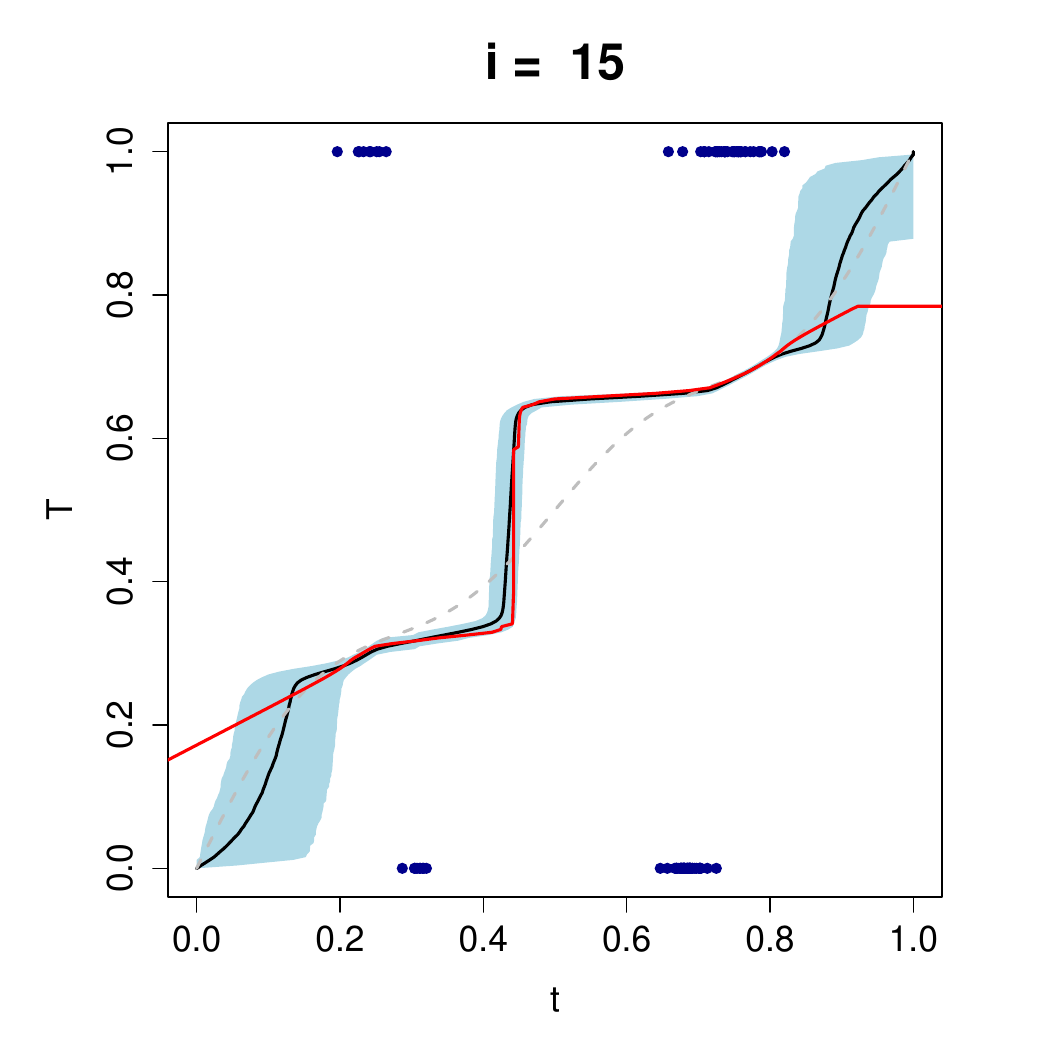}    \\
 \includegraphics[scale=0.180]{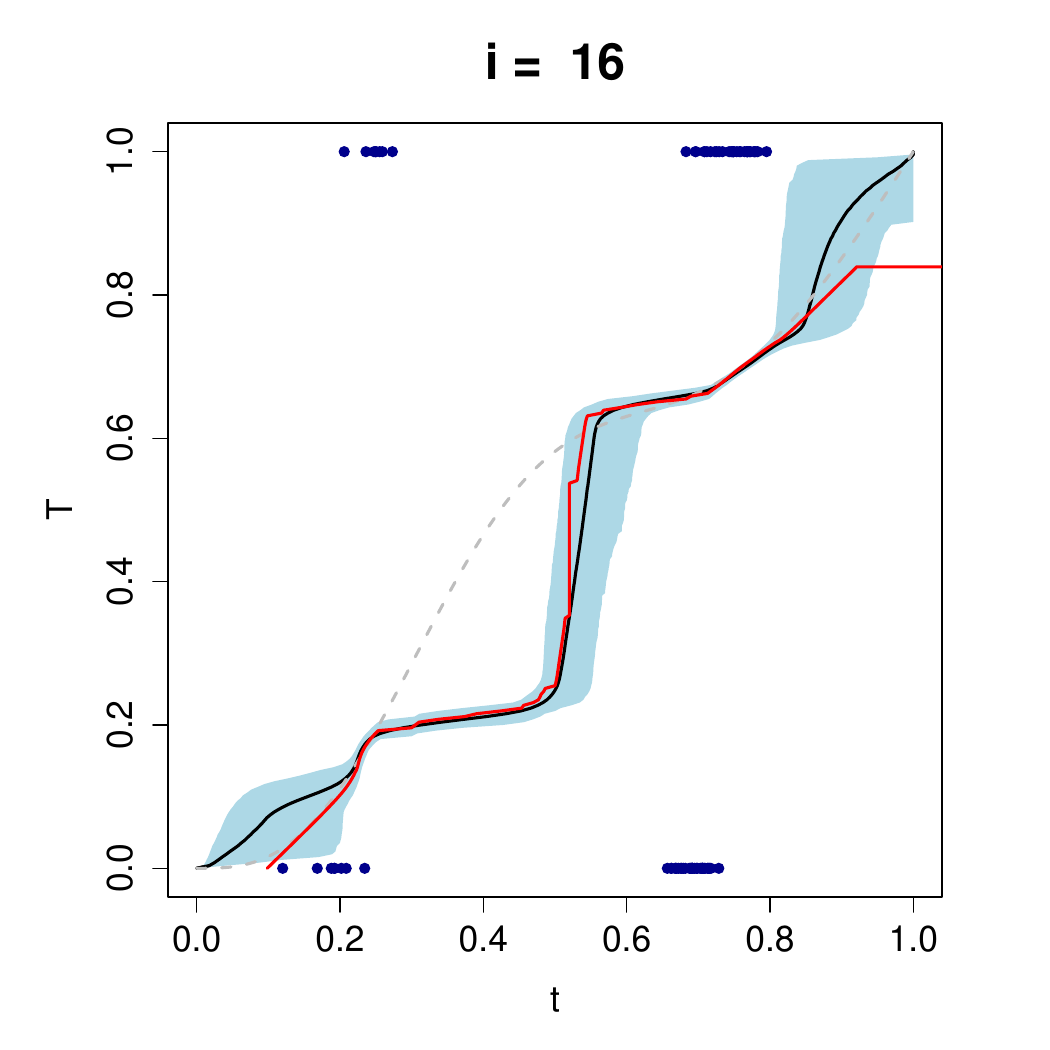}
 \includegraphics[scale=0.180]{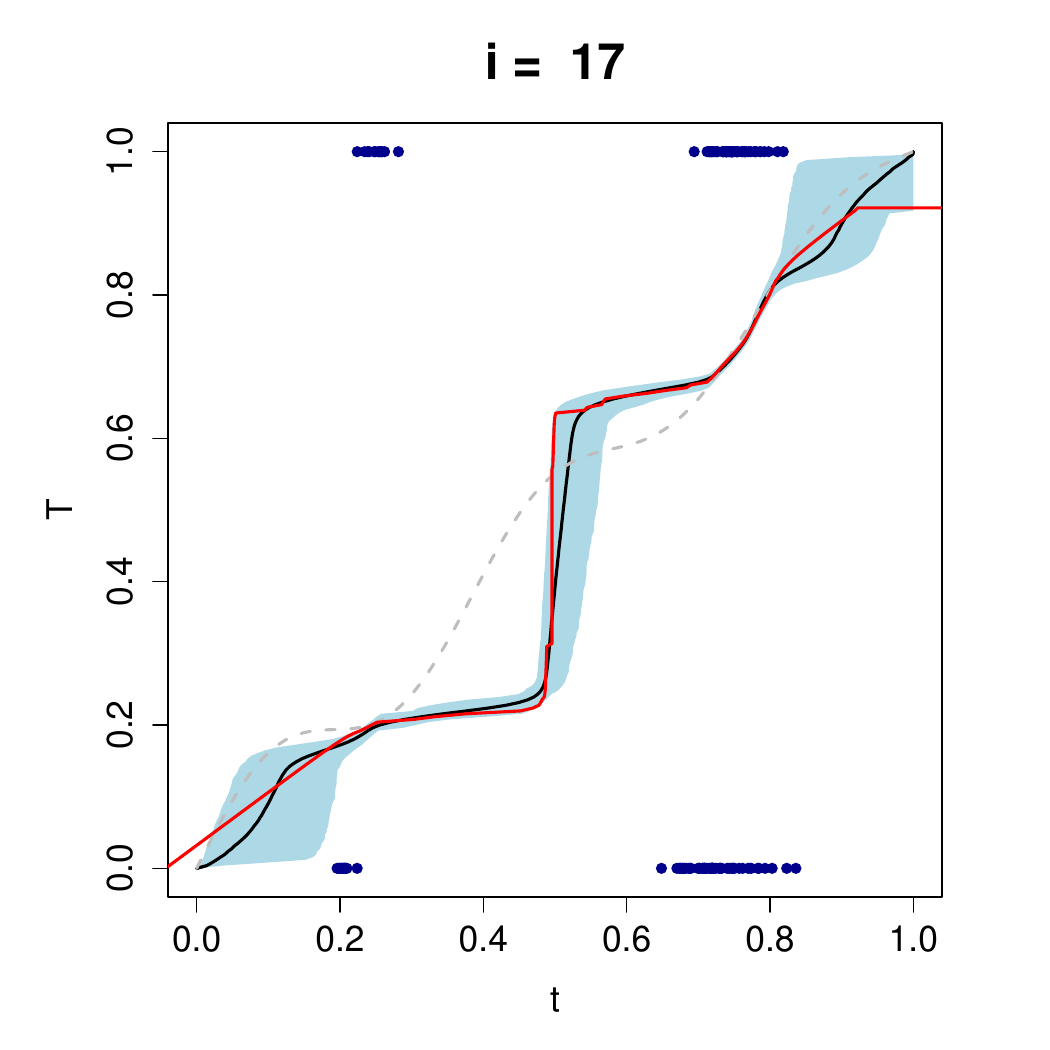}
 \includegraphics[scale=0.180]{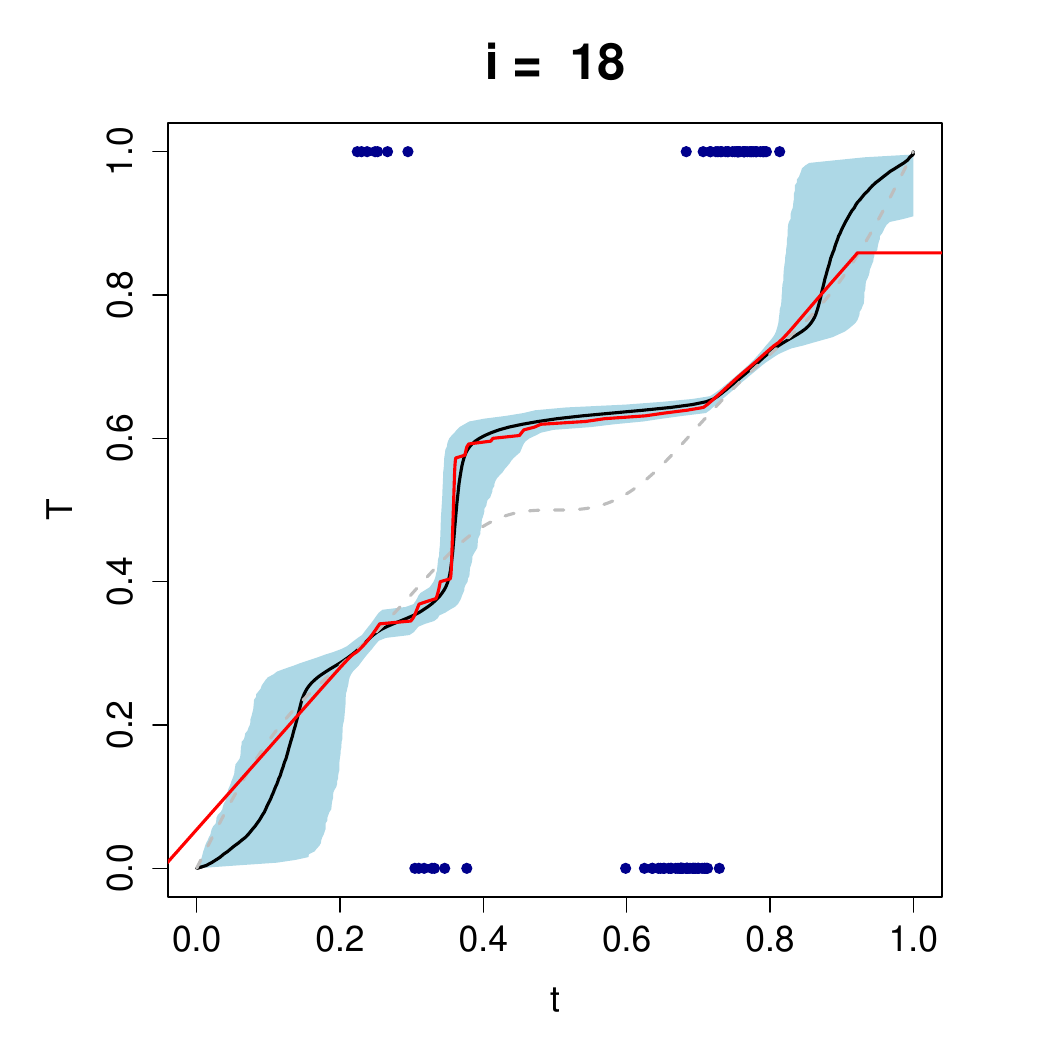}
 \includegraphics[scale=0.180]{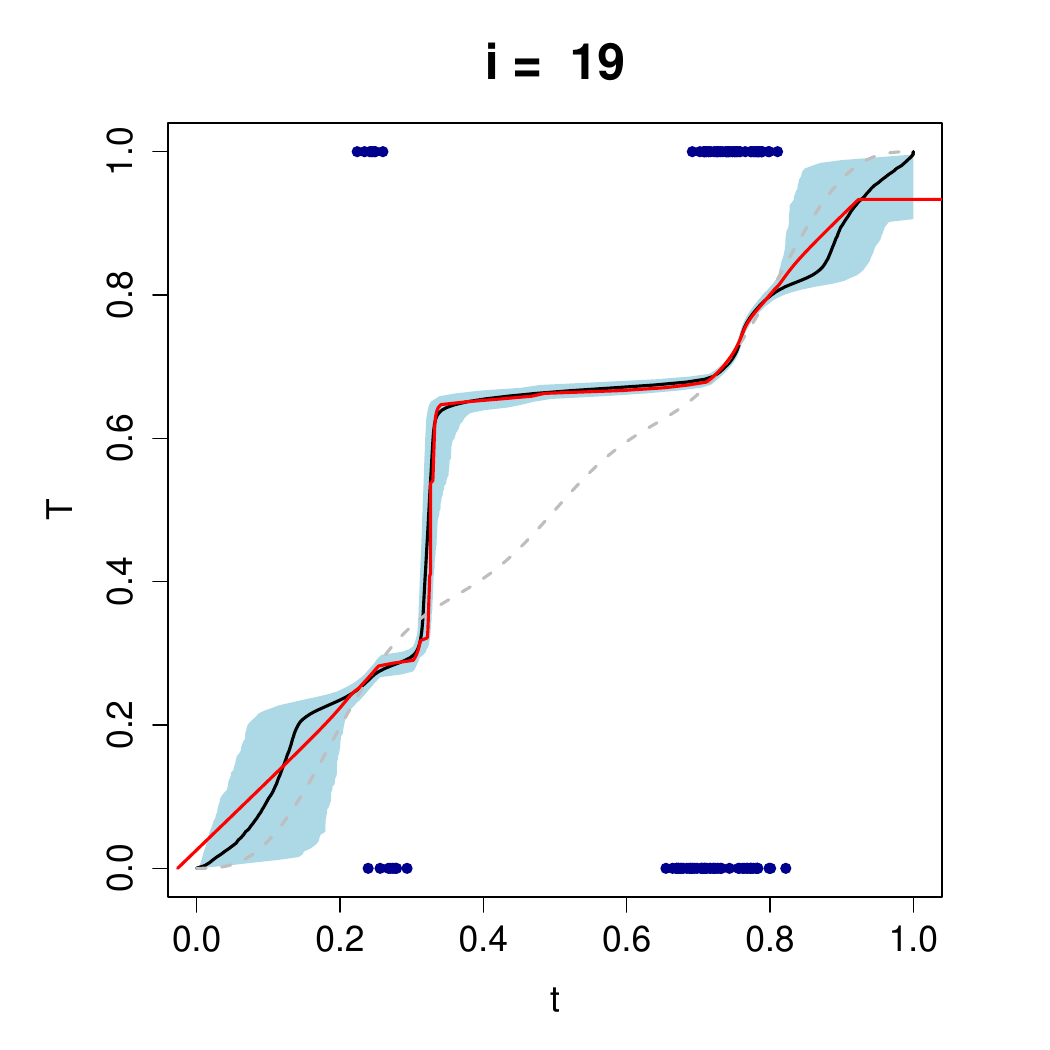}
 \includegraphics[scale=0.180]{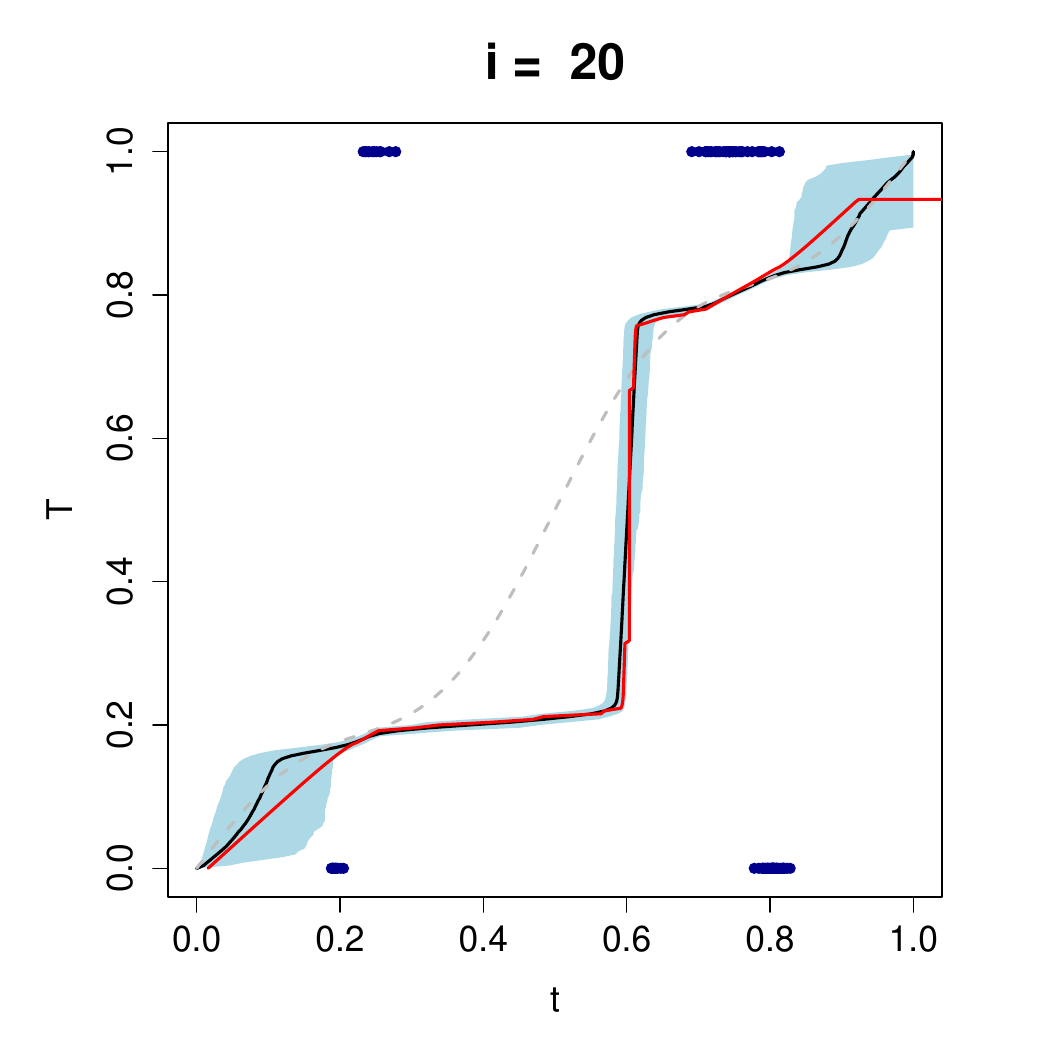}    \\
 \includegraphics[scale=0.180]{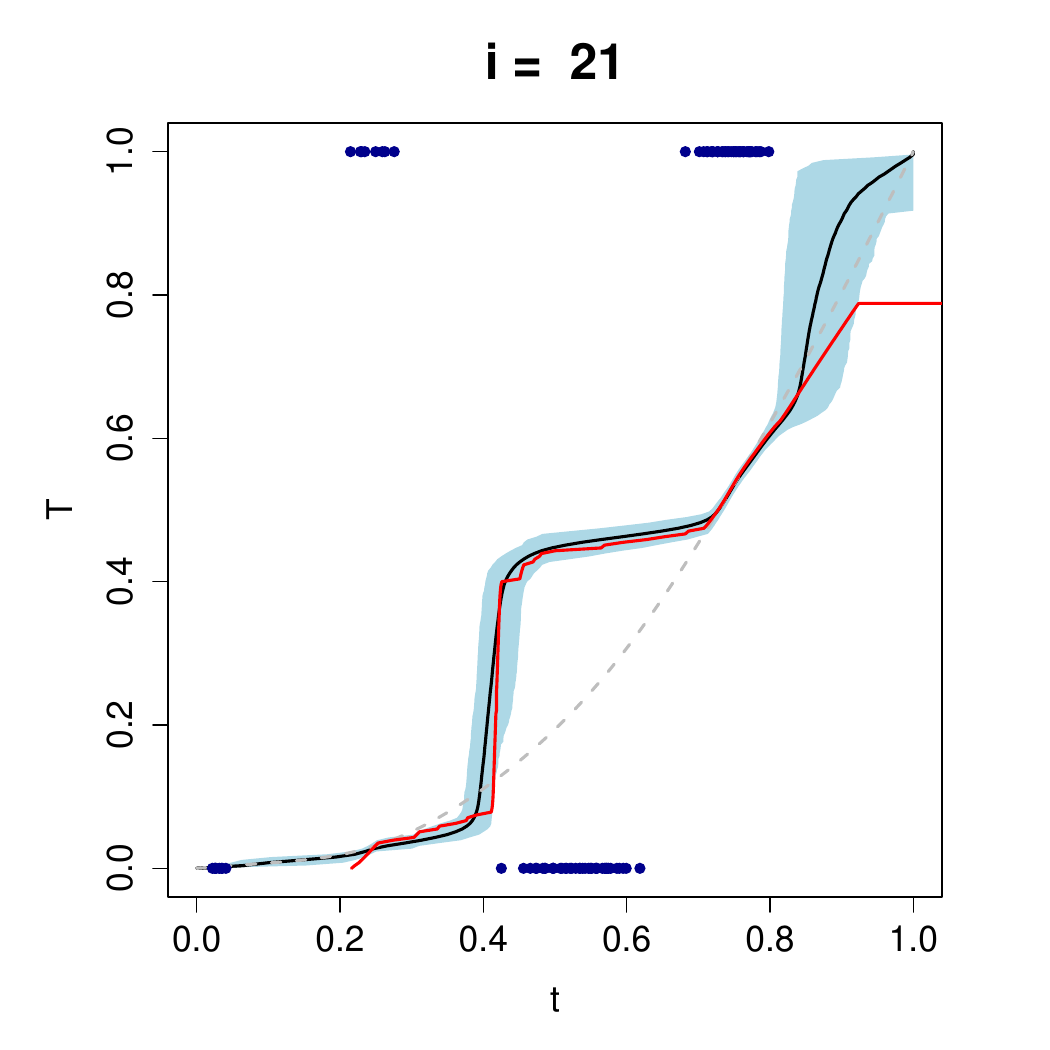}
 \includegraphics[scale=0.180]{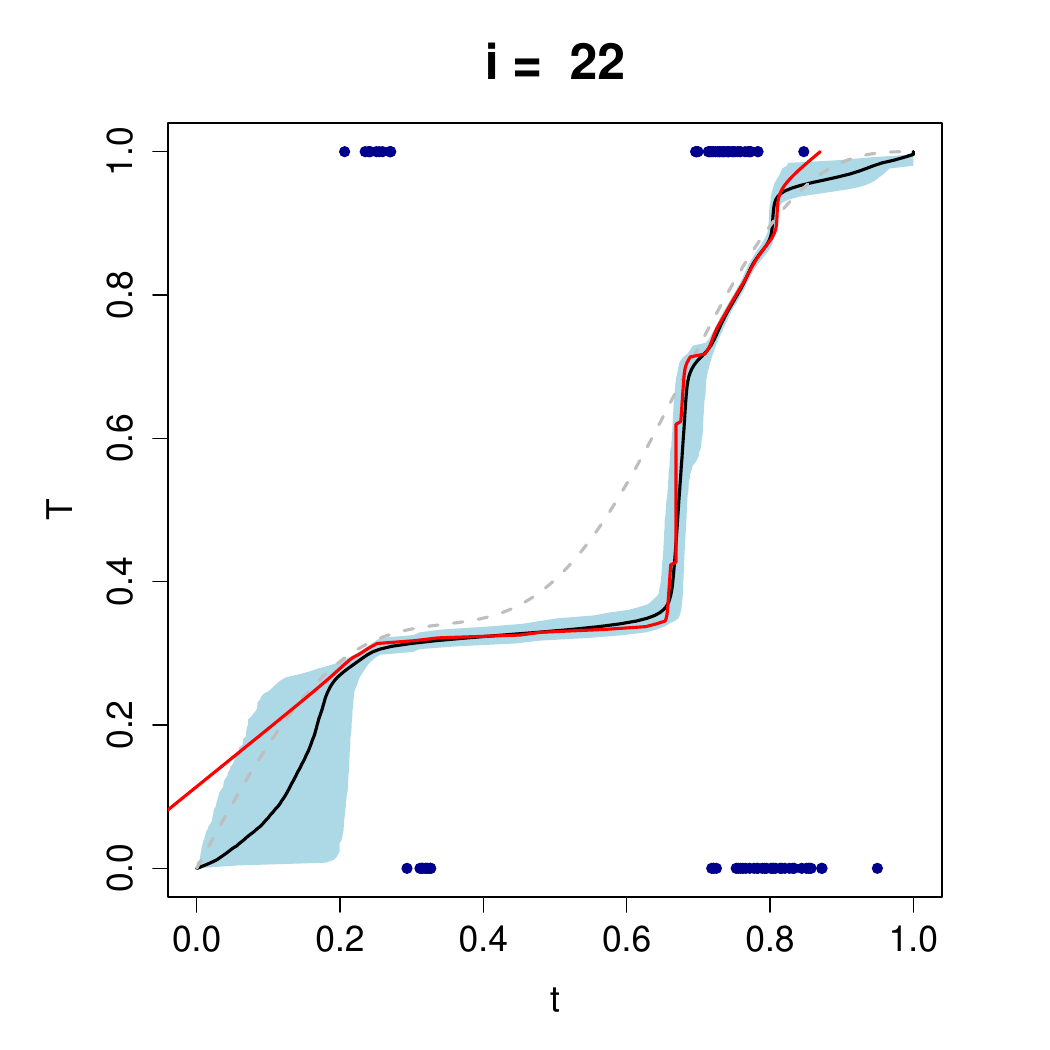}
 \includegraphics[scale=0.180]{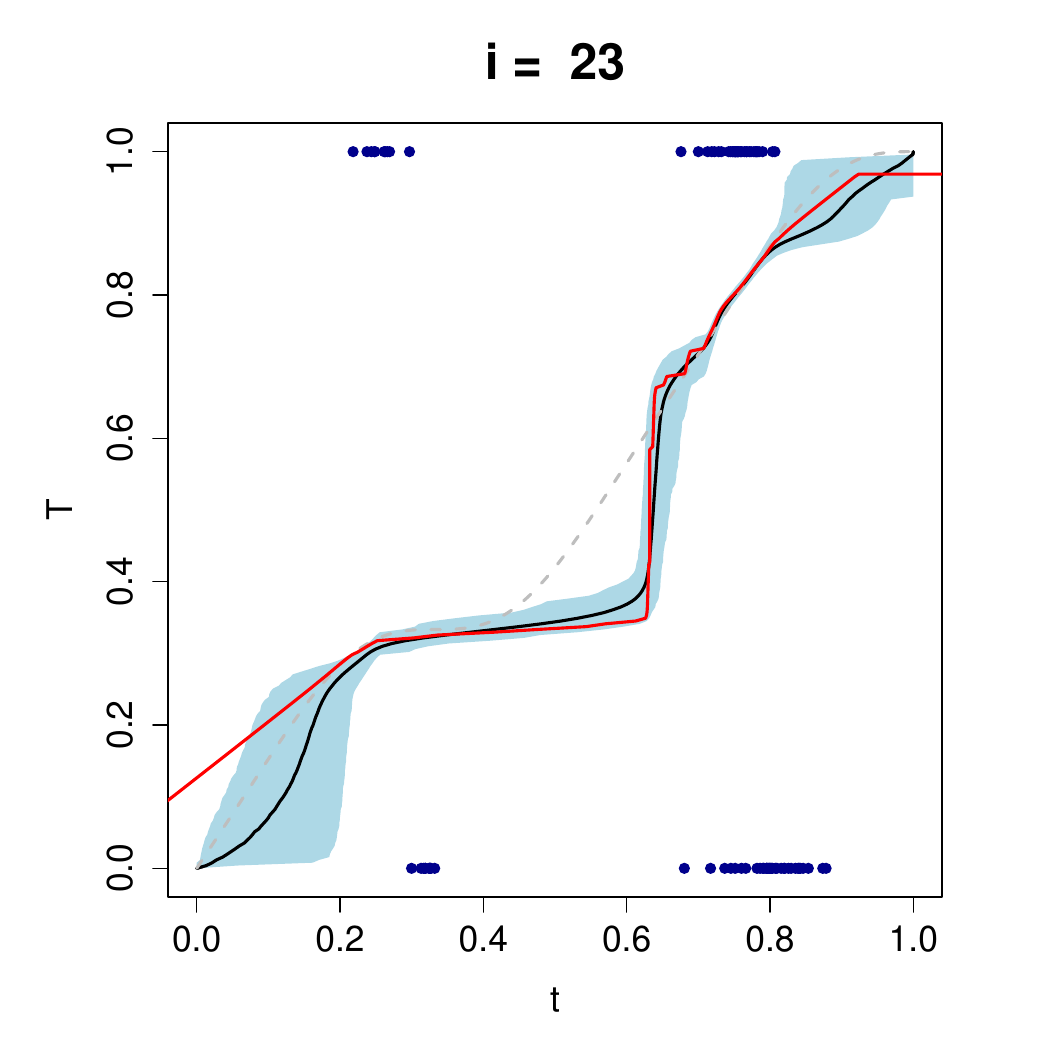}
 \includegraphics[scale=0.180]{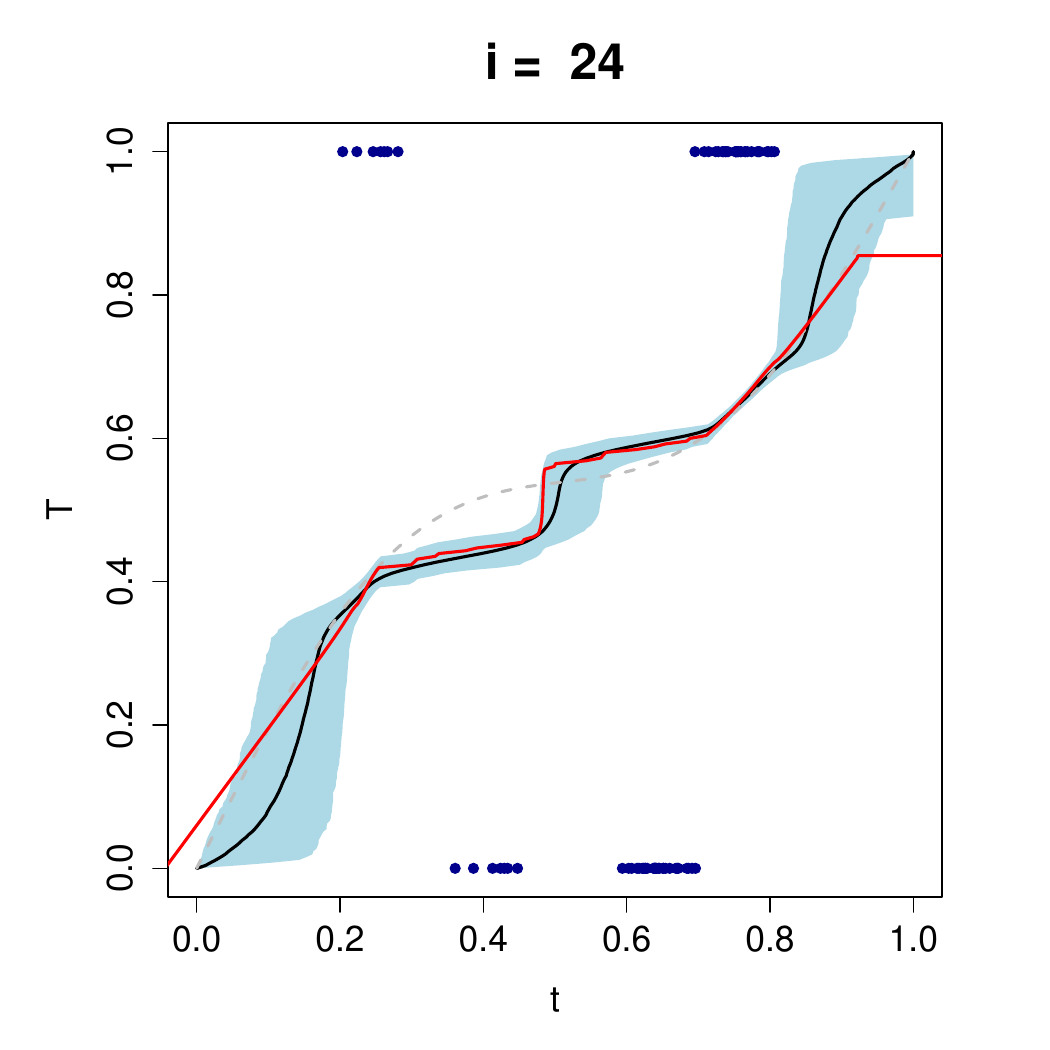}
 \includegraphics[scale=0.180]{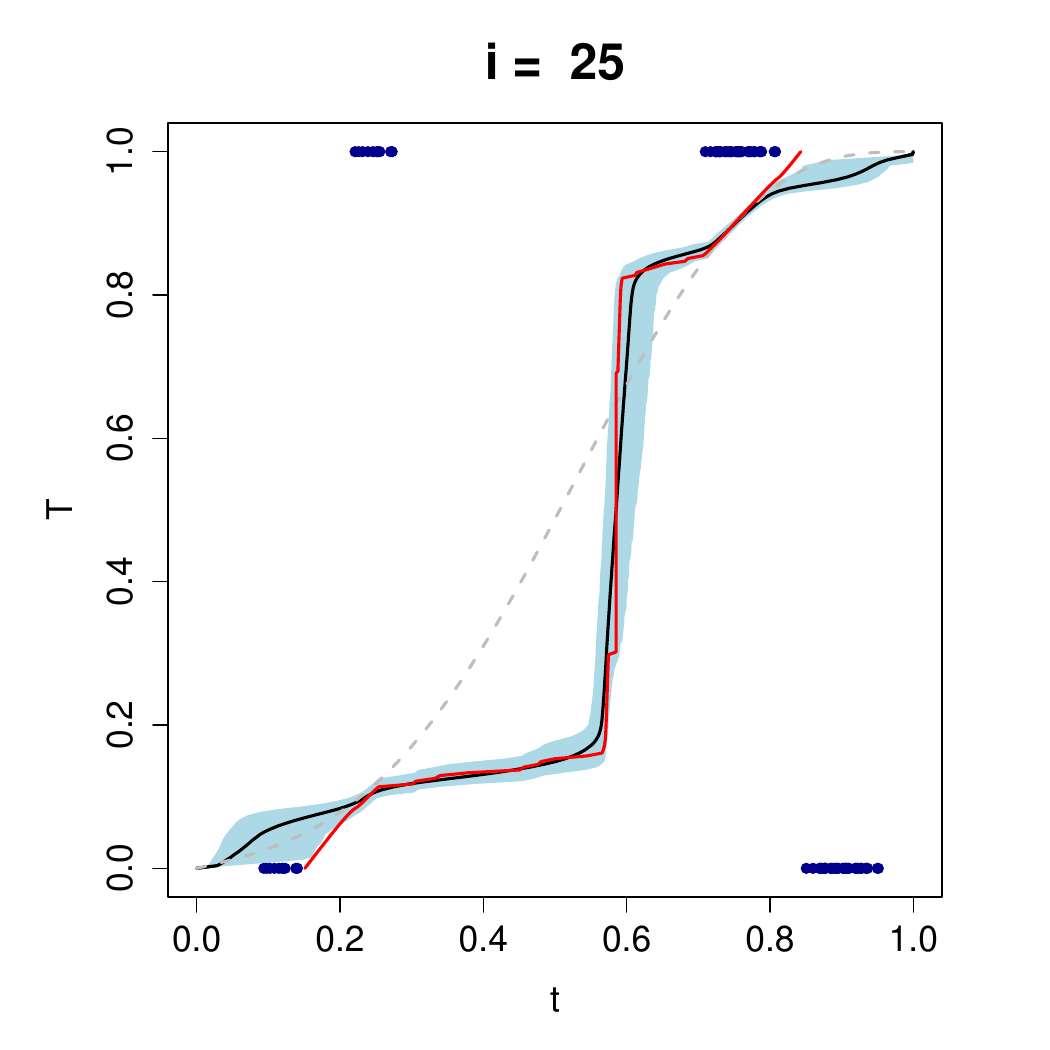}    \\
 \includegraphics[scale=0.180]{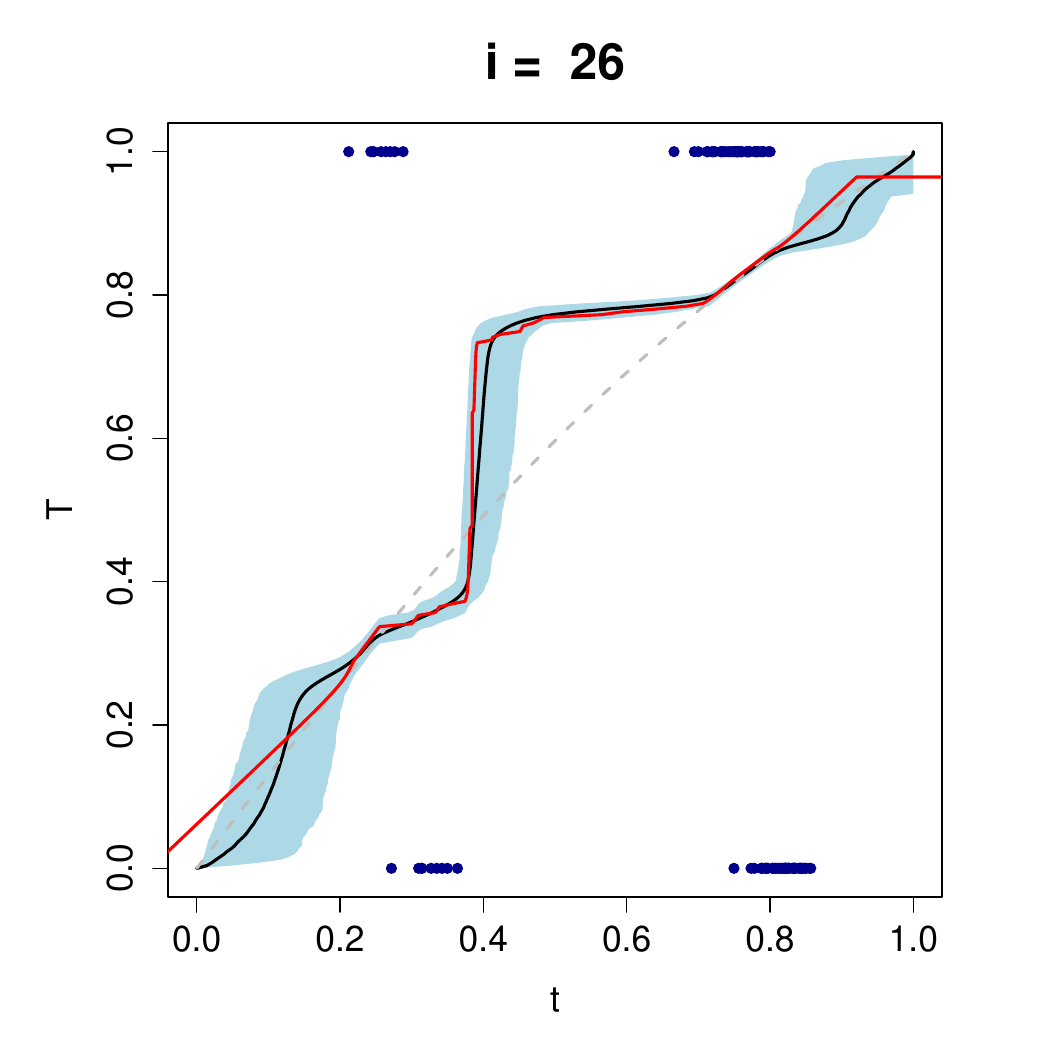}
 \includegraphics[scale=0.180]{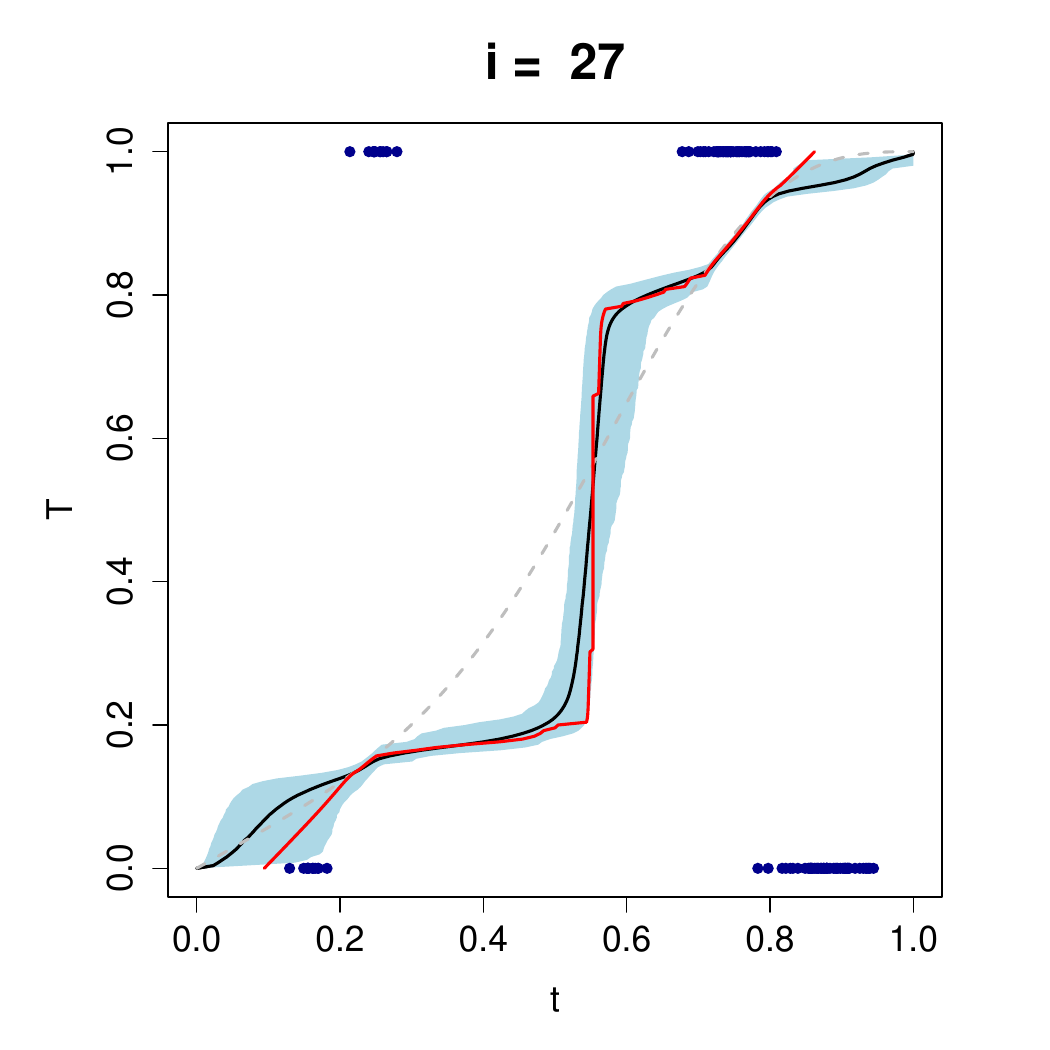}
 \includegraphics[scale=0.180]{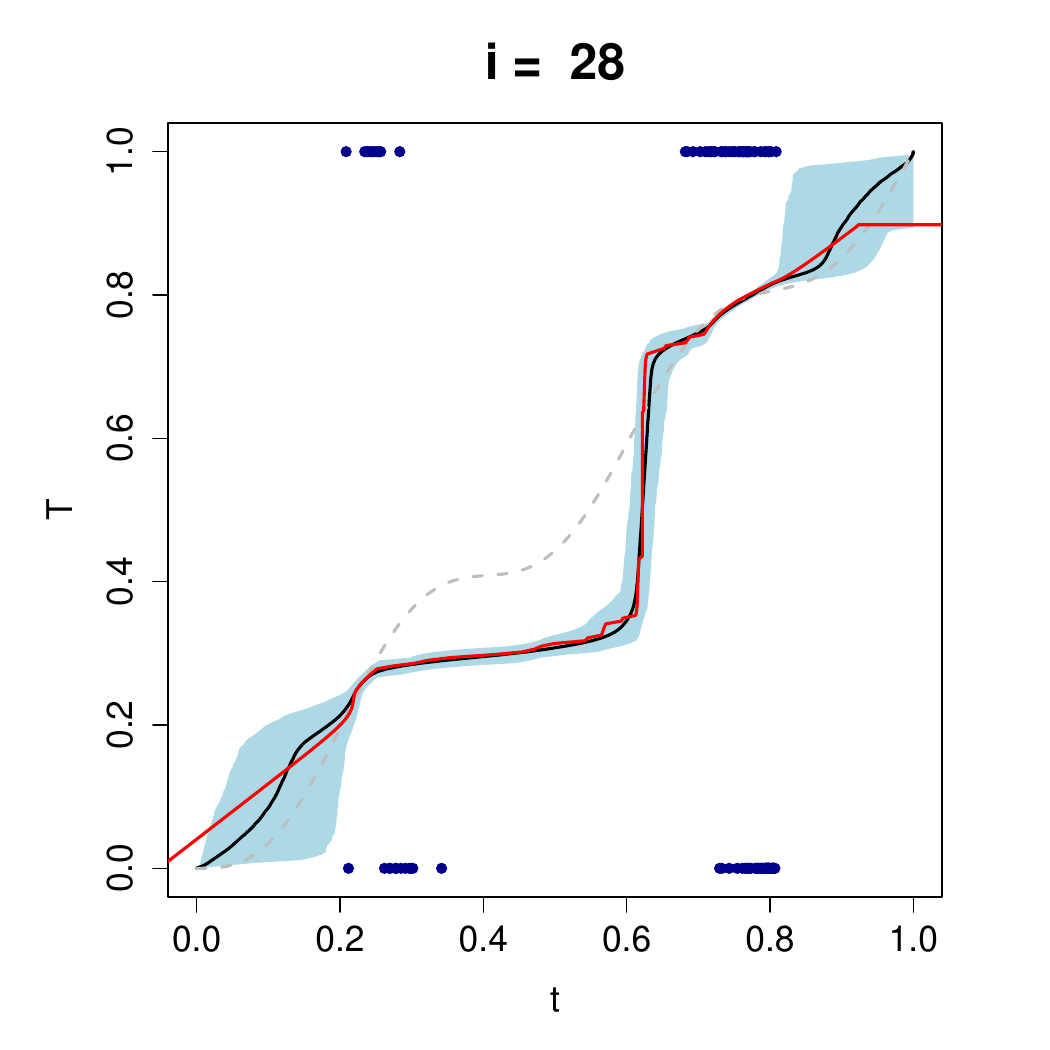}
 \includegraphics[scale=0.180]{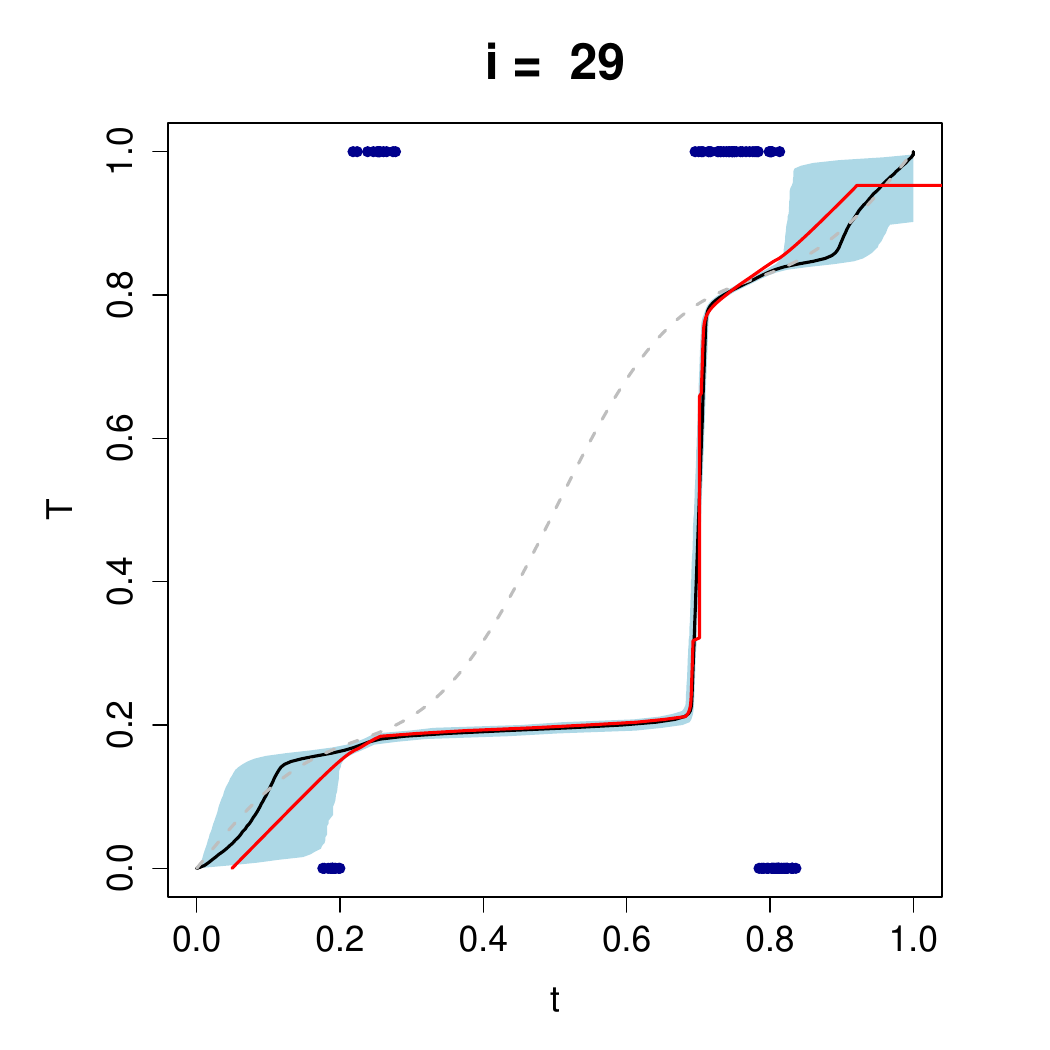}
 \includegraphics[scale=0.180]{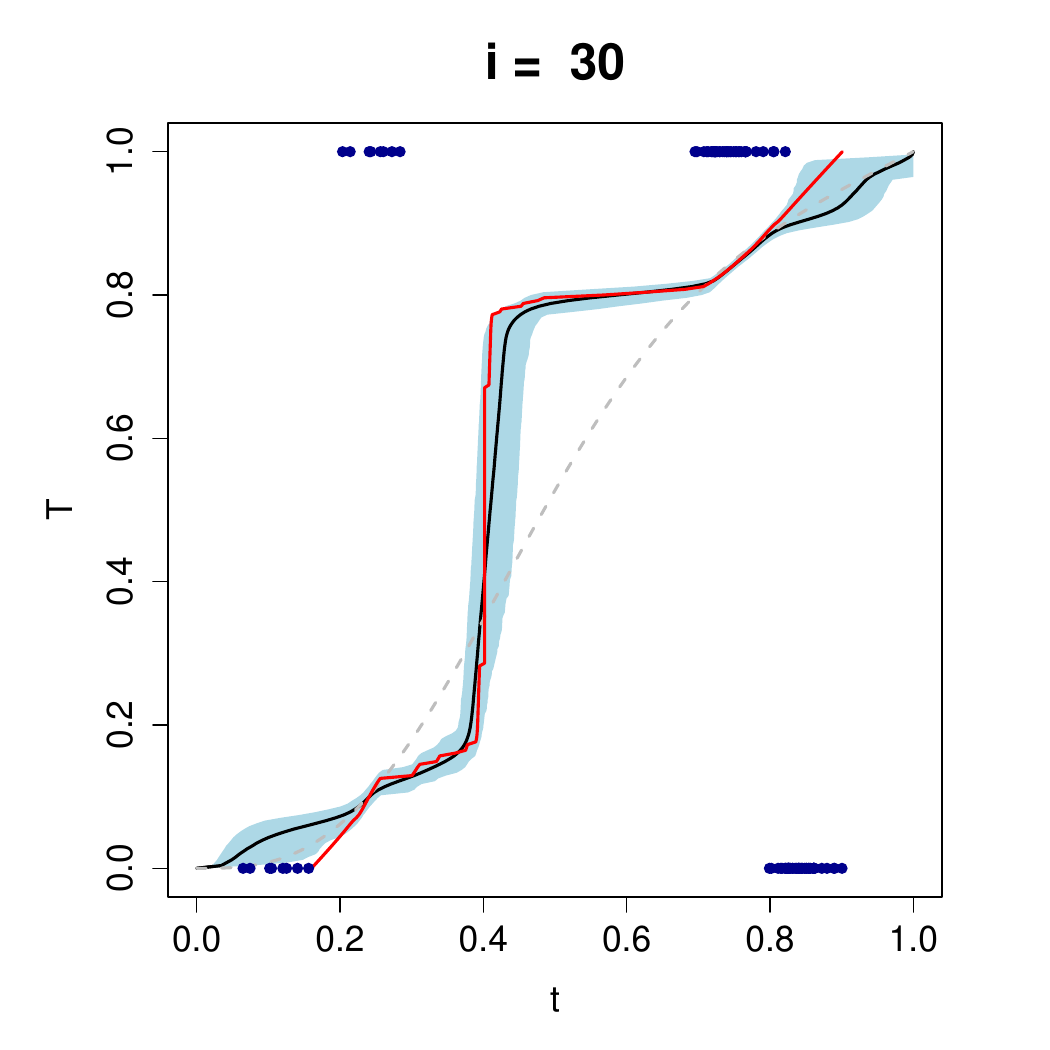}   
\end{center} 
\caption{\footnotesize 30 posterior mean Bernstein polynomial warp functions (solid black) and corresponding credible bands, with their kernel-based counterparts (solid red) and the original warp functions (dashed grey). Warped and original data are in the bottom and top, respectively.}
\label{FNE:3}
\end{figure}

In this section we present some figures which are derived from the simulation studies conducted in Section~\ref{numerics}. In detail, Figure~\ref{FNE:1} refers to results in the simulation study in Subsection~\ref{sc1}, Figure~\ref{FNE:2} refers to the comparison conducted in Subsection~\ref{sc2}, and Figure~\ref{FNE:3} corresponds to Figure~\ref{fig3} (left) but for all warp maps.


\newpage
\subsection{Simulation study under misspecification}\label{supp:misspec}
\begin{figure}\centering
  \begin{minipage}{0.31\linewidth}
    \includegraphics[scale = 0.42]{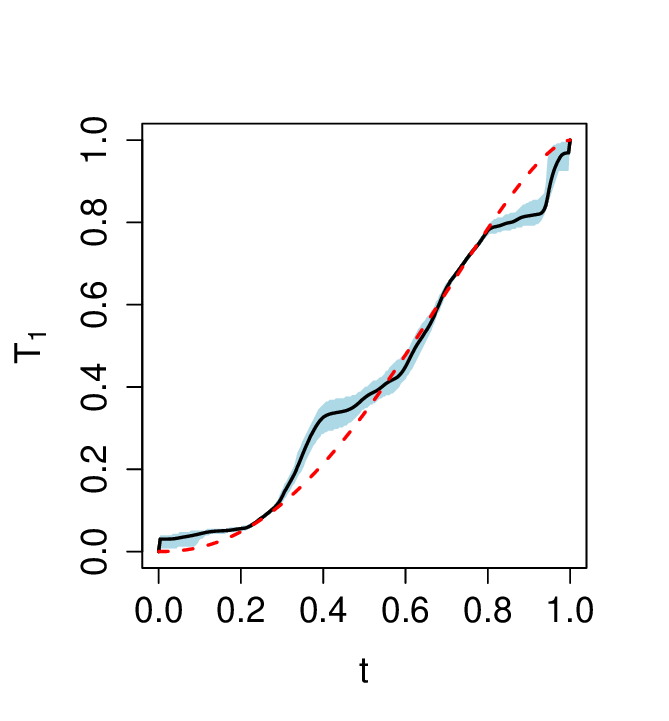}    
    \includegraphics[scale = 0.33]{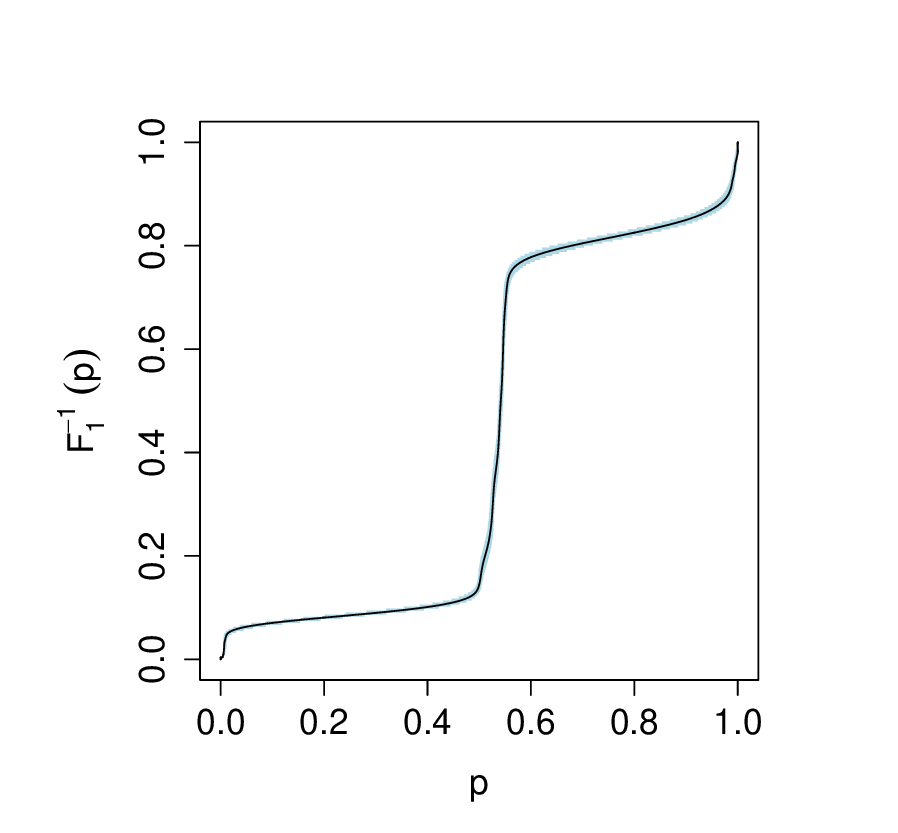}    
  \end{minipage} \hspace{0.2cm}
  \begin{minipage}{0.31\linewidth}
     \includegraphics[scale = 0.42]{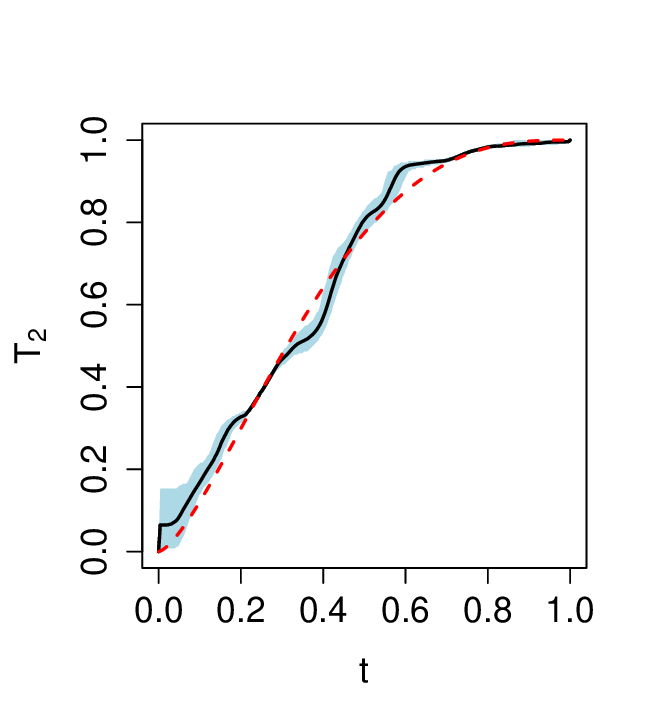}
     \includegraphics[scale = 0.33]{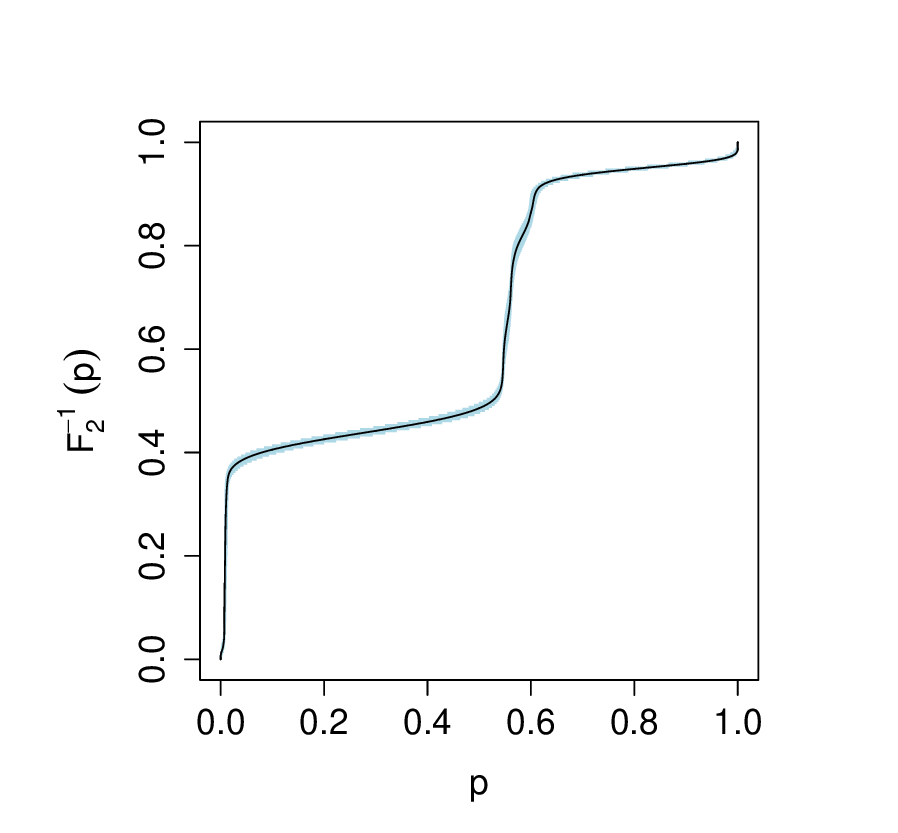}
  \end{minipage} \hspace{0.2cm}
  \begin{minipage}{0.31\linewidth}
    \includegraphics[scale = 0.42]{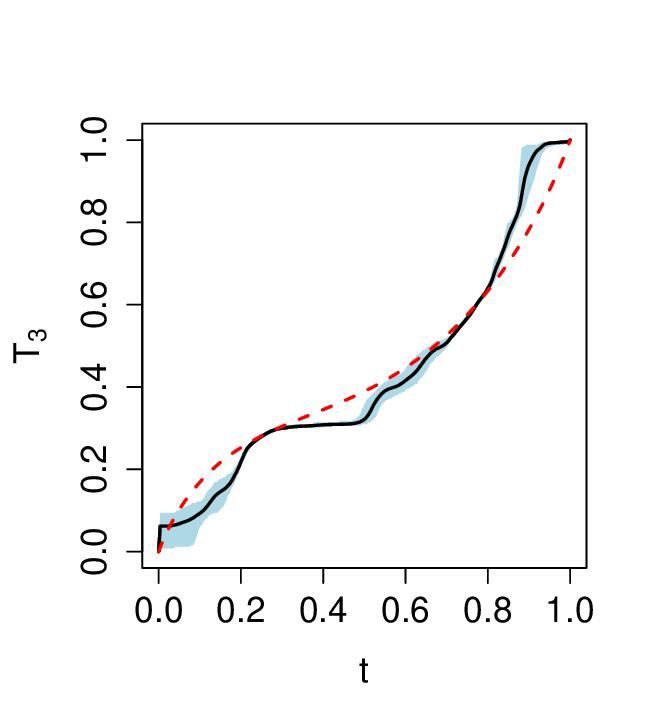}
    \includegraphics[scale = 0.33]{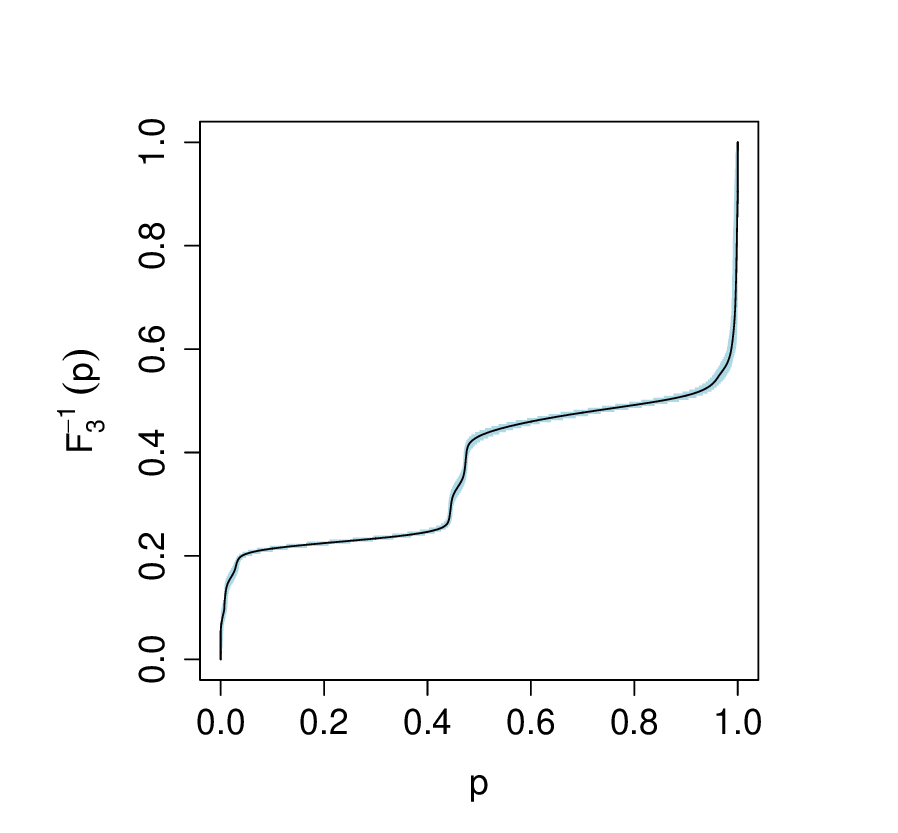}
  \end{minipage}
\caption{\footnotesize Above: True (dashed red) and estimated (solid black) warp functions along with credible bands.  The estimators are constructed as the posterior mean of the induced prior. Below: Corresponding quantile function estimates along with credible bands.}\label{supp:fig2a}
\end{figure}

Here we analyse a simulation scenario similar to that in Subsection~\ref{sc1}, but this time using warp maps $T_i$ which not satisfy $E[T_i(t)] = t$; the goal will be to illustrate the performance of the proposed registration method under misspecification. We generate random samples $x_{i,1},\ldots, x_{i,m_i} \mid m_i$, from
\begin{equation*}
f(t) = 0.45\, \{\phi(t \mid 0.25,0.02^2) +  \phi(t \mid 0.75,0.03^2)\} + 0.1\, \beta(t \mid 1.5,1.5), 
\end{equation*}
where $m_i \sim \text{Poisson}(L)$, for $i = 1,2,3,$ with $L = 150$; here, $\phi(t \mid \mu, \sigma^2)$ denotes the Normal density function and $\beta(t \mid a,b)$ denotes the Beta density. The warped data $\tilde{x}_{i,j} = T_i(x_{i,j})$ are obtained using 
\begin{equation*}
T_i(t) = \int_0^t \beta(y \mid a,b) \, \dif y, \quad i=1,2, \quad T_3(t) = 3t - T_1(t) - T_2(t), \quad a,b \stackrel{\text{iid}}{\sim}  \text{Unif}[1,3].
\end{equation*}
Figure~\ref{supp:fig2a} shows the estimators of each of the three warp maps through the posterior mean of the induced prior defined in Subsection~\ref{bnp_pvp}, along with their credible bands and the true warp maps over a one shot experiment.

\begin{figure}
  \hspace{-.3cm}
  \begin{minipage}{0.32\linewidth}
    \includegraphics[scale = 0.35]{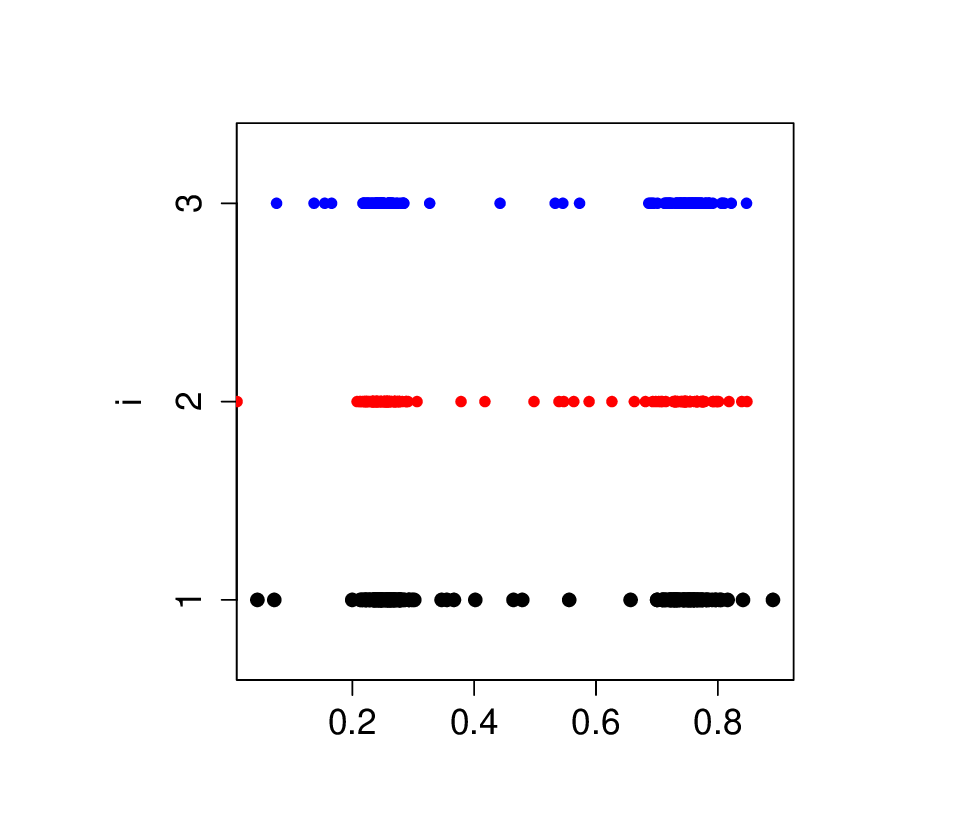}    
  \end{minipage}
  \begin{minipage}{0.32\linewidth}
     \includegraphics[scale = 0.35]{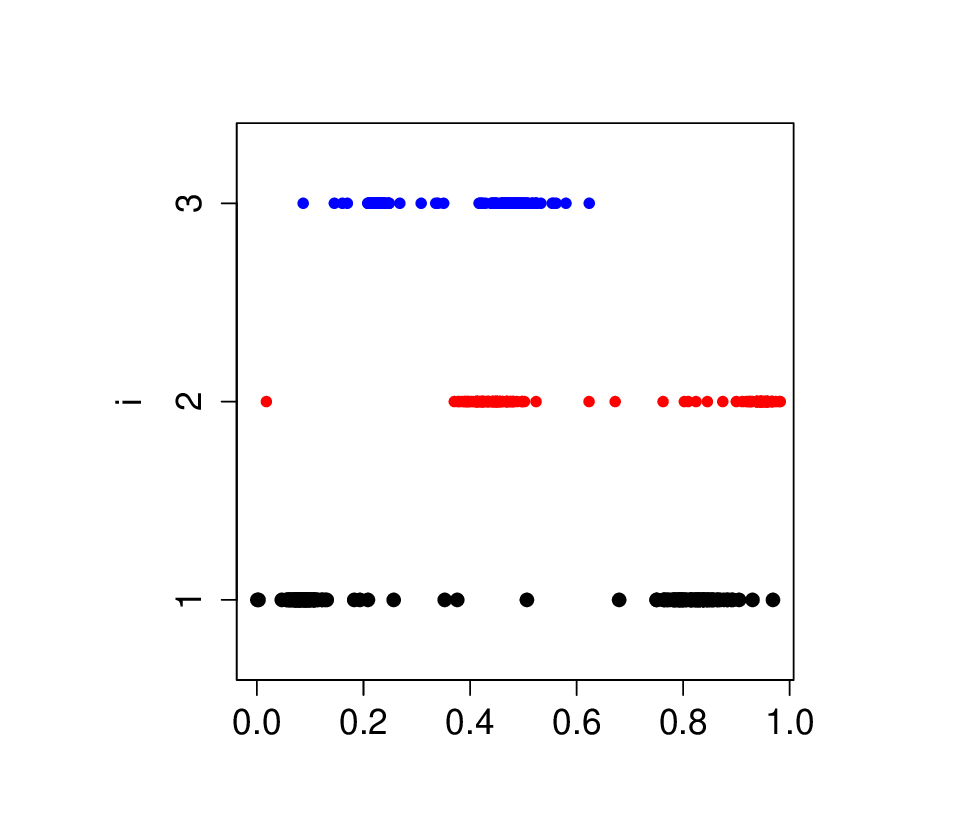}
  \end{minipage}
  \begin{minipage}{0.32\linewidth}
    \includegraphics[scale = 0.35]{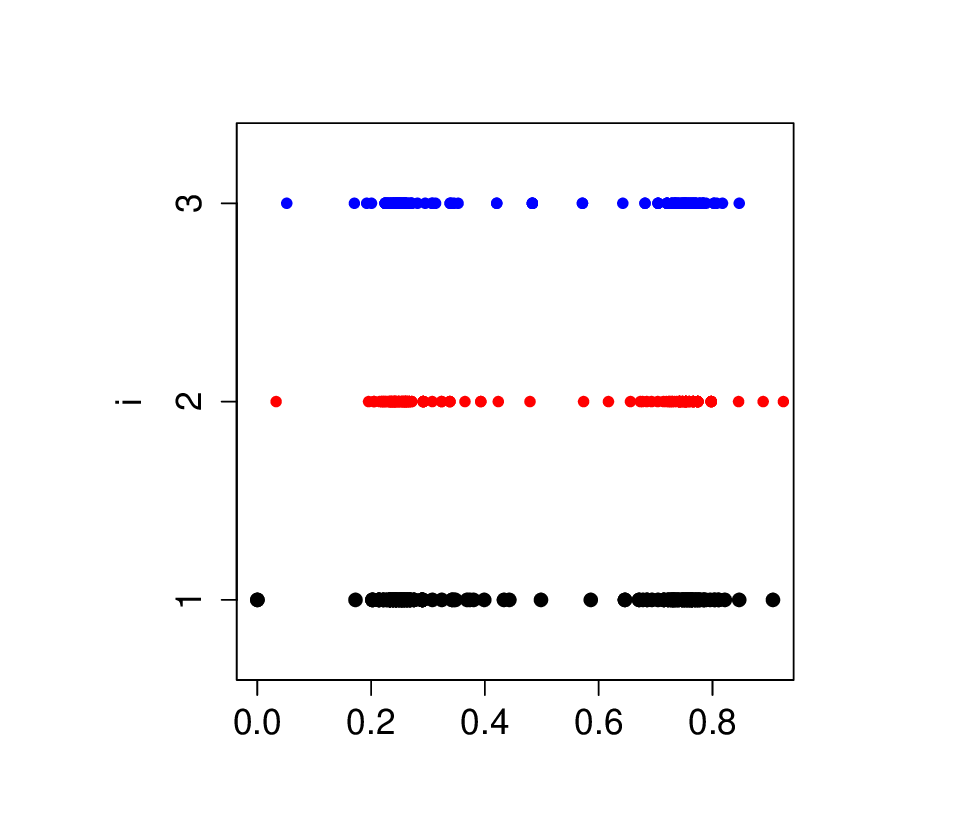}
  \end{minipage}
\caption{\footnotesize Left: Realisations of the original point process from the setup of Section 1 (paper) in the small $n$, large $m$ regime. Middle: Their corresponding phase-varying point process.  Right: Their corresponding registered versions.}\label{supp:fig1a}
\end{figure}

From Figure~\ref{supp:fig2a} it can be noticed that---even under misspecification---our estimators are reasonably in line with the true warp functions, and as a consequence, the method recovers quite well the original point processes, as can be seen when comparing the left and right panels of Figure~\ref{supp:fig1a}.

A Monte Carlo study was conducted based on $B = 50$ simulated datasets. The WDM (Monte Carlo $L^2$-Wasserstein distance mean) defined in Equation~\eqref{wrm} in the paper was at this time $0.041677$---which is of the same order of magnitude as the value obtained in Subsection~\ref{sc1} in the paper under a well-specified setting. For the sake of comparison, the WDM computed using $\widetilde{\Pi}_i$ instead of $\widehat\Pi_i$, is 0.15137, nearly four times larger that that based on $\widehat\Pi_i$.

\section{Additional outputs from application}\label{sec:supp-app}
As in Section~\ref{application}, we analyse the annual peaks over threshold, $\{\widetilde{x}_{i, j}^+ \geq u_j^+\}$, and annual peaks below threshold, $\{\widetilde{x}_{i, j}^- \leq u_j^-\}$; we set the thresholds $u_j^+$ and $u_j^-$ using the 97.5\% and 2.5\% quantiles of temperature over year $j$, and this results in $m_1^+, \dots, m_n^+$  ranging from 10 to 18 and $m_1^-, \dots, m_n^-$ ranging from 10 to 20.

\begin{figure}
  \centering
  \includegraphics[scale=0.35]{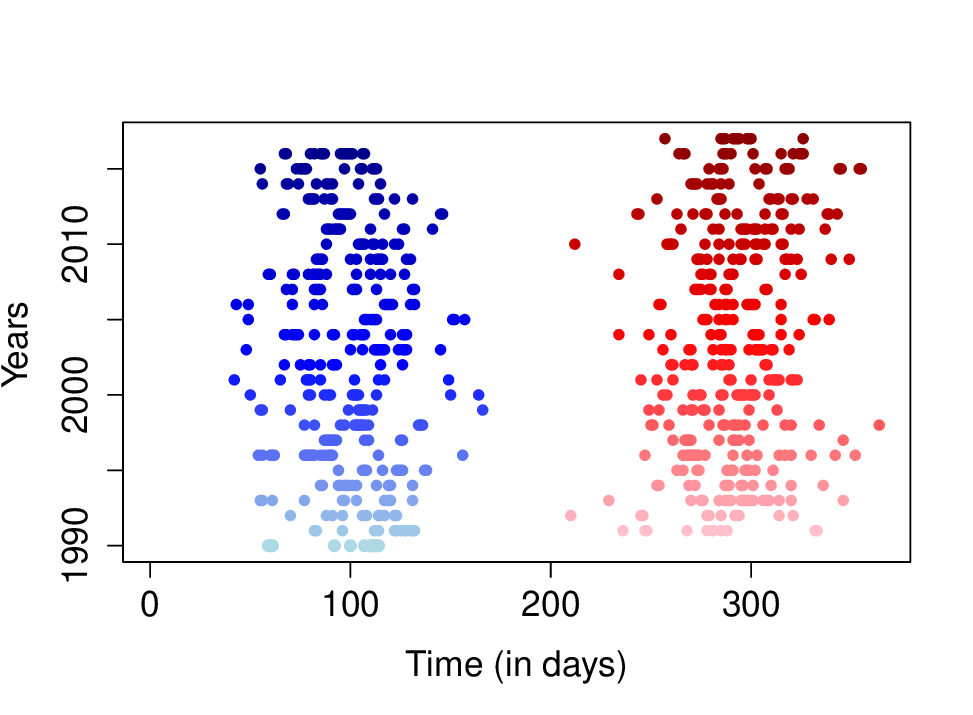}
    \includegraphics[scale=0.35]{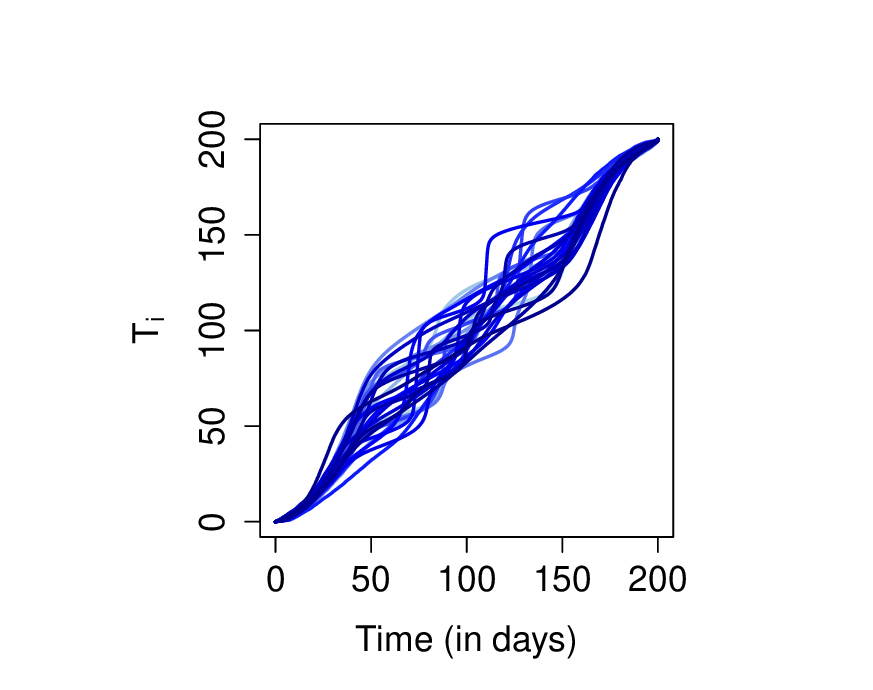}
  \includegraphics[scale=0.35]{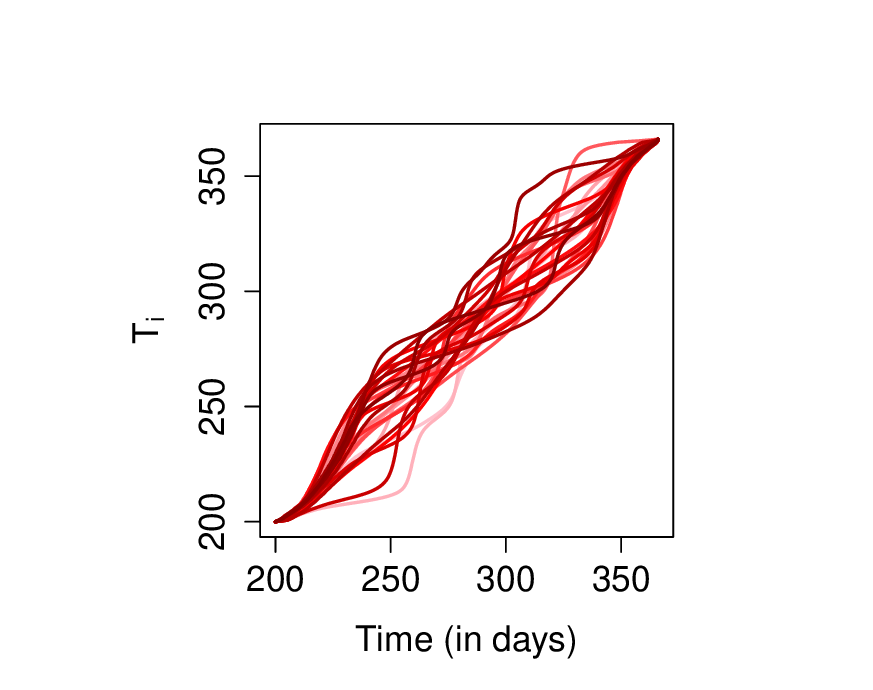}
  \caption{\footnotesize Left: Point processes of annual peaks for peaks above (red) and below (blue) the thresholds. Middle and Right: Corresponding posterior mean warp functions in the same palette of colors for the 2.5\% and 97.5\% quantiles data.}
  \label{fig:supp-pp}
\end{figure}

\begin{figure}\centering 
\includegraphics[scale=0.180]{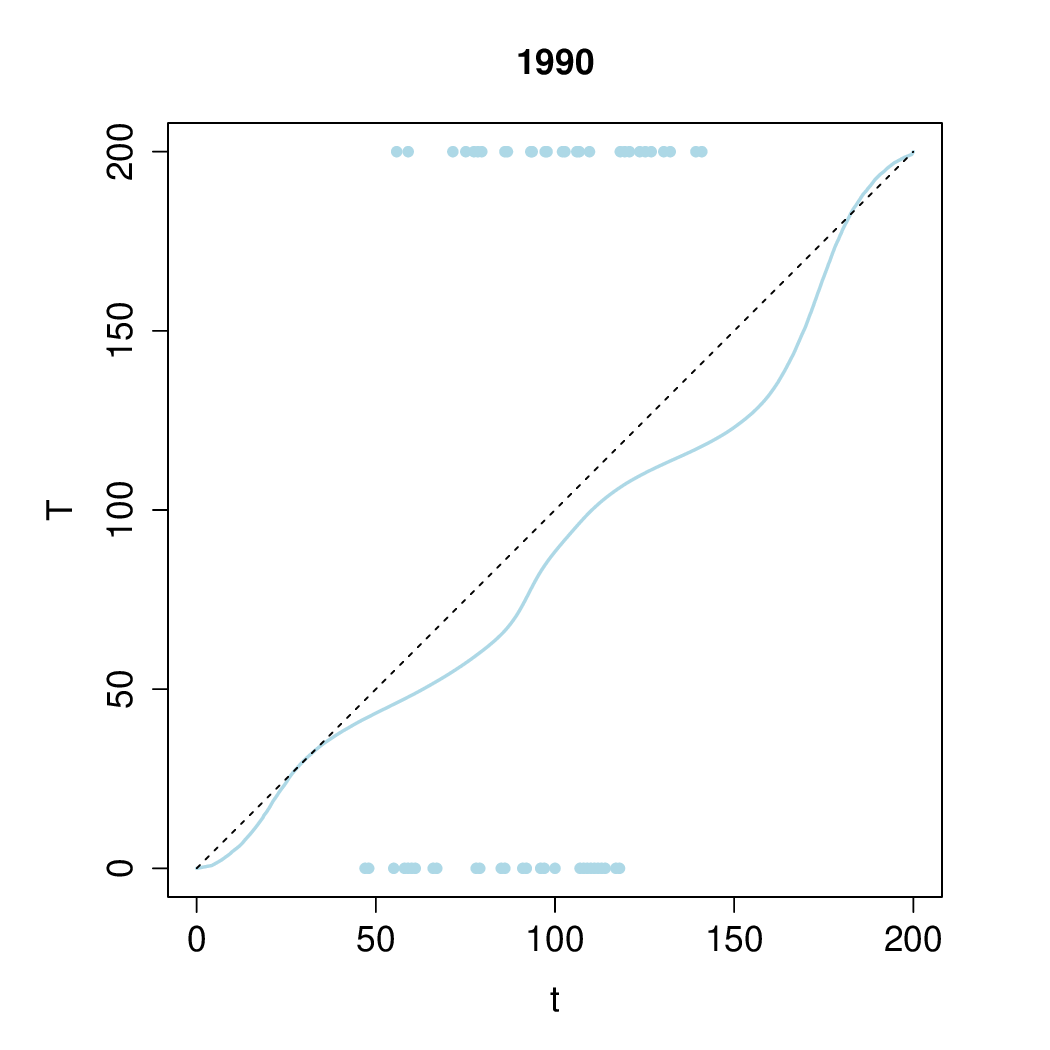}
\includegraphics[scale=0.180]{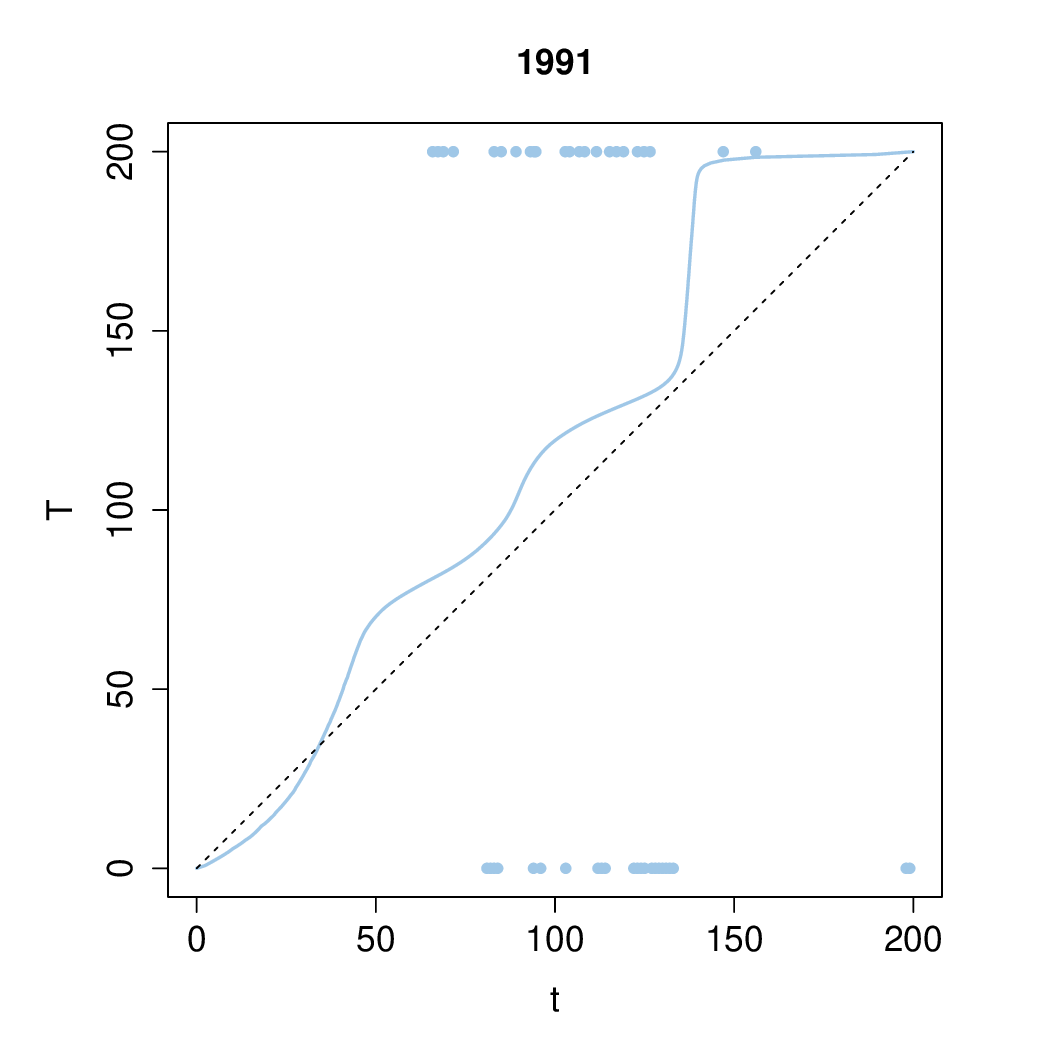}
\includegraphics[scale=0.180]{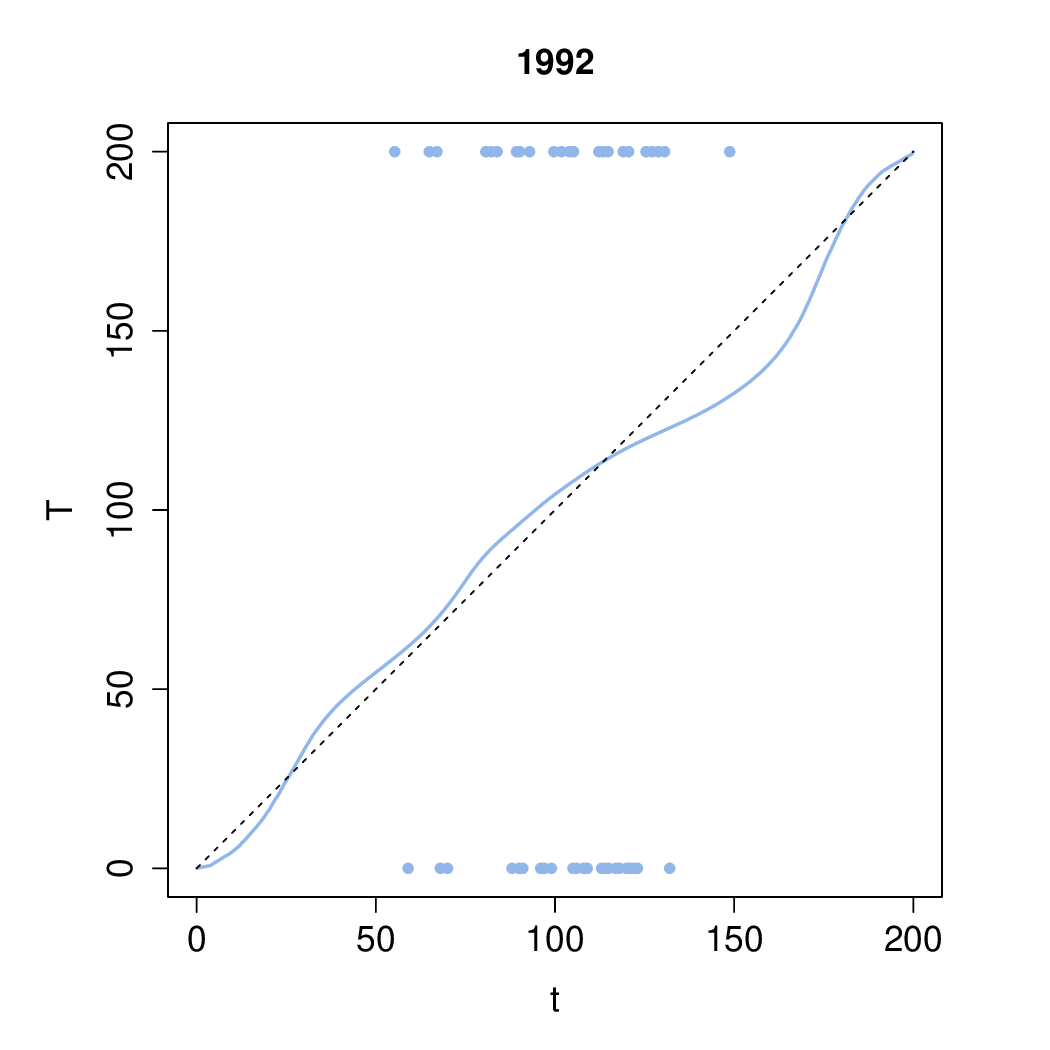}
\includegraphics[scale=0.180]{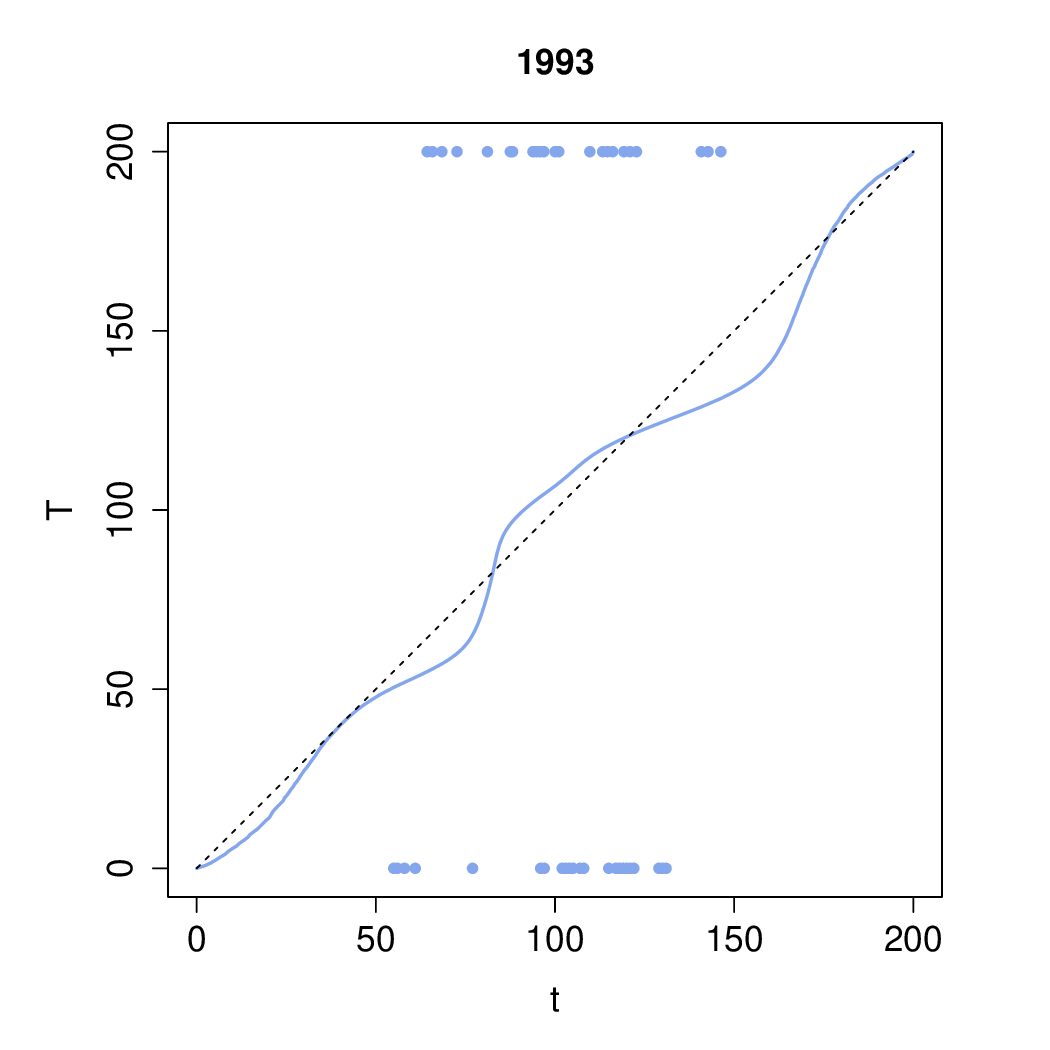}
\includegraphics[scale=0.180]{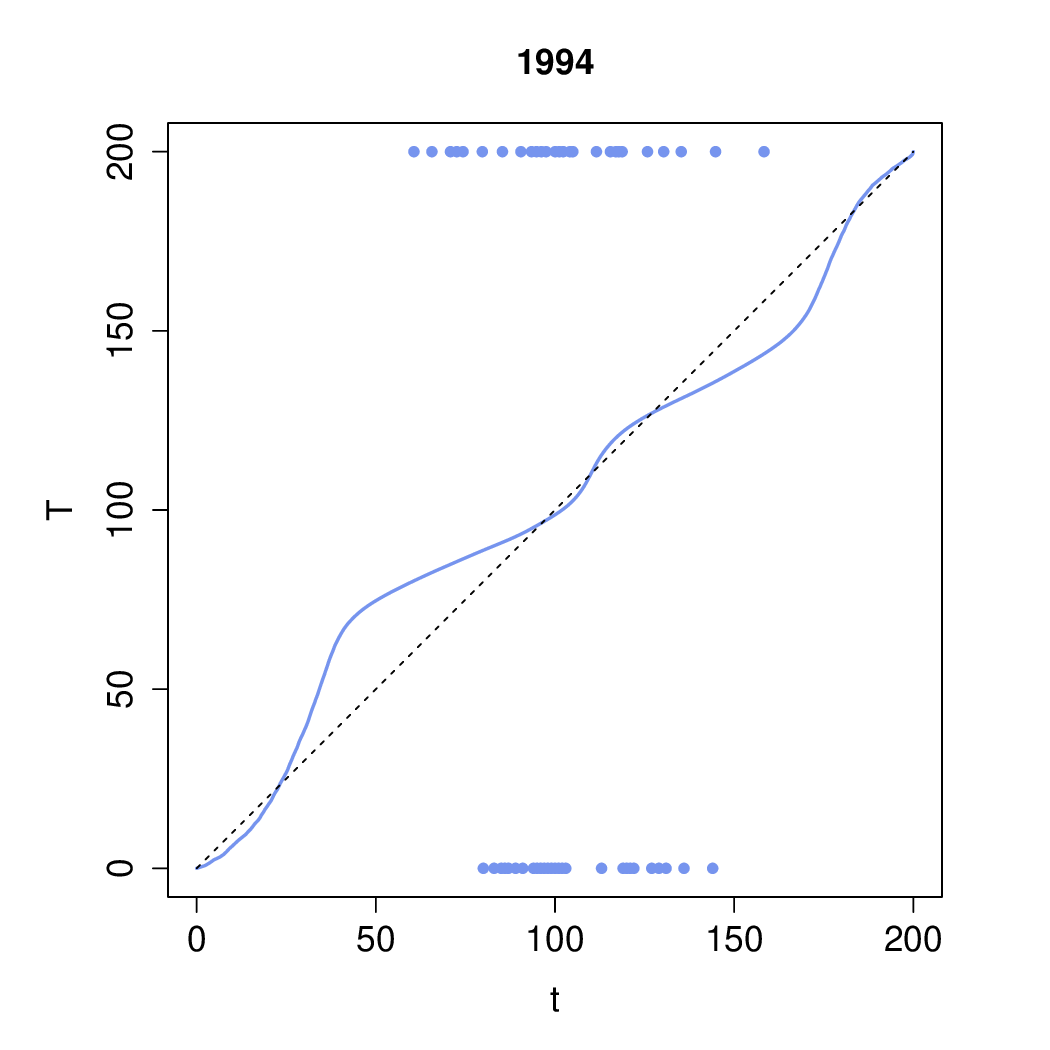}\\
\includegraphics[scale=0.180]{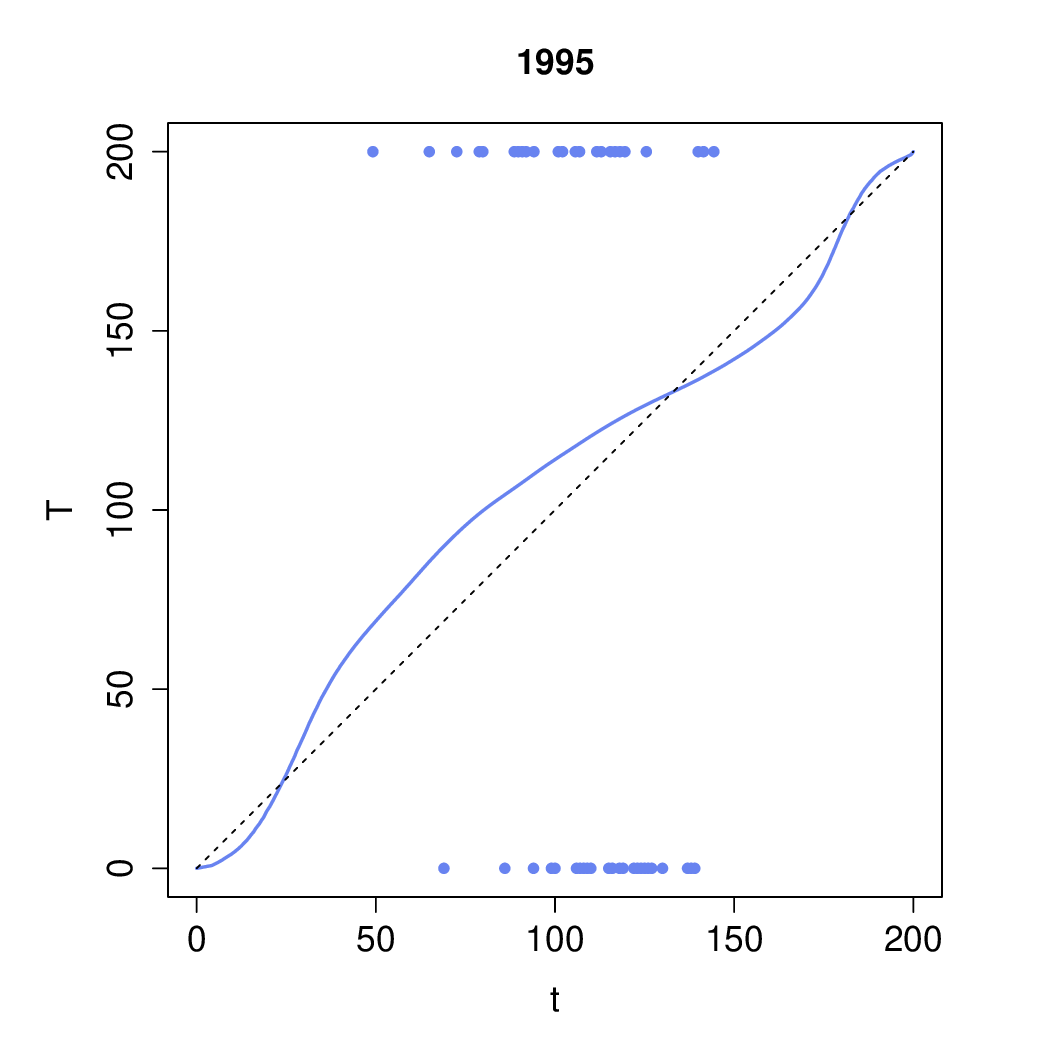}
\includegraphics[scale=0.180]{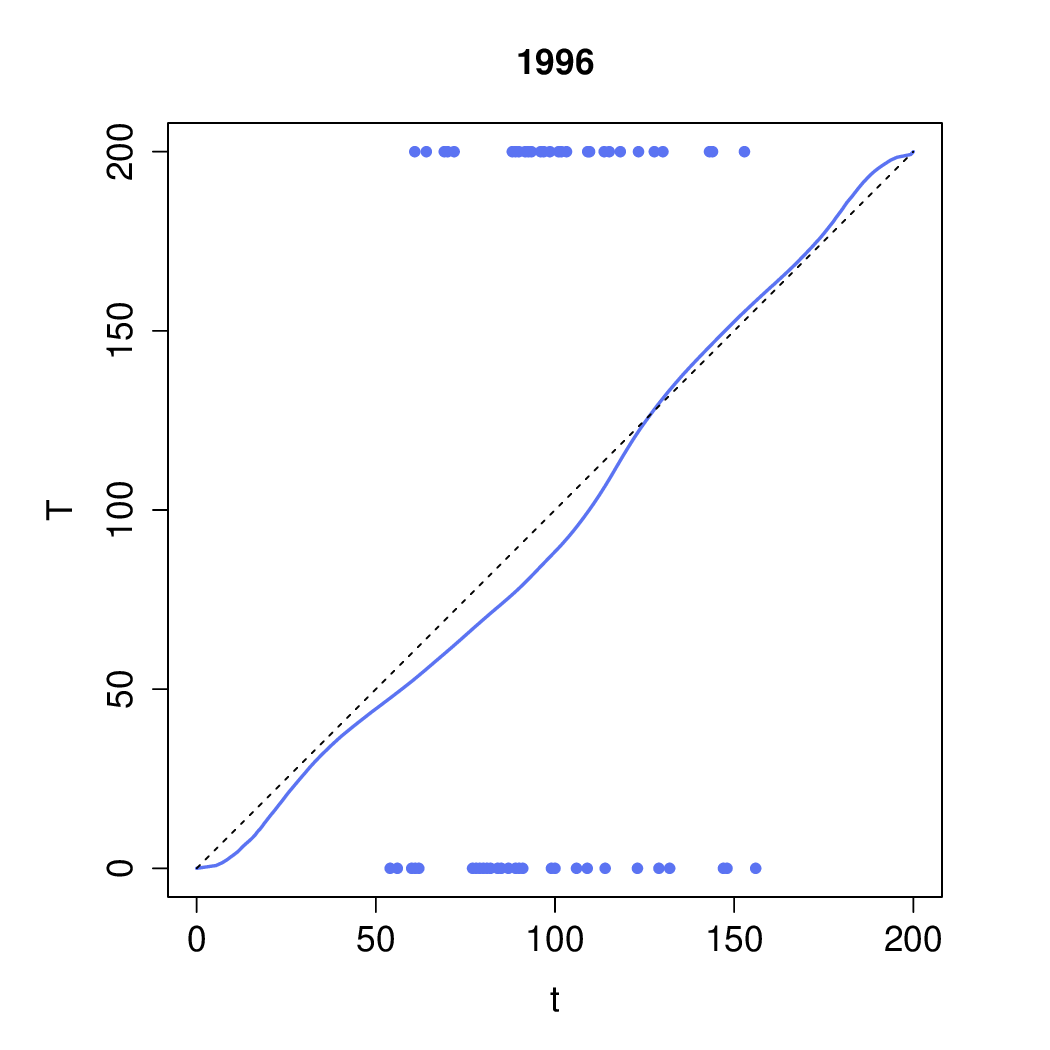}
\includegraphics[scale=0.180]{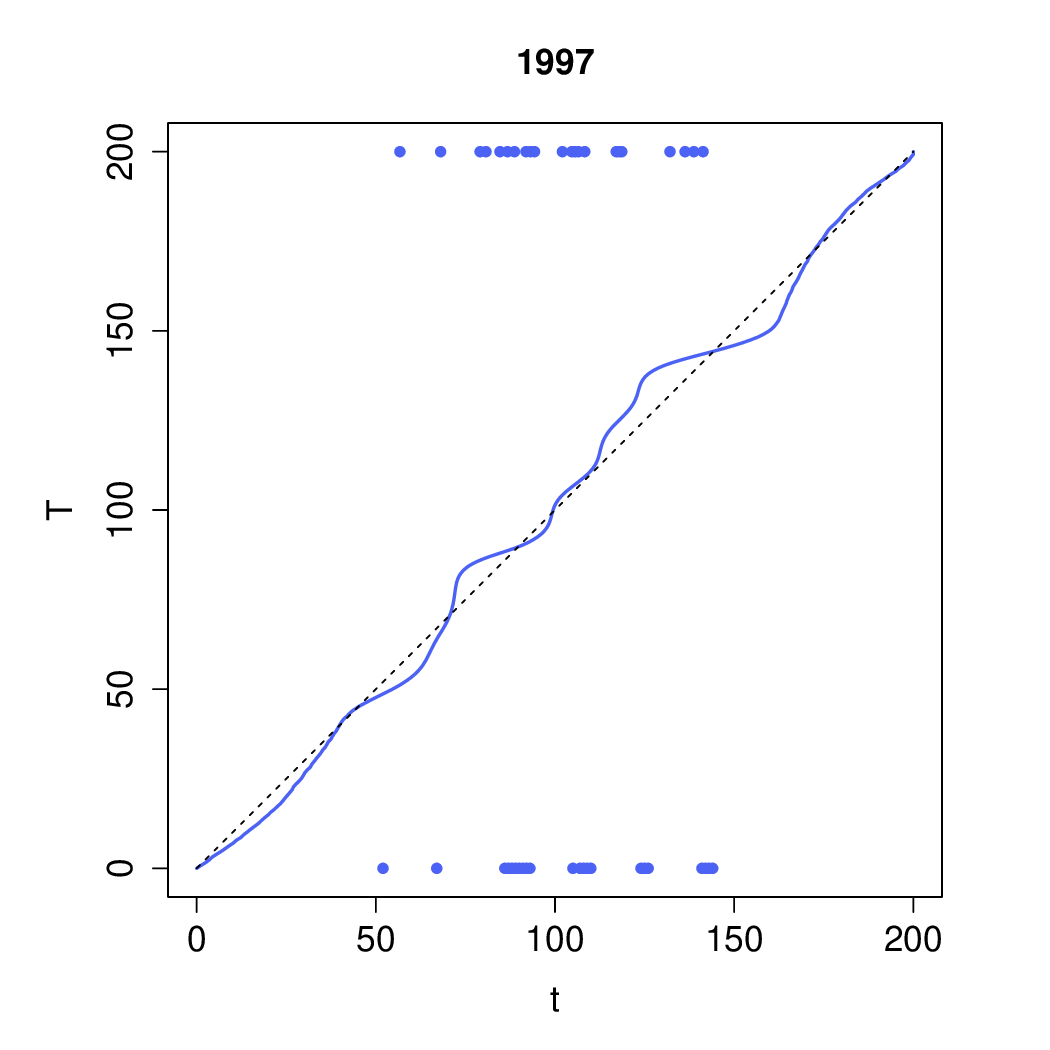}
\includegraphics[scale=0.180]{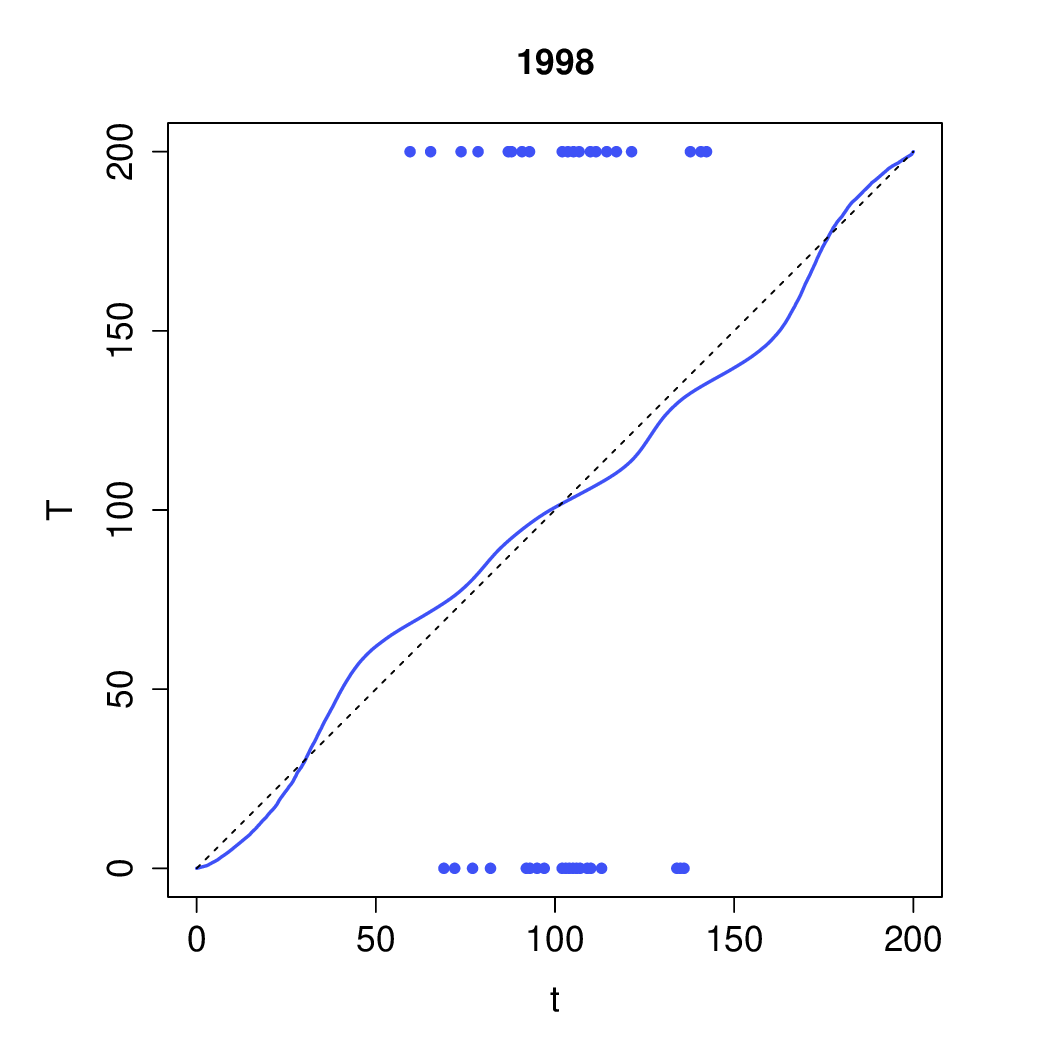}
\includegraphics[scale=0.180]{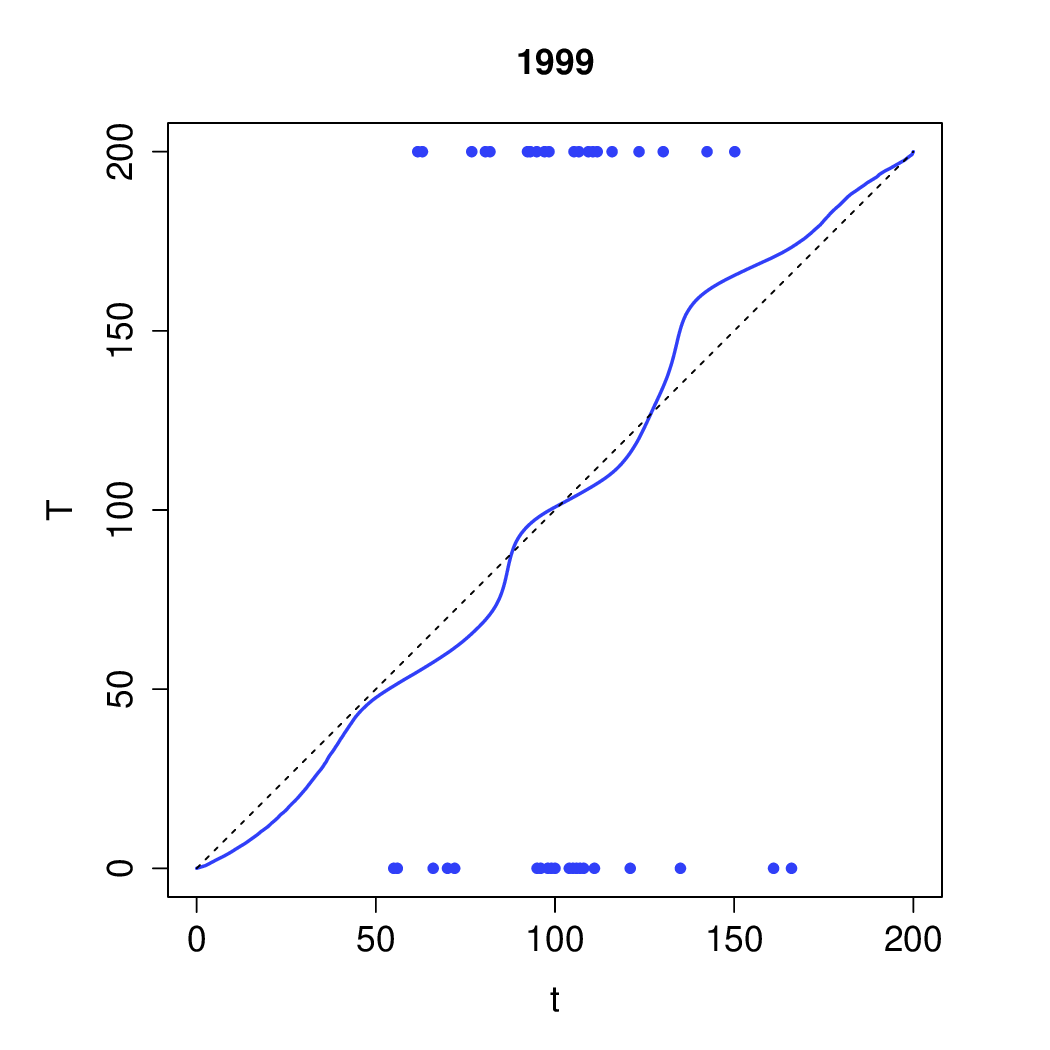}\\
\includegraphics[scale=0.180]{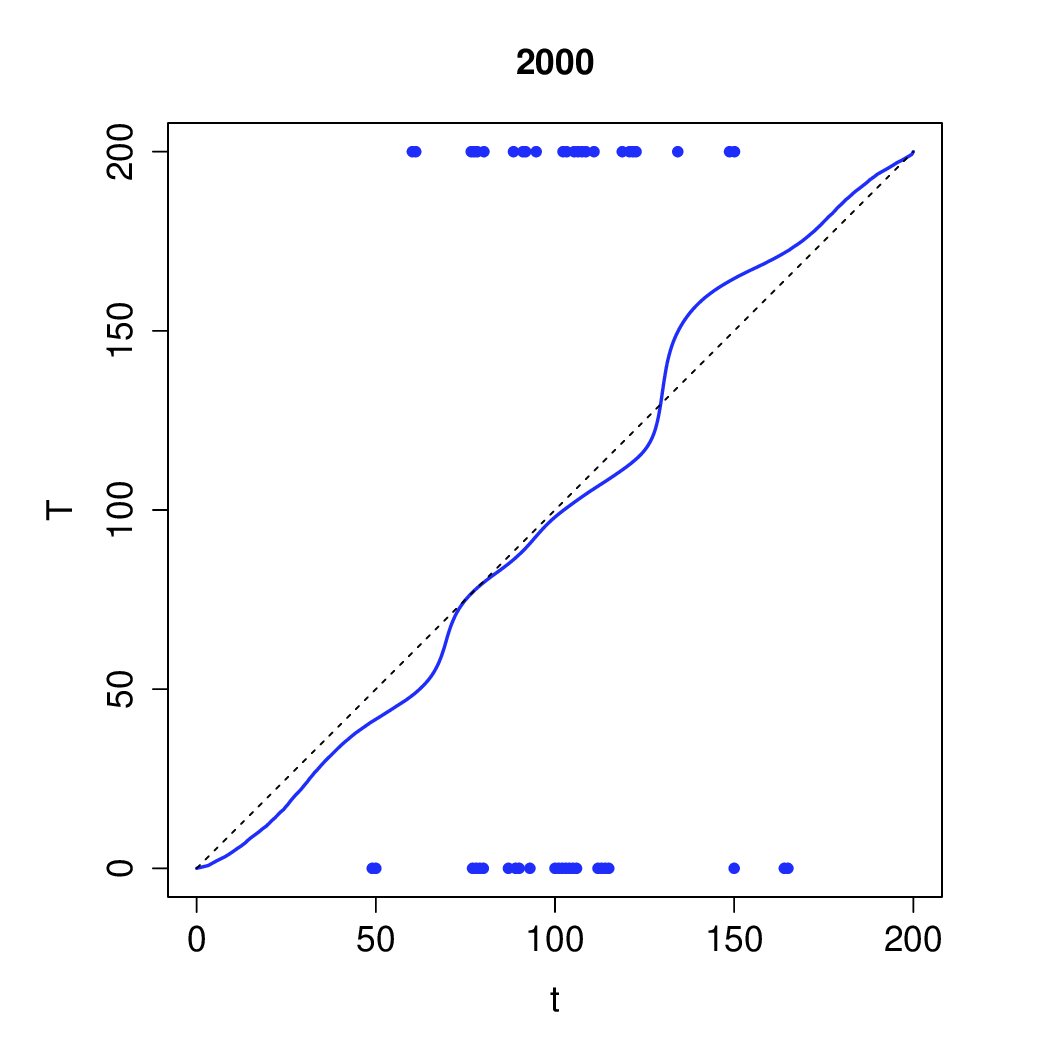}
\includegraphics[scale=0.180]{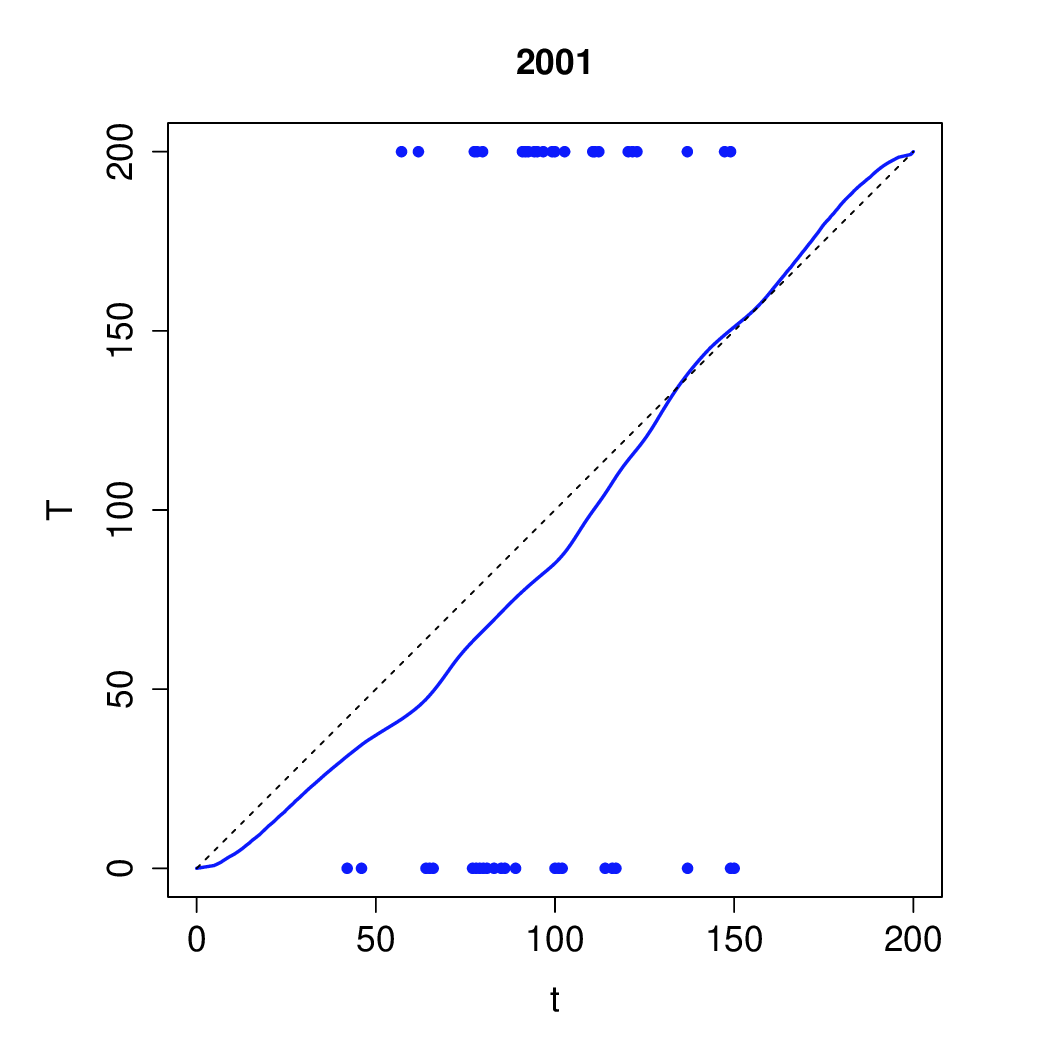}
\includegraphics[scale=0.180]{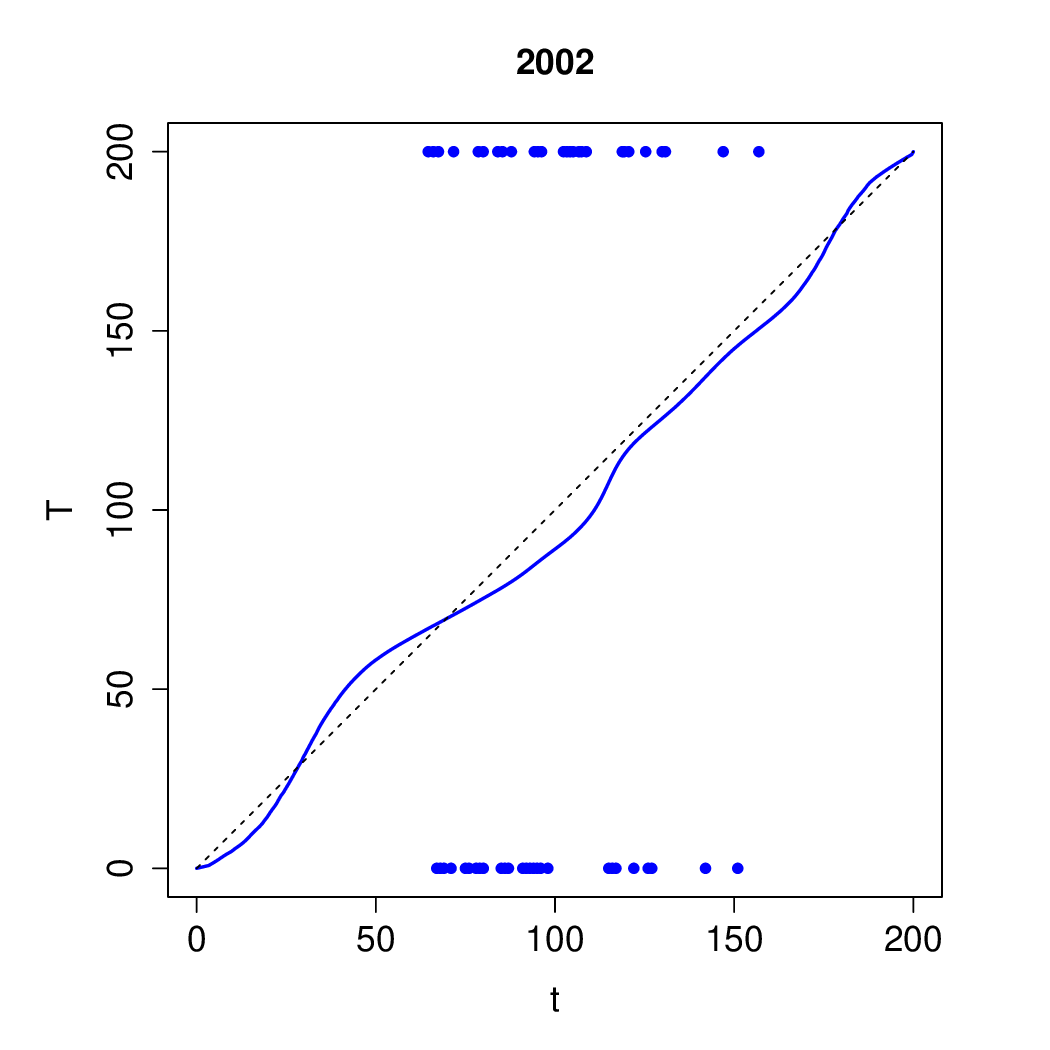}
\includegraphics[scale=0.180]{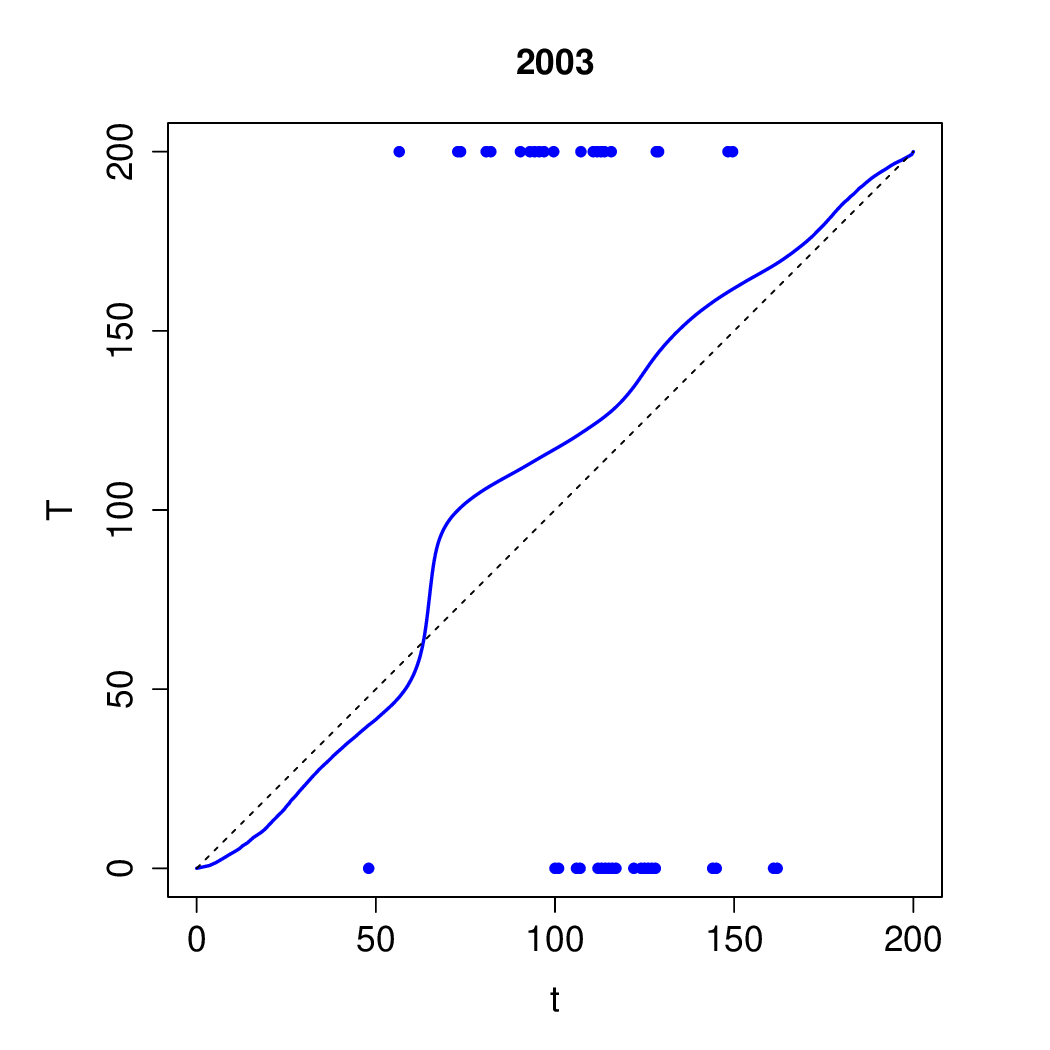}
\includegraphics[scale=0.180]{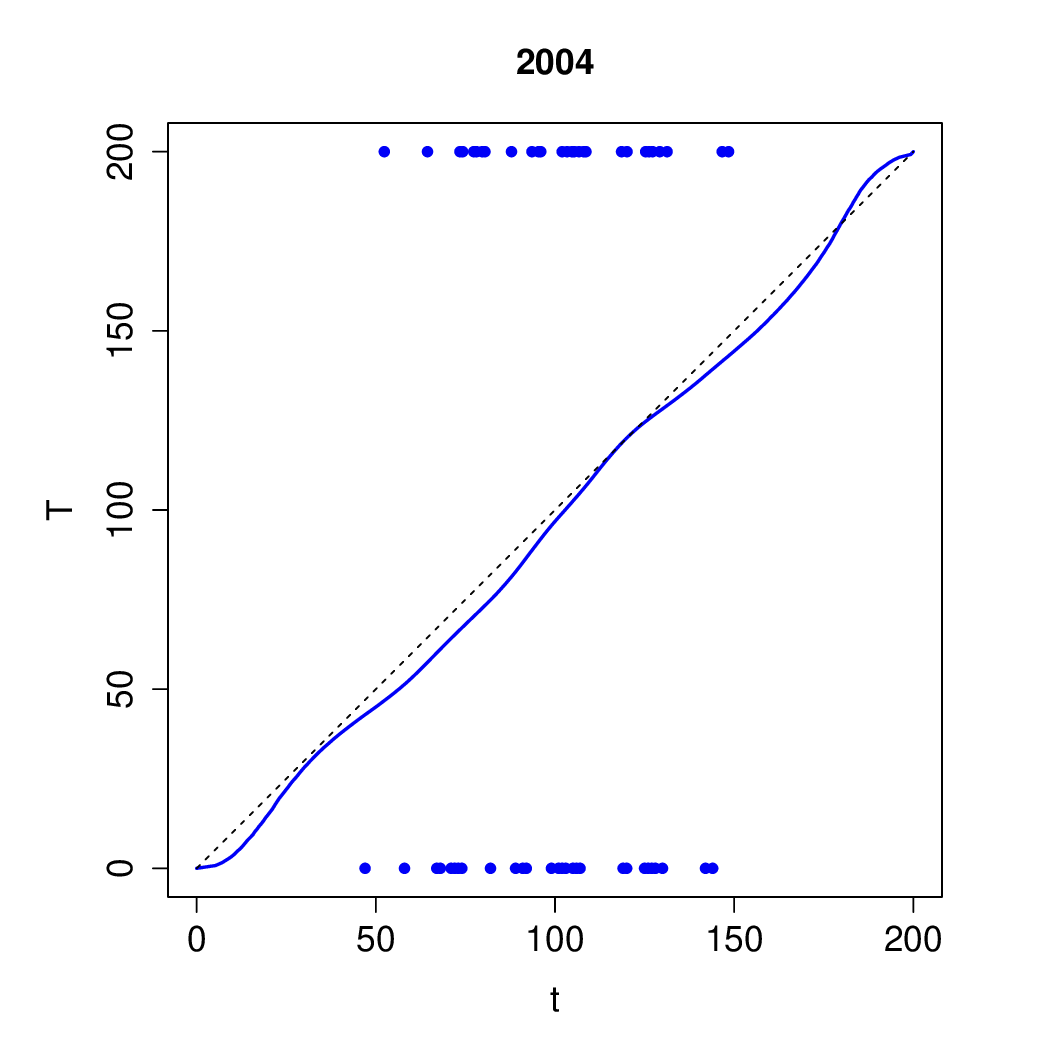}\\
\includegraphics[scale=0.180]{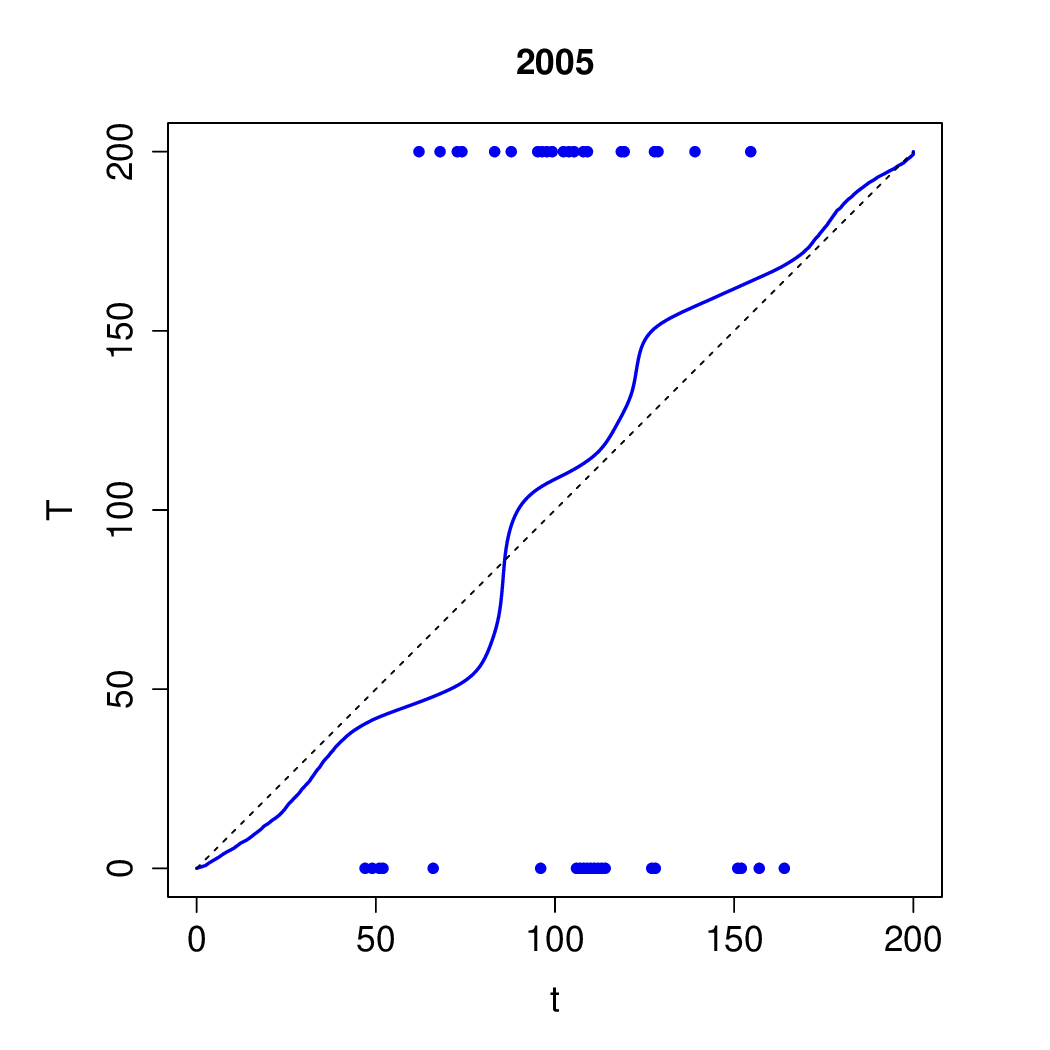}
\includegraphics[scale=0.180]{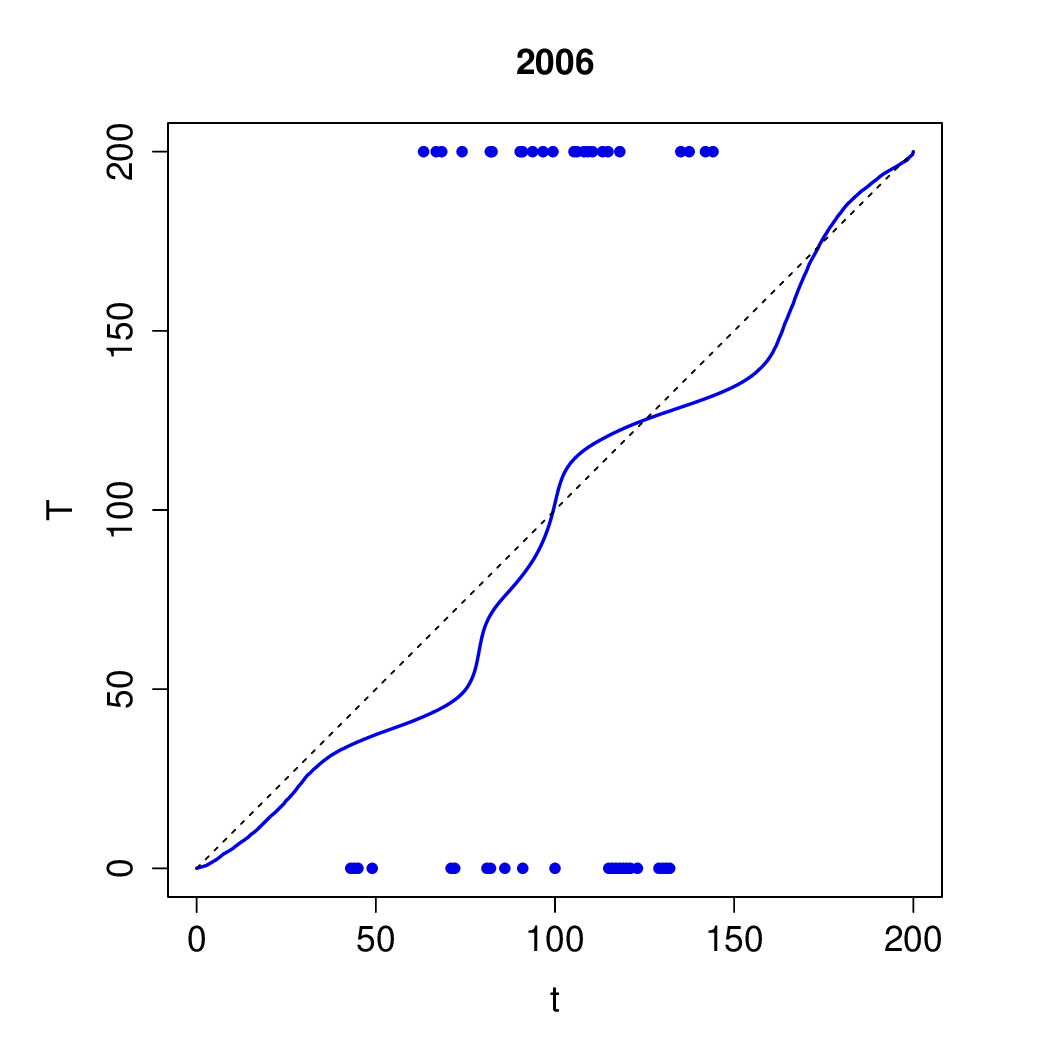}
\includegraphics[scale=0.180]{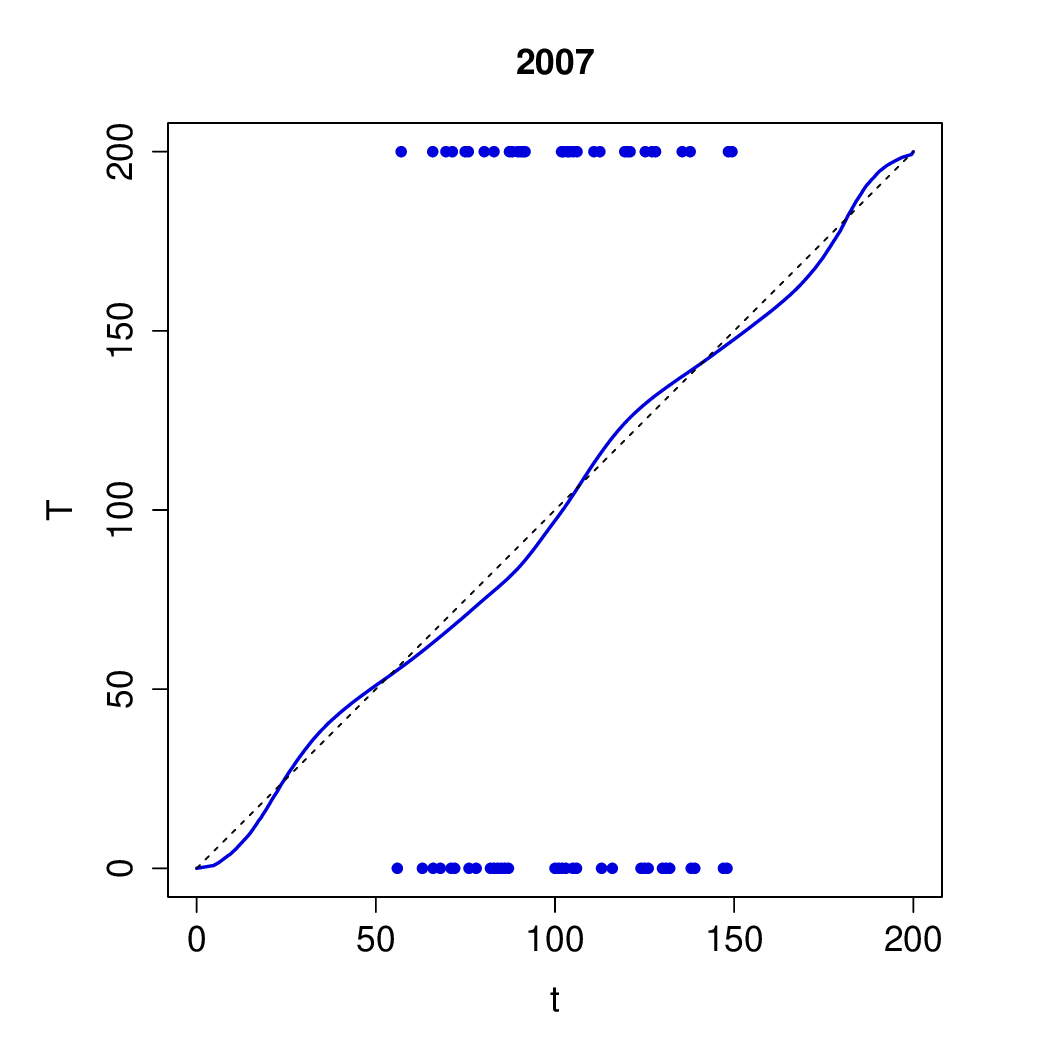}
\includegraphics[scale=0.180]{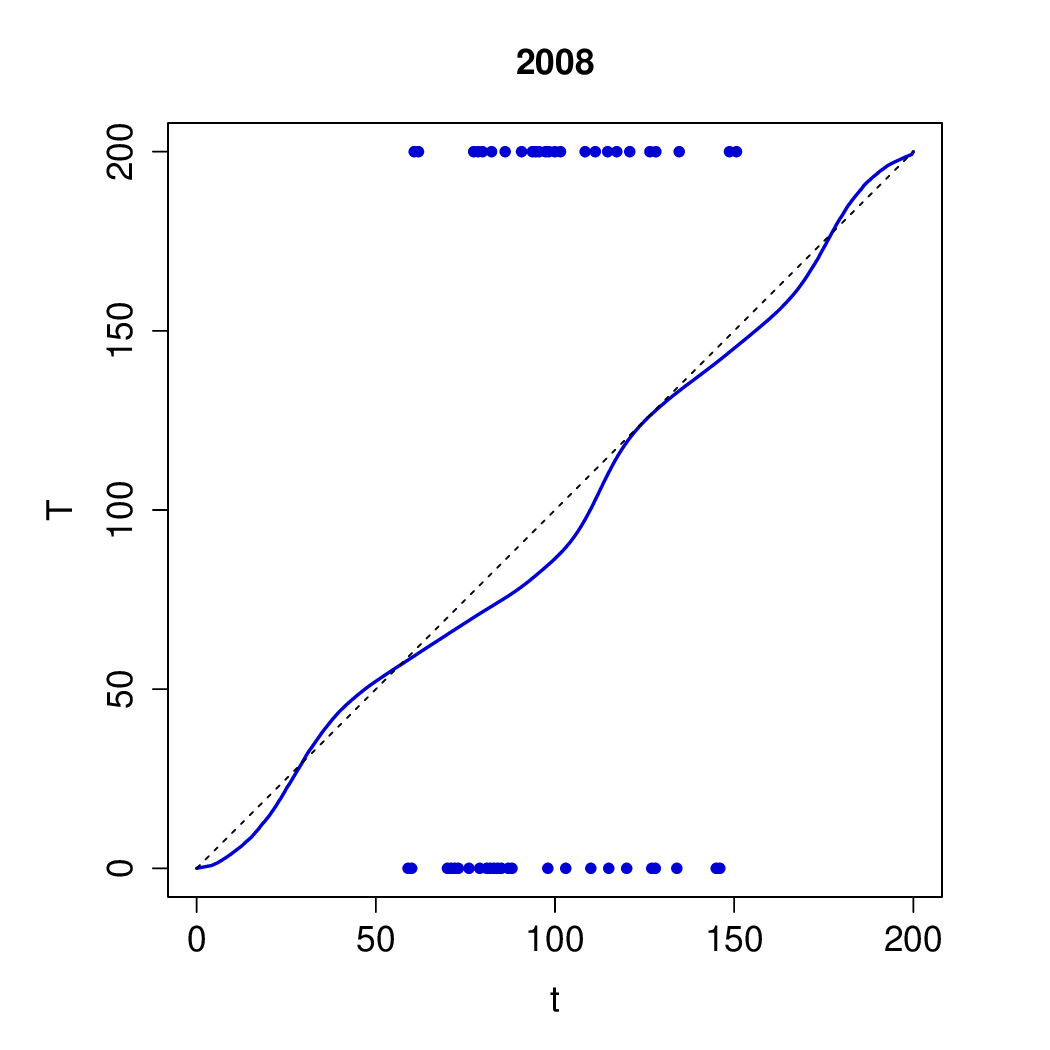}
\includegraphics[scale=0.180]{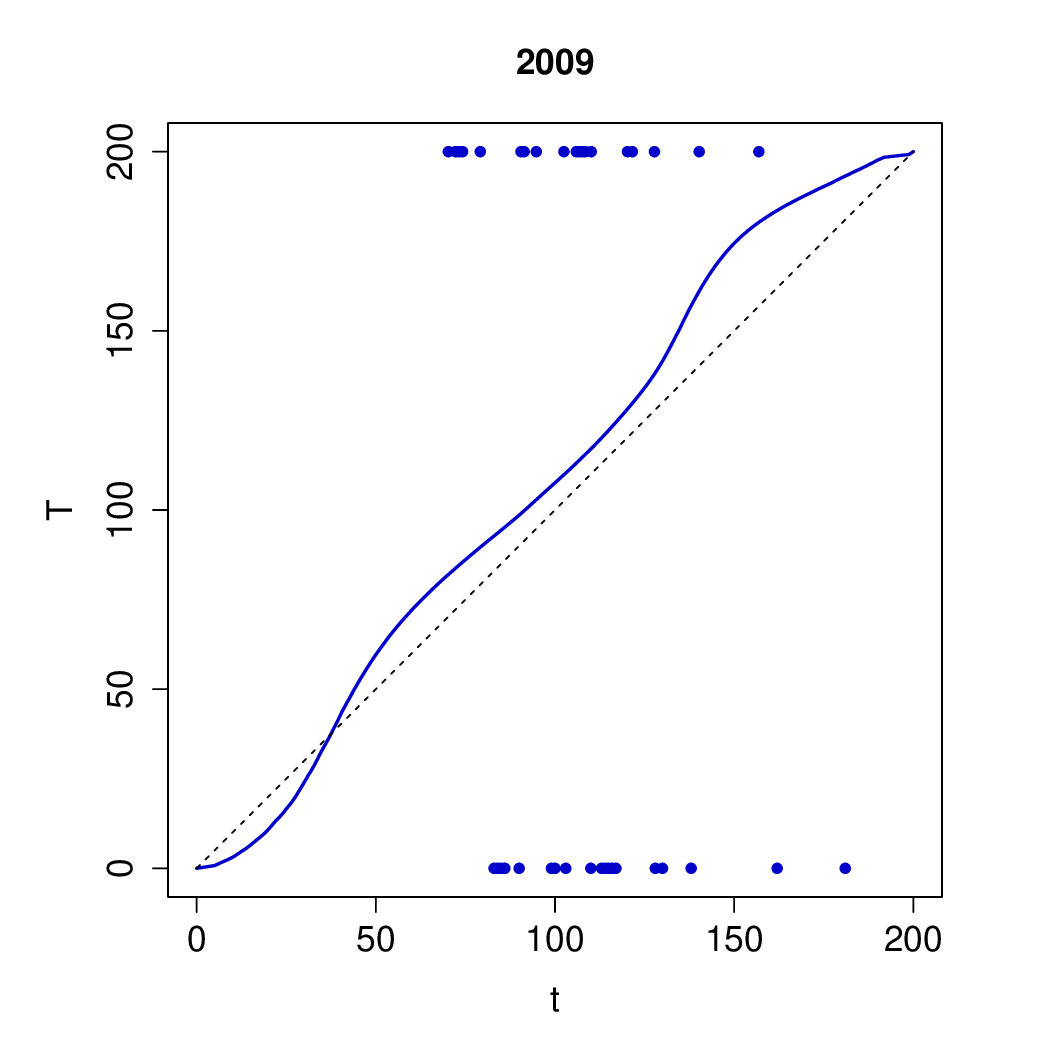}\\
\includegraphics[scale=0.180]{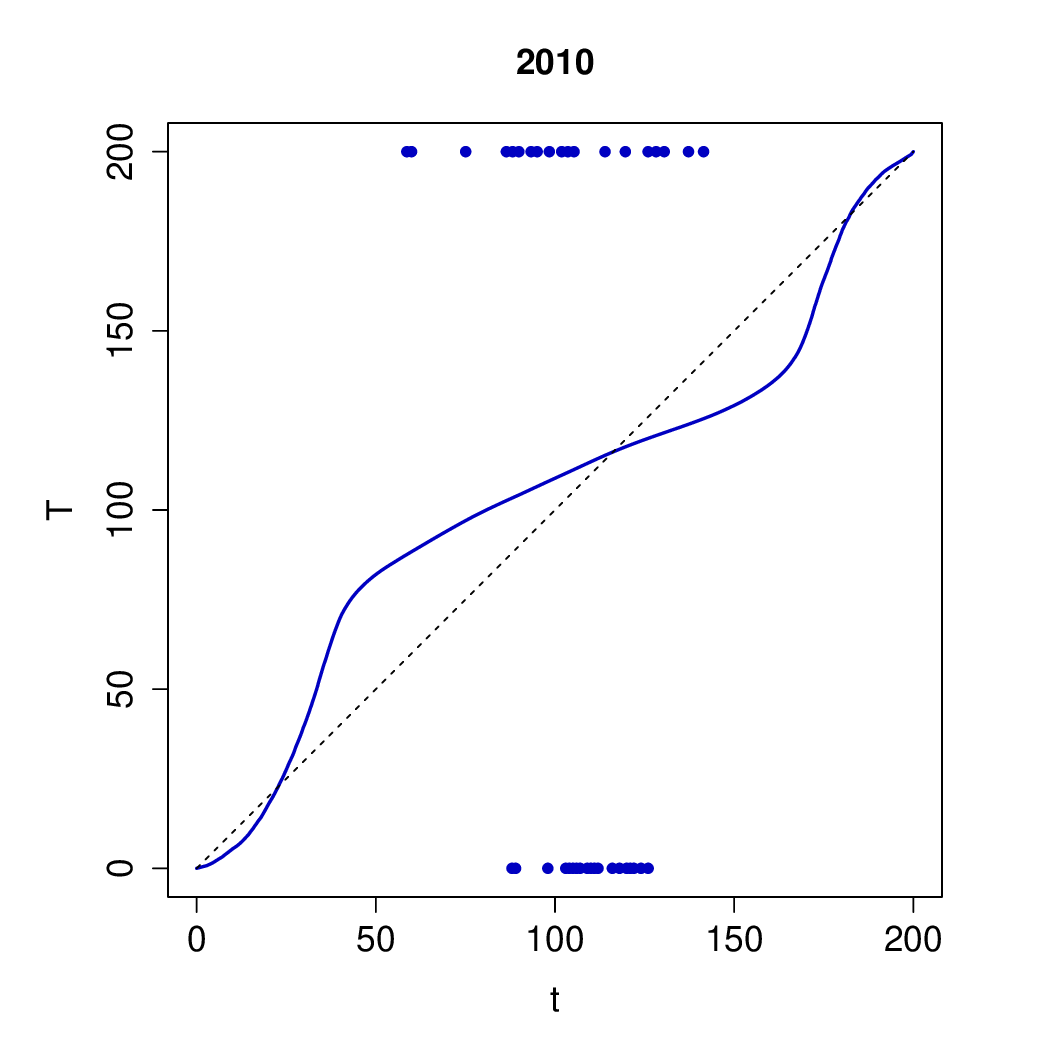}
\includegraphics[scale=0.180]{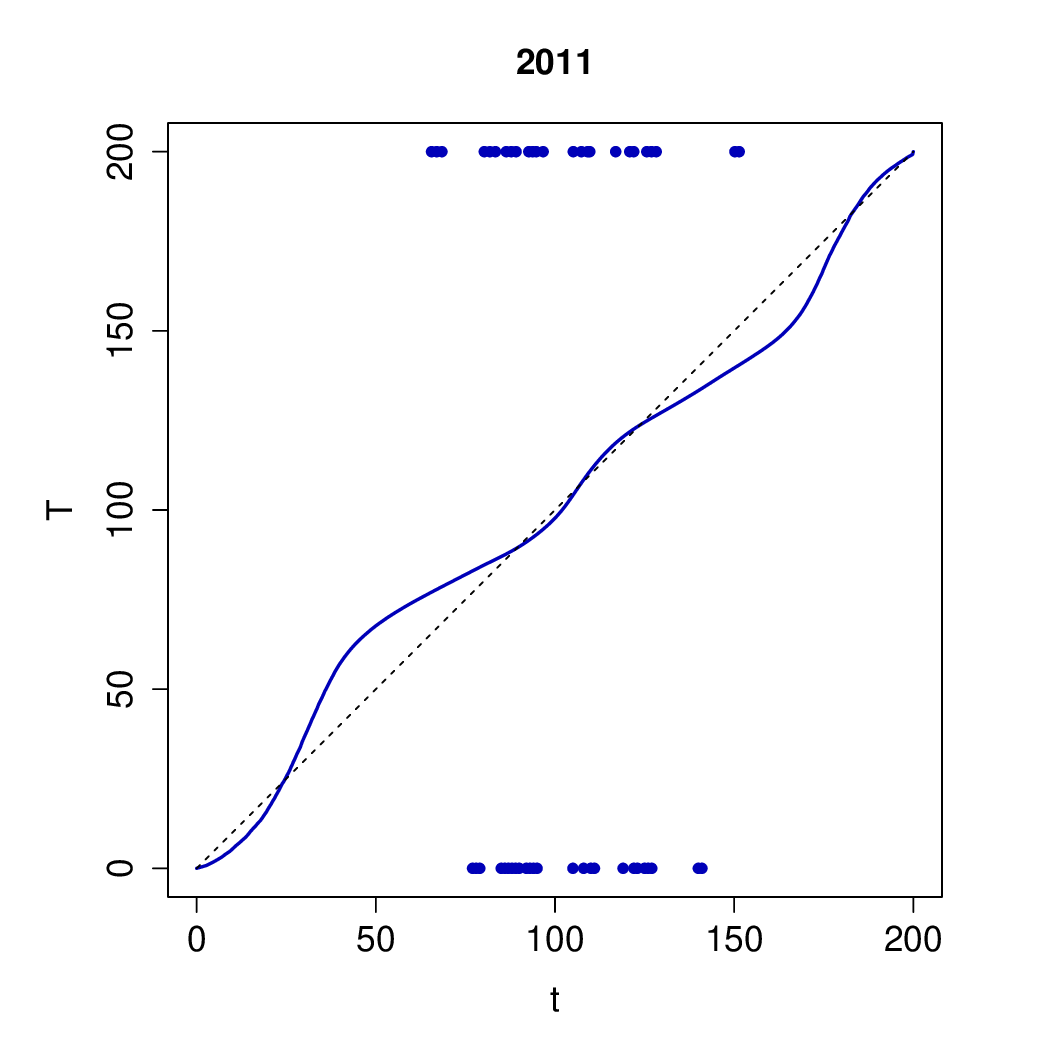}
\includegraphics[scale=0.180]{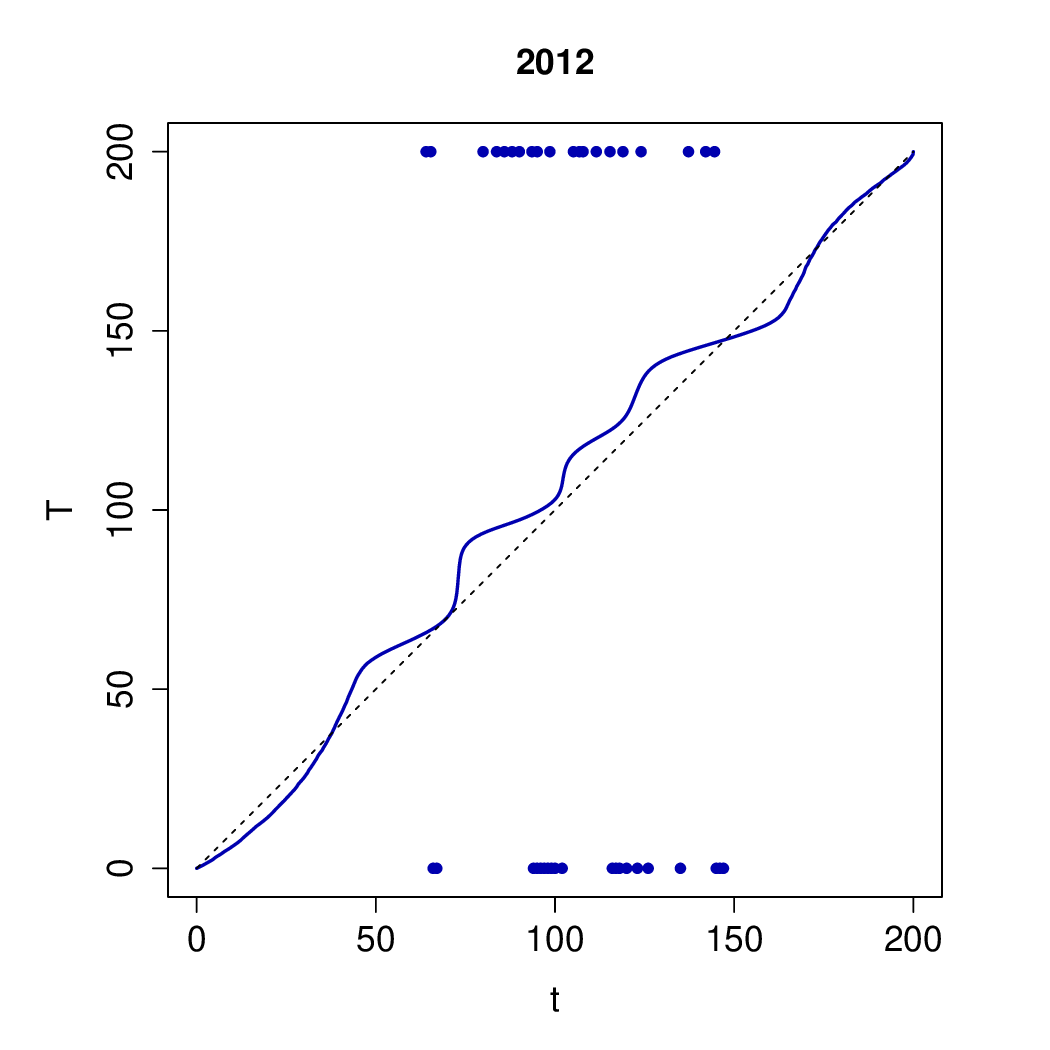}
\includegraphics[scale=0.180]{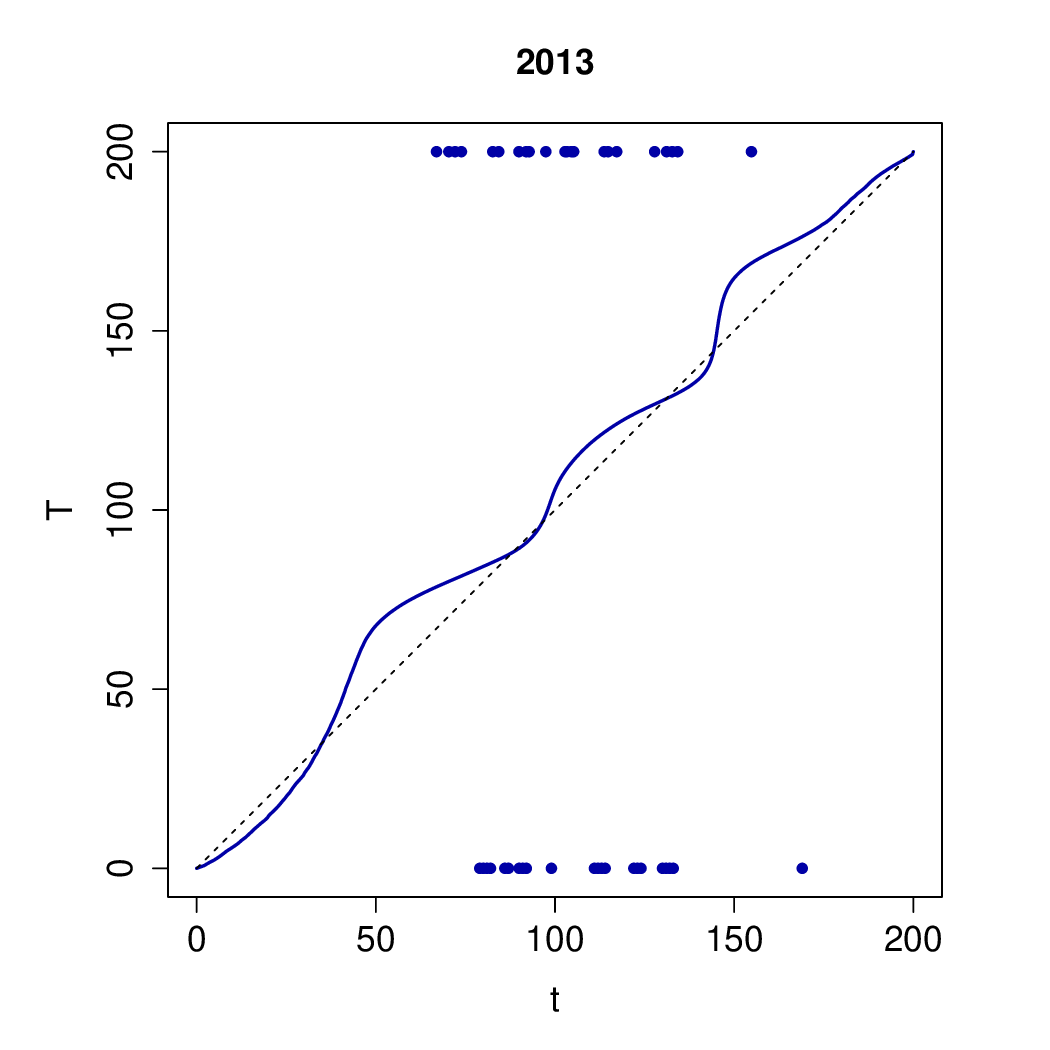}
\includegraphics[scale=0.180]{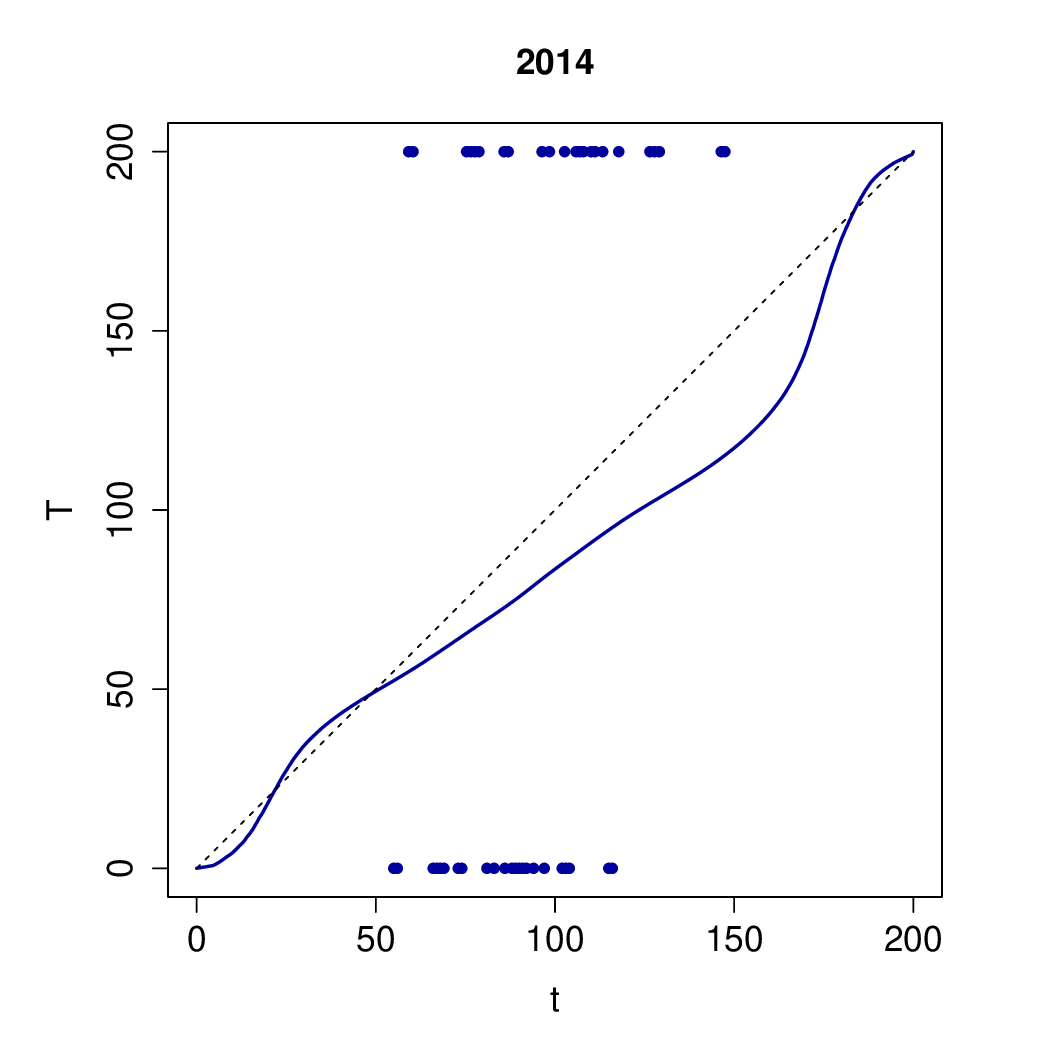}\\
\includegraphics[scale=0.180]{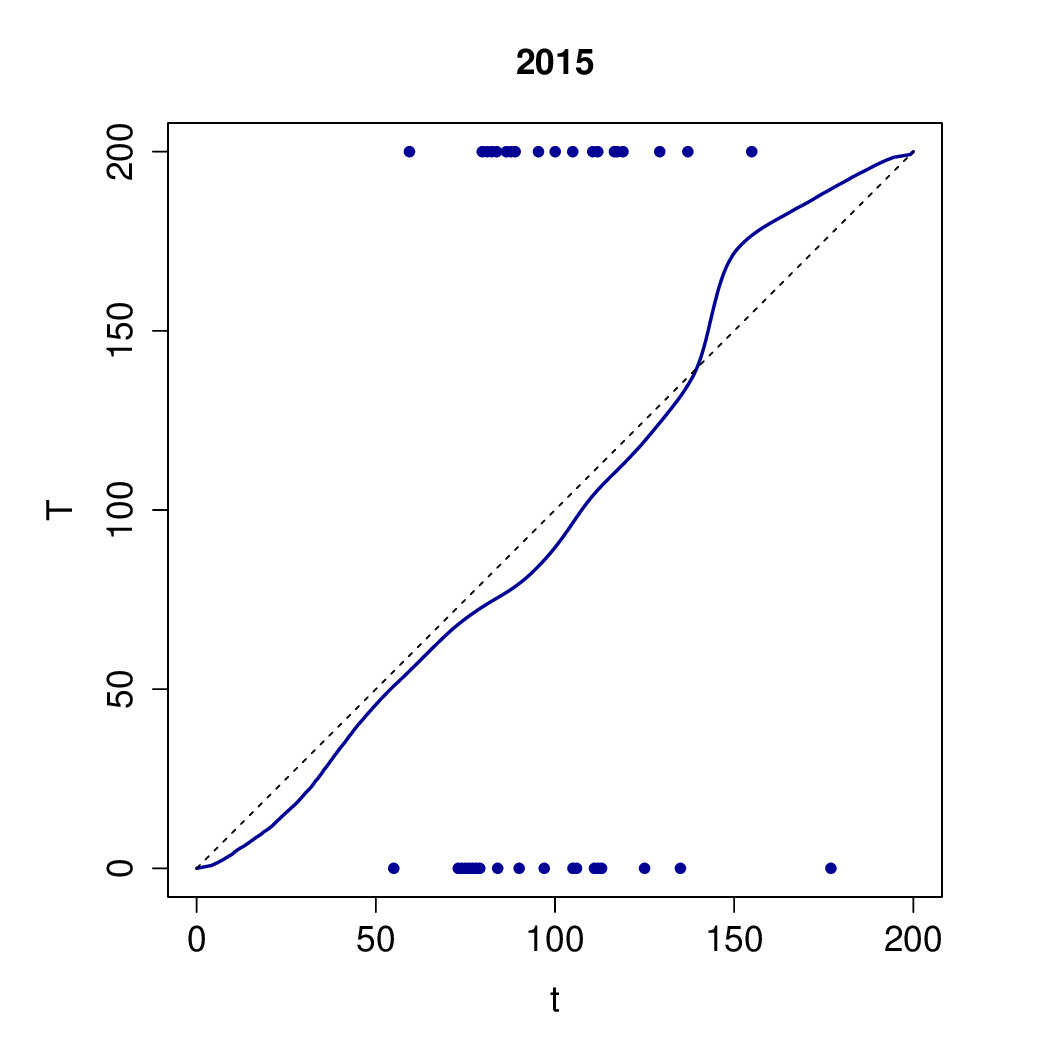}
\includegraphics[scale=0.180]{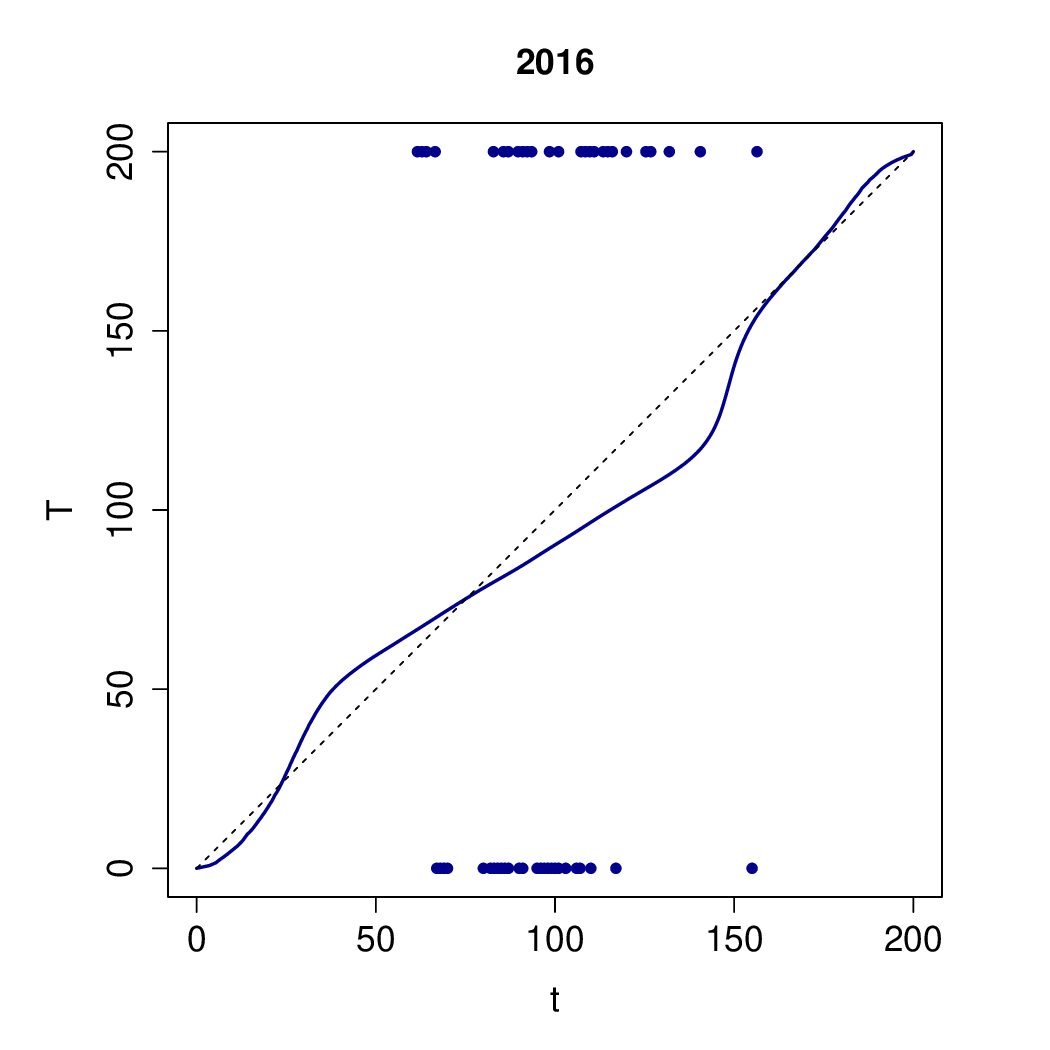}
\caption{\footnotesize Yearly posterior mean Bernstein polynomial warp functions of low-temperatures in the same color palette as data, plotted with raw data (bottom), registered points (top) and the identity function (dashed black).}
\label{warps:cold}
\end{figure}

\begin{figure}\centering 
\includegraphics[scale=0.180]{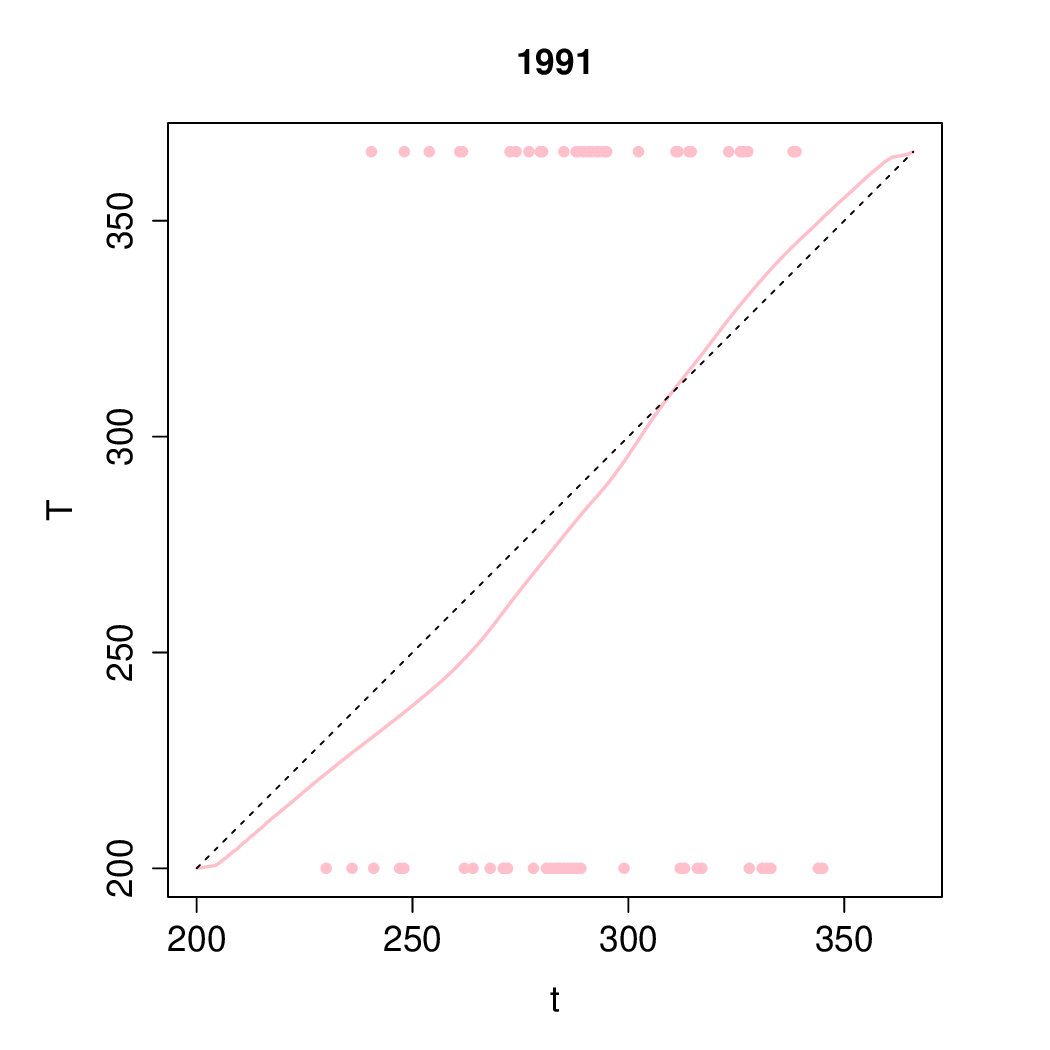}
\includegraphics[scale=0.180]{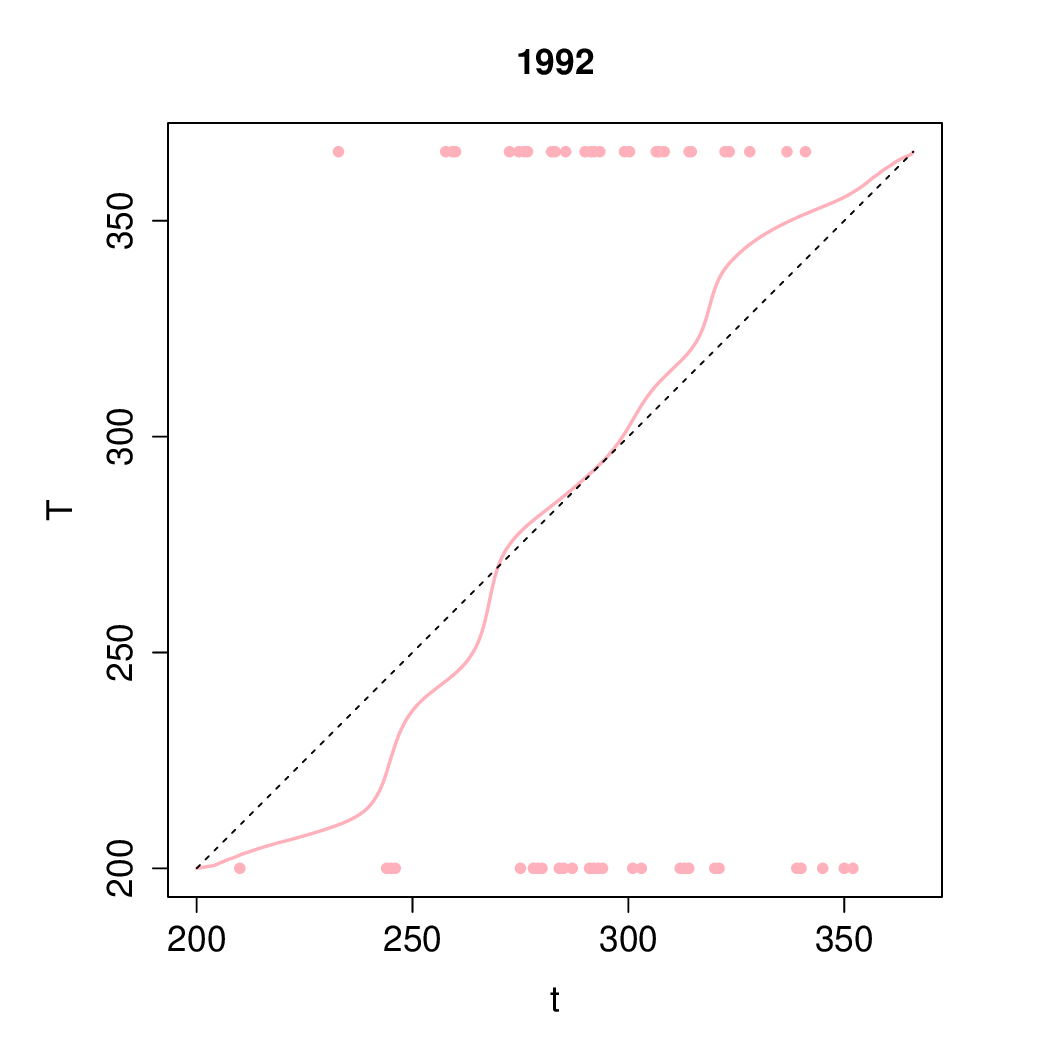}
\includegraphics[scale=0.180]{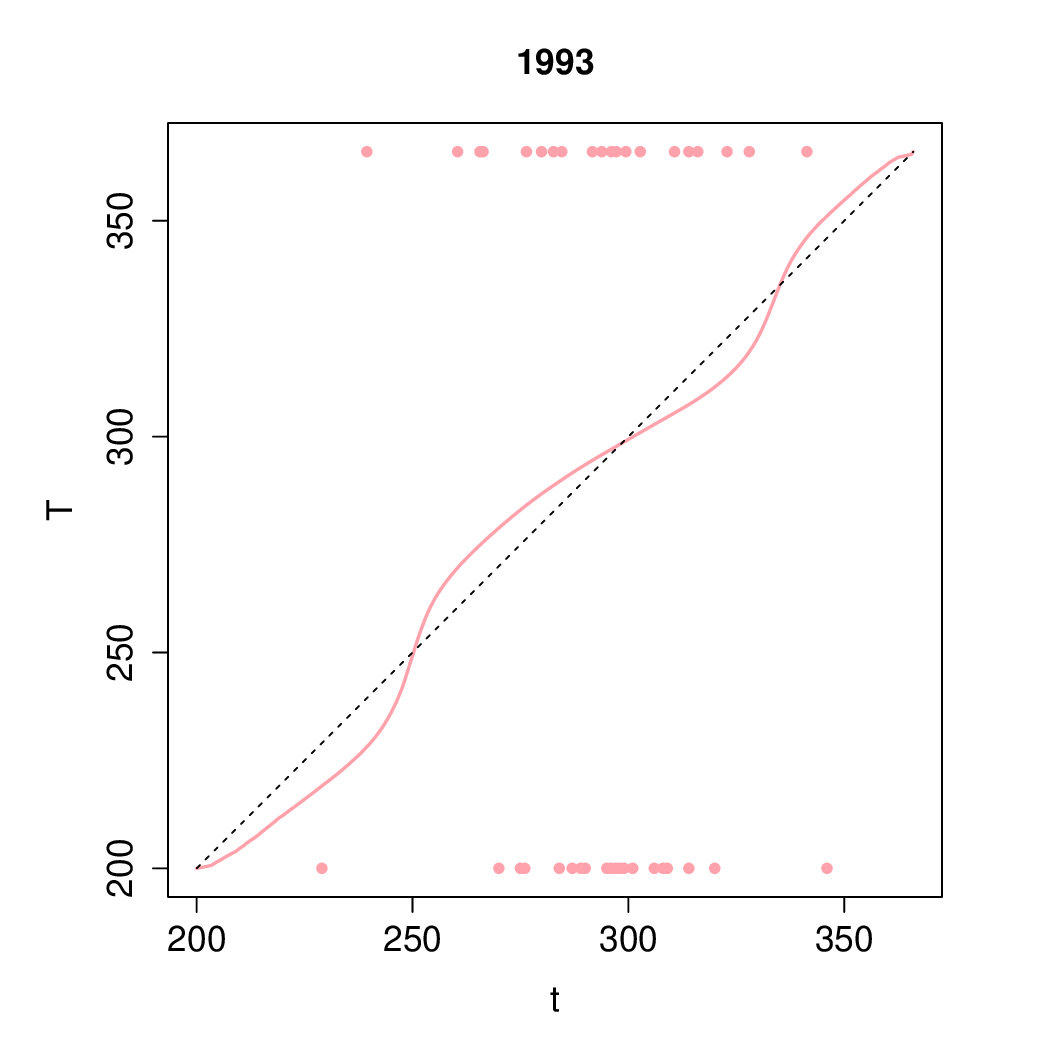}
\includegraphics[scale=0.180]{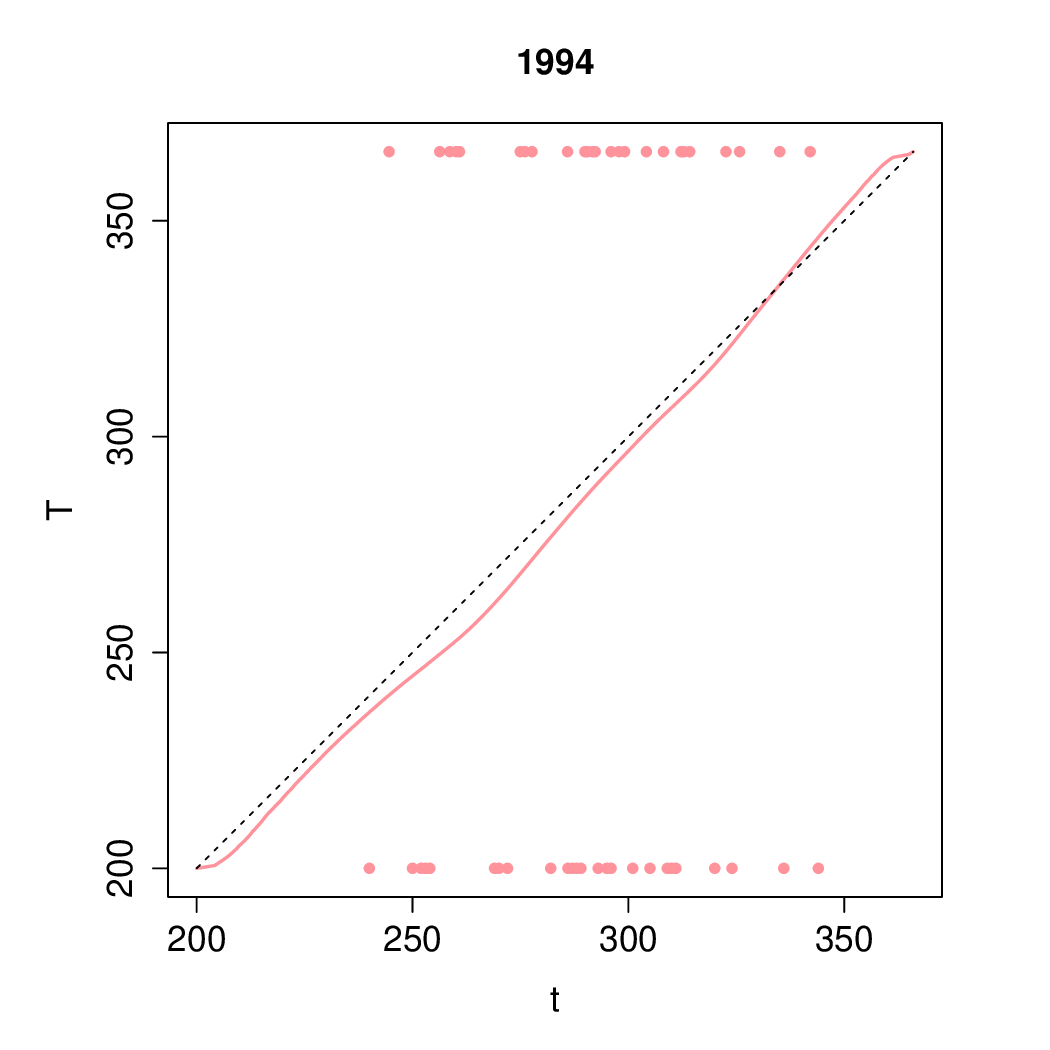}
\includegraphics[scale=0.180]{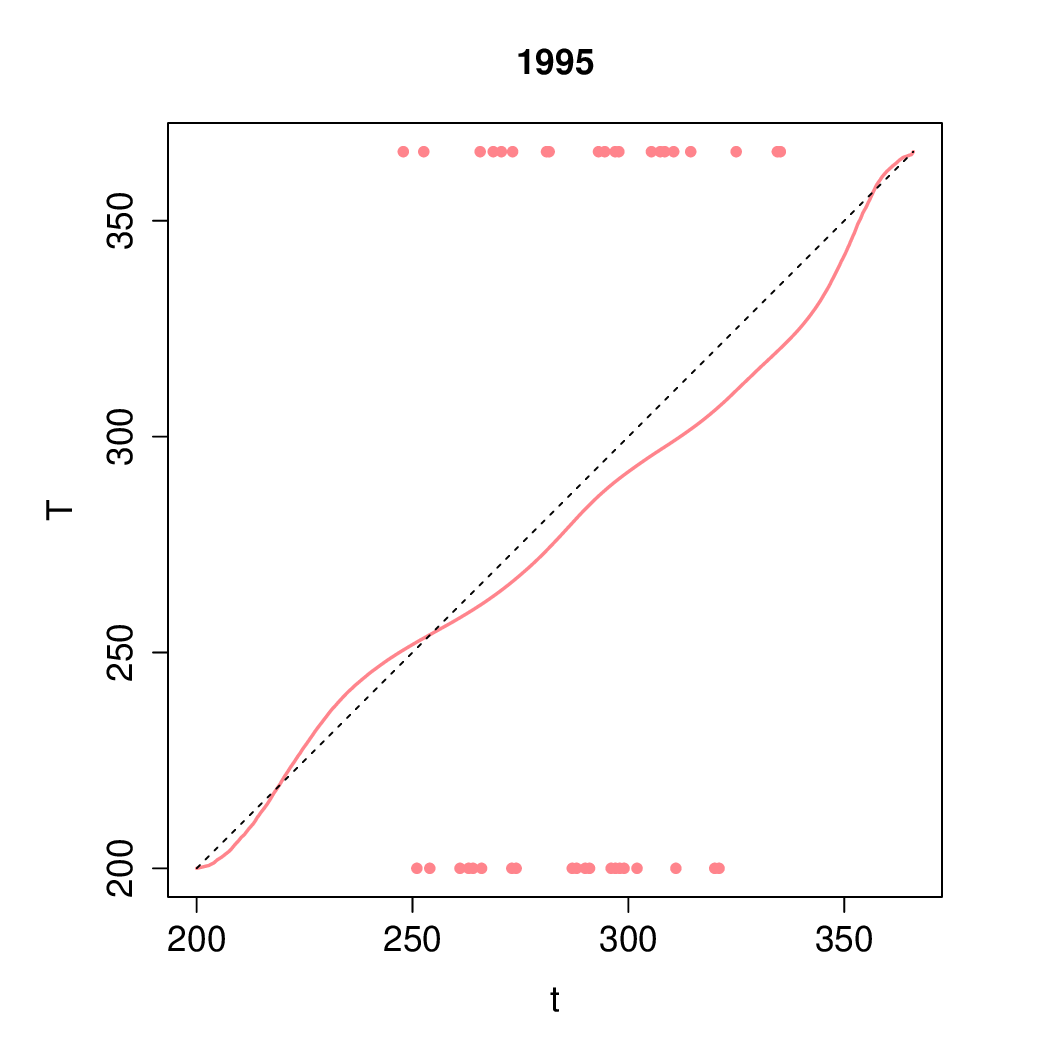}\\
\includegraphics[scale=0.180]{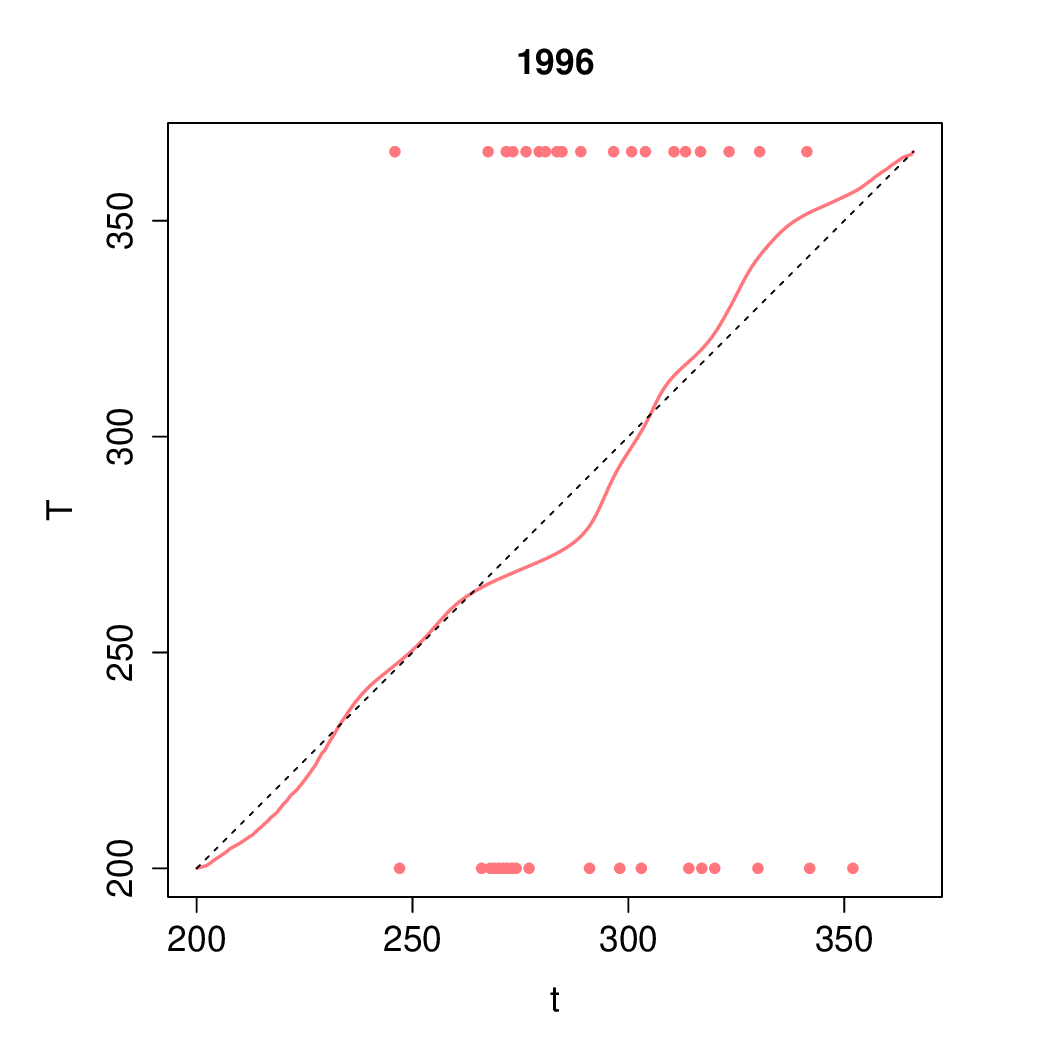}
\includegraphics[scale=0.180]{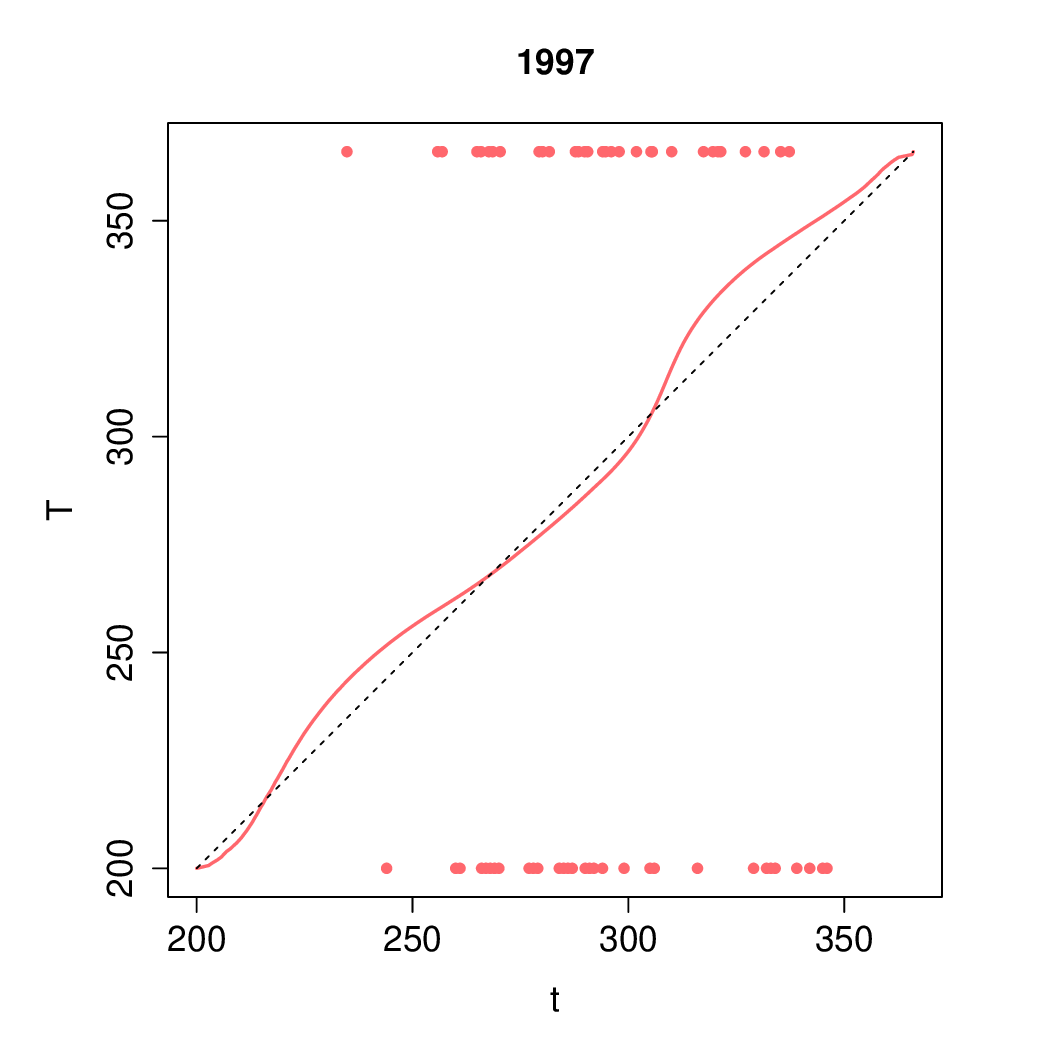}
\includegraphics[scale=0.180]{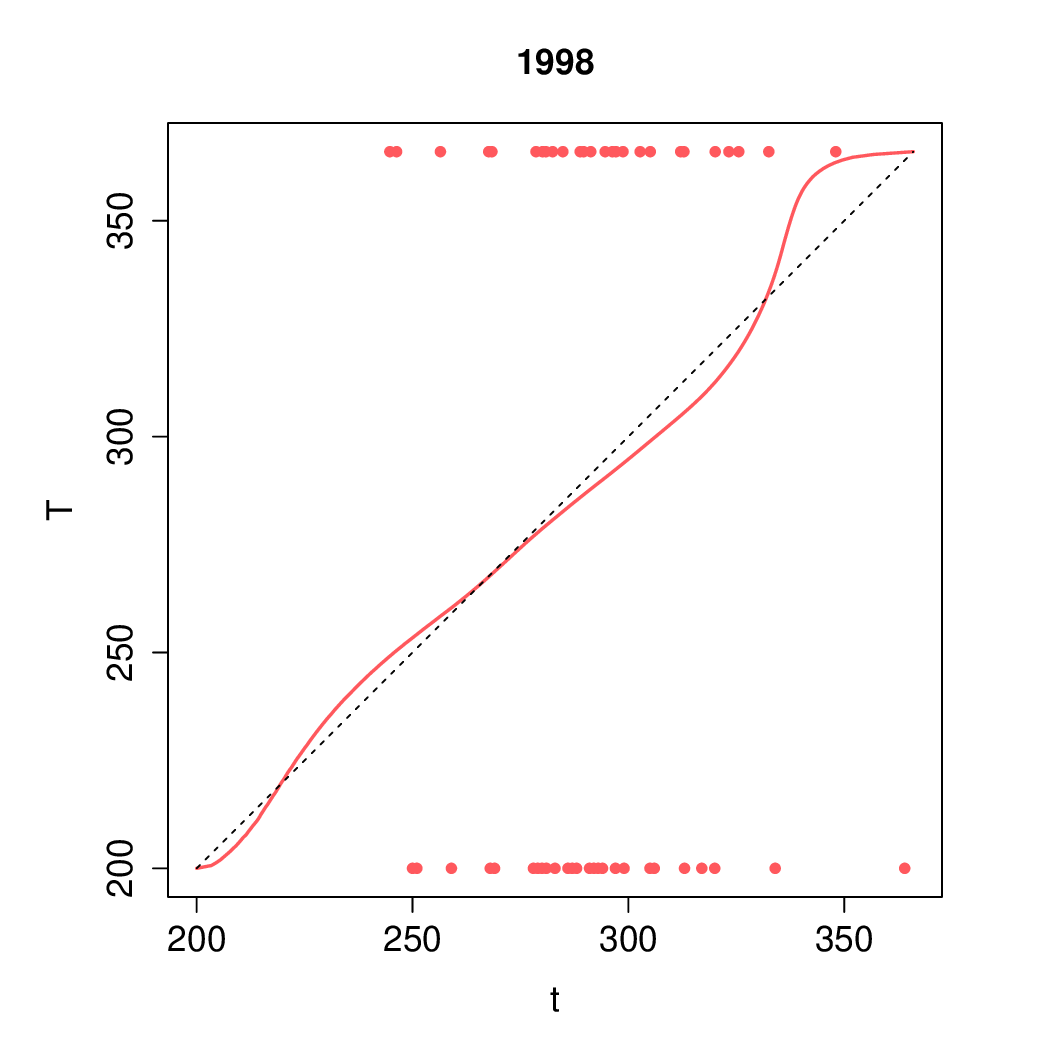}
\includegraphics[scale=0.180]{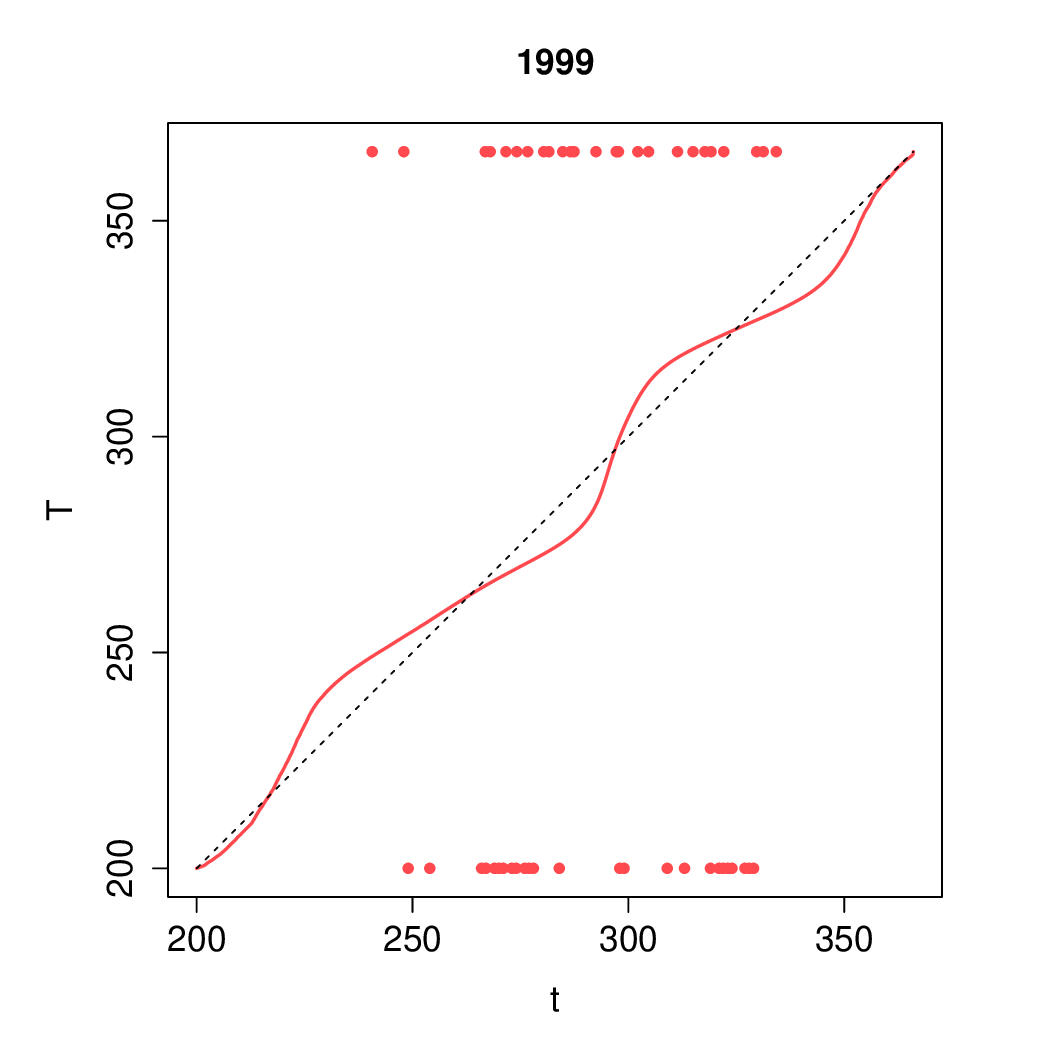}
\includegraphics[scale=0.180]{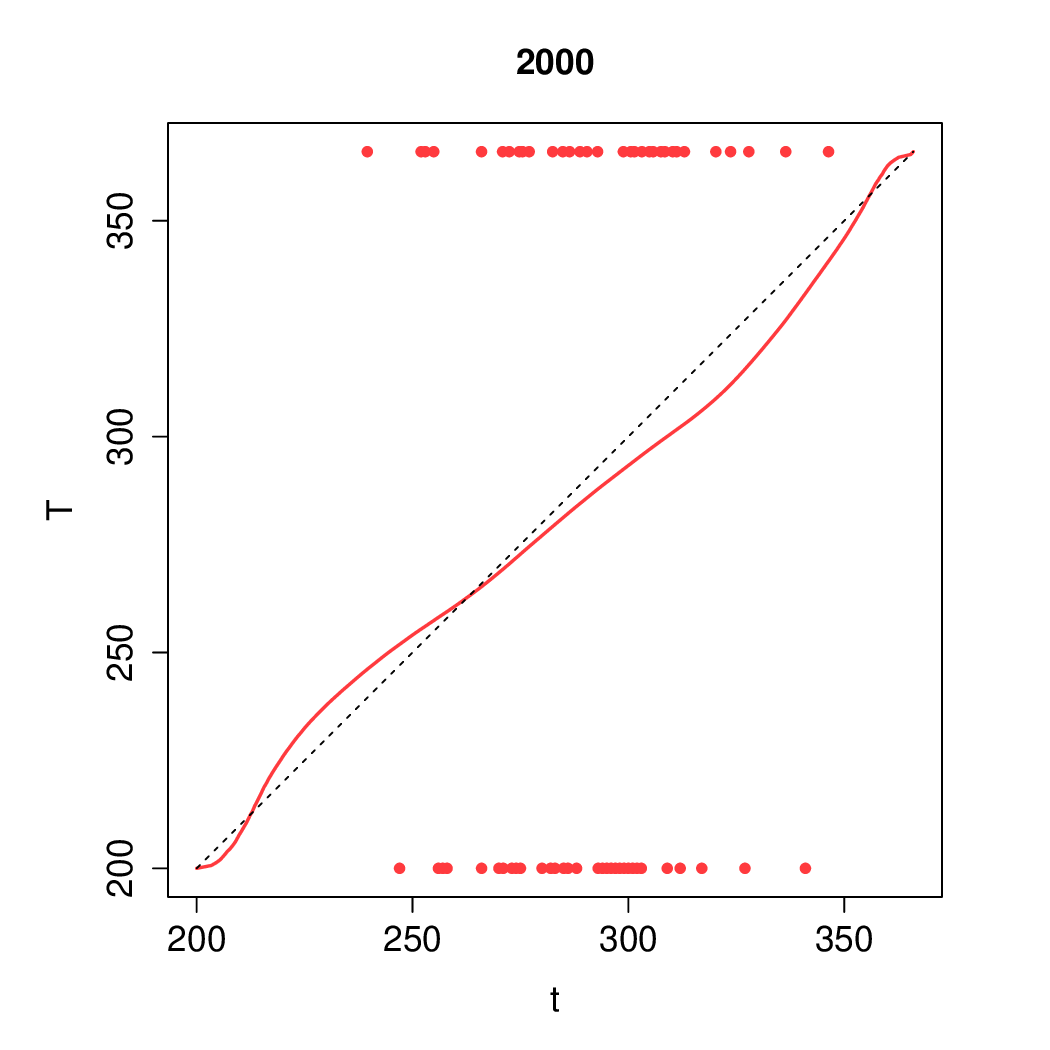}\\
\includegraphics[scale=0.180]{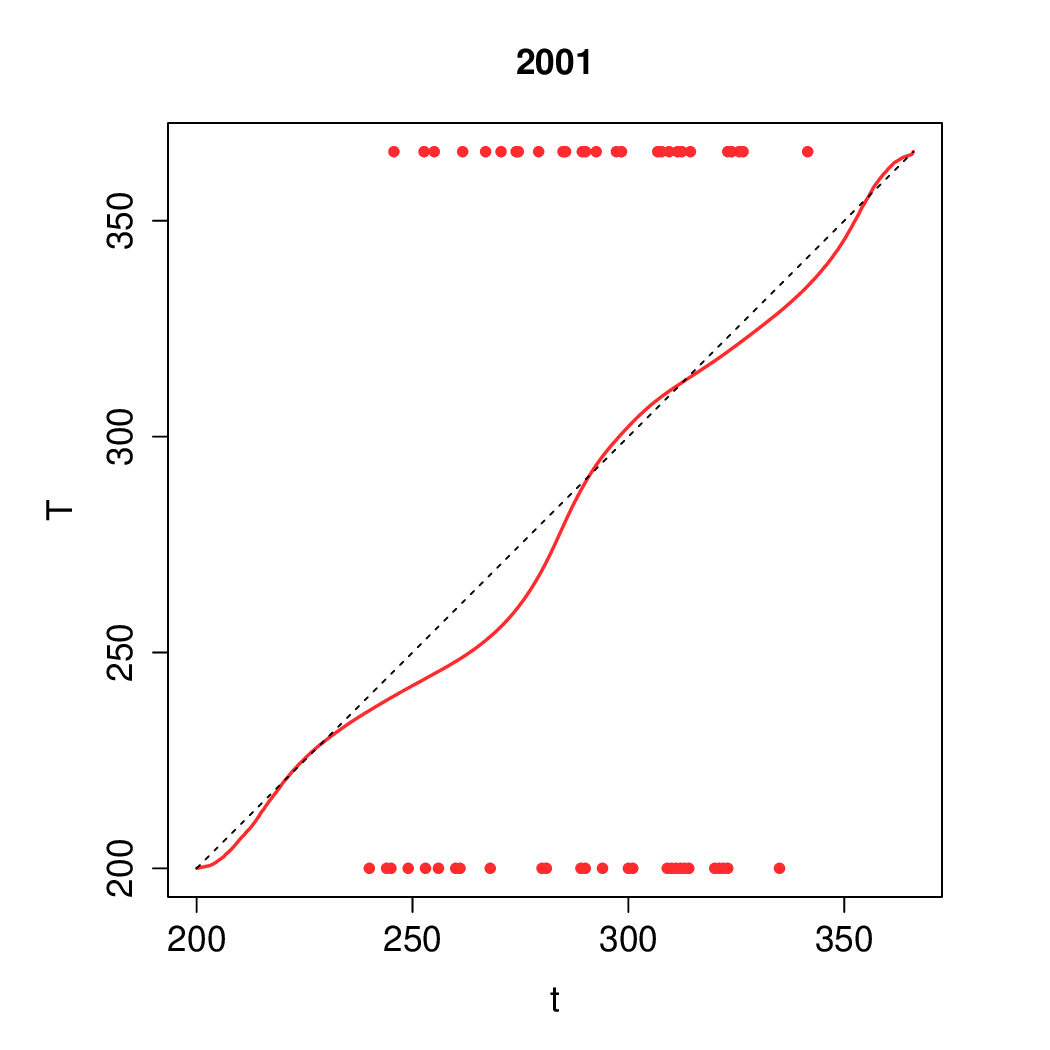}
\includegraphics[scale=0.180]{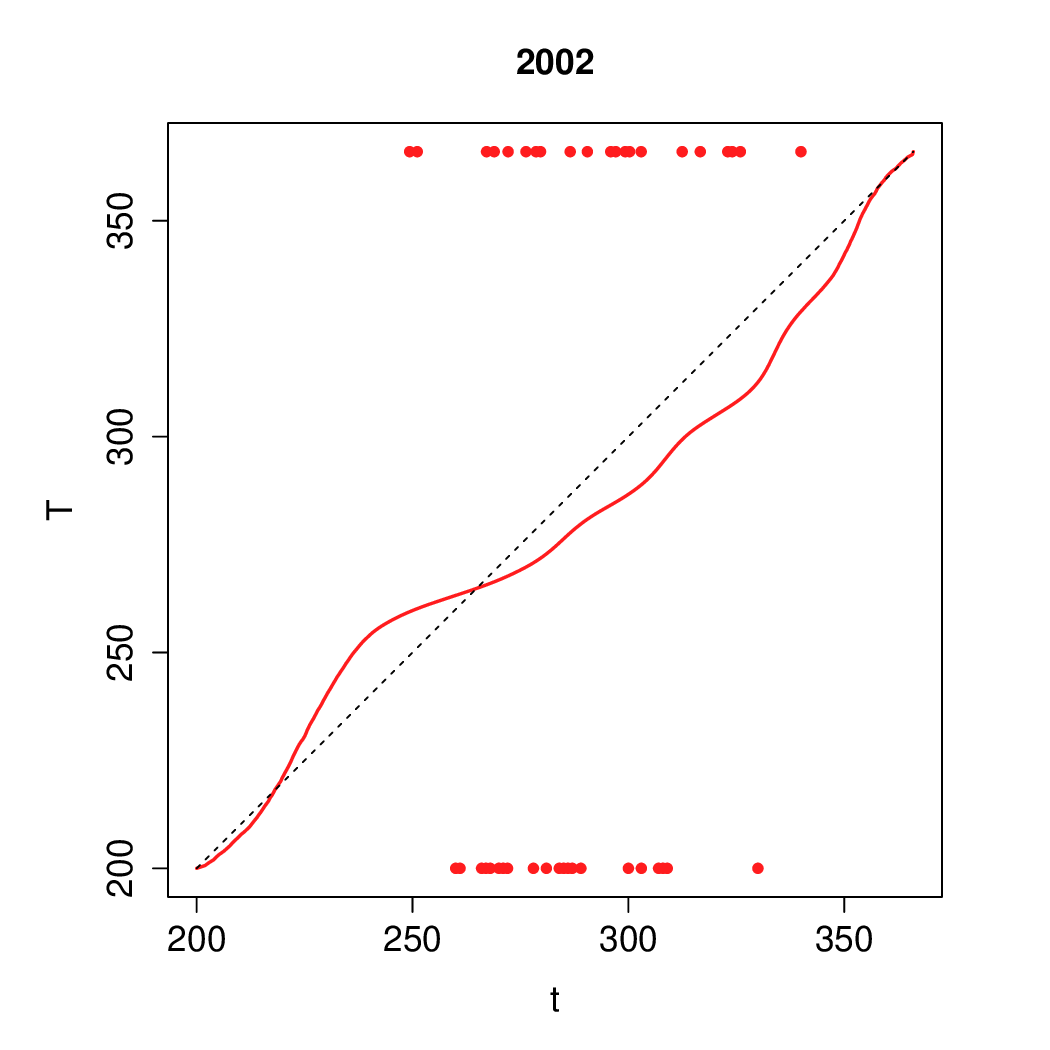}
\includegraphics[scale=0.180]{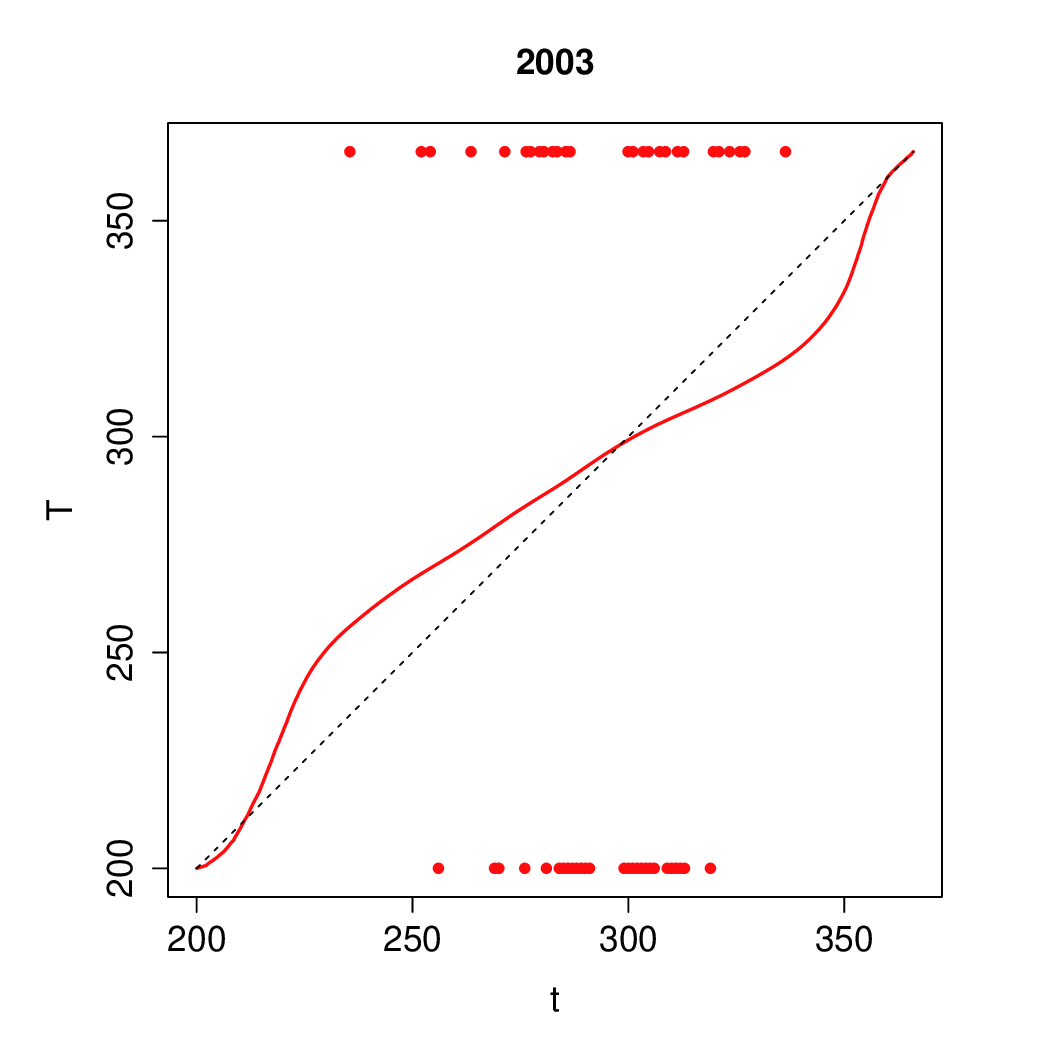}
\includegraphics[scale=0.180]{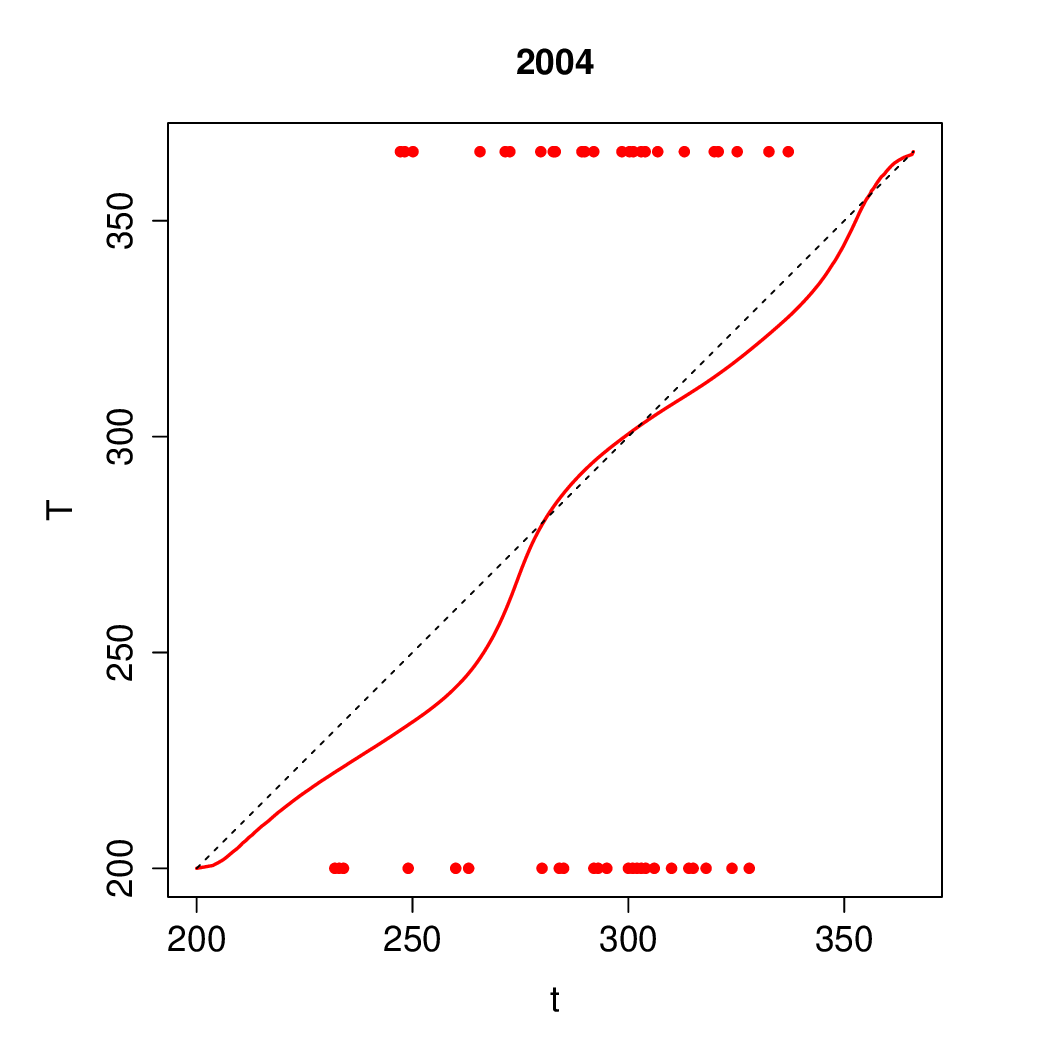}
\includegraphics[scale=0.180]{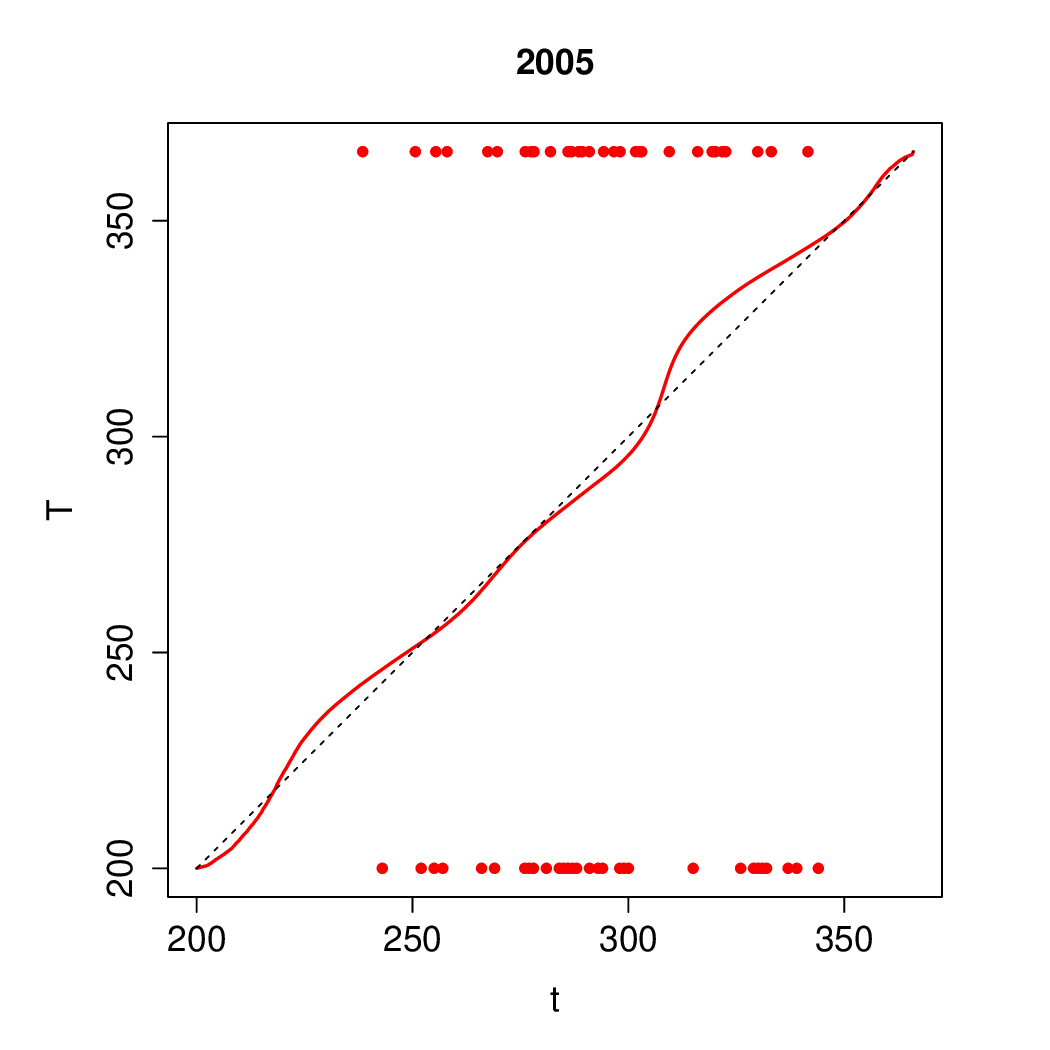}\\
\includegraphics[scale=0.180]{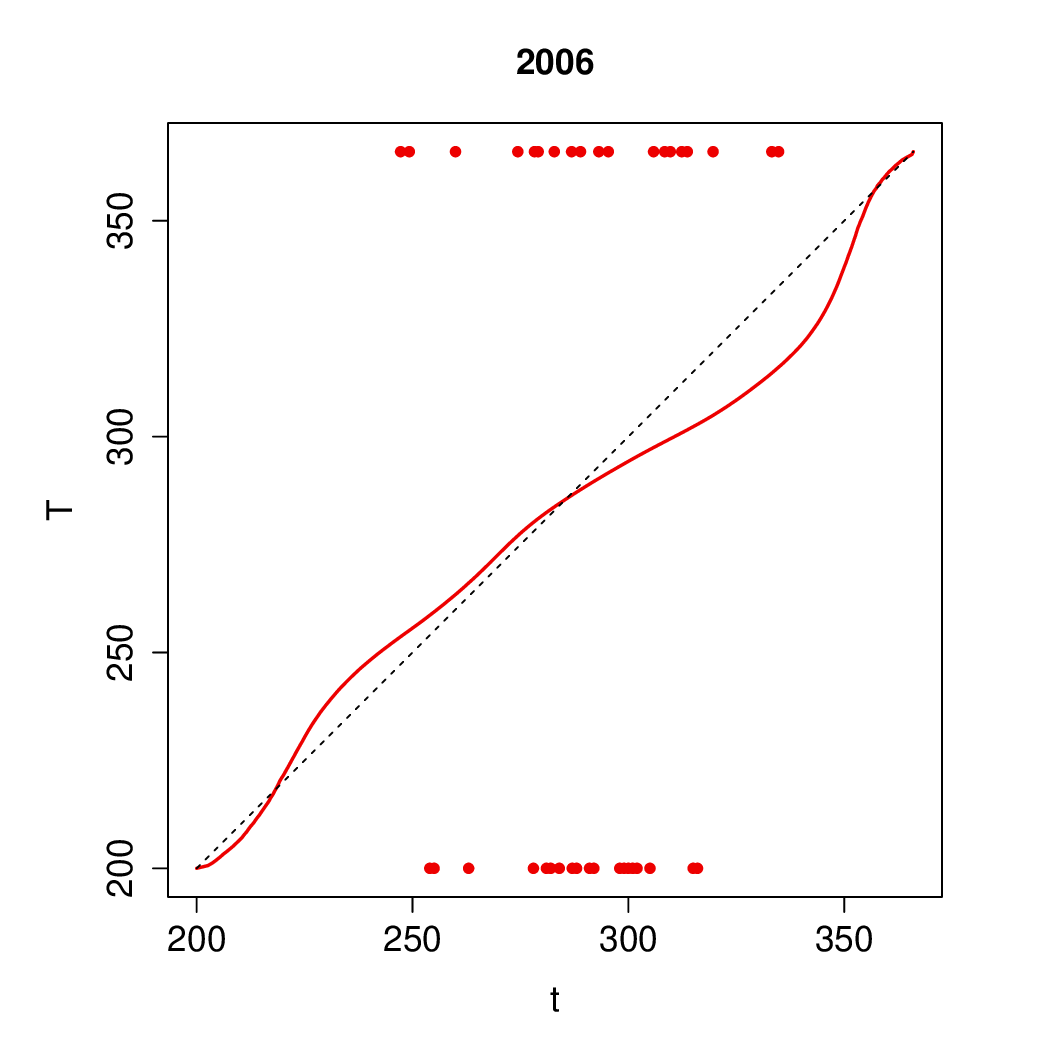}
\includegraphics[scale=0.180]{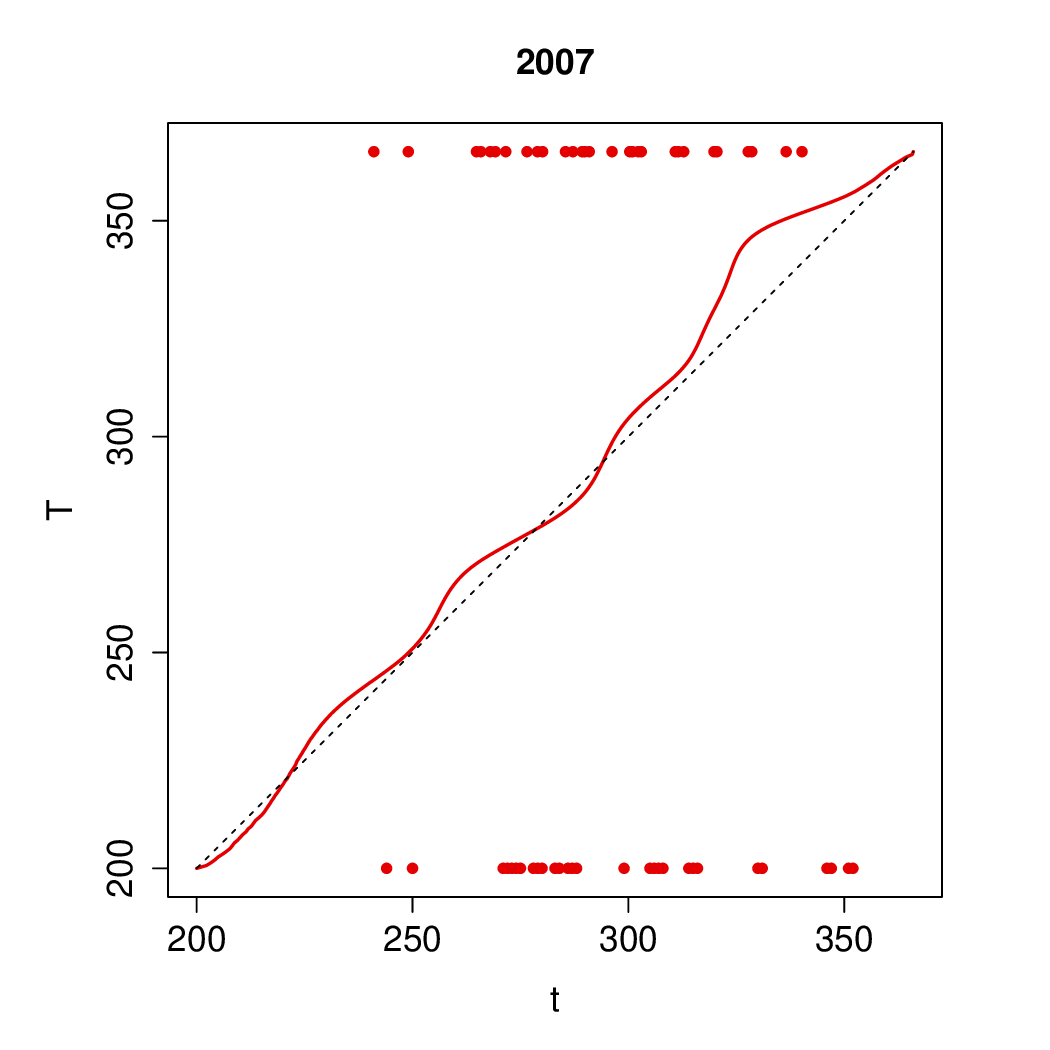}
\includegraphics[scale=0.180]{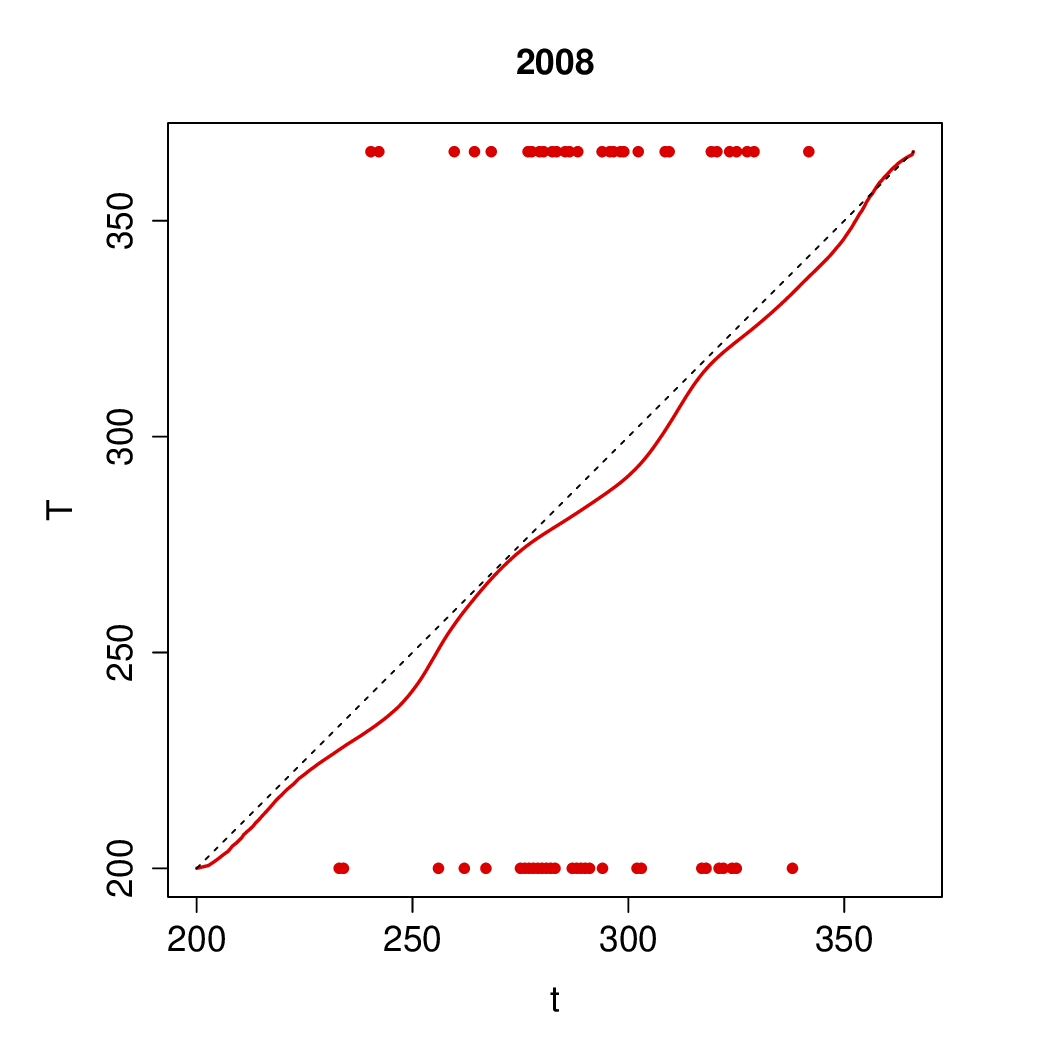}
\includegraphics[scale=0.180]{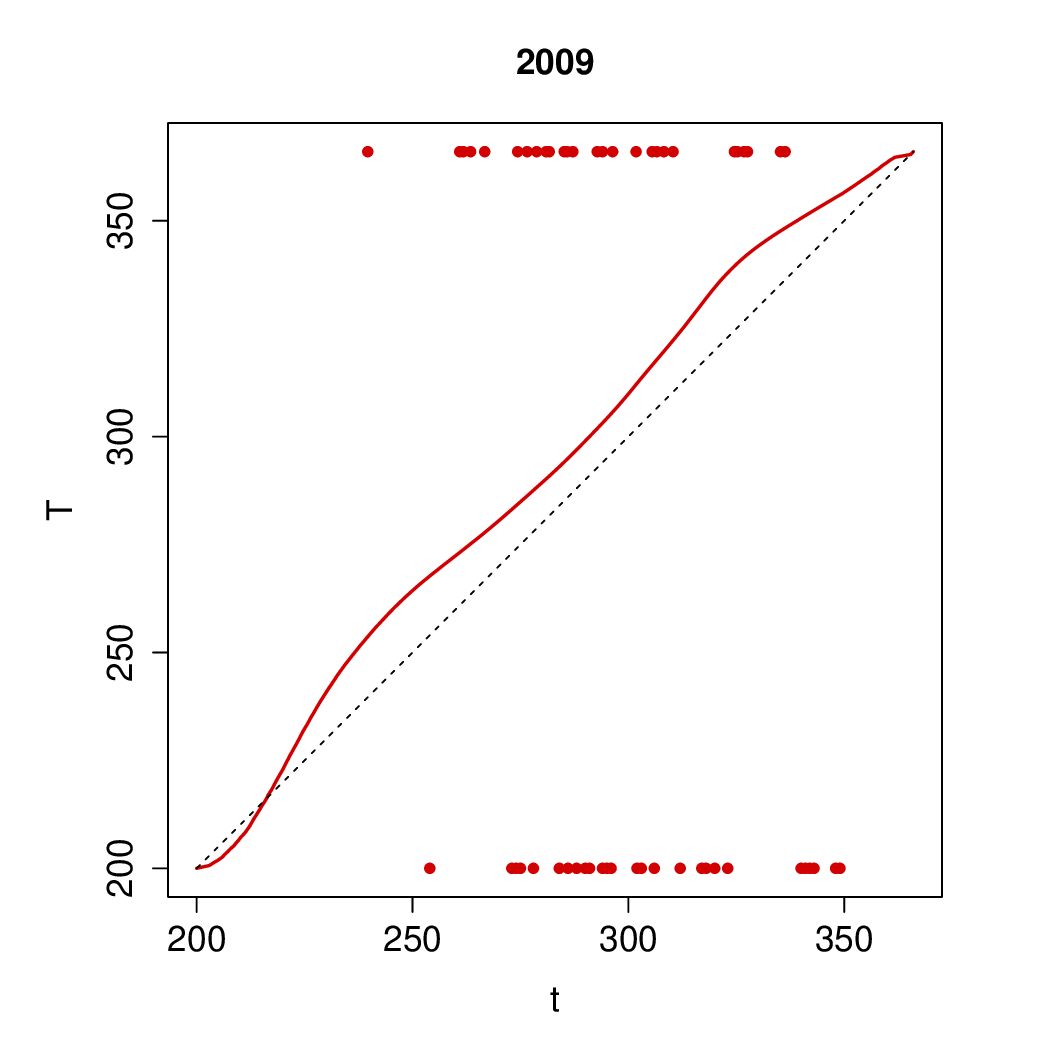}
\includegraphics[scale=0.180]{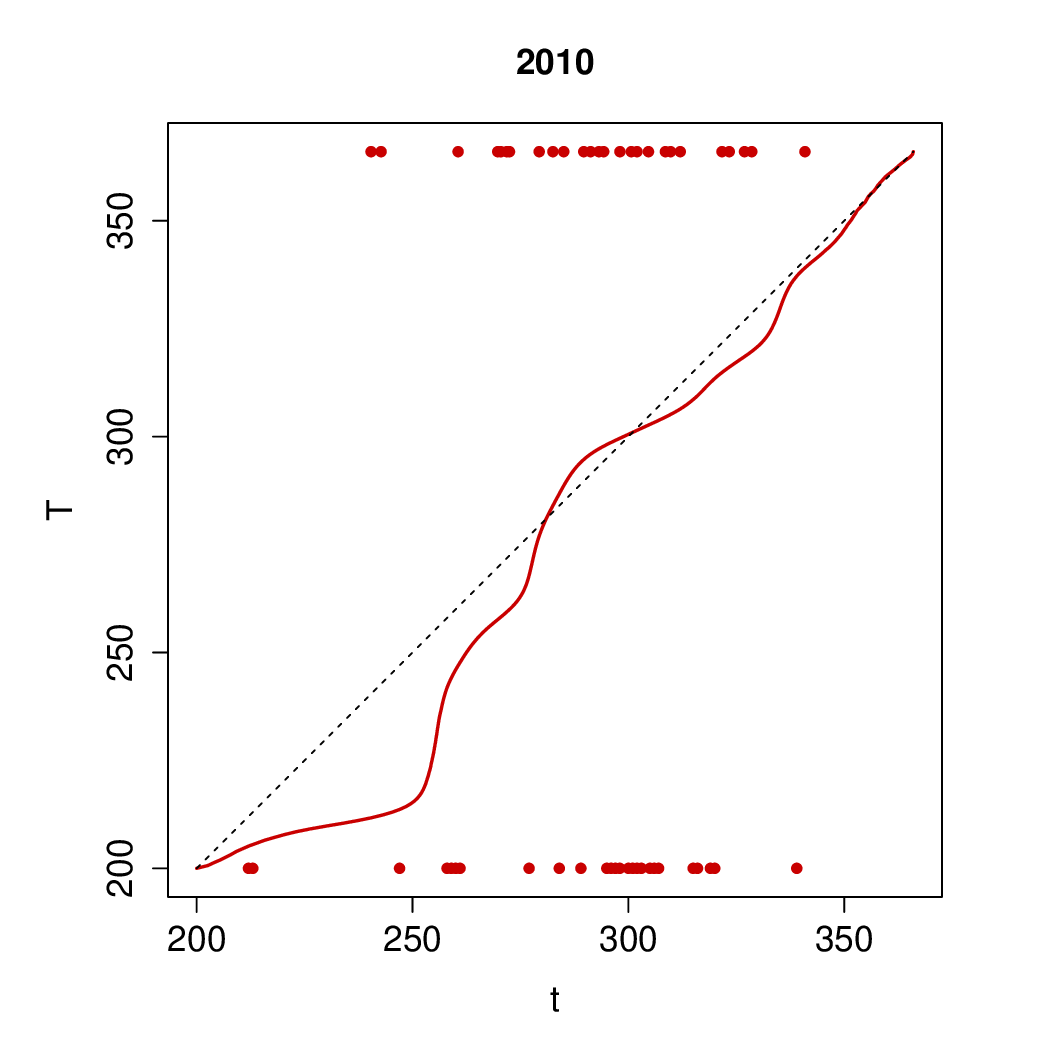}\\
\includegraphics[scale=0.180]{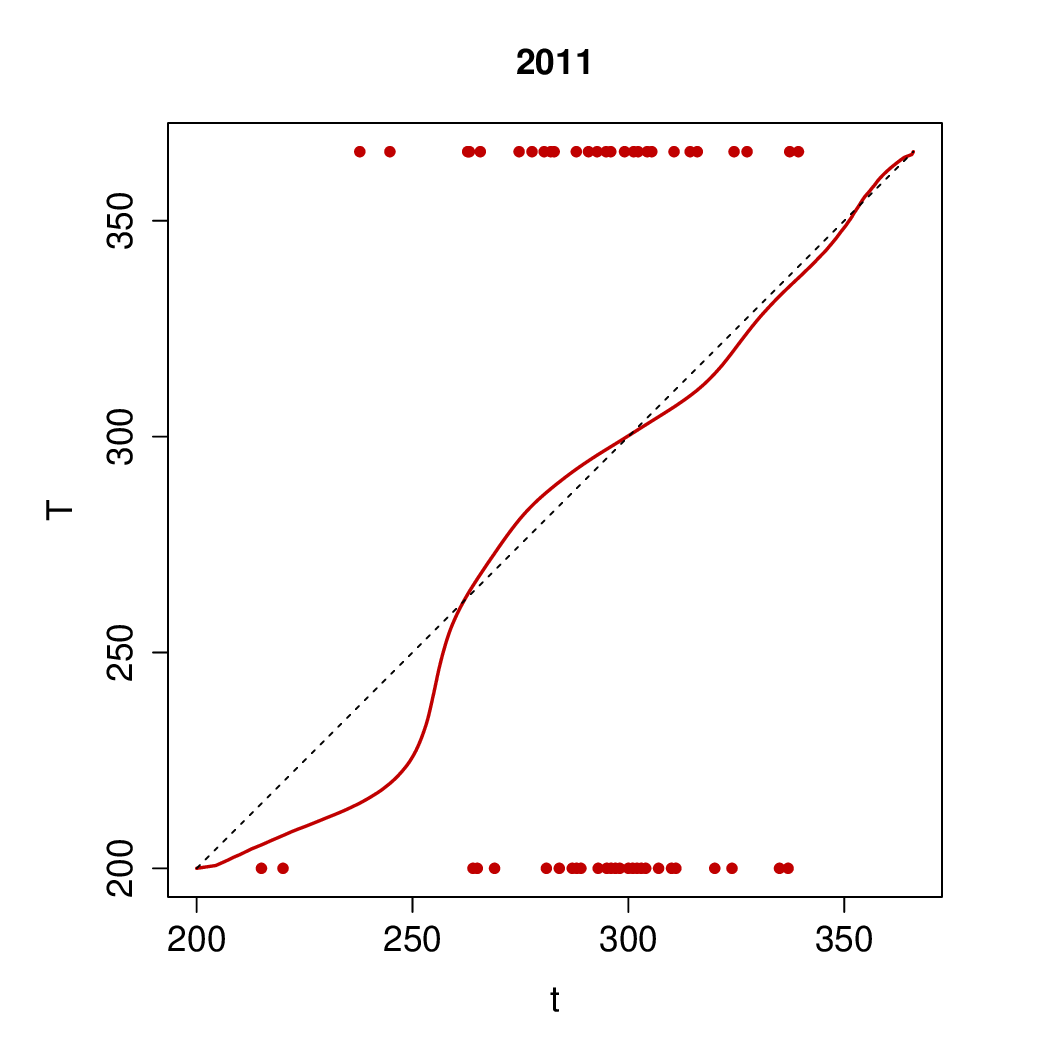}
\includegraphics[scale=0.180]{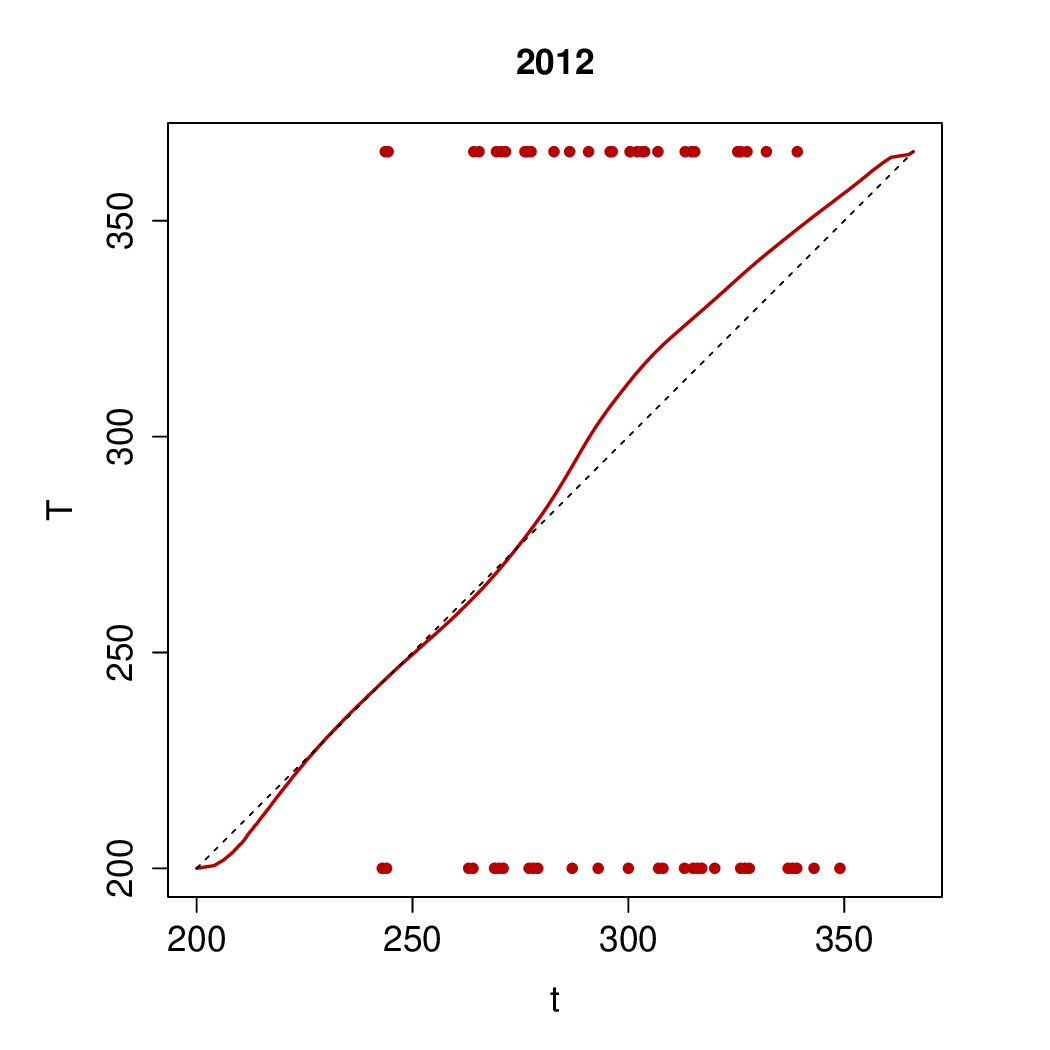}
\includegraphics[scale=0.180]{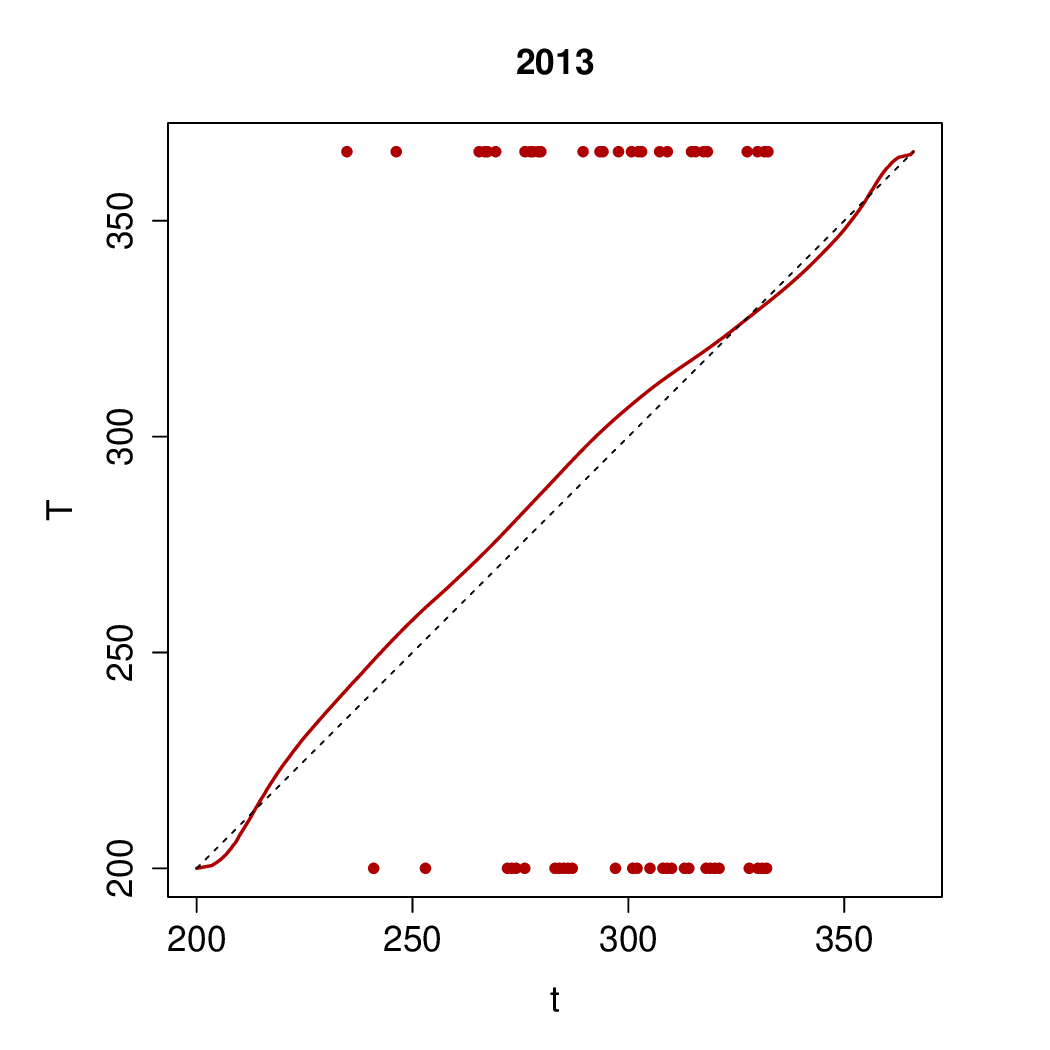}
\includegraphics[scale=0.180]{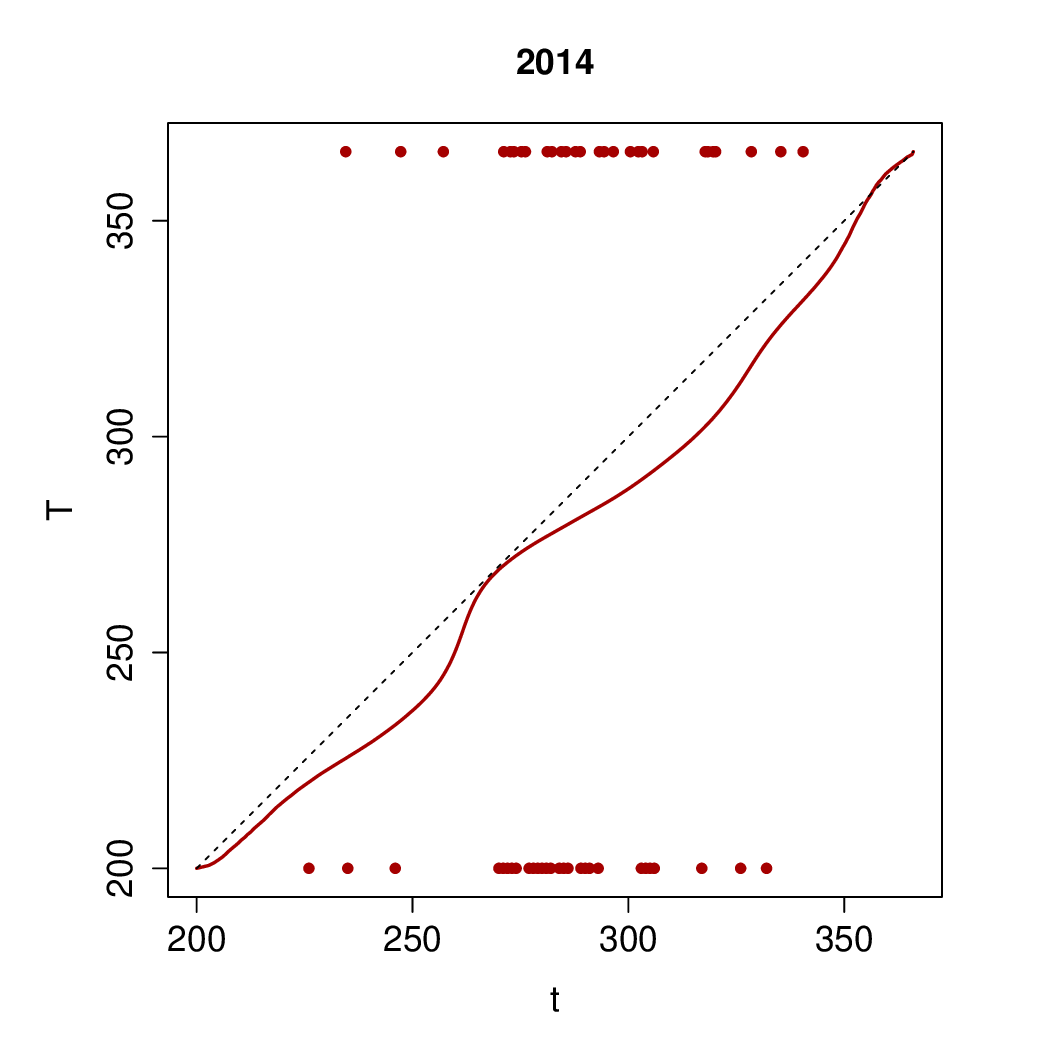}
\includegraphics[scale=0.180]{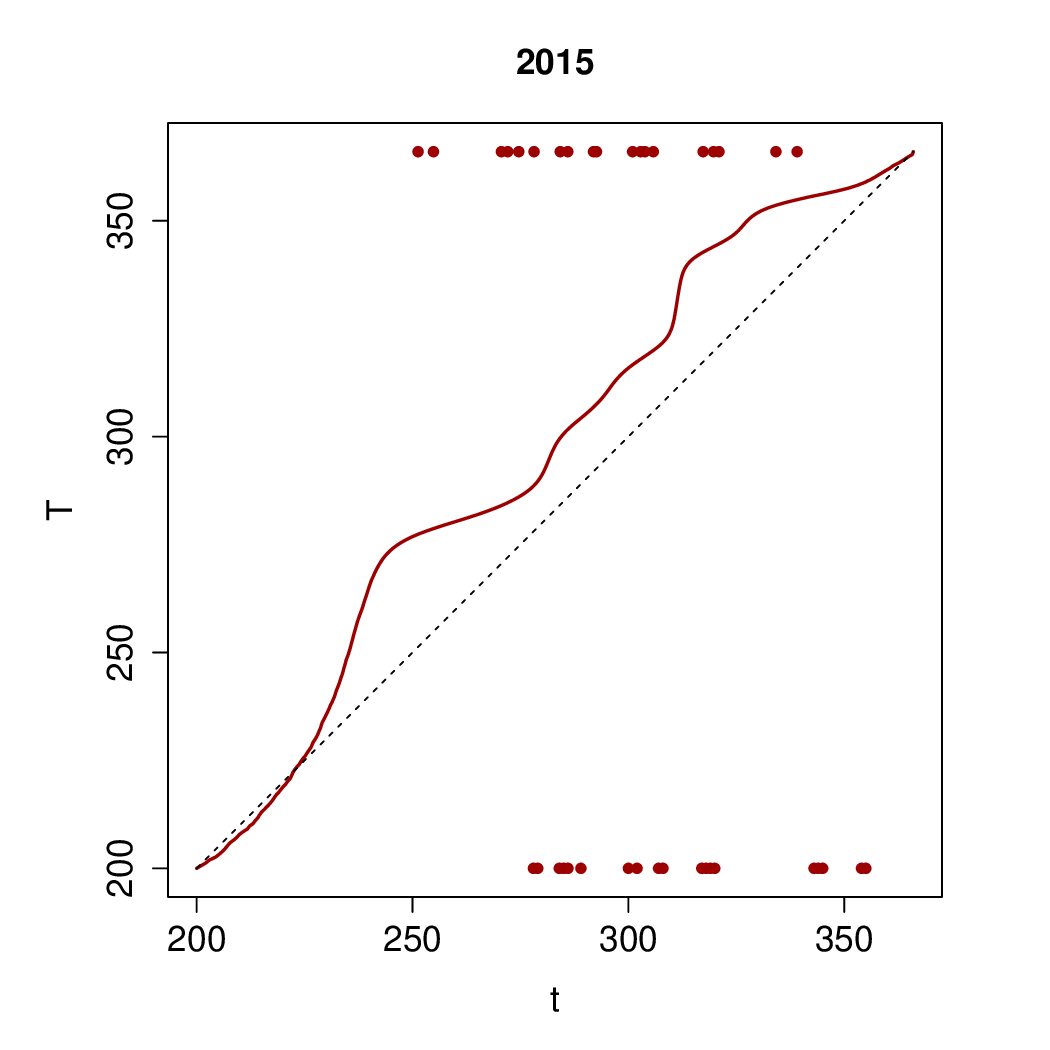}\\
\includegraphics[scale=0.180]{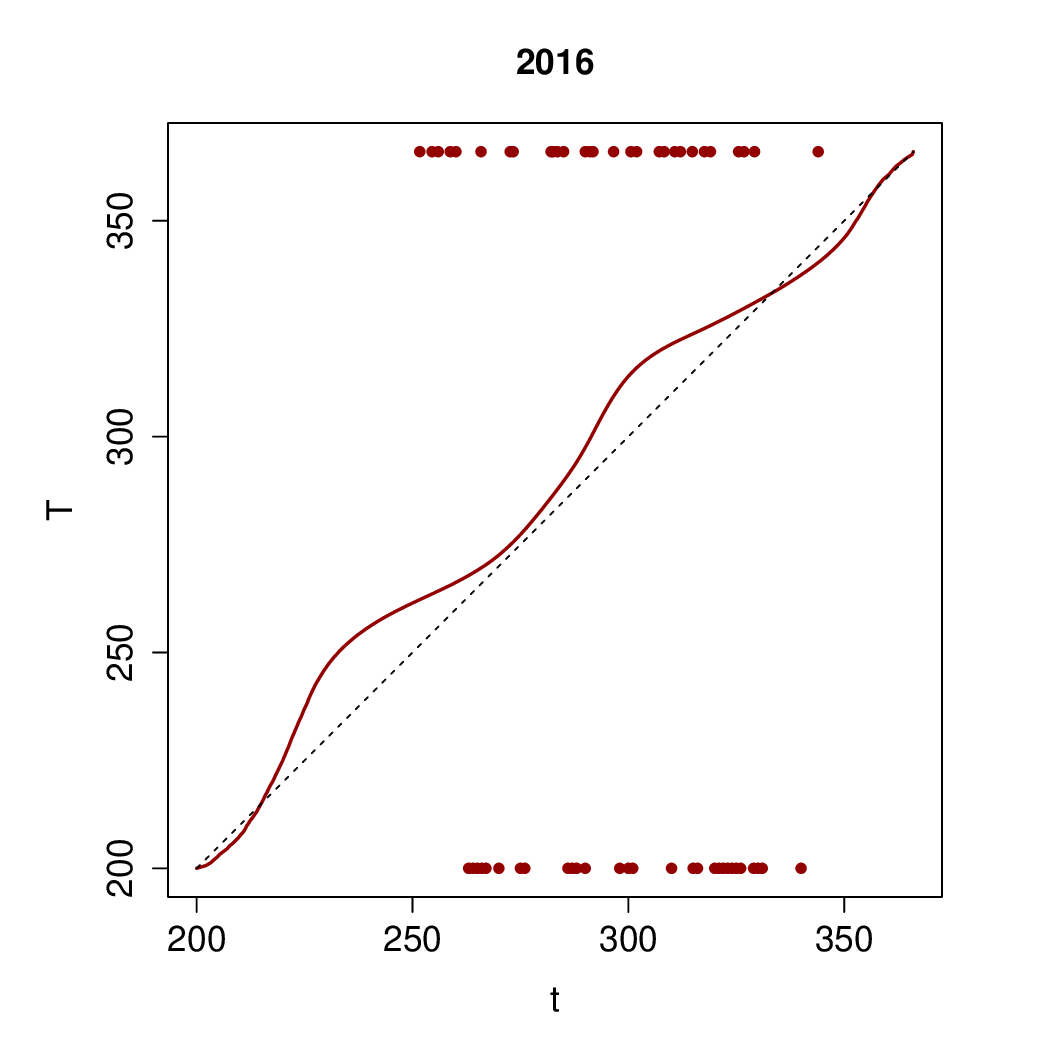}
\includegraphics[scale=0.180]{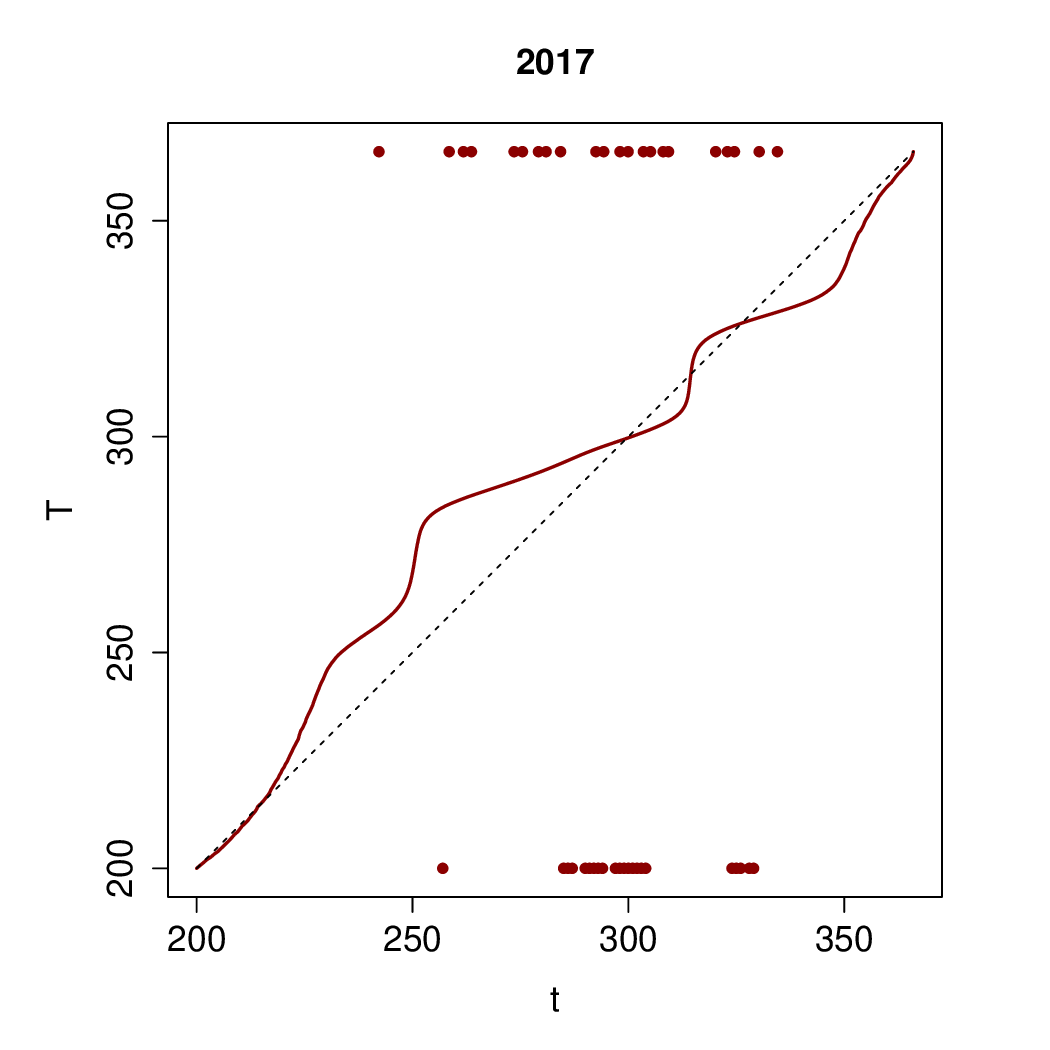}
\caption{\footnotesize Yearly posterior mean Bernstein polynomial warp functions of high-temperatures in the same color palette as data, plotted with raw data (bottom), registered points (top), and the identity function (dashed black). Here the year refers to that of onset of summer.}
\label{warps:hot}
\end{figure}

\begin{figure}\centering
  \hspace{-.8cm}
  \begin{minipage}{0.33\linewidth}
    \includegraphics[scale = 0.43]{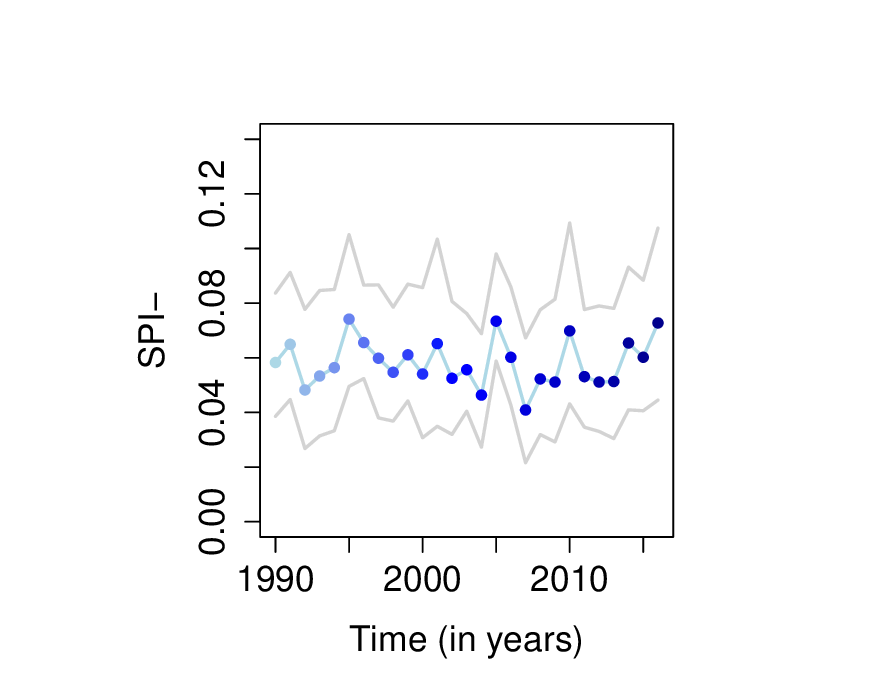}
  \end{minipage}\hspace{0.2cm}
  \begin{minipage}{0.33\linewidth}
    \includegraphics[scale = 0.43]{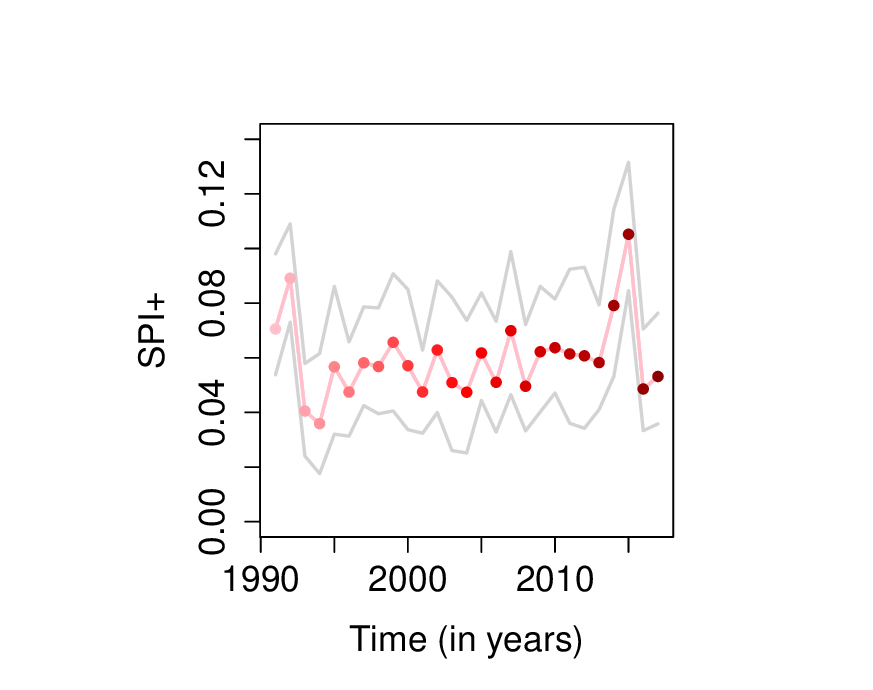}    
  \end{minipage}\hspace{0.2cm}
  \begin{minipage}{0.33\linewidth}
    \includegraphics[scale = 0.43]{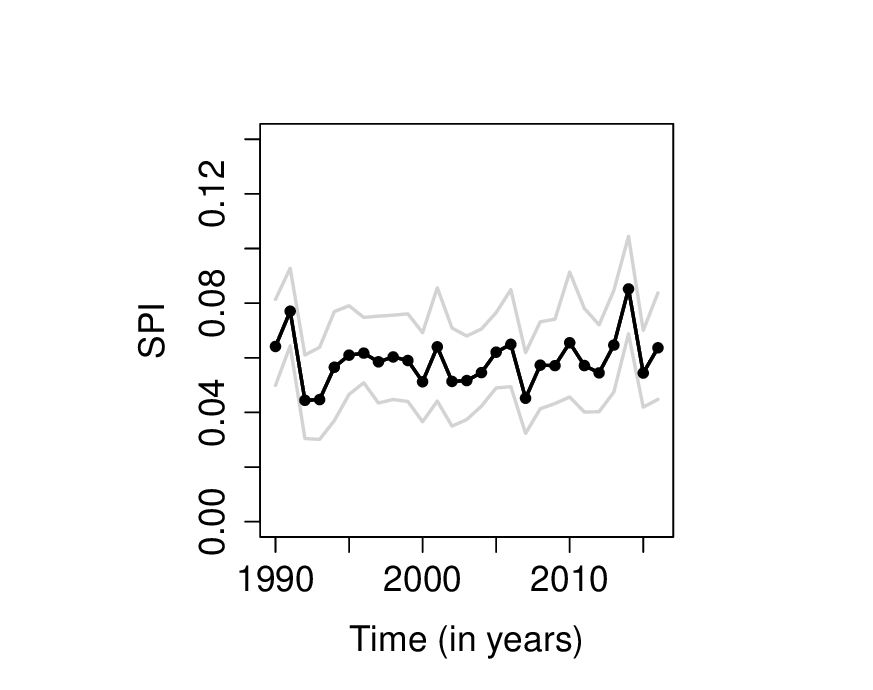}
   \end{minipage}\vspace{-.2cm}
  \caption{\footnotesize Posterior mean \textsc{spi} (scores of peak irregularity), as defined in (11), along with credible intervals, for below threshold (Left), above  threshold (Middle), and global (Right), for the 2.5\% and 97.5\% quantiles data.}
  \label{fig:supp-spi}
\end{figure}

Some comments on the results reported above are in order. Figures~\ref{fig:supp-pp} and~\ref{fig:supp-spi} correspond to Figures~\ref{fig:pp} and \ref{fig:spi} in the paper, respectively. Figures~\ref{warps:cold} and~\ref{warps:hot} depict the fits of the warp maps from the data application in Section~\ref{application}.

\newpage 
\begin{footnotesize}

\end{footnotesize}

\end{document}